\documentclass{jfm}
\usepackage{graphicx,caption,subcaption}
\usepackage{epstopdf, epsfig}
\usepackage{multirow}

\shorttitle{An anisotropic particle in a simple shear flow}
\shortauthor{Mahan Raj Banerjee and Ganesh Subramanian}

\title{An anisotropic particle in a simple shear flow: an instance of chaotic scattering}

\author{Mahan Raj Banerjee
 \and Ganesh Subramanian
 \corresp{\email{sganesh@jncasr.ac.in}}}

\affiliation{Jawaharlal Nehru Centre for Advanced Scientific Research, Jakkur, Bangalore}

\begin{document}

\maketitle

\begin{abstract}
In the Stokesian limit, the streamline topology around a single neutrally buoyant sphere is identical to the topology of pair-sphere pathlines, both in an ambient simple shear flow. In both cases there are fore-aft symmetric open and closed trajectories spatially demarcated by an axisymmetric separatrix surface. This topology has crucial implications for both scalar transport from a single sphere, and for the rheology of a dilute suspension of spheres. We show that the topology of the fluid pathlines around a neutrally buoyant freely rotating spheroid, in simple shear flow, is profoundly different, and will have a crucial bearing on transport from such particles in shearing flows. To the extent that fluid pathlines in the single-spheroid problem and pair-trajectories in the two-spheroid problem, are expected to bear a qualitative resemblance to each other, the non-trivial trajectory topology identified here will also have significant consequences for the rheology of dilute suspensions of anisotropic particles.

An inertialess non-Brownian spheroid in a simple shear flow rotates indefinitely in any one of a one-parameter family of Jeffery orbits. The parameter is the orbit constant $C$, with $C = 0$ and $C = \infty$ denoting the limiting cases of a spinning(log-rolling) spheroid, and a spheroid tumbling in the flow-gradient plane, respectively. The streamline pattern around a spinning spheroid is qualitatively identical to that around a sphere regardless of its aspect ratio. For a spheroid in any orbit other than the spinning one ($C >0$), the velocity field being time dependent in all such cases, the fluid pathlines may be divided into two categories. Pathlines in the first category extend from upstream to downstream infinity without ever crossing the flow axis; unlike the spinning case, these pathlines are fore-aft asymmetric, suffering a net displacement in both the gradient and vorticity directions. The second category includes primarily those pathlines that loop around the spheroid, and to a lesser extent those that cross the flow axis, without looping around the spheroid, reversing direction in the process. The residence time, in the neighbourhood of the spheroid, is a smooth function of upstream conditions for pathlines belonging to the first category. In sharp contrast, the number of loops, and thence the residence time associated with the pathlines in the second category, is extremely sensitive to upstream conditions. Plots of the residence time as a function of the upstream co-ordinates of these pathlines reveal a fractal structure with singularities distributed on a Cantor-like set, suggesting the existence of a chaotic saddle in the vicinity of the spheroid.
\end{abstract}


\section{Introduction}
Characterizing the rheological behaviour of a dilute Stokesian suspension of hydrodynamically interacting particles requires the solution of the two-body problem at the microscale. Einstein(1906) first obtained the effective viscosity($\mu_e$) of a dilute suspension of non-interacting rigid spherical particles(the one-body problem) in terms of the viscosity of the suspending fluid$(\mu)$ as $\mu_e = \mu\left(1+\frac{5}{2}\phi\right)$, $\phi$ being the volume fraction of the suspended spheres; the factor $5/2$ referred to as the Einstein coefficient. Much later, Batchelor and Green\citep{batchelor_stress,batchelor_2sphere} investigated the effect of pair-interactions
between spherical particles, in an ambient linear flow, in an attempt to calculate the $O(\phi^2)$ correction to the effective viscosity. In obtaining this correction, they examined the pair-sphere pathlines in simple shear flow, and showed that these pathlines had the same character as the streamlines around a single sphere in a simple shear flow \citep{cox1968}. In both instances, the trajectories may be classified into two groups. The first group consists of fore-aft symmetric open trajectories that extend to upstream and downstream infinity. The second group consists of closed orbits around the test sphere. The two groups are demarcated by a separatrix surface consisting of open trajectories that asymptote to the flow axis infinitely far away in both the upstream and downstream directions. The surface itself is axisymmetric, the axis of symmetry being the gradient direction of the ambient simple shear. Both the fore-aft symmetry of open trajectories, and the existence of closed ones, arise from the reversibility of the Stokes equations. Closed streamlines in the single-sphere problem are known to profoundly affect heat and mass transfer, leading to diffusion-limited transport at large Peclet  numbers \citep{acrivos,subramanian_2006PRL,subramanian_2006POF,deepak_2018_1, deepak_2018_2}. Likewise, for the pair-sphere problem, Batchelor and Green\citep{batchelor_2sphere} showed that  the pair-distribution function is rendered indeterminate, in the region of closed pathlines, in the pure hydrodynamic limit. This in turn leads to an indeterminate $O(\phi^2)$ coefficient for the suspension stress in all linear flows with regions of closed pair-pathlines, simple shear flow being a special case\citep{kao}.

In this paper, we analyze the fluid pathlines around a torque-free neutrally buoyant spheroid of an arbitrary aspect ratio (regarded as a canonical anisotropic particle) in simple shear flow. Keeping in mind the scenario for suspensions of spherical particles detailed above, we expect our findings to, on one hand, shed some light on the transport of heat or mass from a single spheroid immersed in a shearing flow. On the other hand, our findings will also be relevant to the nature of pair-spheroid interactions, and thence, to the pair-level microstructure of Stokesian suspensions of interacting anisotropic particles. In any event, analyzing the fluid motion around a single spheroid is a natural first step towards an understanding of the more complicated two-body problem. The results reported here suggest that the diffusion-limited scalar transport at $O(\phi)$, and the indeterminate rheology at $O(\phi^2)$, encountered for the case of spherical particles above, are both likely to be resolved, to a significant extent, by the effects of shape anisotropy.

We characterize the topology of the fluid pathlines around a single prolate or oblate spheroid in an ambient simple shear flow. The exact disturbance velocity field required for this purpose is available in closed form from earlier efforts \citep{dabade1,dabade2,marath_linear} that have used a spheroidal harmonics formalism for this purpose. As is well known, a neutrally buoyant torque-free spheroid in simple shear rotates along Jeffery orbits, and the disturbance velocity field is therefore time dependent in all cases except when the spheroid axis is aligned with the ambient vorticity (the log-rolling or spinning mode). Our results show that the departure from sphericity has a profound effect on the nature of the fluid pathlines around a spheroid in a precessional (non-spinning) orbit. The obvious changes happen for the open pathlines which are still open, but unlike the case of a sphere, are no longer fore-aft symmetric. Such pathlines undergo a lateral displacement in both the gradient and vorticity directions as they proceed from upstream to downstream infinity. The unexpected aspect concerns the originally closed pathlines. A fraction of these pathlines opens up, but unlike the pathlines above, the resulting open pathlines do not head directly downstream. Instead, these pathlines loop around the rotating spheroid with the number of loops being extremely (indeed, infinitely) sensitive to the upstream coordinates of the particular pathline. The sensitivity with regard to the upstream coordinates has the signatures of chaotic scattering \citep{bleher,aref-pomphrey-letter,aref-pomphrey,aref-stremler,aref}. The presence of a chaotic saddle suggested by the aforementioned scattering pattern implies that both closed (periodic) and bounded pathlines continue to exist for a spheroid in a generic precessional  orbit, but only constitute a set of vanishingly small measure, and one that is therefore numerically inaccessible. 

The paper is organized as follows. In Section \ref{sec:canonical} we discuss the problem formulation. Calculation of the fluid pathlines requires an expression for the disturbance velocity field due to a torque-free spheroid in simple shear flow, and as mentioned above, this is obtained in terms of the appropriate vector spheroidal harmonics \citep{kushch1997,kushch1998,kushch_book}, after resolving the ambient simple hear into simpler canonical components in body-fixed coordinates \citep{dabade1,dabade2,marath_linear}. In Section \ref{sec:spinning} we first analyse the fluid pathlines (which are the same as streamlines) for the simpler steady case of a spinning spheroid. In this case, the topology of the streamline pattern is shown to be identical to that around a torque-free sphere and a circular cylinder\citep{kao,torza,powell}, with fore-aft symmetric open and closed streamlines being demarcated by a (non-axisymmetric) separatrix surface. For a prolate spheroid, the streamline pattern (including the separatrix in particular) transitions from that for a sphere to a cylinder, with increasing spheroid aspect ratio. In Sections \ref{sec:open} and \ref{sec:closed}, we investigate the topology of the fluid pathlines for the general time dependent case when the spheroid is not in its spinning orbit. Section \ref{sec:open} concerns regular open pathlines that do not loop around the spheroid, while heading from upstream to downstream infinity. Unlike both a sphere and a spinning spheroid, these pathlines are fore-aft asymmetric. This asymmetry is best characterized in terms of the net gradient and vorticity displacements suffered by the pathline \citep{subramanian_brady}. The numerically determined displacements are then compared to  the far-field analytical predictions, as a function of the upstream pathline coordinates, for the spheroid rotating over a range of Jeffery orbits. In Section \ref{sec:closed} we analyze the second category of pathlines which includes the singular open pathlines that, although open, loop around the spheroid before heading off to downstream infinity. For this case, we first show that the number of loops around the spheroid, and therefore, the residence time of the pathline in a sufficiently large neighborhood of the spheroid, are sensitively dependent on the initial pathline coordinates(chosen to lie on the negative flow axis). Plots of the residence time reveal an extremely fine-scaled fractal-like structure, suggesting the existence of a chaotic saddle in the vicinity of the spheroid. We examine the nature of this sensitive dependence as a function of the spheroid aspect ratio, and more importantly, as a function of the orbit constant for a given aspect ratio; the latter reveals the singular nature of the approach to the integrable limit ($C = 0$) of a log-rolling or a spinning spheroid. In section \ref{sec:transition},  we analyze the fluid pathlines from the chaotic scattering perspective that has been gainfully used in other scenarios \citep{bleher,faisst,janosi,aref}, by focussing on the nature of the pathlines as a function of their upstream coordinates and the Jeffery phase angle. Herein, we examine the transition from regular to chaotic scattering with varying spheroid aspect ratio, and again, as a function of the Jeffery orbit constant. There exists a critical upstream offset corresponding to the onset of chaotic scattering, and for offsets less than this value, intervals of regular and chaotic scattering appear interlaced down to the smallest scales; the interlaced pattern is strongly dependent on aspect ratio. This critical offset also varies with the Jeffery phase angle, and one may therefore characterize the scattering in terms of a residence time surface plotted as a function of the upstream offset and the Jeffery phase angle, with a critical curve that separates regions of regular and irregular dependence. The extent of chaotic scattering is nevertheless shown to decrease for the largest aspect ratios, with the relevant signatures being eventually undetectable (numerically). We also examine the probability density of residence times, the emphasis being on the functional form of the tail corresponding to large residence times. Section 6 presents a summary of our findings, together with the implications for the nature of fluid pathlines in a more general setting - the one-parameter family of hyperbolic linear flows. This is followed by a more detailed explanation of the relevance of chaotic scattering for transport in the in disperse.

\section{Single spheroid in a simple shear flow}\label{sec:canonical}
\subsection{Jeffery orbits}
\begin{figure}
  \centering
  \includegraphics[scale=0.15]{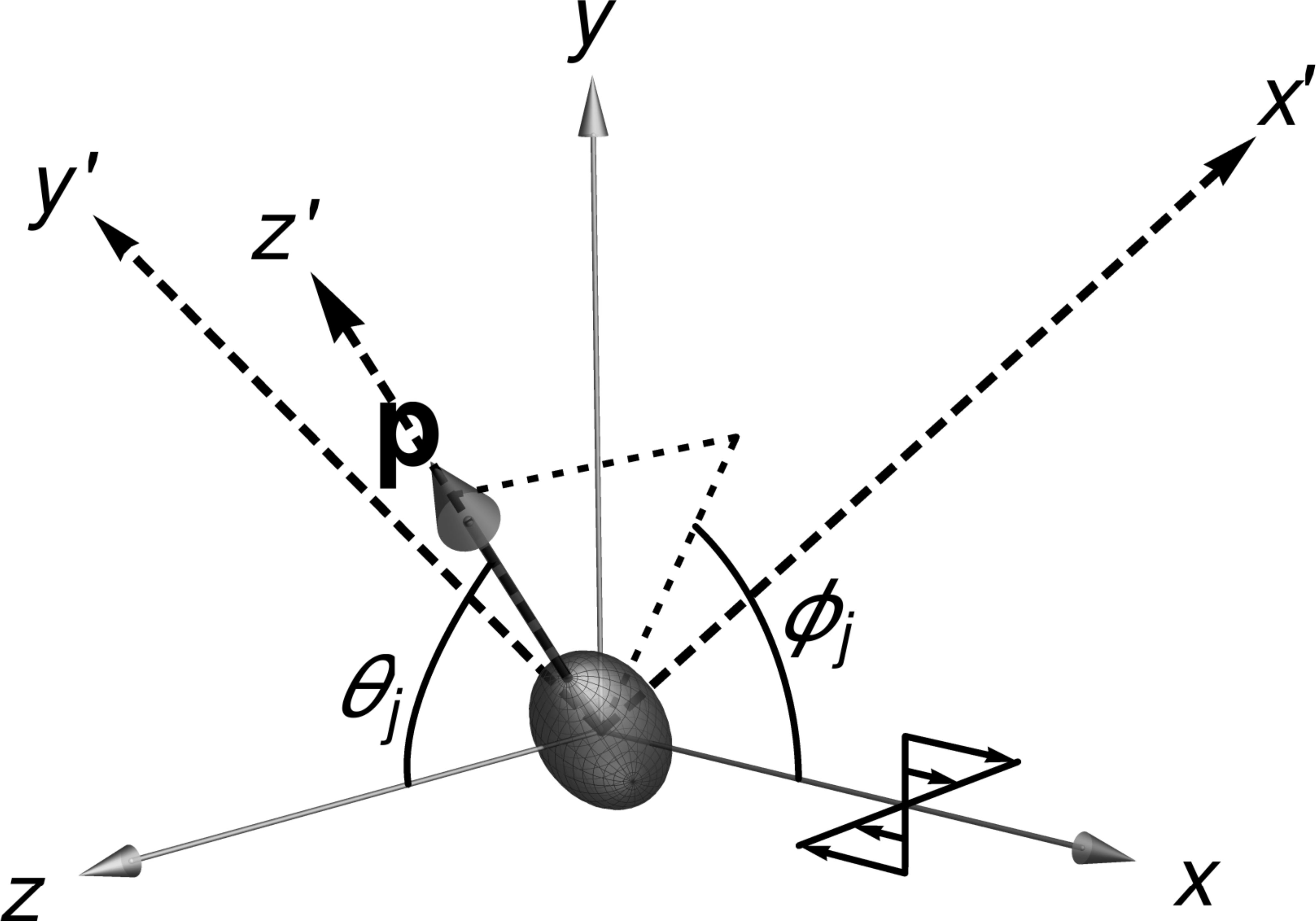}
  \includegraphics[scale=0.15]{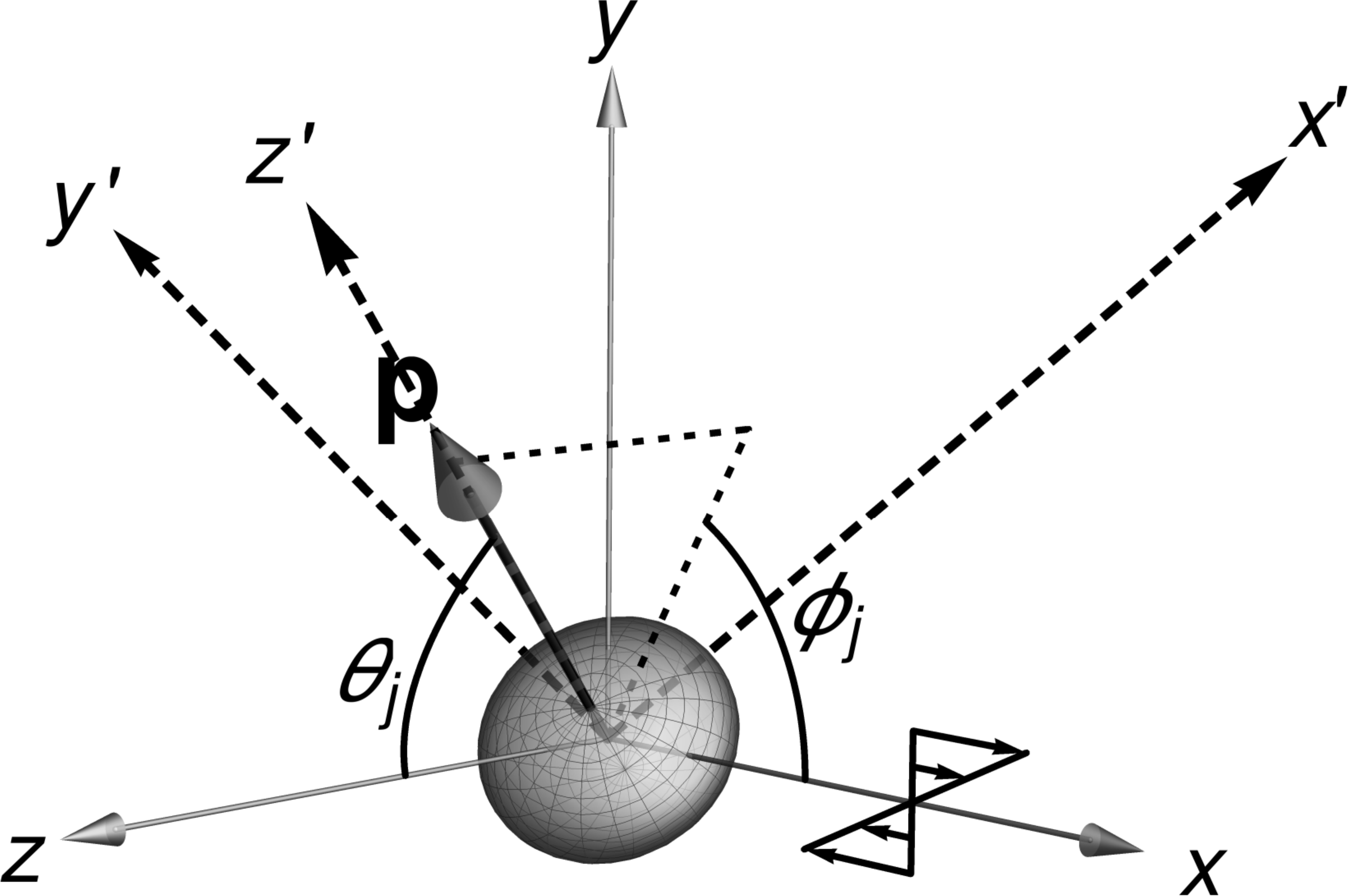}
  \caption{A prolate and an oblate spheroid in simple shear flow. The primed($X',Y',Z'$) and unprimed ($X,Y,Z$) coordinates denote the body and the space fixed axes, respectively. $\theta_j$ and $\phi_j$ are polar and azimuthal Jeffery angles, respectively.}
\label{fig:spheroid}
\end{figure}
\begin{figure}
  \begin{center}
  \includegraphics[scale=2.5]{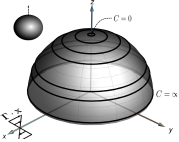}
  \includegraphics[scale=2.5]{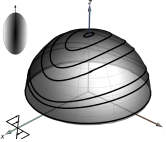}
  \includegraphics[scale=2.5]{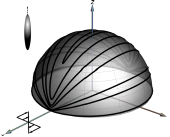}
  \\
  \includegraphics[scale=2.5]{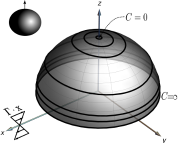}
  \includegraphics[scale=2.5]{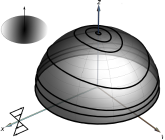}
  \includegraphics[scale=2.5]{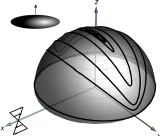}
  \end{center}
  \caption{Jeffery orbits for prolate(top)($\kappa=1.25$, $2$, $4$) and oblate spheroids($\kappa=0.8$, $0.5$, $0.25$)}
\label{fig:orbit}
\end{figure}

The motion of spheroidal particles in linear flows in the absence of inertia is known since the work of \cite{jeffery}. In simple shear flow, a single torque-free spheroid rotates in any one of a single parameter family of closed trajectories now known as Jeffery orbits. The motion along a Jeffery orbit may be characterized in terms of the polar  $(\theta_j)$ and azimuthal $(\phi_j)$ angles of the spheroid axis (see Figure \ref{fig:spheroid}), and these are given by the following functions of time:
\begin{equation}\label{eq:jeffery}
\phi_j=\cot^{-1}\left[\kappa\,\tan\left(\frac{\kappa\dot{\gamma}t}{1+\kappa^2}\right)\right],\;
\theta_j=\tan^{-1}\left[\frac{C\kappa}{\kappa^2 \sin^2\phi_j+\cos^2\phi_j}\right].
\end{equation}
Here, $\dot{\gamma}$ is the shear rate, $\kappa$ is the spheroid aspect ratio and $C$ is the orbit constant, ranging from $0$ to $\infty$, that parameterizes the Jeffery orbits; $\kappa>1(<1)$ for prolate(oblate) spheroids.  The generic orbit is a spherical ellipse, with its major axis along the flow (gradient) direction for a prolate(an oblate) spheroid. The limiting cases $C = 0$ and $C = \infty$ correspond to a log-rolling motion of the prolate spheroid (spinning for the oblate one) about the vorticity axis, and a tumbling motion in the flow-gradient plane, respectively. The period of rotation is independent of $C$, being equal to $T_j = 2\pi/\dot{\gamma}(\kappa + \kappa^{-1})$. Figure \ref{fig:orbit} shows the Jeffery orbits for prolate and oblate spheroids of different aspect ratios. The eccentricity of the orbits can be seen to increase with  increasing (decreasing) aspect ratio of the prolate (oblate) spheroid. For slender fibers ($\kappa = \infty$) and flat disks ($\kappa = 0$), the Jeffery orbits have a meridional character with the end points of the meridians, the poles, corresponding to the intersections of the flow and gradient axes with the unit sphere, respectively.

\subsection{The spheroid velocity field in simple shear flow}

The determination of the velocity field due to an inertialess torque-free spheroid in an ambient simple shear flow involves solving the Stokes equations in spheroidal coordinates\citep{kushch1997,kushch1998,kushch_book}:
\begin{equation}\label{eq:stokes}
\mu\nabla^2\boldsymbol{u} = \bnabla p,
\end{equation}
with the following boundary conditions in place
\begin{equation}\label{eq:bc}
\boldsymbol{u}(\xi=\xi_0,\eta,\phi)=\boldsymbol{\omega}\times\boldsymbol{x}\quad\textit{and}\quad\boldsymbol{u}(\xi\to\infty,\eta,\phi)=\boldsymbol{\Gamma}\bcdot\boldsymbol{x}.
\end{equation}
Here, $\mu$ is the fluid viscosity, $p$ is the hydrodynamic pressure, $\boldsymbol{x} \equiv (x,y,z)$ is the position vector, and $\boldsymbol{\Gamma}$ is the (transpose of the) velocity gradient tensor; $\boldsymbol{\Gamma} = \boldsymbol{1}_{x}\boldsymbol{1}_{y}$  in the space-fixed coordinate system ($X,Y,Z$) shown in Figure \ref{fig:spheroid}, with $\boldsymbol{1}_{x}$, $\boldsymbol{1}_{y}$ and  $\boldsymbol{1}_{z}$ denoting the unit vectors along the flow, gradient and vorticity directions, respectively. The spheroid angular velocity, $\boldsymbol{\omega}$,  in (\ref{eq:bc}) is determined from a torque-free constraint. The component of $\boldsymbol{\omega}$ normal to the spheroid axis, is obtained from the Jeffery relations (\ref{eq:jeffery}), while that along the axis is just half the projected ambient vorticity. In (\ref{eq:stokes}) and (\ref{eq:bc}), ($\xi,\eta,\phi$) are prolate spheroidal coordinates which, for a prolate spheroid, are related to the Cartesian ones as:  $x'=d\bar{\xi}\bar{\eta}\cos\phi$, $y'=d\bar{\xi}\bar{\eta}\sin\phi$, $z'=d\xi\eta$, where $\bar{\xi}=\sqrt{\xi^2-1}$, and $\bar{\eta}=\sqrt{1-\eta^2}$ (for an oblate spheroid, the corresponding relations are  obtained by setting $d\rightarrow -\mathrm{i}d$ and $\xi\rightarrow\mathrm{i}\bar{\xi}$). Here, the constant-$\xi$ surfaces represent a family of confocal prolate(oblate) spheroids with the interfoci distance(diameter of the focal circle) being $2d$; $\xi=\xi_0$ denotes the surface of the spheroidal particle. The constant-$\eta$ surfaces denote a family of confocal two-sheeted(single-sheeted) hyperboloids, and the constant-$\phi$ surfaces are planes passing though the axis of symmetry. 

The total velocity field $\boldsymbol{u}$ may be written as $\boldsymbol{ u} = \boldsymbol{\Gamma}\cdot\boldsymbol{x} + \boldsymbol{u'}$ where $\boldsymbol{u'}$ is the disturbance velocity field that is a function of the instantaneous spheroid orientation and vanishes in the far-field. The solution for $\boldsymbol{u'}$ is best accomplished in a body-fixed coordinate system. Defining ($X',Y',Z'$) as the body-fixed coordinate system (see Figure \ref{fig:spheroid}), the polar angle $\theta_j$ denoting the angle between the $Z$ and $Z'$ axes, and the azimuthal angle $\phi_j$ denoting the dihedral angle between $X-Z$  and $X'-Z'$ planes, with the $Y'$ axis constrained to lie in the flow gradient ($X-Y$)  plane, the rate of strain tensor, $\boldsymbol{E} = (\boldsymbol{\Gamma} + \boldsymbol{\Gamma}^T)/2$, is given by:
\begin{equation}
\setlength{\arraycolsep}{0pt}
\renewcommand{\arraystretch}{1.3}
\mathsfbi{E'} = \left[
\begin{array}{ccc}
  \frac{\cos^2\theta_j\sin 2\phi_j}{2} & \frac{\cos\theta_j\cos 2\phi_j}{2} & \frac{\sin 2\theta_j\sin 2\phi_j}{4} \\
  \frac{\cos\theta_j\cos 2\phi_j}{2} & -\frac{\sin 2\phi_j}{2} & \frac{\sin\theta_j\cos 2\phi_j}{2} \\
  \frac{\sin 2\theta_j\sin 2\phi_j}{4} & \frac{\sin\theta_j\cos 2\phi_j}{2} & \frac{\sin^2\theta_j\sin 2\phi_j}{2}
\end{array}  \right] 
\label{strain}
\end{equation}
in body-fixed coordinates. As shown by  \cite{subramanian_koch_2006} in the context of a nearly spherical particle, the rate of strain tensor above can be resolved into five elementary components, each corresponding to a canonical linear flow. The corresponding disturbance velocity fields may be expressed in terms of the appropriate decaying vector spheroidal biharmonics. Following \cite{dabade1,dabade2}, the disturbance velocity field for an arbitrary aspect ratio spheroid may similarly be written as:
\begin{equation}
\boldsymbol{u'(x')} = \sum_{i=1}^5\boldsymbol{u}_{is}(\boldsymbol{x'}),
\end{equation}
where the $\boldsymbol{u}_{is}(\boldsymbol{x'})$'s refer to the component disturbance fields, and are defined in Table \ref{tab:canonical}.
\begin{table}
\begin{tabular}{|c|c|}
 Axisymmetric Extension                  &  $\boldsymbol{u}_{1s}=-\frac{d\bar{\xi}_0}{2(Q_1^1(\xi_0)-\xi_0Q_2^1(\xi_0))}\sin^2{\theta_j}\sin{2\phi_j}\boldsymbol{S}_{20}^{(3)}$ \\ \hline
\multirow{2}{*}{Transverse planar extension} & $\boldsymbol{u}_{2s}= -\frac{d\bar{\xi}_0}{2(3Q_1^1(\xi_0)-\xi_0Q_2^1(\xi_0))}\sin{2\phi_j}\left(1+\cos^2{\theta_j}\right)\left[\boldsymbol{S}_{22}^{(3)}+\boldsymbol{S}_{2,-2}^{(3)}\right]$  \\
                  					&  $\boldsymbol{u}_{3s}=\frac{\mathrm{i}\;d\bar{\xi}_0}{3Q_1^1(\xi_0)-\xi_0Q_2^1(\xi_0)}\cos{\theta_j}\cos{2\phi_j}\left[\boldsymbol{S}_{22}^{(3)}-\boldsymbol{S}_{2,-2}^{(3)}\right]$   \\ \hline
\multirow{2}{*}{Longitudinal planar extension} & $\boldsymbol{u}_{4s}=\frac{d\xi_0\bar{\xi_0}}{2Q_2^1(\xi_0)(2\xi_0^2-1)}\sin{2\theta_j}\sin{2\phi_j}\left[\boldsymbol{S}_{21}^{(3)}-\boldsymbol{S}_{2,-1}^{(3)}\right]$   \\
                  				 	 &  $\boldsymbol{u}_{5s}=-\frac{\mathrm{i}\,d\xi_0\bar{\xi_0}}{2Q_2^1(\xi_0)(\xi_0^2-1)}\sin{\theta_j}\cos{2\phi_j}\left[\boldsymbol{S}_{21}^{(3)}+\boldsymbol{S}_{2,-1}^{(3)}\right]$ \\ \hline
\end{tabular}
\caption{The five component disturbance velocity fields in a body-fixed coordinate system}
\label{tab:canonical}
\end{table}
$\boldsymbol{u}_{1s}$ corresponds to an axisymmetric extension with the axis of symmetry coincident with the spheroid axis; $\boldsymbol{u}_{2s}$ and $\boldsymbol{u}_{3s}$ correspond to a pair of planar extensions in the plane orthogonal to the spheroid axis, with their principal axes rotated by an angle of $\pi/4$ relative to each other, and $\boldsymbol{u}_{4s}$ and $\boldsymbol{u}_{5s}$ correspond to planar extensions in a pair of orthogonal planes containing the spheroid axis. The decaying spheroidal biharmonics in the expressions for the disturbance velocity fields in Table \ref{tab:canonical} are of the general form $\boldsymbol{S}^{(3)}_{ts}$, and are defined as:
\begin{eqnarray}
\boldsymbol{S}^{(3)}_{ts}&=&\boldsymbol{e}_1\left\{-(x'-\mathrm{i}y')D_2F_{t-1}^{s-1}-\bar{\xi}_0^2dD_1F_{t}^{s}+(t+s-1)(t+s)\beta_{-(t+1)}F_{t-1}^{s-1}\right\}+ \nonumber \\
&& \boldsymbol{e}_2\left\{(x'+\mathrm{i}y')D_1F_{t-1}^{s+1}-\bar{\xi}_0^2dD_2F_{t}^{s}-(t-s-1)(t-s)\beta_{-(t+1)}F_{t-1}^{s+1}\right\}+ \nonumber \\
&& \boldsymbol{1}_{z'}\left\{z'D_3F_{t-1}^s-\xi_0^2dD_3F_t^s-C_{-(t+1),s}F_{t-1}^s\right\},
\end{eqnarray}
where,
\begin{eqnarray}
&& F_t^s(\boldsymbol{r},d)=Q_t^s(\xi)Y_t^s(\eta,\phi), \quad Y_t^s(\eta,\phi)=\frac{(t-s)!}{(t+s)!}P_t^s(\eta)\exp(\mathrm{i}\,s\phi)\\
&& \beta_t=\frac{t+3}{(t+1)(2t+3)}, \quad C_{t,s}=(t+s+1)(t-s+1)\beta_t, \quad |s|\leq t\\
&& \boldsymbol{e}_1=\frac{\boldsymbol{1}_{x'}+\mathrm{i}\,\boldsymbol{1}_{y'}}{2}, \quad \boldsymbol{e}_2=\frac{\boldsymbol{1}_{x'}-\mathrm{i}\,\boldsymbol{1}_{y'}}{2}\\
&& D_1=(\partial_{x'}-i\,\partial_{y'}), \quad D_2=(\partial_{x'}+i\,\partial_{y'}), \quad D_3=\partial_{z'}.
\end{eqnarray}
Here, $Y_t^s$ are the scalar surface harmonics, with $P_t^s$ and $Q_t^s$ being the associated Legendre functions of the first and second kind, respectively; the index $t$ in $\boldsymbol{S}^{(3)}_{ts}$  denotes the rapidity of the decay for large $\xi$(or $r$), with $\lim_{r\to\infty}\boldsymbol{S}^{(3)}_{ts}\propto r^{-t}$. Thus, all of the $\boldsymbol{S}^{(3)}_{2s}$'s appearing in the disturbance velocity fields (Table \ref{tab:canonical}) decay as $1/r^2$, as is appropriate since the force-free freely rotating spheroid appears as a fluctuating force-dipole in the far-field.

The complete dynamical system governing the evolution of the fluid pathlines in terms of the position vector($\boldsymbol{x}$) and Jeffery angles($\theta_j,\phi_j$) may now be written as:
\begin{eqnarray}\label{eq:system}
 && \frac{d\boldsymbol{x}}{dt}=\boldsymbol{\Gamma}\bcdot\boldsymbol{x} + \boldsymbol{\mathcal{R}}^T(\theta_j,\phi_j)\cdot\sum_{i=1}^5\boldsymbol{u}_{is}(\boldsymbol{x'}),\label{eq:sys1}\\
 && \frac{d\phi_j}{dt}=-\frac{1}{\kappa^2+1}\left(\kappa^2\sin^2{\phi_j}+\cos^2{\phi_j}\right), \label{eq:sys2}\\
 && \frac{d\theta_j}{dt}=\frac{1}{4}\left(\frac{\kappa^2-1}{\kappa^2+1}\right)\sin{2\theta_j}\sin{2\phi_j}. \label{eq:sys3}
\end{eqnarray} 
Here, $\boldsymbol{\mathcal{R}}(\theta_j,\phi_j)$ is the rotation tensor relating the space and body-fixed coordinates.
\section{A spheroid in the spinning or log-rolling orbit($C=0$)}\label{sec:spinning}
To begin with, we analyze the simpler case of a spheroid in its spinning(log-rolling) orbit. The results for the streamlines around an arbitrary aspect ratio spheroid, aligned with the vorticity axis, haven't been reported before. This simple case also serves as a partial validation for our numerical integration, based on prior knowledge of the streamline patterns for the limiting cases of a cylinder and a sphere. The spinning (log-rolling) orbit corresponds to $\theta_j = 0$, and is obtained from (\ref{eq:jeffery}) by setting the orbit constant $C = 0$. For this configuration, the only non-zero component in the general disturbance field is $\boldsymbol{u}_{3s}$, corresponding to a transverse planar extension with the extensional axis oriented at 45 degrees in the flow-gradient plane.

\subsection{The streamline pattern for the spinning spheroid}
\begin{figure}
  \centering
\begin{subfigure}[b]{0.45\textwidth}
\centering
  \includegraphics[scale=0.2]{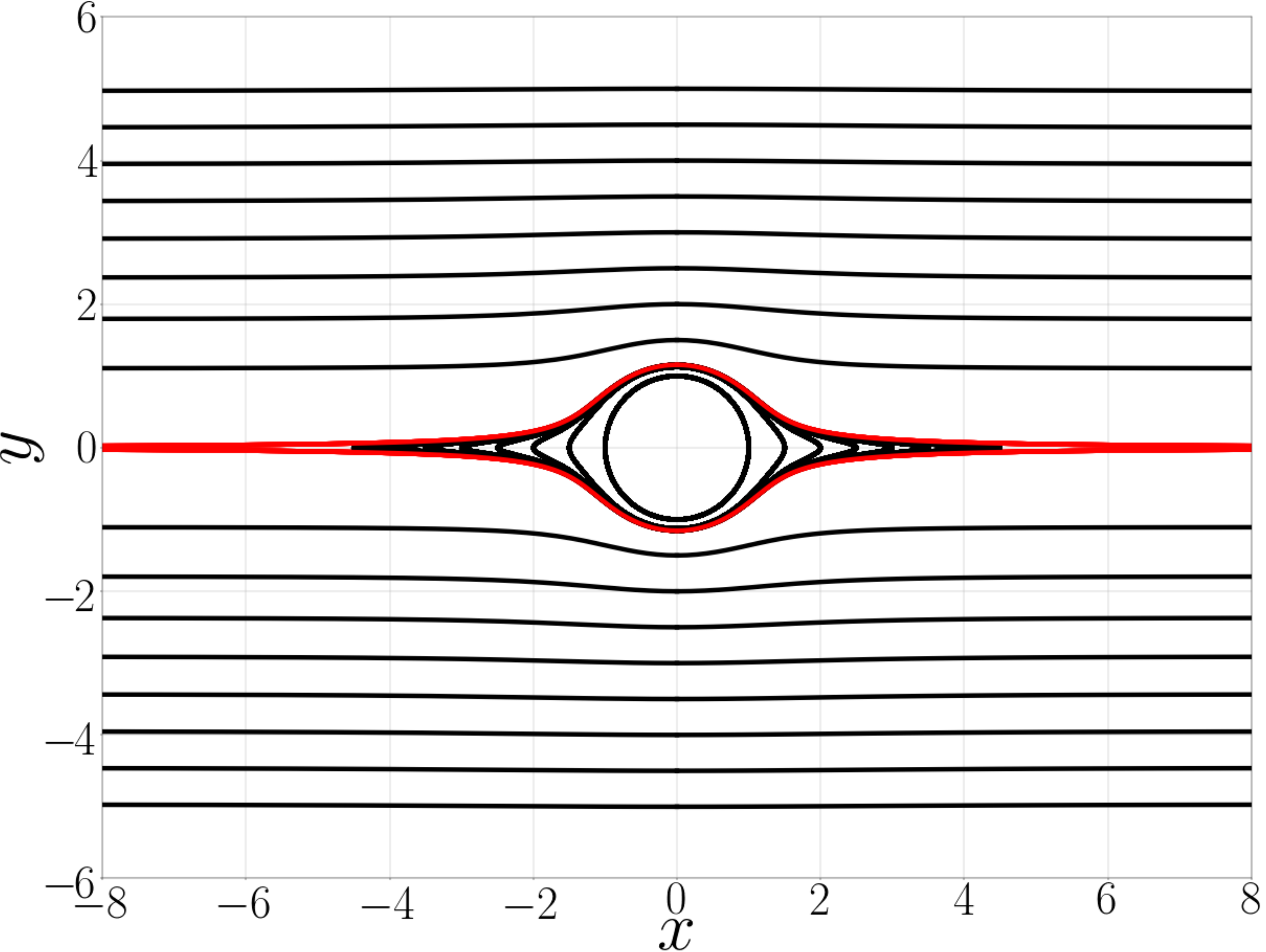}
\caption{}
\end{subfigure}
\quad
\begin{subfigure}[b]{0.45\textwidth}
\centering
  \includegraphics[scale=0.2]{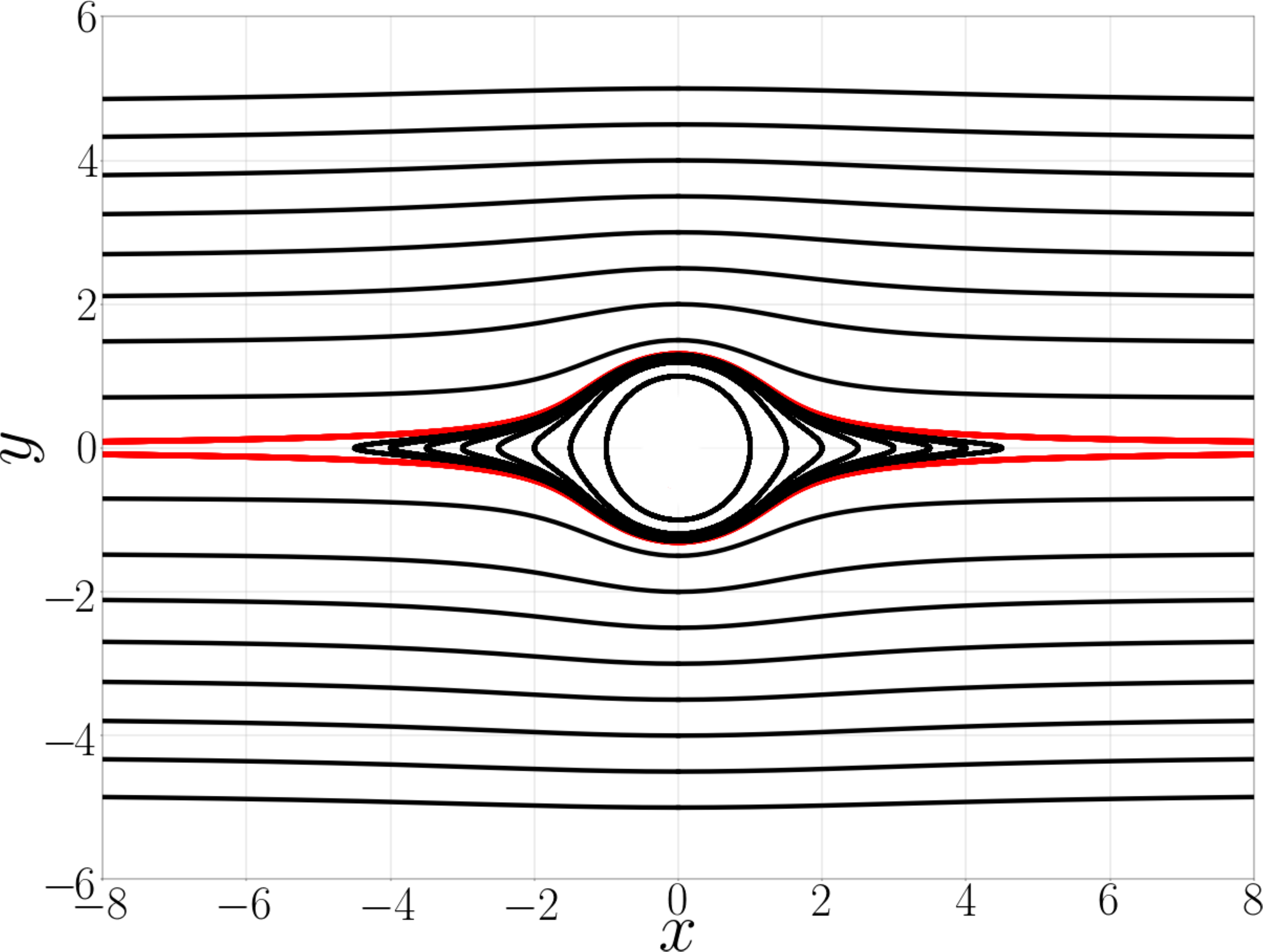}
\caption{}
\end{subfigure}
\vfill
\begin{subfigure}[b]{0.45\textwidth}
\centering
  \includegraphics[scale=0.2]{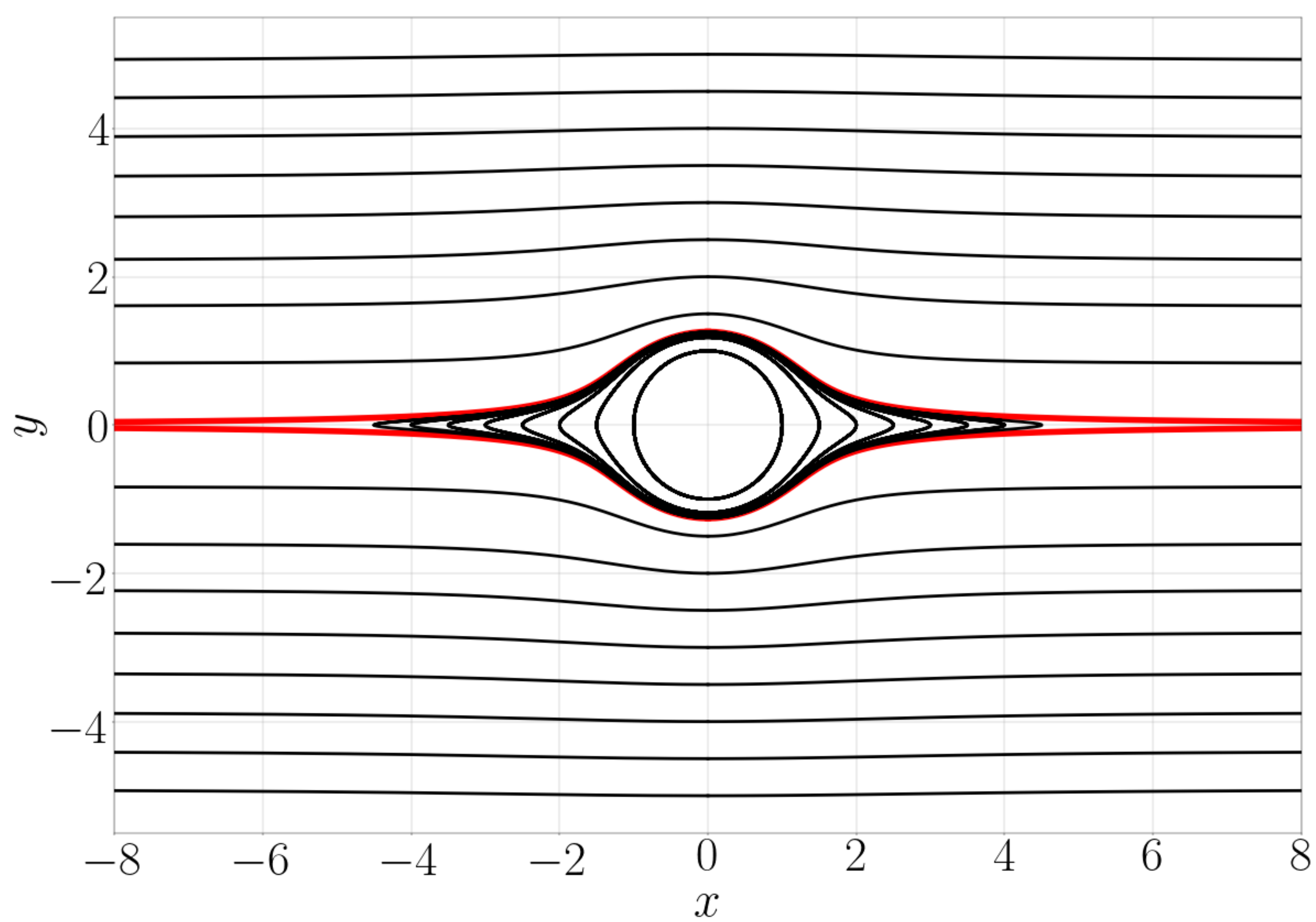}
\caption{}
\end{subfigure}
  \caption{The in-plane fore-aft symmetric streamline pattern for (a) a sphere, (b) a cylinder and (c) a prolate spheroid with $\kappa=3.28(\xi_0=1.05)$. In each case, fore-aft open streamlines are separated from the closed ones by separatrices(shown in red).}
\label{fig:spin_pathline1}
\end{figure}

\begin{figure}
  \centering
\begin{subfigure}[b]{0.45\textwidth}
\centering
  \includegraphics[scale=0.25]{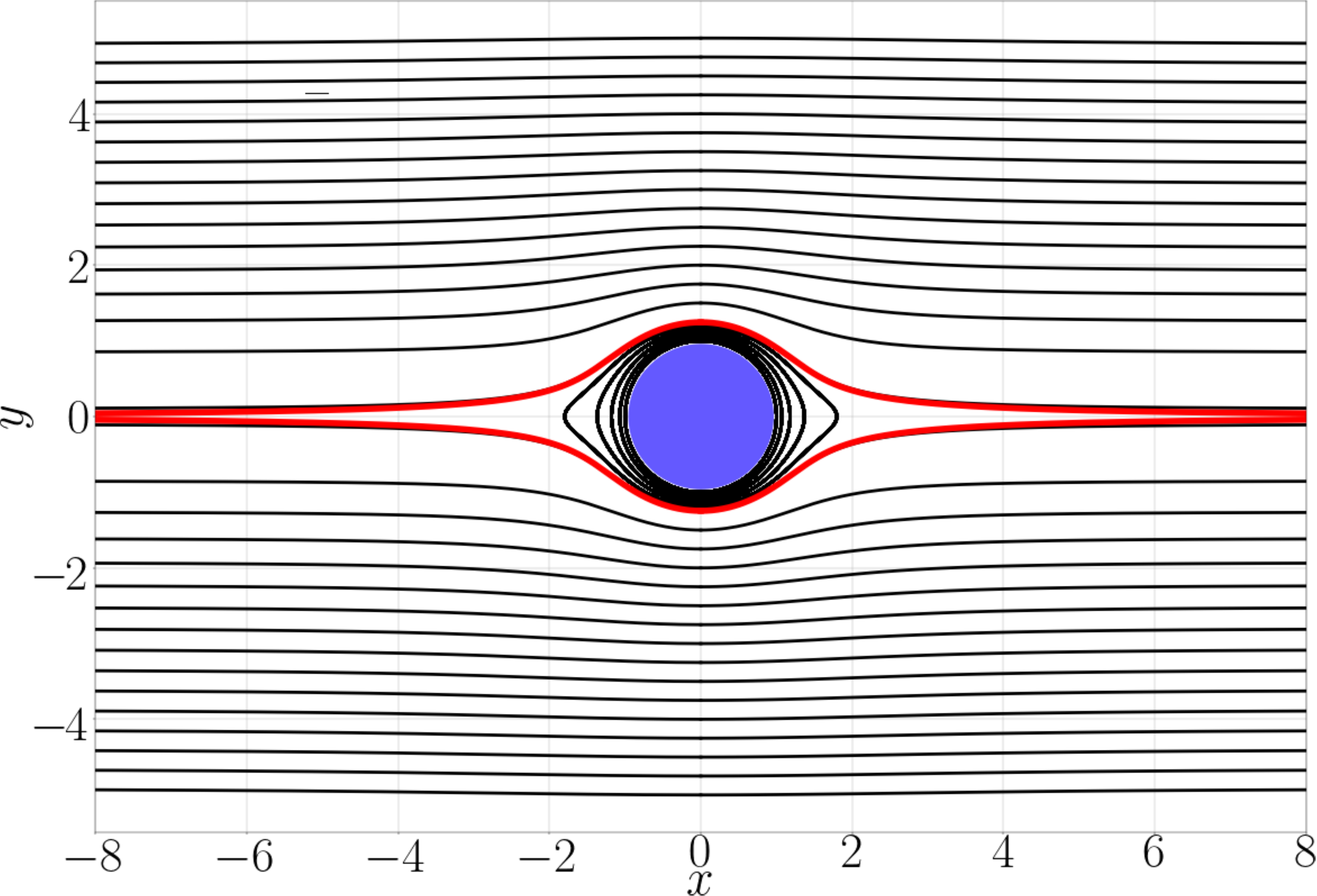}
\caption{}
\label{spin-x-y}
\end{subfigure}
\quad
\begin{subfigure}[b]{0.45\textwidth}
\centering
  \includegraphics[scale=0.3]{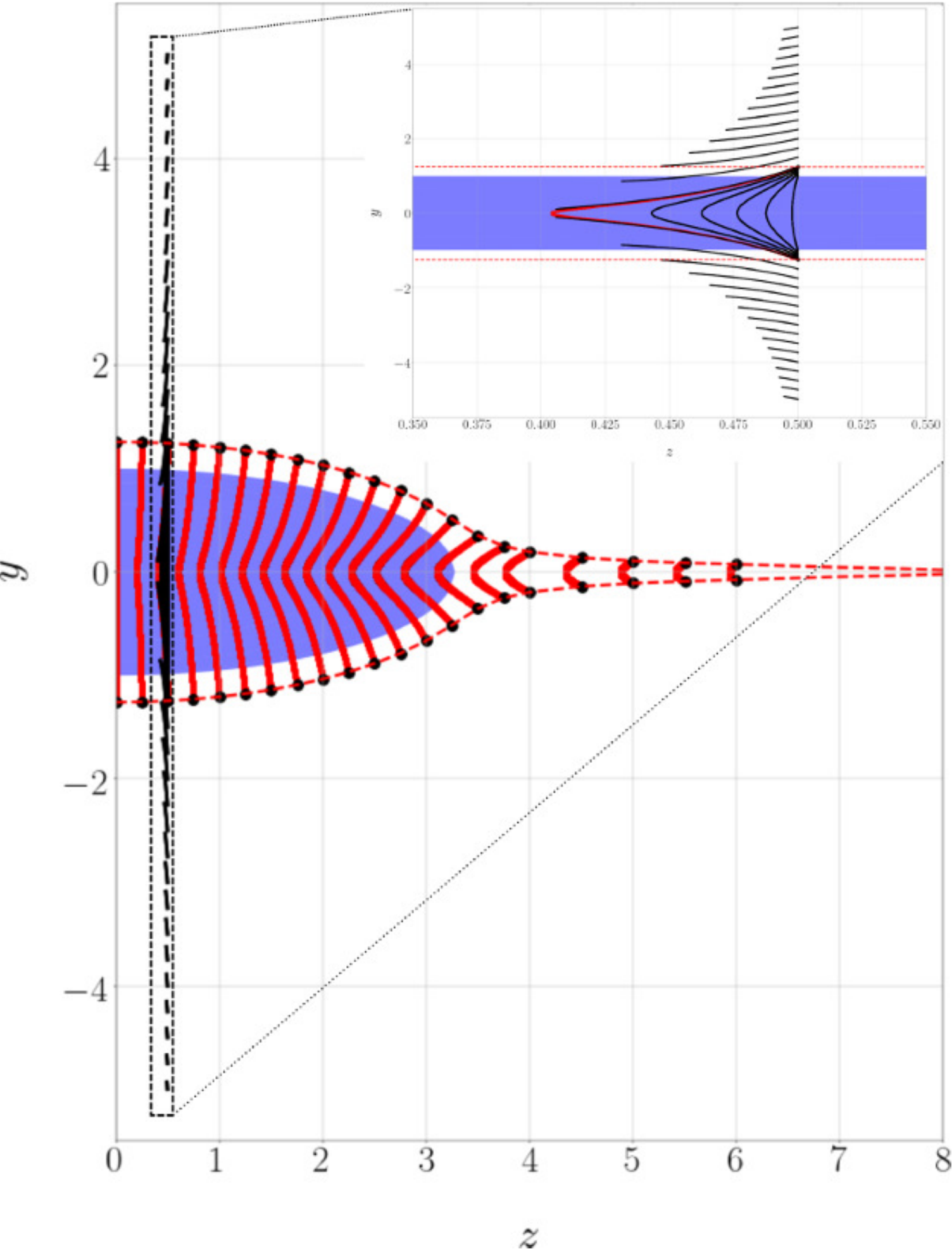}
\caption{}
\label{spin-y-z}
\end{subfigure}
\vfill
\begin{subfigure}[b]{0.85\textwidth}
\centering
  \includegraphics[scale=.4]{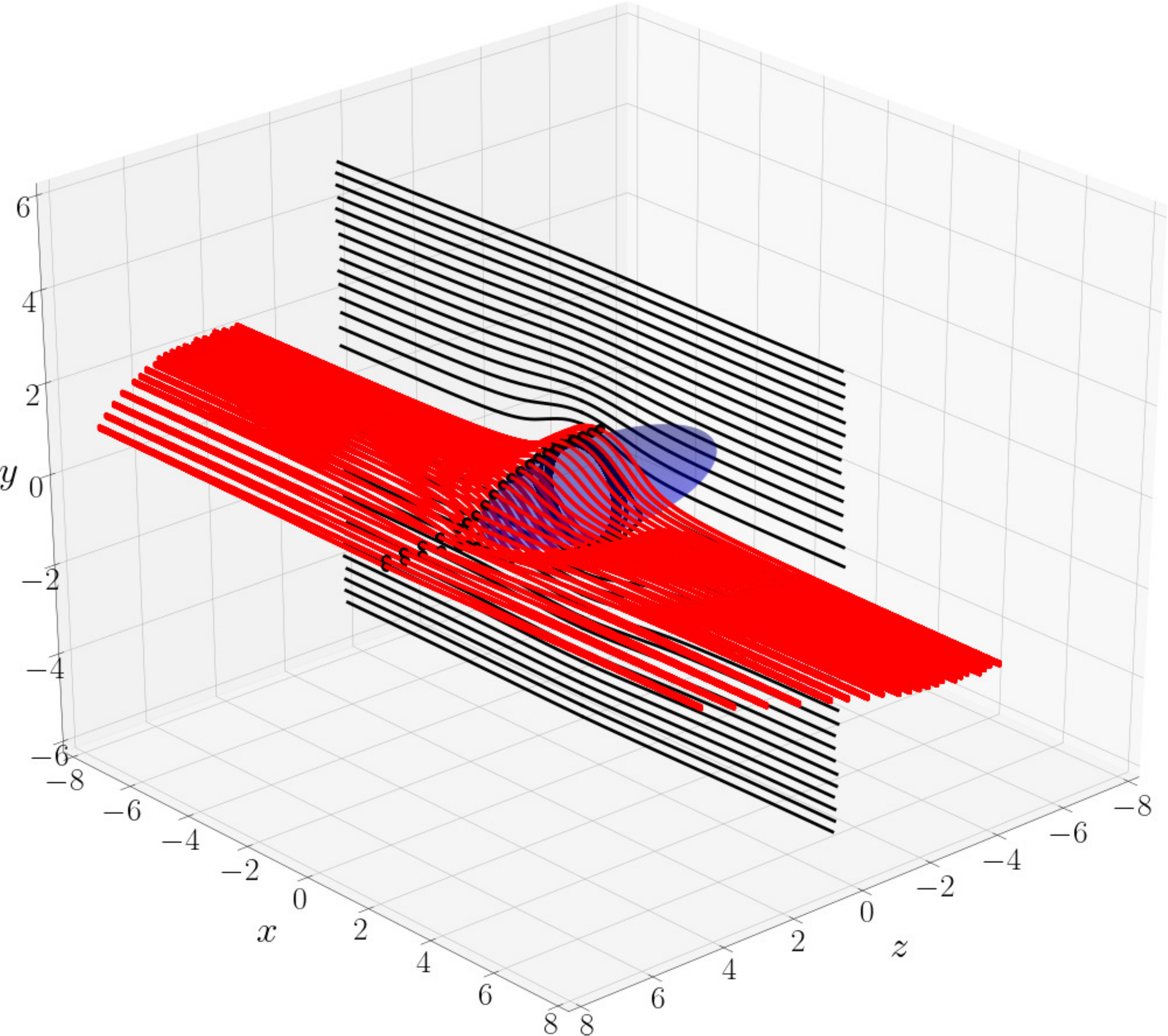}
\caption{}
\label{spin-3d}
\end{subfigure}
  \caption{The three-dimensional streamline pattern around a log-rolling prolate spheroid with $\kappa = 3.28 (\xi_0 = 1.05)$ corresponding to $z = 0.5$(on the gradient-vorticity plane). The subfigures include projections on the flow-gradient plane (a), the gradient-vorticity plane (b), and a three dimensional view (c). The projection of the separatrix surface, and the constituent separatrix streamlines, in each of these figures are shown in red. In figure 4c, black dots denote the intersection of the separatrix envelope with the gradient-vorticity plane.}
\label{fig:spin_pathline2}
\end{figure}

\begin{figure}
  \centering
\begin{subfigure}[b]{0.48\textwidth}
\centering
  \includegraphics[scale=0.2]{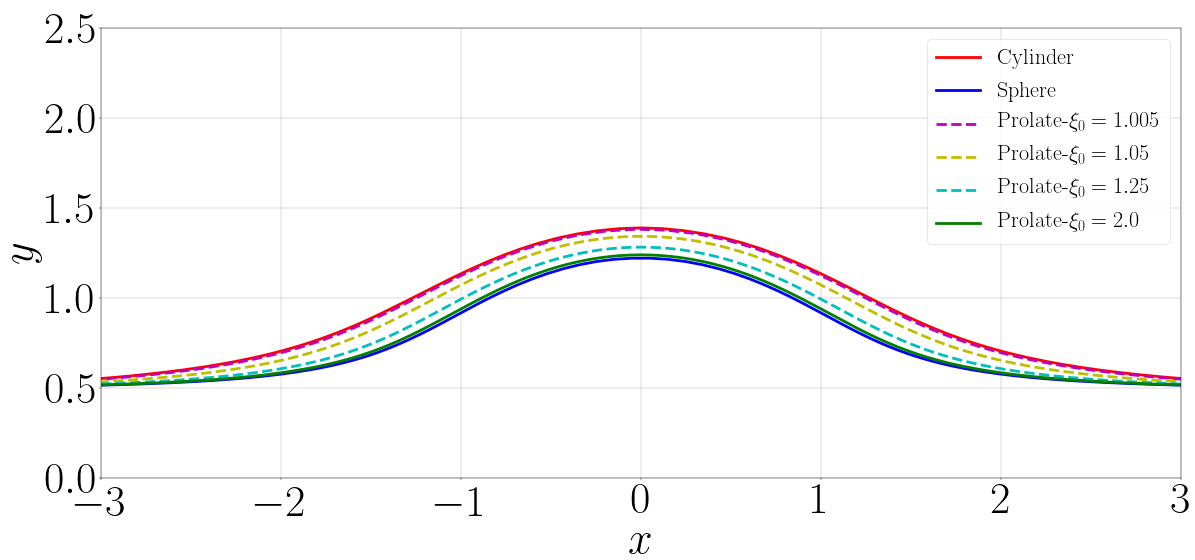}
\caption{}
\label{fig:spin_open}
\end{subfigure}
\hfill
\begin{subfigure}[b]{0.48\textwidth}
\centering
  \includegraphics[scale=0.17]{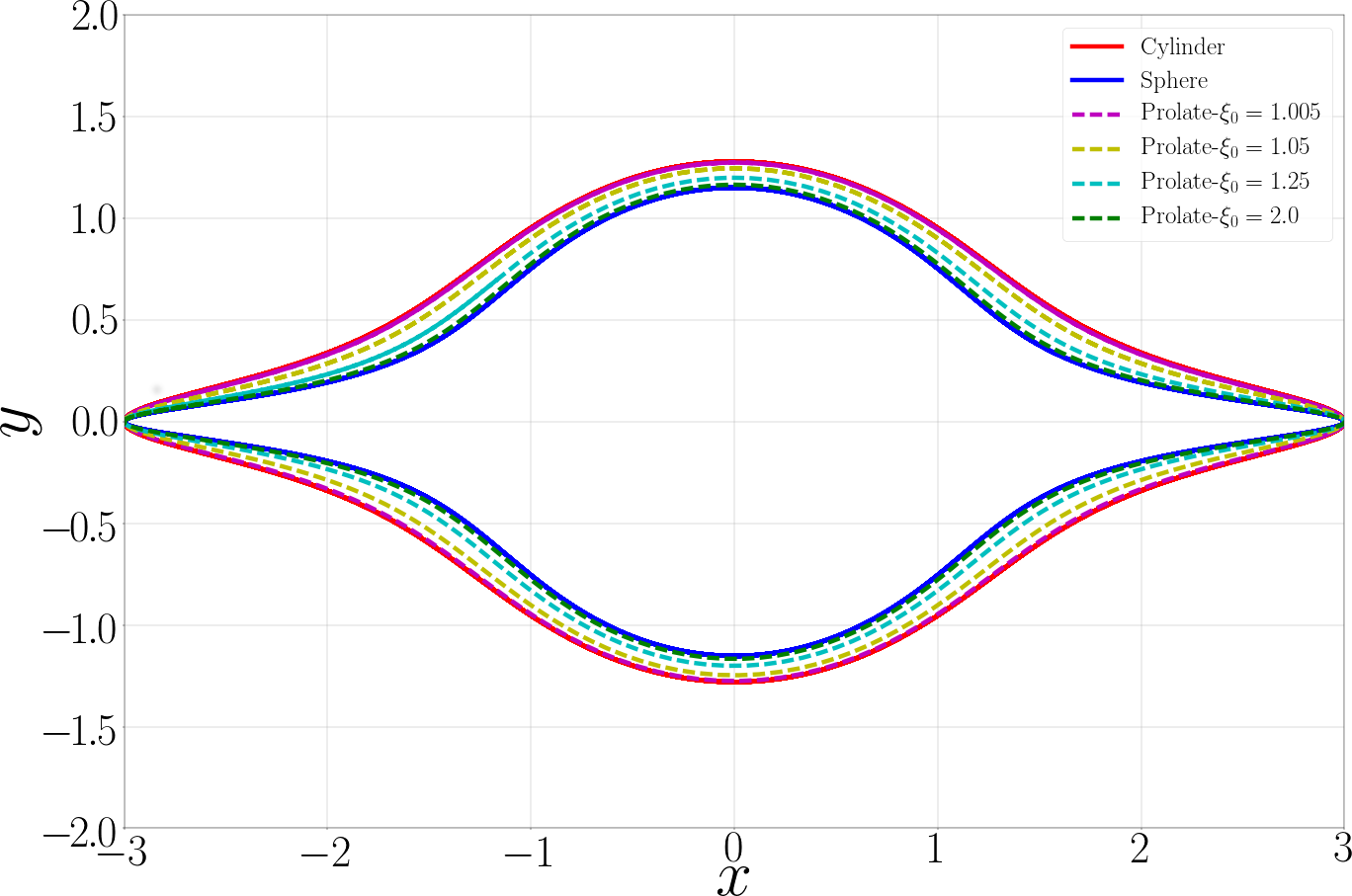}
\caption{}
\label{fig:spin_closed2}
\end{subfigure}
\begin{subfigure}[b]{0.85\textwidth}
\centering
  \includegraphics[scale=0.23]{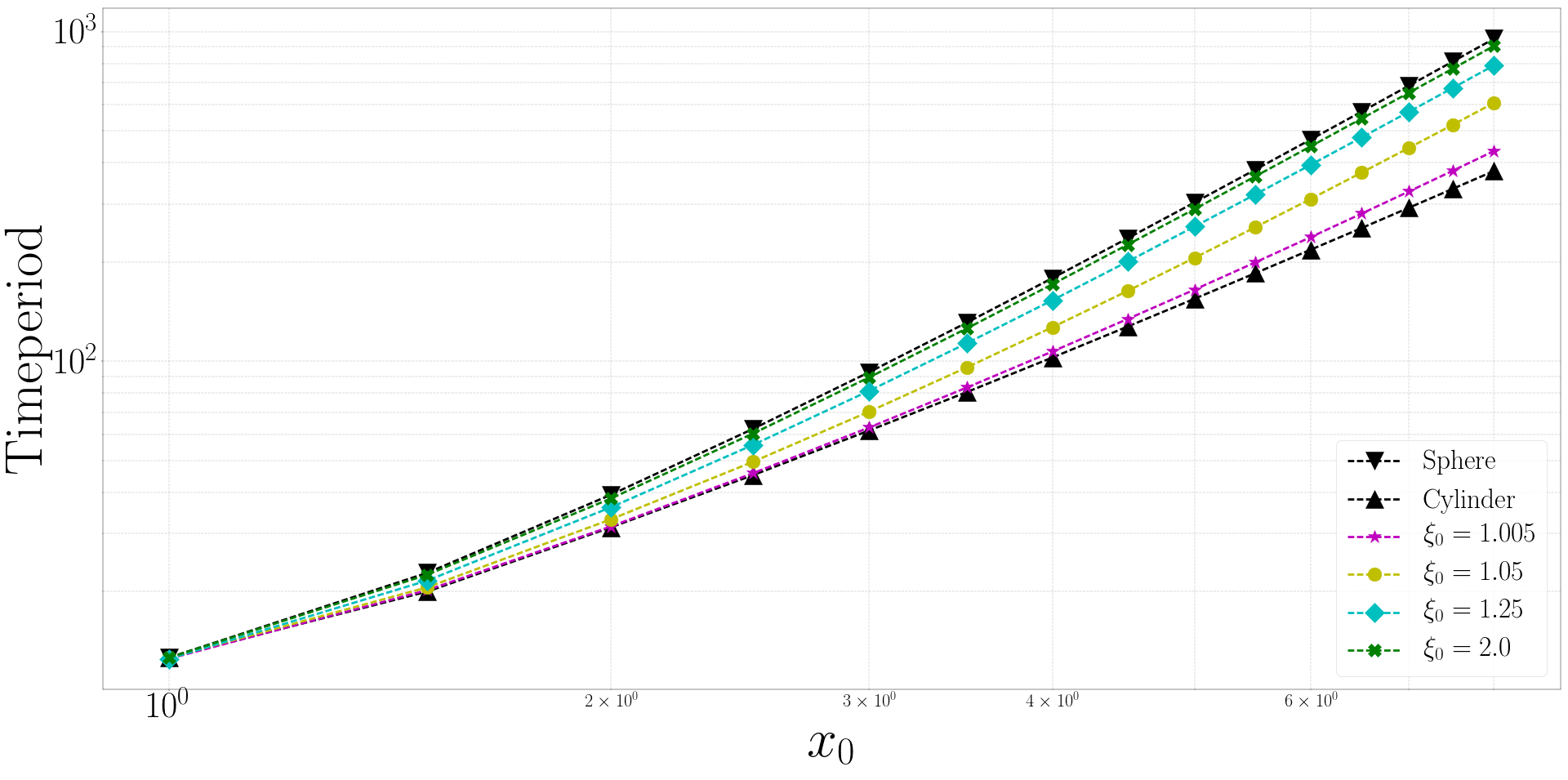}
\caption{}
\label{fig:timeperiod}
\end{subfigure}
  \caption{The shapes of the open (a) and closed (b) streamlines for prolate spheroids with $\kappa(\xi_0)=10.04(1.005),3.25(1.05),1.67(1.25),1.15(2)$; the limiting cases of a sphere($\kappa=1$) and a cylinder($\kappa=\infty$) are shown. The open streamlines coincide at upstream and downstream infinity, while the closed streamlines start from the same location on the flow axis ($-3.0,0$); (c) A comparison of closed-streamline time periods for the aforementioned aspect ratios, including the limiting sphere and cylinder cases.}
\label{fig:spin_closed}
\end{figure}

Numerically integrating (\ref{eq:system}) with only the disturbance velocity field $\boldsymbol{u}_{3s}$ one may obtain the streamlines around a spinning spheroid. Figures \ref{fig:spin_pathline1} and \ref{fig:spin_pathline2} show the topology of the streamlines both within and outside of the flow gradient plane. Figure \ref{fig:spin_pathline1} depicts the in-plane streamline patterns for a sphere, a cylinder and a prolate spheroid with $\xi_0=1.05(\kappa = 3.28)$. The qualitative resemblance between the patterns is readily evident, the streamline topology being identical in all three cases. The only perceptible difference is an increase in the size of the closed streamline region as one moves from the sphere to the cylinder, this  being consistent with the increasing strength of the velocity disturbance, as is also evident from the far-field behavior of the disturbance field in the two cases - the $O(1/r^2)$ decay for the sphere as opposed to the $O(1/r)$ decay for the circular cylinder. Figure \ref{fig:spin_pathline2} shows the off-plane streamlines, which intersect the gradient-vorticity plane at $z = 0.5$. The three-dimensional view (figure \ref{spin-3d}), and the projections of these streamlines on the flow-gradient (figure \ref{spin-x-y}) and gradient-vorticity (figure \ref{spin-y-z}) planes are included. The separatrix surface is non-axisymmetric for all cases except that of a sphere, and this is evident from the differing projections of the separatrix envelope in the aforementioned planes; note that the envelope in the flow-gradient plane is a single streamline, but that in the gradient-vorticity plane is constructed from multiple separatrix streamlines (also shown in red in figure \ref{spin-y-z}). 
Figures \ref{fig:spin_open} and \ref{fig:spin_closed2} show the shapes of both an open and a closed streamline around log-rolling prolate spheroids with different aspect ratios ($\kappa =10.04,3.25,1.67,1.15 $); also included are limiting cases of a sphere ($\kappa = 1$) and a cylinder ($\kappa = \infty$). The open streamline deviates the most (least) from the corresponding ambient streamline, and the closed streamline is largest (smallest) in size, for $\kappa = \infty (1)$. Figure \ref{fig:timeperiod} shows  the time periods for closed streamlines, starting from a given location on the flow axis, increasing monotonically as one goes from a cylinder to a sphere. The limiting values for a sphere and cylinder are available for earlier efforts\citep{torza,powell}. All of these features are again consistent with the larger disturbance field associated with longer prolate spheroids (the cross-sectional diameter is fixed as $\kappa$ varies), with the largest disturbance corresponding to a cylinder. In summary, the streamline topology associated with a spinning spheroid is identical to that around a sphere and a circular cylinder, with the various measures associated with the streamline pattern around a log-rolling prolate spheroid, including the time periods of the closed streamlines, the width of the closed streamline region etc, bounded between those for a sphere and a cylinder.

\subsection{The far-field separatrix for a spinning spheroid}\label{sec:farfield}
The analytical form of the separatrix for a sphere is known\citep{batchelor_2sphere}, and is given by $y=\sqrt{1/3}x^{-3/2}$; an implication of the rather slow approach of the separatrix towards the flow axis is that an infinite volume of fluid is contained within the axisymmetric separatrix envelope. To derive the far-field separatrix for an arbitrary aspect ratio spheroid, we start from the multipole expansion for the velocity field induced by a spheroid in an ambient linear flow, given by\citep{brenner3,brenner4,brenner5}: 
\begin{eqnarray}\label{eq:multipole}
\boldsymbol{u}(\boldsymbol{x})&=&\boldsymbol{\Gamma}\bcdot\boldsymbol{x}+\mathsfbi{S}\bcdot\bnabla\frac{3}{D}\frac{\partial}{\partial D}\left[\frac{\sinh D}{D}\right]\frac{\boldsymbol{\cal G(\boldsymbol{x})}}{8\upi\mu}\\ \nonumber
 &=&\boldsymbol{\Gamma}\cdot\boldsymbol{x}+\mathsfbi{S}\bcdot\bnabla\left[1+\frac{3.4}{5!}D^2+\frac{3.6}{7!}D^4+\cdots\right]\frac{\boldsymbol{\cal G(\boldsymbol{x})}}{8\upi\mu},
\end{eqnarray}
where $D^2=d^2\bar{\xi_0}^2\left[\nabla^2+(\kappa^2-1)\boldsymbol{pp\colon\nabla\nabla}\right](D^2=d^2\xi_0^2\left[\nabla^2+(\kappa^2-1)\boldsymbol{pp\colon\nabla\nabla}\right])$ for a prolate(an oblate) spheroid. In (\ref{eq:multipole}), $\boldsymbol{p}=\boldsymbol{1}_z$, $\boldsymbol{\mathcal{G}}$ is the Oseen-Burger's tensor given by $\mathcal{G}_{ij}=\delta_{ij}/r+x_ix_j/r^3$, and $\mathsfbi{S}$ is the stresslet induced by a force and torque-free spinning spheroid in an ambient shear flow, which may be written as:
\begin{equation}\label{eq:stresslet1}
\mathsfbi{S}=\frac{20\upi\mu d^3\xi_0^3}{3}K_{tp/o}\,(\delta_{il}-p_ip_l)E_{lk}(\delta_{kj}-p_kp_j)+\frac{1}{2}(E_{kl}p_kp_l)(\delta_{ij}-p_ip_j).
\end{equation}
In (\ref{eq:stresslet1}), the stresslet coefficient $K_{tp/o}$ is a different function of aspect ratio for prolate and oblate spheroids(see Table \ref{tab:coeff}), and characterizes the magnitude of the stresslet response for a tranverse planar extensional flow.

\begin{figure}
  \centerline{
  \includegraphics[scale=0.4]{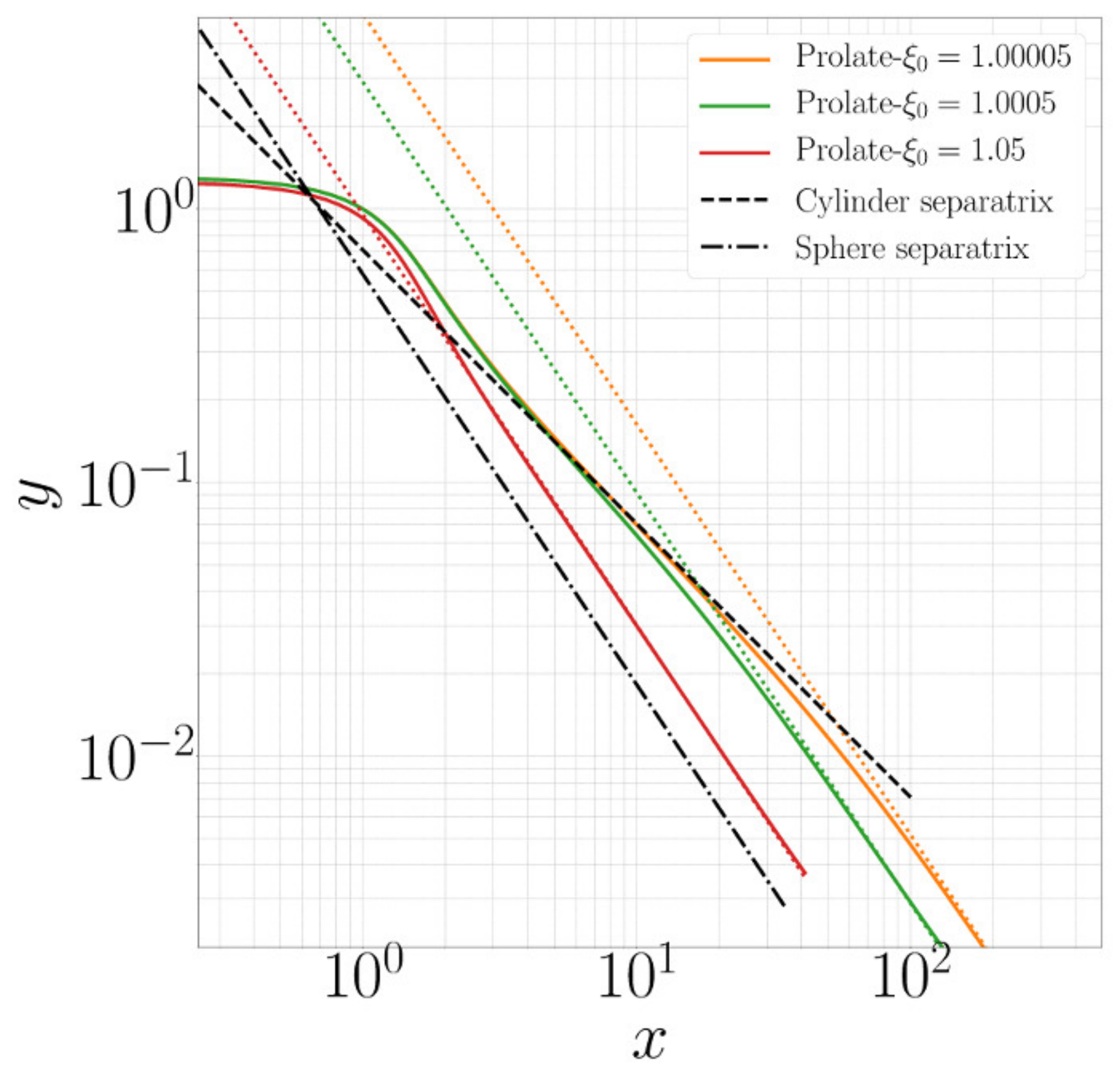}
  }
  \caption{Separatrices in the flow-gradient plane, for spinning prolate spheroids($\kappa=100.04,31.63,10.04,3.28$) in a simple shear flow. The dashed lines represent the predictions of the asymptotic analysis in the text.}
\label{fig:spin_sep}
\end{figure}

The dipole($\boldsymbol{u}^d$) and octupole($\boldsymbol{u}^o$) contributions correspond to the first and second terms in (\ref{eq:multipole}) and, for a prolate spheroid, are given by:
\begin{eqnarray}\label{eq:dipole_octupole}
 u^d_i&=&\frac{1}{8\upi\mu}S_{jk}\partial_k\mathcal{G}_{ij}=-\frac{3}{8\pi\mu}\frac{S_{jk}n_jn_k}{r^2}n_i,\\
 u^o_i&=& \frac{d^2\bar{\xi}_0^2}{80\pi\mu\,r^4}\left[-12S_{ij}n_j+30n_i(S_{jk}n_jn_k)+(\kappa^2-1)\left\{-3\left(4p_i(S_{jk}n_jp_k)+2n_i(S_{jk}p_kp_k)\right)\right.\right. \nonumber \\
 && \left.\left. +15\left(4n_i(p_ln_lS_{jk}n_jp_k)+2p_i(p_ln_lS_{jk}n_jn_k)+n_i(S_{jk}n_jn_k)\right)\right.\right. \nonumber \\
 && \left.\left. -105n_i(p_ln_l)^2(S_{jk}n_jn_k)\right\}\right],
\end{eqnarray}
where the unit radial vector $\boldsymbol{n}=\boldsymbol{x}/r$.
To derive the asymptotic form of the separatrix, one needs to account for the higher-order octupole contribution, owing to the dipole field being purely radial, and thereby, lacking a transverse ($y$) component along the flow axis. Using the above velocity field in $d\boldsymbol{x}/dt = \boldsymbol{\Gamma.x} + \boldsymbol{u}^d + \boldsymbol{u}^o$, and considering the limit $y \rightarrow 0$, $x \rightarrow \infty$, one obtains the following equations at leading order:
\begin{eqnarray}
&& \frac{dx}{dt} = \lim_{y\to 0}u_x=y,\label{eq:spin_sepsys_1}\\ 
&& \frac{dy}{dt} =\lim_{y\to 0}u_y=-\frac{5y^2d^3\xi_0^3\,K_p}{2x^4}-\frac{d^5\xi_0^3\bar{\xi_0}^2}{2x^4}K_{p}.\label{eq:spin_sepsys_2}
\end{eqnarray}
It is evident that the $x$-component of the velocity in (\ref{eq:spin_sepsys_1}) is just the ambient flow ($y$) at leading order, while the dominant contribution in (\ref{eq:spin_sepsys_2}) is the octupolar contribution; that this term of $O(1/x^4)$ is dominant over the dipole contribution of $O(y^2/x^4)$ is easily seen since $y$ for the separatrix is asymptotically small in the far-field. With the far-field forms in place, the form of the separatrix may be obtained by taking the ratio of (\ref{eq:spin_sepsys_1}) and (\ref{eq:spin_sepsys_2}), and integrating the resulting differential equation. One obtains $y = 1/\sqrt{3}\,d^{5/2}\xi_0^{3/2}\bar{\xi_0}K_p^{1/2}x^{-3/2}$  which exhibits the same far-field decay as the sphere separatrix. Figure \ref{fig:spin_sep} shows the separatrices for log-rolling prolate spheroids of different aspect ratios, and compares their far-field forms with the analytical prediction above (shown in dotted lines). For sufficiently large aspect ratios ($\kappa \rightarrow \infty$ or $\xi_0 \rightarrow 1$), the separatrix exhibits an intermediate asymptotic form, $1/\sqrt{2} x^{-1/2}$,  in the interval $1 \ll x \ll \kappa$, corresponding to that of a circular cylinder, before eventually transitioning to the more rapid $O(x^{-3/2})$ decay for $x \gg \kappa$.

\section{A spheroid in non-spinning orbits($C>0$)}
\subsection{ Regular open pathlines}\label{sec:open}
The results presented thus far show that the streamline pattern around spinning spheroids is steady, and analogous to that around a sphere. In this section, we consider spheroids in orbits other than the spinning one ($C > 0$). For all such cases, the velocity field is time dependent. The focus from hereon will therefore be on the fluid pathlines rather than the (time periodic) sequence of instantaneous streamline patterns associated with the rotating spheroid. In our examination of fluid pathlines, it is natural to start from a spheroid tumbling in the flow-gradient plane. The polar Jeffery angle is now a constant ($\theta_j = \pi/2$), with the time dependence of the velocity field arising only due to $\phi_j$ varying with time. Thus, one only needs to integrate (2.11) and (2.12) in order to characterize the pathlines. Further, the disturbance velocity field only involves three of the five canonical components in the disturbance velocity fields in Table \ref{tab:canonical} viz. the axisymmetric extension($\boldsymbol{u}_{1s}$), the first transverse extension($\boldsymbol{u}_{2s}$) and the second longitudinal extension($\boldsymbol{u}_{5s}$). An additional symmetry-induced simplification is that fluid elements in the flow-gradient plane are confined to this plane for all times, despite the time dependence of the flow field. This makes it easier to visualize the changes in the pathline topology relative to the case of a spheroid rotating in a generic precessional orbit.

\begin{figure}
  \centering
\begin{subfigure}[b]{0.48\textwidth}
\centering
  \includegraphics[scale=0.2]{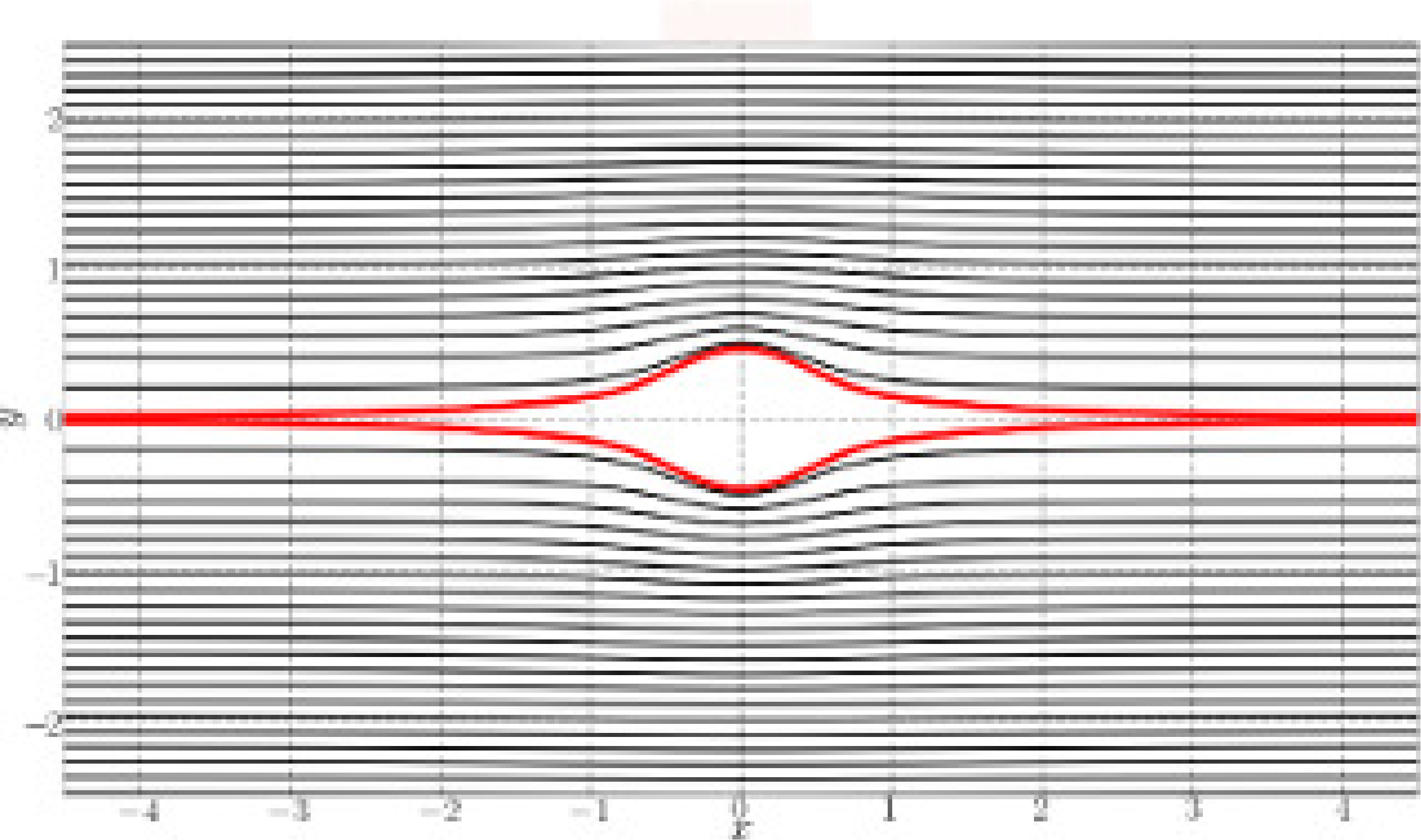}
\caption{$\phi_{j0}=0$}
\end{subfigure}
\hfill
\begin{subfigure}[b]{0.48\textwidth}
\centering
  \includegraphics[scale=0.22]{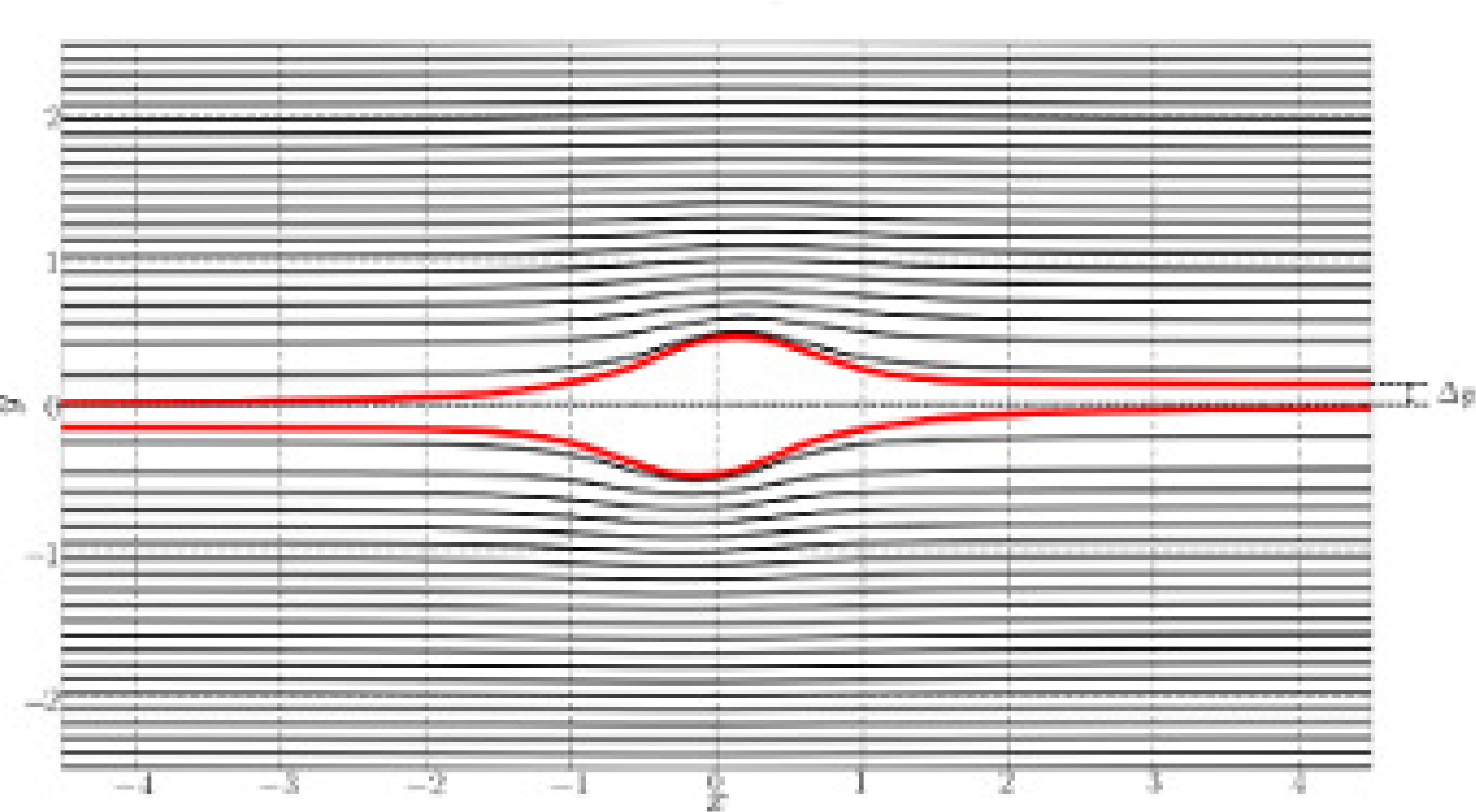}
\caption{$\phi_{j0}=\pi/6$}
\end{subfigure}
\begin{subfigure}[b]{0.48\textwidth}
\centering
  \includegraphics[scale=0.2]{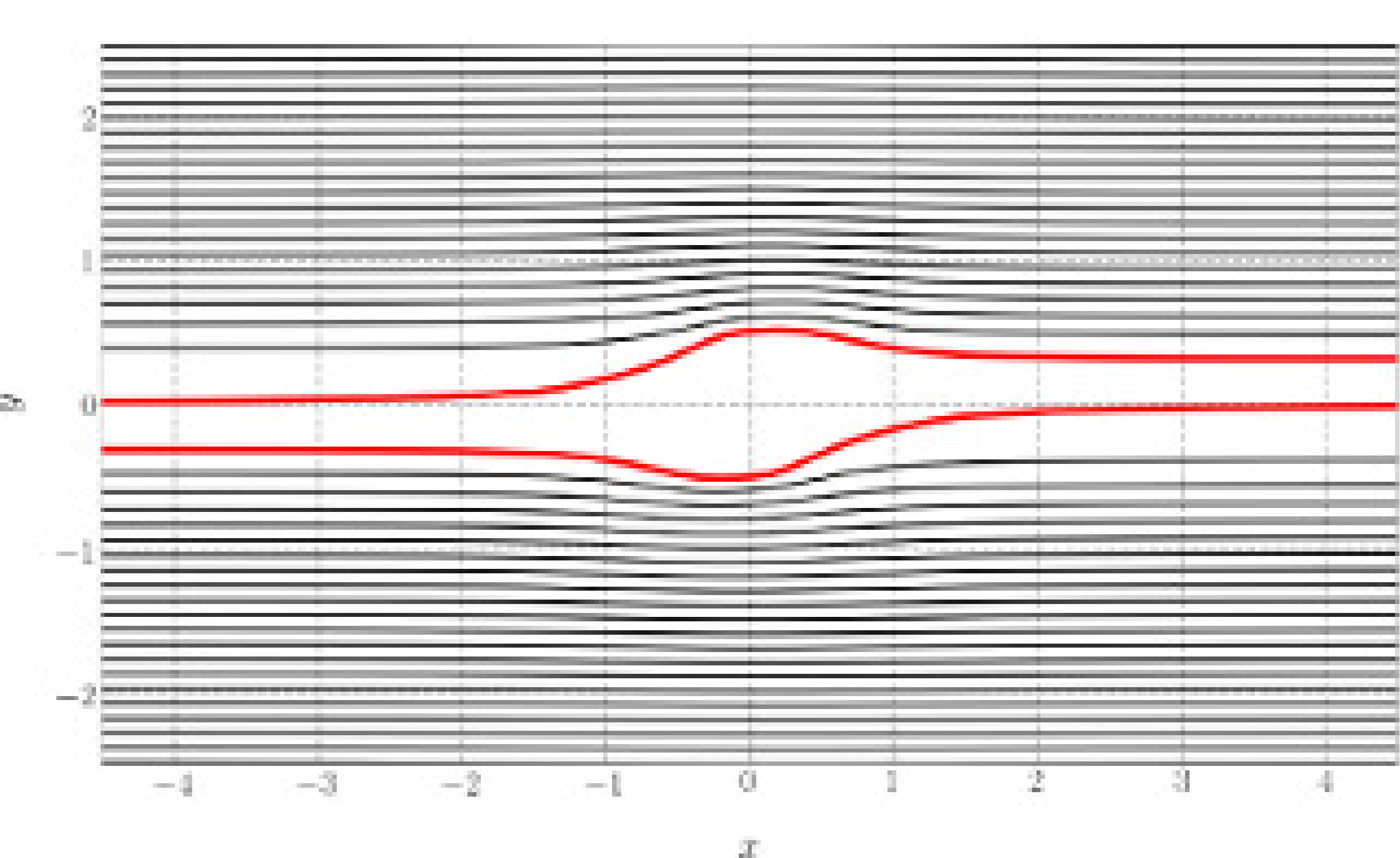}
\caption{$\phi_{j0}=\pi/4$}
\end{subfigure}
\hfill
\begin{subfigure}[b]{0.48\textwidth}
\centering
  \includegraphics[scale=0.22]{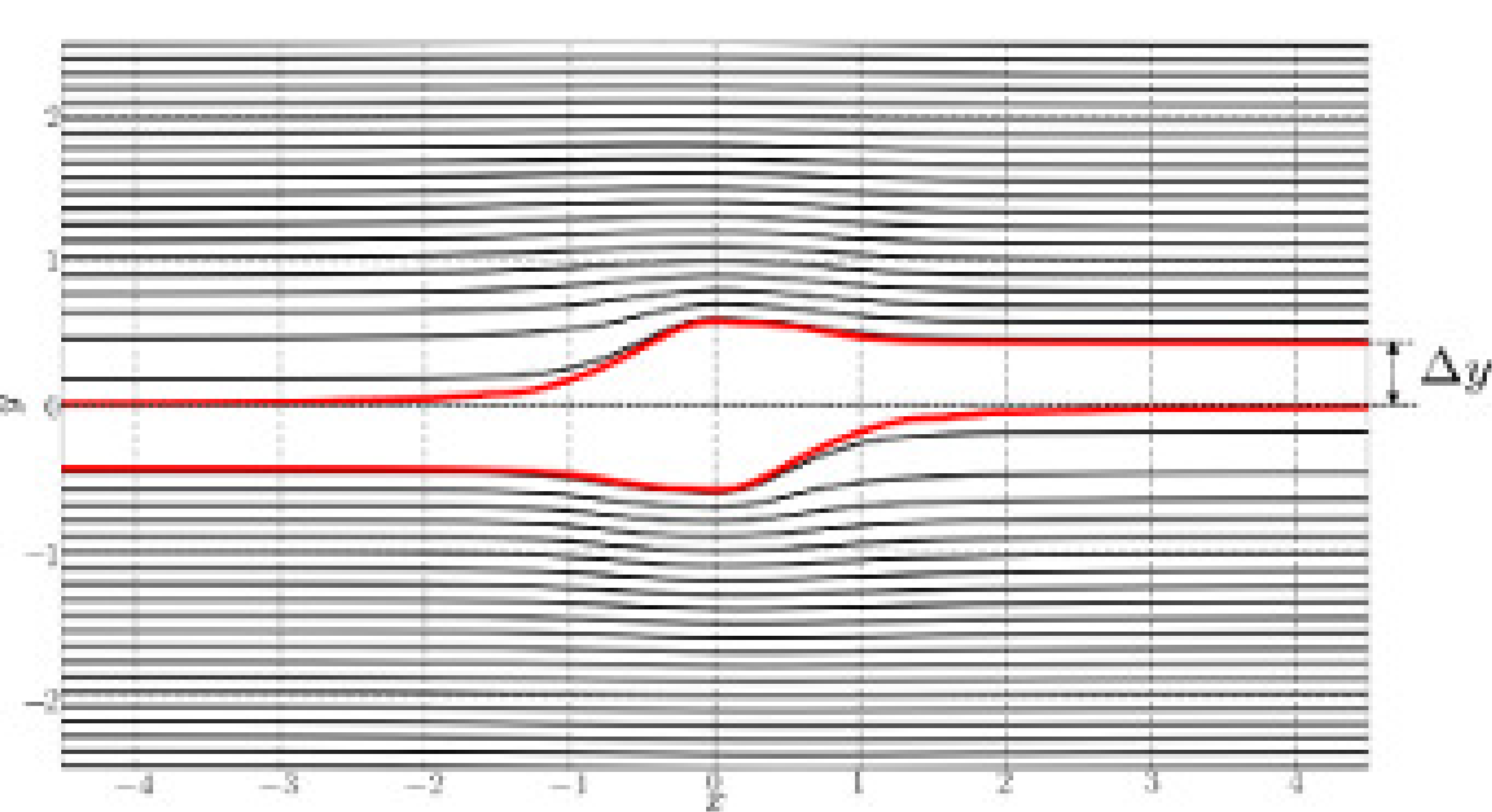}
\caption{$\phi_{j0}=\pi/3$}
\end{subfigure}
\begin{subfigure}[b]{0.9\textwidth}
\centering
  \includegraphics[scale=0.2]{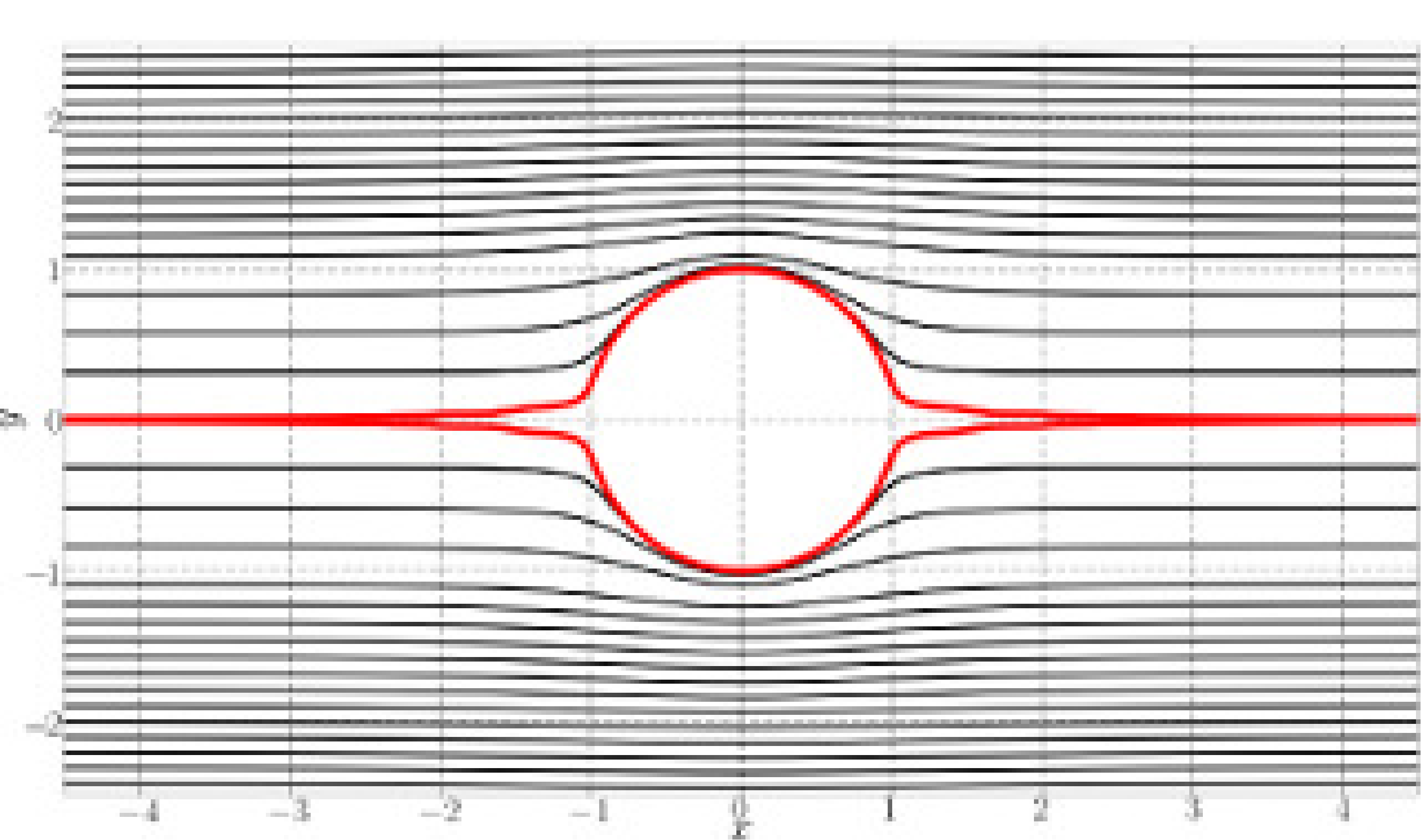}
\caption{$\phi_{j0}=\pi/2$}
\end{subfigure}
  \caption{The configuration of the inplane regular pathlines for a tumbling prolate spheroid with $\kappa=3.28(\xi_0=1.05)$. The different configurations correspond to $\phi_{j0}$ values of  (a)$0$, (b)$\pi/6$, (c)$\pi/4$, (d)$\pi/3$, and (e)$\pi/2$. All configurations are bounded below by separatrices shown in red. The $\Delta y$ marked in (b) and (d) denotes the lateral displacement in the gradient direction.}
\label{fig:latdisp_phi}
\end{figure}

The changing orientation of the tumbling spheroid implies that, unlike the spinning case, one needs to specify an `initial' orientation in order to fix the fluid pathlines. A convenient choice is the spheroid orientation at the instant that the fluid element crosses the gradient-vorticity plane ($x =0$); the coordinates of the element at this instant may be taken as ($y_0,z_0$). This orientation is fixed by the azimuthal Jeffery angle $\phi_{j0}$. The fore-aft symmetry of the spheroid implies that $\phi_{j0}$ may be restricted to the interval ($0,\pi$). Further, from reversibility considerations, one only need consider $\phi_{j0}$'s in the interval ($0,\pi/2$), since the pathline configuration for $\pi-\phi_{j0}$ may be obtained from that for $\phi_{j0}$ via a reflection transformation with respect to the gradient-vorticity plane. Each  $\phi_{j0}$ in ($0,\pi/2$) leads to a distinct configuration of pathlines, an individual pathline being obtained by integrating forward and backward in time starting from the initial position of the fluid element in the gradient-vorticity plane. The pathline configurations for fluid elements in the flow-gradient plane, and for five different values of $\phi_{j0}$ in ($0,\pi/2$), are shown in figure \ref{fig:latdisp_phi}. While the pathlines for the cases $\phi_{j0} = 0$ and $\pi/2$ are evidently fore-aft symmetric, all other values of $\phi_{j0}$ lead to fore-aft asymmetric open pathlines that suffer a net displacement ($\Delta y$) in the gradient direction as they head from upstream to downstream infinity. By analogy with the spinning spheroid, the `limiting' open pathlines are again shown in red for all configurations in figure \ref{fig:latdisp_phi}. The expectation is that these red curves now serve as fore-aft asymmetric separatrices (for $\phi_{j0} \neq 0,\pi/2$), asymptoting to the flow axis far upstream or downstream. These separatrices would therefore seem to demarcate pathlines that extend to infinity in the upstream and downstream directions from those that do not, being forced to cross the flow axis at a finite $x$, either positive or negative (in a manner similar to the spinning spheroid or sphere above). It will be seen in section 5, where we consider pathlines that loop around the spheroid as a function of their upstream coordinates, that a subset of these separatrices mark the onset of chaotic scattering. This subset must correspond to the interval $\phi_{j0} \in (\pi/2,\pi)$ since, as implied by figure 7, it is separatrices with $\phi_{j0}$'s in the above interval that begin from finite offsets at negative $x$'s, and are therefore accessible from upstream infinity.

\begin{figure}
  \centering
\begin{subfigure}[b]{0.3\textwidth}
\centering
  \includegraphics[scale=0.17]{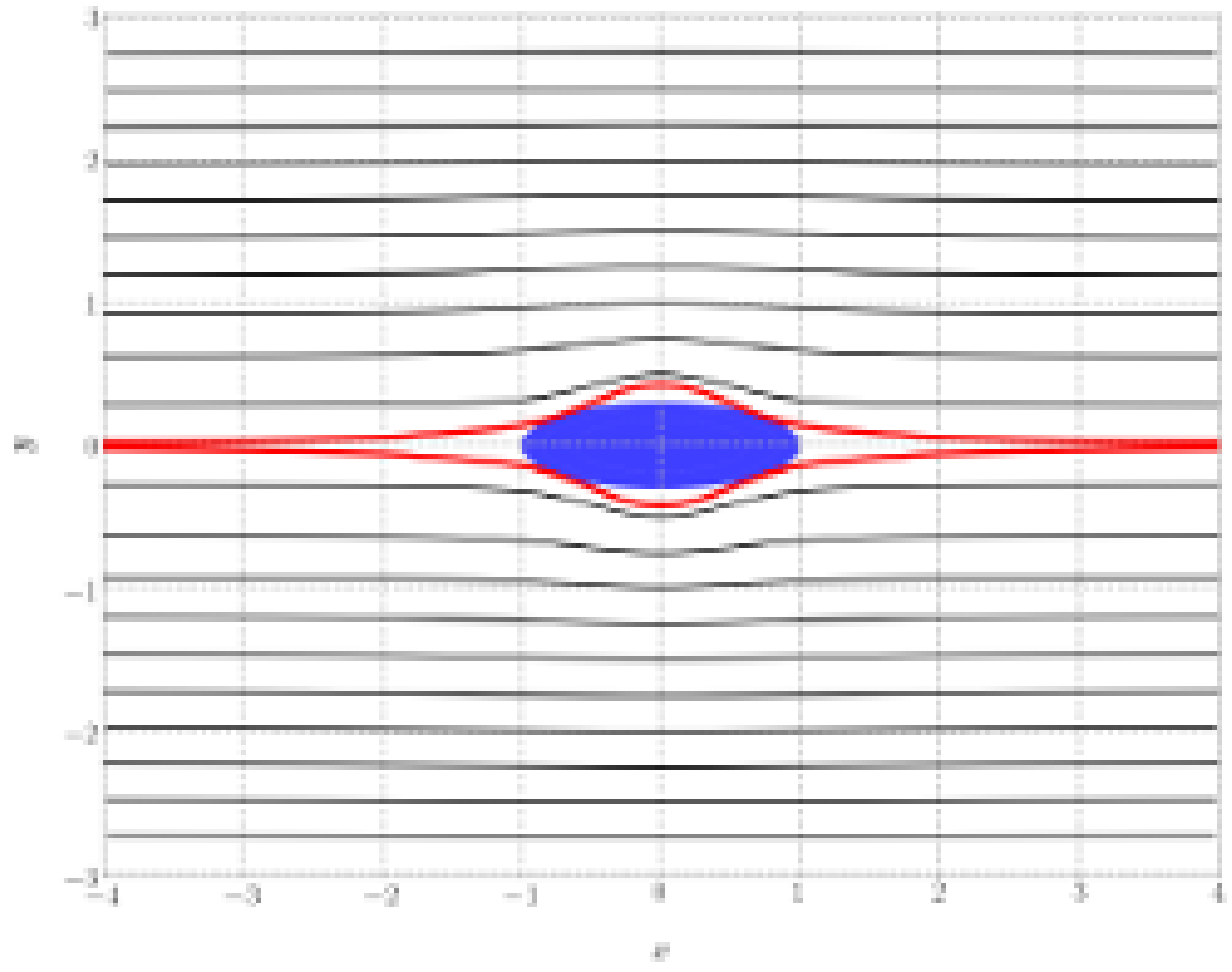}
\caption{}
\end{subfigure}
\hfill
\begin{subfigure}[b]{0.3\textwidth}
\centering
  \includegraphics[scale=0.17]{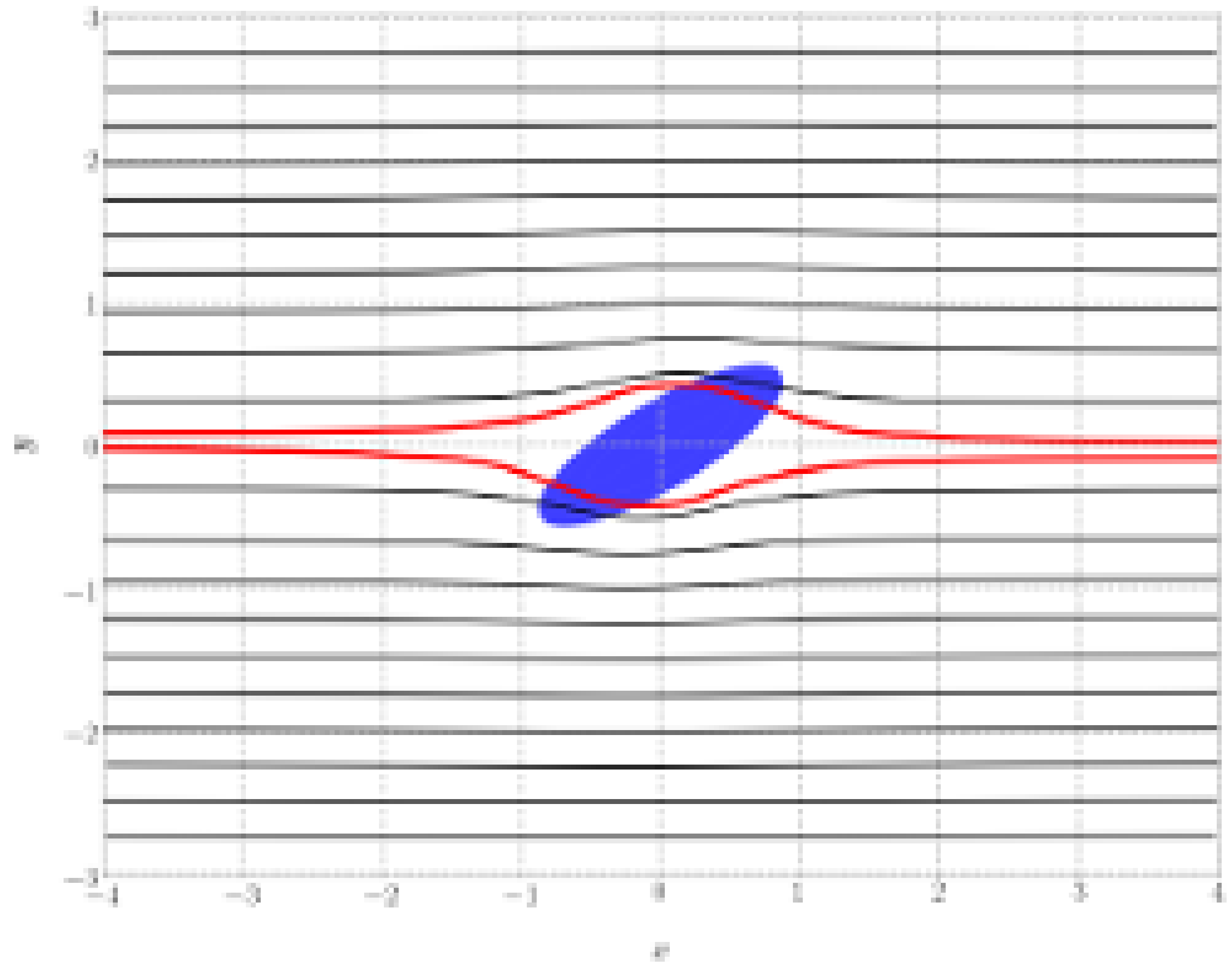}
\caption{}
\end{subfigure}
\hfill
\begin{subfigure}[b]{0.3\textwidth}
\centering
  \includegraphics[scale=0.17]{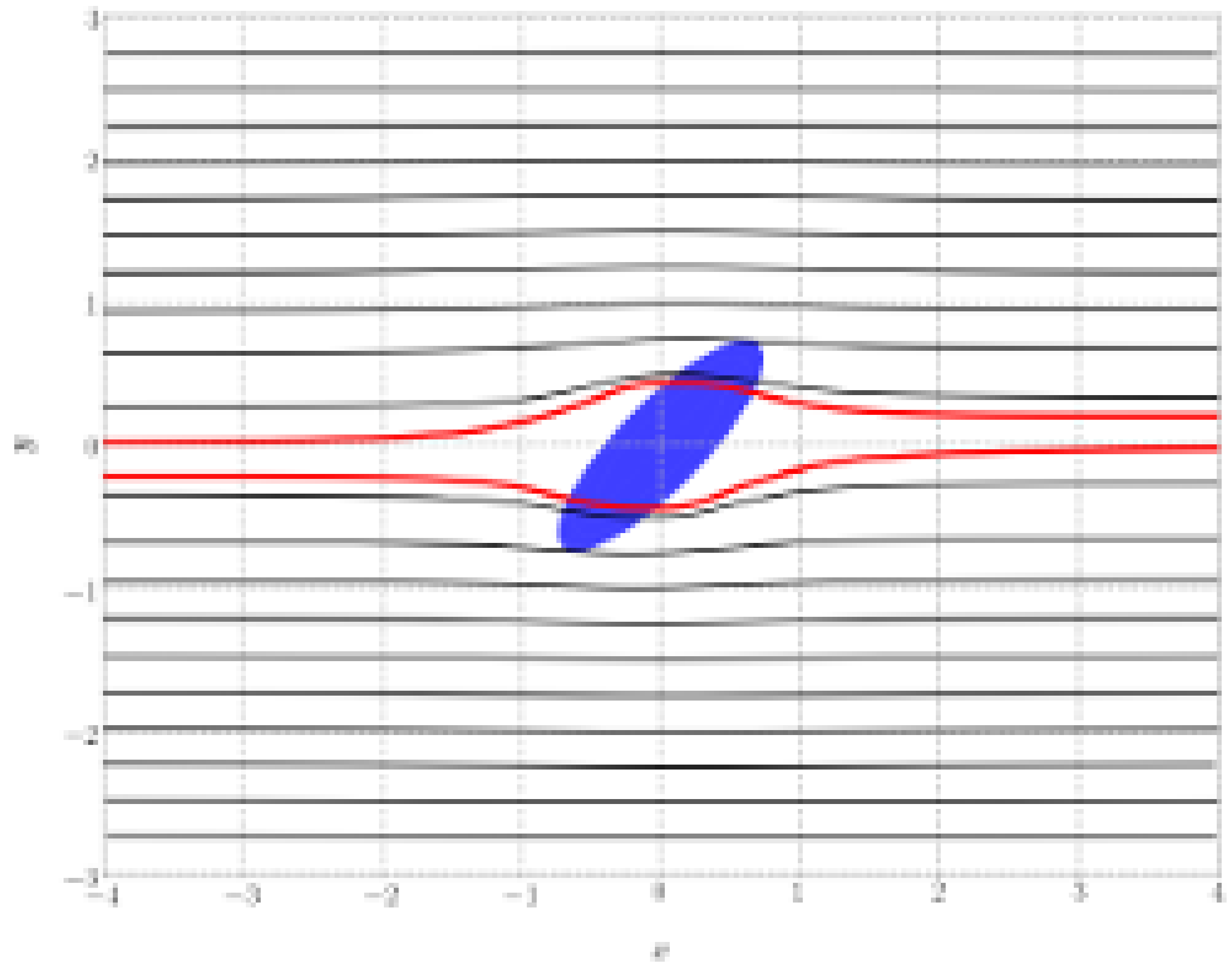}
\caption{}
\end{subfigure}
\begin{subfigure}[b]{0.48\textwidth}
\centering
  \includegraphics[scale=0.18]{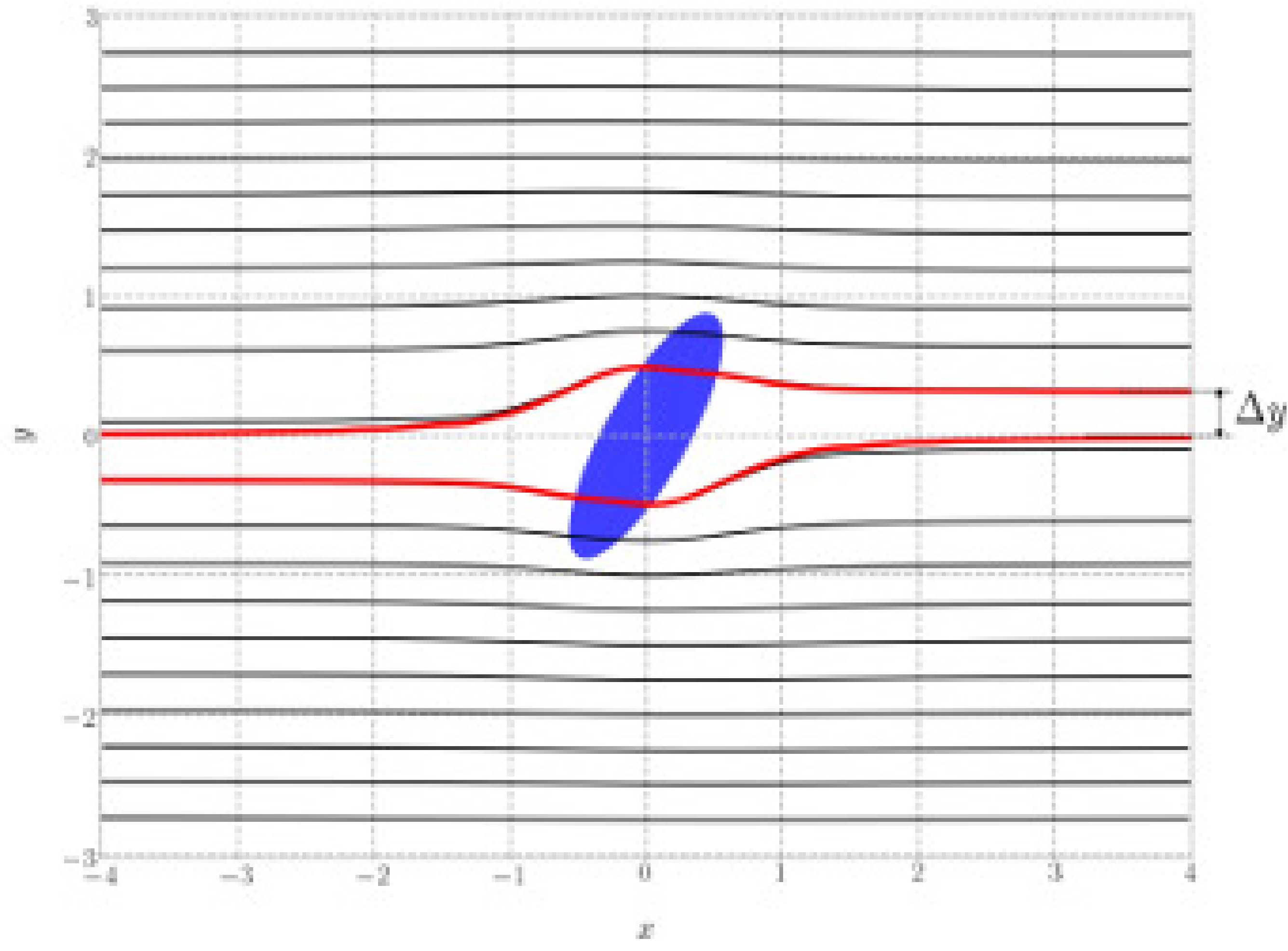}
\caption{}
\end{subfigure}
\hfill
\begin{subfigure}[b]{0.48\textwidth}
\centering
  \includegraphics[scale=0.18]{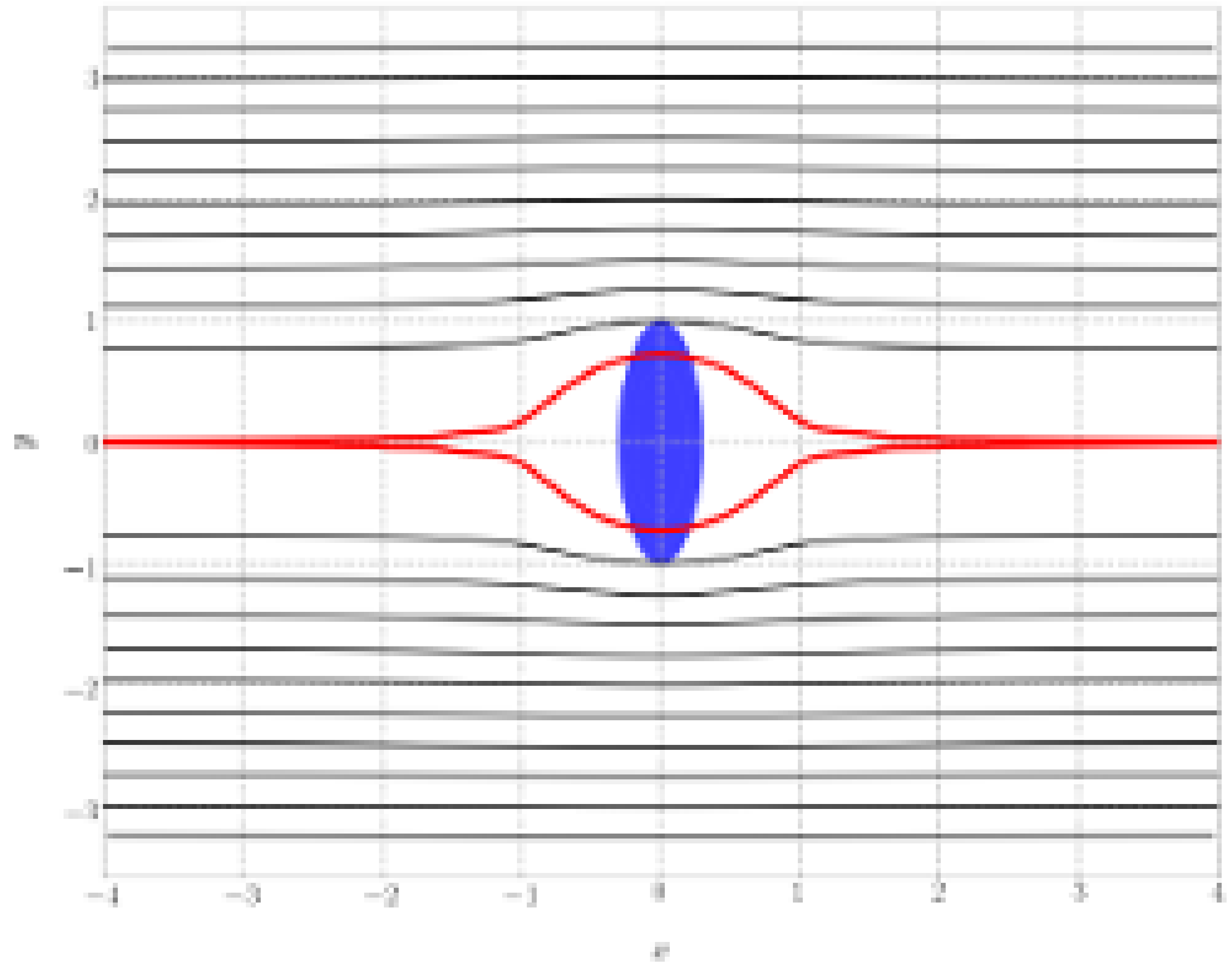}
\caption{}
\end{subfigure}
\begin{subfigure}[b]{0.3\textwidth}
\centering
  \includegraphics[scale=0.2]{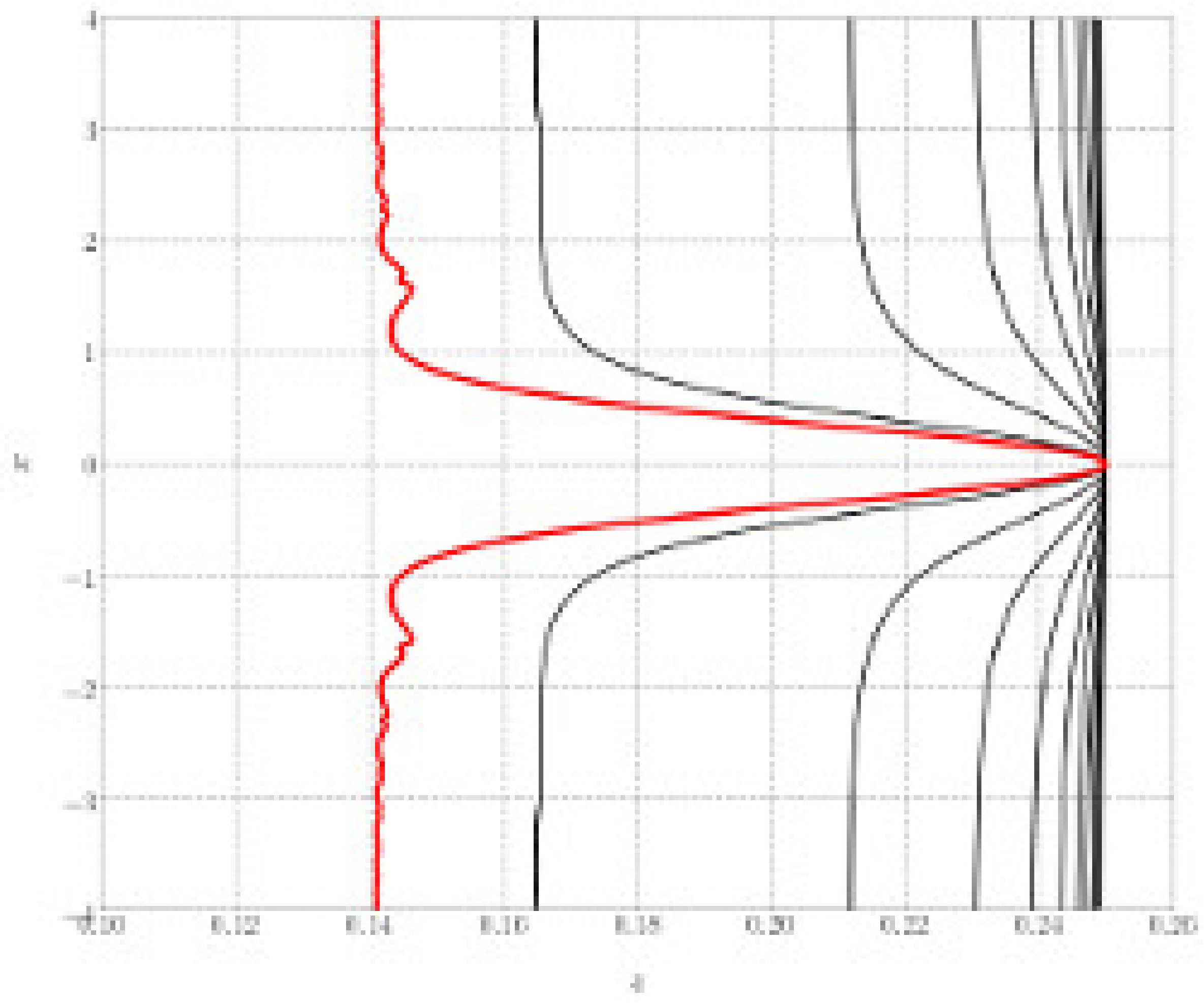}
\caption{}
\end{subfigure}
\hfill
\begin{subfigure}[b]{0.3\textwidth}
\centering
  \includegraphics[scale=0.2]{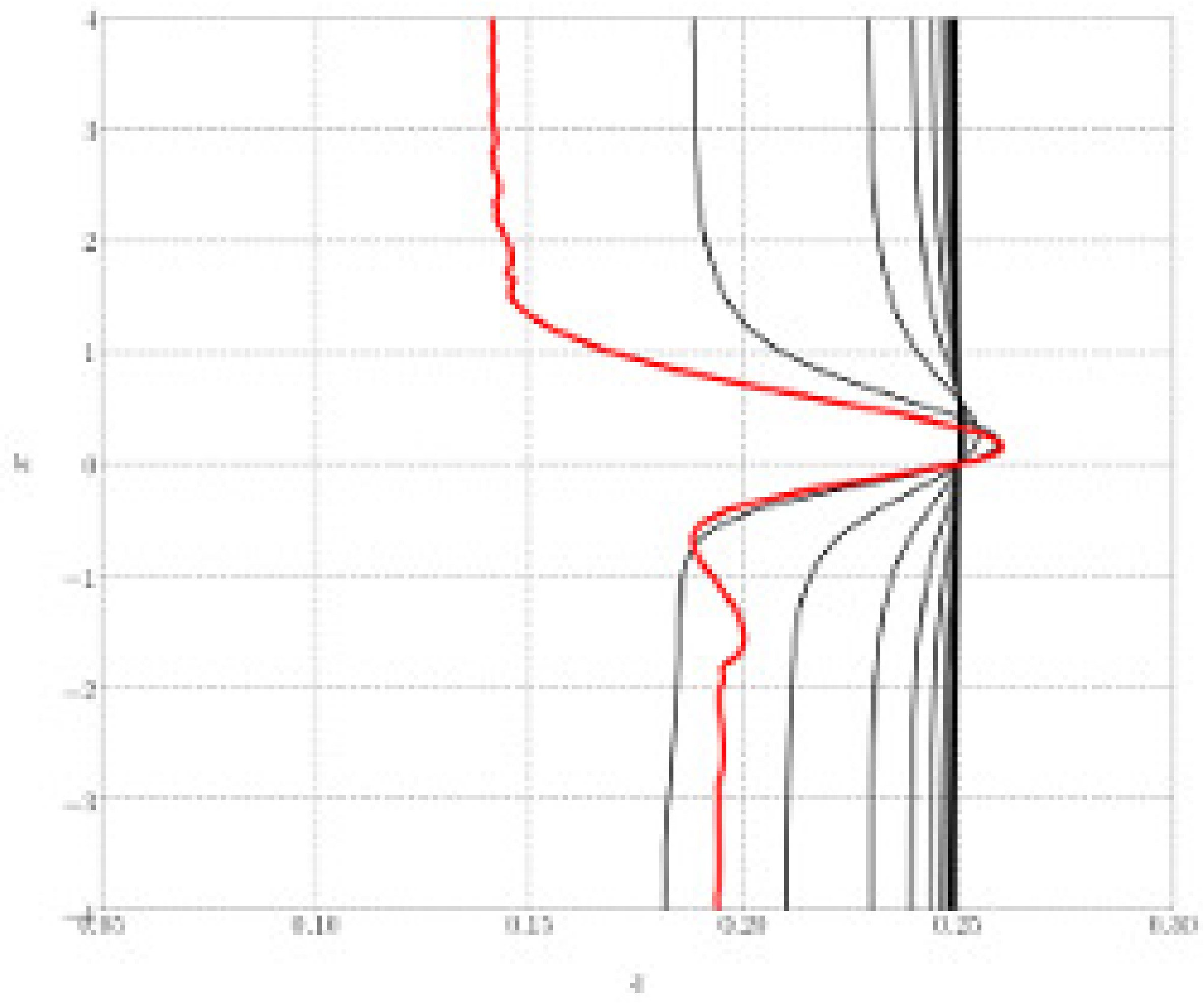}
\caption{}
\end{subfigure}
\hfill
\begin{subfigure}[b]{0.3\textwidth}
\centering
  \includegraphics[scale=0.2]{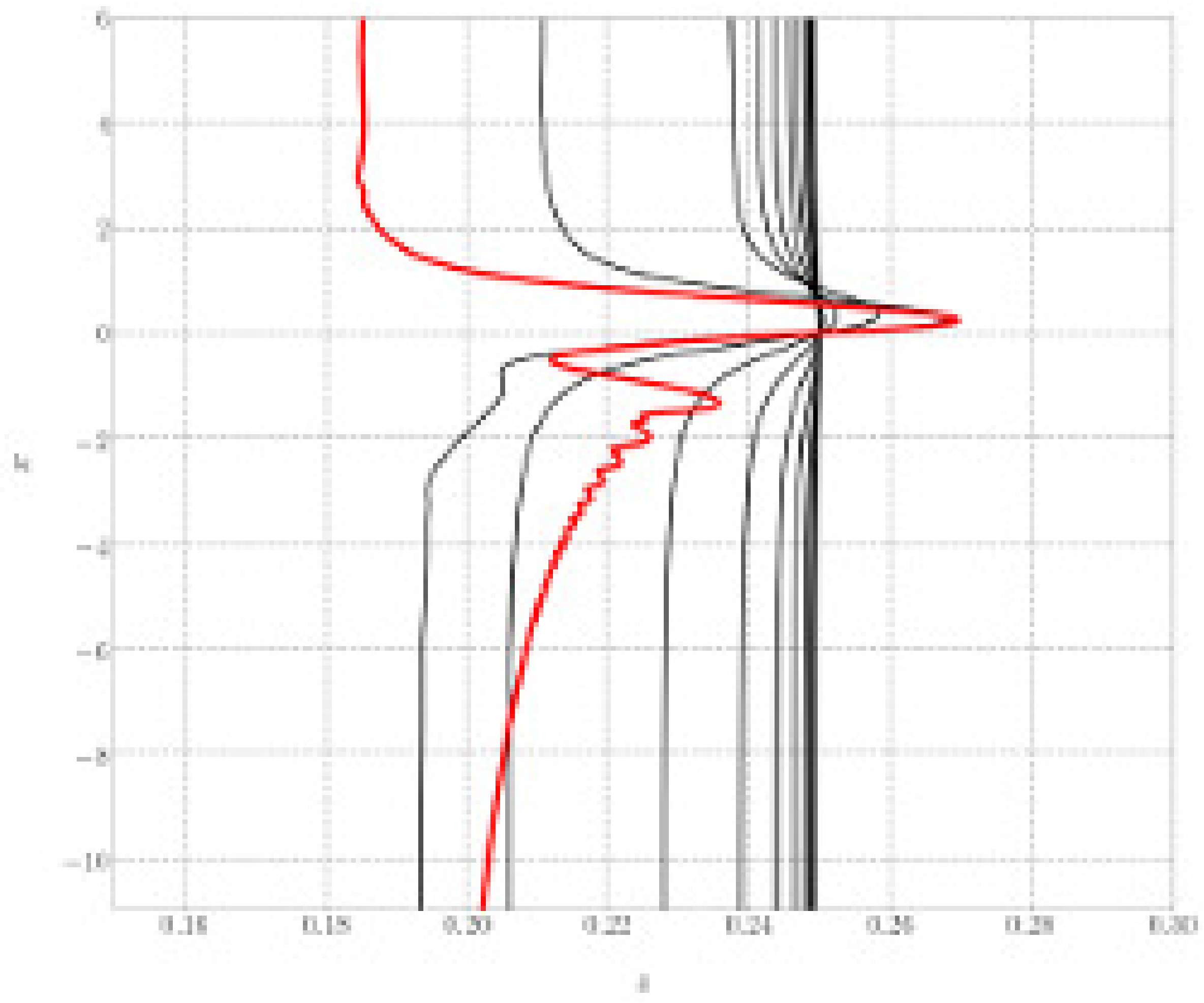}
\caption{}
\end{subfigure}
\begin{subfigure}[b]{0.48\textwidth}
\centering
  \includegraphics[scale=0.21]{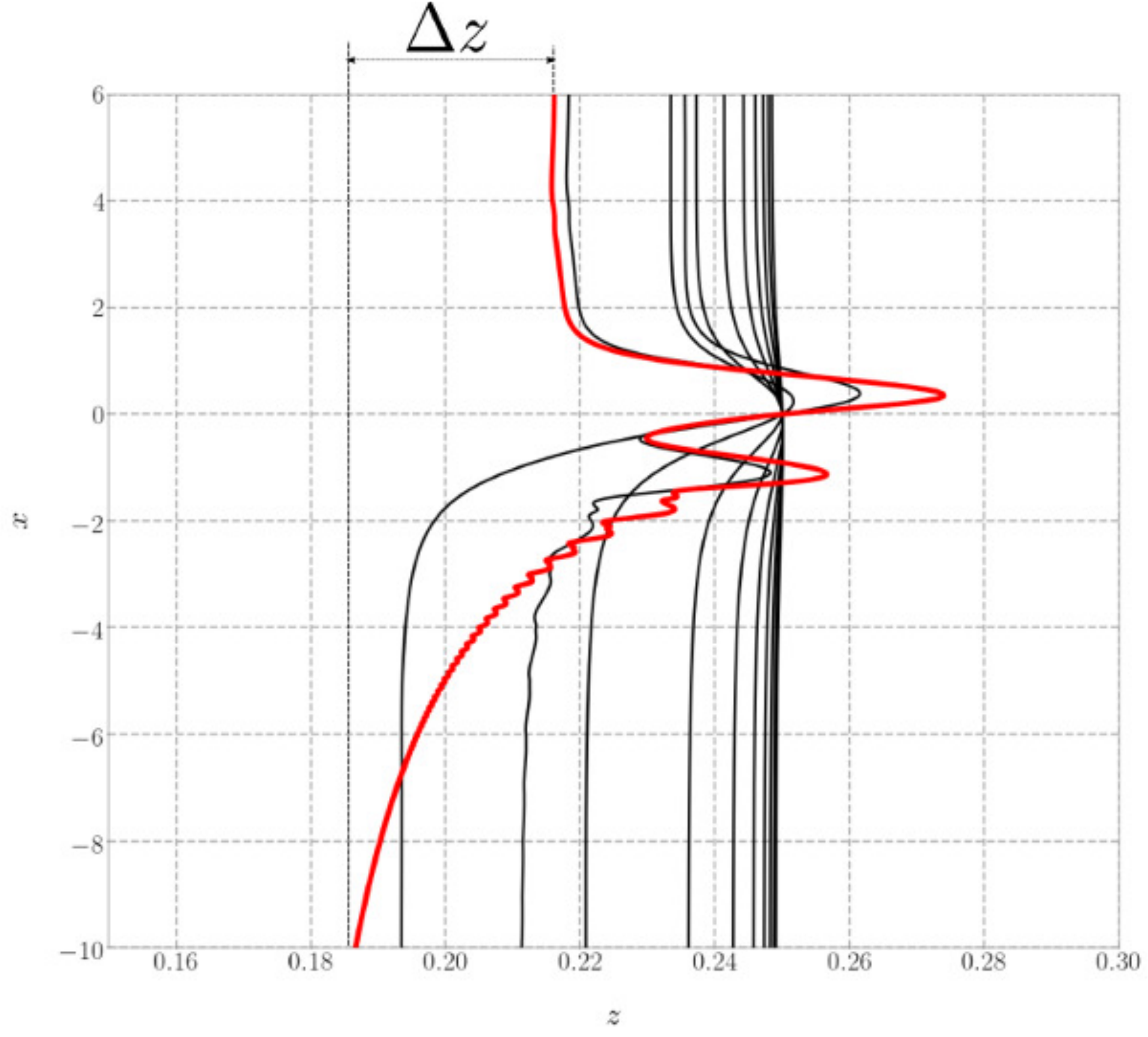}
\caption{}
\end{subfigure}
\hfill
\begin{subfigure}[b]{0.48\textwidth}
\centering
  \includegraphics[scale=0.2]{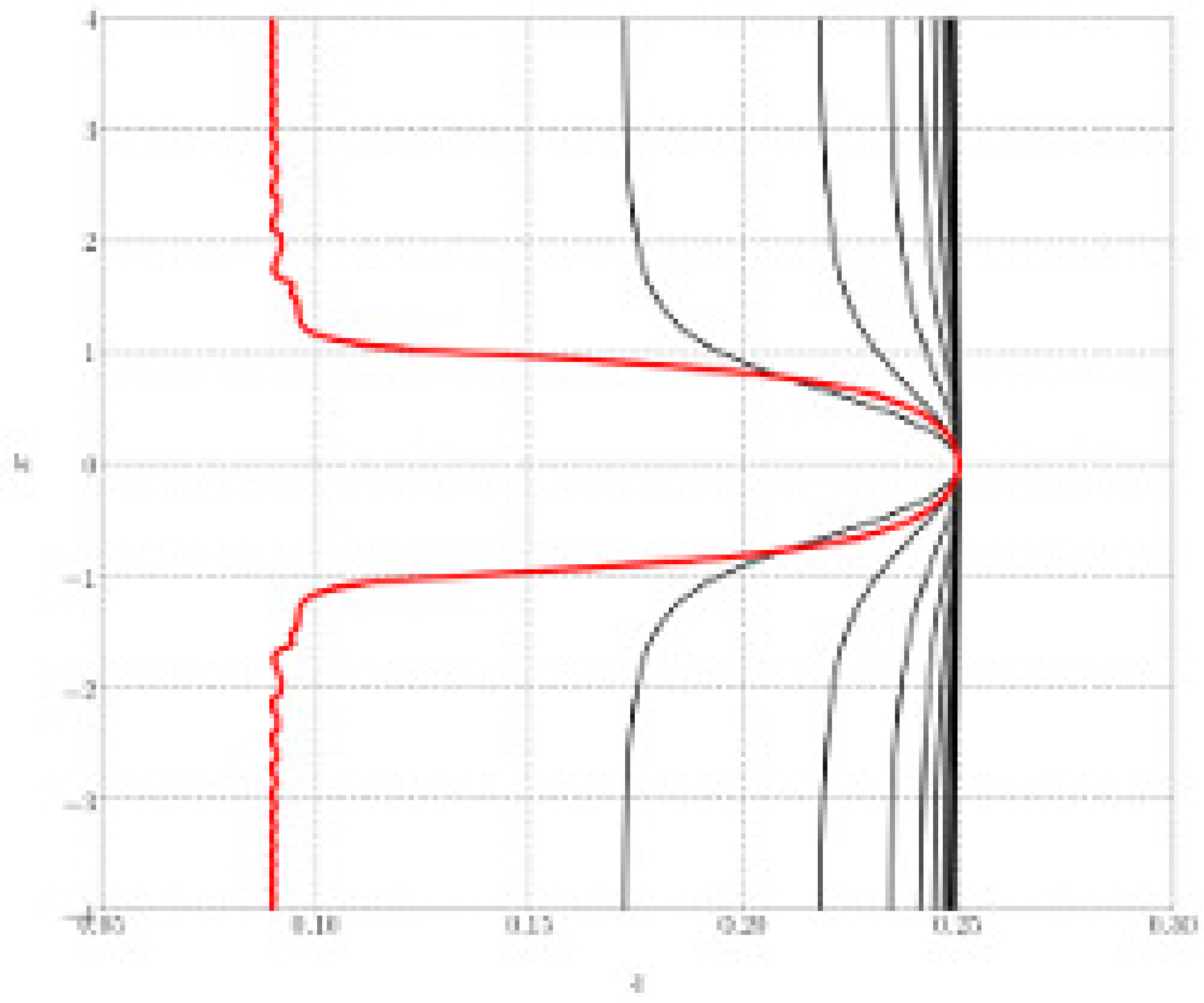}
\caption{}
\end{subfigure}
  \caption{The configurations of the off-plane(vorticity offset $z = 0.25$) regular open pathlines projected onto the flow-gradient(a,b,c,d,e) and flow-vorticity(f,g,h,i,j) planes for a tumbling prolate spheroid with $\kappa=3.28(\xi_0=1.05)$; off-plane separatrices are shown in red. The different configurations correspond to $\phi_{j0}$ values of  (a,f)$0$, (b,g)$\pi/6$, (c,h)$\pi/4$, (d,i)$\pi/3$, and (e,j)$\pi/2$. $\Delta y$ and $\Delta z$ denote the lateral displacements in the gradient and vorticity directions, respectively.}
\label{fig:latdisp_phi_z}
\end{figure}

Figure \ref{fig:latdisp_phi_z} shows the off-plane pathline configurations, again for a tumbling prolate spheroid. As for the inplane case, the pathlines correspond to fluid elements that cross the gradient-vorticity plane ($x= 0$) at a fixed vorticity offset of $z = 0.25$; the $\phi_{j0}$ values are the same as  in figure \ref{fig:latdisp_phi}. Expectedly, the off-plane pathlines are three-dimensional trajectories, and their fore-aft asymmetry manifests as net displacements in both the gradient and vorticity directions; these displacements($\Delta y$ or $\Delta z$) are indicated in the relevant pathline projections. Note that the `separatrices' in particular have small scale wiggles superposed on a slower large-scale variation, on account of the short-scale tumbling dynamics of the spheroid.

\subsubsection{Lateral displacements of the regular open pathlines: $0<C\leq\infty$}
\begin{figure}
  \centering
\begin{subfigure}[b]{0.48\textwidth}
\centering
  \includegraphics[scale=0.3]{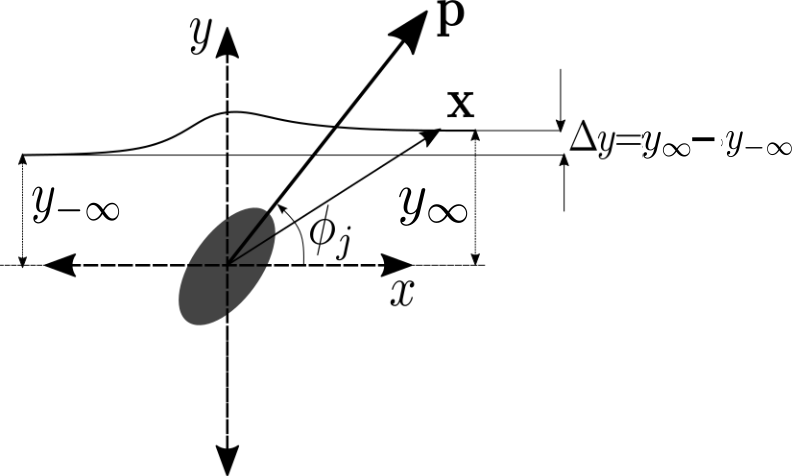}
\caption{}
\label{schematic_1}
\end{subfigure}
\hfill
\begin{subfigure}[b]{0.48\textwidth}
\centering
  \includegraphics[scale=0.2]{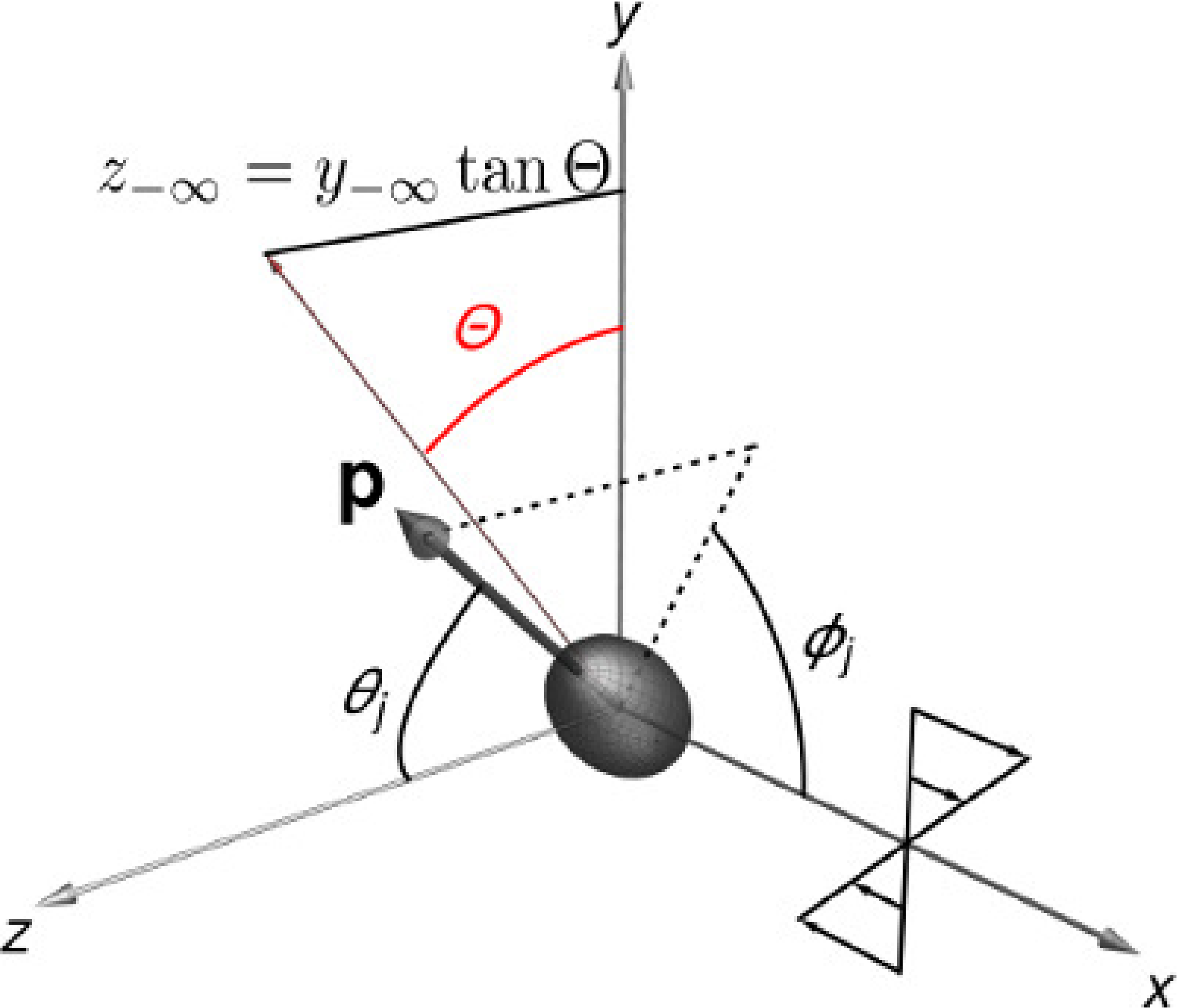}
\caption{}
\label{schematic_2}
\end{subfigure}
  \caption{Figure (a) shows the lateral displacement of a regular open fluid pathline in the flow-gradient plane. $\Theta$ in (b) characterizes the initial upstream location of a fluid element outside the flow-gradient plane.}
\label{fig:latdisp}
\end{figure}

The upstream and downstream coordinates, along the gradient and vorticity directions, are identical for all open streamlines associated with the spinning spheroid or a sphere. An obvious measure of fore-aft asymmetry, as already implied in figures \ref{fig:latdisp_phi} and \ref{fig:latdisp_phi_z}, is therefore the difference between the upstream and downstream gradient and vorticity coordinates of a given open pathline.  These lateral displacements in the gradient ($\Delta y$) and vorticity ($\Delta z$) directions have been used earlier to characterize the asymmetry of finite-$St$ pair-sphere trajectories($St$ here being the Stokes number, and a measure of particle inertia; see \cite{subramanian_brady}). The definition of the lateral displacement in the gradient direction($\Delta y$) is indicated in figure \ref{schematic_1}, along with a schematic in figure \ref{schematic_2} highlighting the coordinates used for a general off-plane pathline.
Both $\Delta y$ and $\Delta z$ may be calculated numerically, in a straightforward manner, for all pathlines down to the `separatrices' shown in figures \ref{fig:latdisp_phi} and \ref{fig:latdisp_phi_z}. For pathlines that approach the spheroid to within a distance of order its major axis, the lateral displacements will be comparable to the respective upstream coordinates: that is, $\Delta y \sim O(y_{-\infty})$ and $\Delta z \sim O(z_{-\infty})$. For far-field pathlines, however, the displacements are much smaller, allowing for an analytical approach involving a perturbation about the upstream coordinates. The analysis requires the expression for the far-field disturbance velocity field due to a spheroid rotating in a precessional orbit. In contrast to a spinning spheroid discussed in section \ref{sec:farfield}, a torque-free spheroid rotating in a generic Jeffery orbit appears, at leading order, as a time-dependent stresslet. The stresslet is now the sum of three distinct contributions, and may be written in the form:
\begin{equation}\label{eq:stresslet01}
\mathsfbi{S}=\frac{20\upi\mu d^3\xi_0^3}{3}\left(K_{ap/o}\,\mathsfbi{E_a}+K_{tp/o}\,\mathsfbi{E_t}+K_{lp/o}\,\mathsfbi{E_l}\right).
\end{equation}
where the subscripts $p$ and $o$ denote prolate and oblate spheroids, respectively. $\mathsfbi{E_a}$, $\mathsfbi{E_t}$ and $\mathsfbi{E_l}$ are the rate of strain tensors associated with axisymmetric extension, transverse and longitudinal planar extensions, respectively, that, in invariant form, are given by: 
\begin{eqnarray}\label{eq:stresslet2}
&& \quad E_{a\,ij}=(E_{kl}p_kp_l)\left[p_ip_j-\frac{\delta_{ij}}{3}\right],\\
&& \quad E_{t\,ij}=(\delta_{il}-p_ip_l)E_{lk}(\delta_{kj}-p_kp_j)+\frac{1}{2}(E_{kl}p_kp_l)(\delta_{ij}-p_ip_j),\\
&& \quad E_{l\,ij}=p_ip_lE_{lk}(\delta_{kj}-p_kp_j)+(\delta_{il}-p_ip_l)E_{kl}p_kp_j.
\end{eqnarray}
There are only three distinct stresslet coefficients, despite the five independent canonical velocity fields in Table \ref{tab:canonical}, since the responses of the spheroid are identical for the pair of planar and transverse longitudinal extensions (owing to its circular cross-section; see \cite{marath2018}). The stresslet coefficients $K_{ap/o}$, $K_{tp/o}$ and $K_{lp/o}$ are functions of aspect ratio, and the relevant expressions for both prolate and oblate spheroids appear in Table \ref{tab:coeff}. Note that the time dependence of the stresslet in (\ref{eq:stresslet01}) arises from the spheroid orientation vector $(\boldsymbol{p})$ being time dependent; $\boldsymbol{p} = \sin{\theta_j}\cos{\phi_j}\,\boldsymbol{1}_x+\sin{\theta_j}\sin{\phi_j}\,\boldsymbol{1}_y+\cos{\theta_j}\,\boldsymbol{1}_z$, where the time evolution of the Jeffery angles is given by (\ref{eq:sys2}) and (\ref{eq:sys3}). The lateral displacements may now be determined, using (\ref{eq:stresslet01}), as:
\begin{equation}\label{eq:latdisp0}
 \Delta y=\int_{-\infty}^{\infty}\boldsymbol{1}_y\cdot\boldsymbol{u}^d\;dt, \quad\quad \Delta z = \int_{-\infty}^{\infty}\boldsymbol{1}_z\cdot\boldsymbol{u}^d\;dt,
\end{equation}
where $\boldsymbol{u}^d = -(3/(8\pi\mu))\mathsfbi{S:}\boldsymbol{nnn}/r^2$ is the leading order dipolar term in the multipole expansion (see (\ref{eq:dipole_octupole})), with $\mathsfbi{S}$ being defined by (\ref{eq:stresslet01}). The distance $r$, between the fluid element and the spheroid center, is a function of time, and may be approximated as $r = [(y_{-\infty} t)^2 + y_{-\infty}^2 (1 + \tan^2\Theta)]^{1/2}$, where the gradient and vorticity coordinates have been replaced by their upstream values, and we have used that $z_{-\infty} = y_{-\infty} \tan{\Theta}$(see figure \ref{schematic_2}). Thus, $r$ changes at leading order only due to convection by the ambient shear, with the instant $t = 0$ corresponding to the fluid element being in the gradient-vorticity plane, with $\phi_j = \phi_{j0}$ (see  (\ref{eq:sys2}) and (\ref{eq:sys3})). The unit radial vector is given by $\boldsymbol{n}=(t\,\boldsymbol{1}_x+\boldsymbol{1}_y+\tan{\Theta}\,\boldsymbol{1}_z)/(1+t^2+\tan^2{\Theta})^{1/2}$ and its projections along the gradient and vorticity directions, appearing in (\ref{eq:latdisp0}), may be written as $n_y=\boldsymbol{n\cdot}\boldsymbol{1}_y=1/(1+t^2+\tan^2{\Theta})^{1/2}$ and $n_z = n_y\tan\Theta$, respectively. Using the expressions for $\mathsfbi{S}$, $n_y$ and $n_z$ above, the detailed expressions for the lateral displacements associated with the far-field pathlines are given by:
\begin{eqnarray}\label{eq:latdisp}
&&\Delta y = \frac{K_{ap/o}}{8y_{-\infty}^2}\int_{-\infty}^{\infty}\frac{dt}{\left(t^2+\sec^2\Theta\right)^{5/2}}\left[\sin ^2\theta_j \left(\tan ^2\Theta (3 \cos 2\theta_j+1) \sin 2\phi_j\right.\right. \\ \nonumber 
&& \left.\left. +6 \tan \Theta \sin 2\theta_j \sin 2\phi_j (\sin \phi_j+t  \cos \phi_j)+12 \sin ^2\theta_j \sin \phi_j \cos \phi_j \left(t ^2 \cos ^2\phi_j+\sin ^2\phi_j\right)\right.\right. \\ \nonumber
&& \left.\left. +6 t  \sin ^2\theta_j \sin ^22\phi_j-2 \left(t ^2+1\right) \sin 2\phi_j\right)\right] \\ \nonumber
&& +\frac{K_{tp/o}}{4y_{-\infty}^2}\int_{-\infty}^{\infty}\frac{dt}{\left(t^2+\sec^2\Theta\right)^{5/2}}\left[\frac{1}{2} \sin 2\theta_j \left(\left(-2 \tan ^2\Theta+t ^2+1\right) \sin 2\theta_j \sin 2\phi_j \right.\right. \\ \nonumber
&& \left.\left. +4 \tan \Theta (t  \sin \phi_j+\cos \phi_j)\right)-8 \tan \Theta \sin ^3\theta_j \cos \theta_j \sin 2\phi_j (\sin \phi_j+t  \cos \phi_j)\right. \\ \nonumber 
&& \left. +\sin ^4\theta_j \left(2 t  \cos 4\phi_j-\left(t ^2-1\right) \sin 4\phi_j\right)+t  \sin ^2\theta_j (\cos 2\theta_j+3)\right] \\ \nonumber
&& + \frac{K_{lp/o}}{8y_{-\infty}^2}\int_{-\infty}^{\infty}\frac{dt}{\left(t^2+\sec^2\Theta\right)^{5/2}}\left[\sin ^2\theta_j \sin 2\phi_j \left(\tan ^2\Theta (\cos 2\theta_j+3)-2 \left(t ^2+1\right)\right)\right. \\ \nonumber
&& \left. +4 \tan \Theta \sin ^3\theta_j \cos \theta_j \sin 2\phi_j (\sin \phi_j+t  \cos \phi_j)-4 \tan \Theta \sin 2\theta_j (t  \sin \phi_j+\cos \phi_j) \right. \\ \nonumber
&& \left. +2 \sin ^4\theta_j \sin 2\phi_j (\sin \phi_j+t  \cos \phi_j)^2+8 t  \cos ^2\theta_j\right] \\
&& \nonumber \\
&& \Delta z=\Delta y\tan\Theta.
\end{eqnarray}
The integral in (4.6) is evaluated numerically, and is a function of the spheroid aspect ratio ($\kappa$), the orbit constant ($C$) and $\phi_{j0}$; the case of a tumbling spheroid may be obtained by setting $\theta_j = \pi/2(C = \infty)$. On account of the radial nature of the dipolar field, the dominant displacement of the pathline occurs when the fluid element is at a distance of O($y_{-\infty}$) or smaller from the gradient-vorticity plane. Since this corresponds to a time interval of O($\dot{\gamma}^{-1}$), independent of $y_{-\infty}$, the far-field lateral displacements exhibit a decay of ($y_{-\infty}^{-2}$) which is the same as that of the dipolar field. In light of this decay, in figure 10, we plot the rescaled lateral displacements, $y_{-\infty}^2\Delta y$, for pathlines in the flow-gradient plane, for tumbling prolate and oblate spheroids. Owing to the simple relationship between the two lateral displacements in (\ref{eq:latdisp}), in figure \ref{fig:latdisp_tumb_z_plot} we plot the re-scaled total lateral displacement, $y_{-\infty}^2 \Delta r =  y_{-\infty}^2\sqrt{\Delta y^2 + \Delta z^2} = y_{-\infty}^2\Delta y\,\sec{\Theta}$, for pathlines outside the flow-gradient plane (corresponding to $\Theta = \pi/9$), again for tumbling prolate and oblate spheroids.\begin{table}
\begin{tabular}{|c|c|}
\hline
$ K_{ap}=\frac{4}{15 \xi_0^3 \left(\left(3 \xi_0^2-1\right) \coth ^{-1}(\xi_0)-3 \xi_0\right)}$ & $ K_{ao}=-\frac{4}{15 \xi_0^3 \left(3 \sqrt{\xi_0^2-1}+\left(2-3 \xi_0^2\right) \csc ^{-1}(\xi_0)\right)}$\\
$ K_{tp}=-\frac{4 \left(\xi_0^2-1\right)}{5 \xi_0^2 \left(2 \xi_0^2-1\right) \left(-3 \xi_0^2+3 \left(\xi_0^2-1\right) \xi_0 \coth ^{-1}(\xi_0)+2\right)}$ & $ K_{to}=-\frac{4 \sqrt{\xi_0^2-1}}{5 \left(2 \xi_0^3-\xi_0\right) \left(-3 \xi_0^2+3 \sqrt{\xi_0^2-1} \xi_0^2 \csc ^{-1}(\xi_0)+1\right)}$ \\
$ K_{lp}=\frac{8 \left(\xi_0^2-1\right)}{5 \xi_0^3 \left(-3 \xi_0^3+3 \left(\xi_0^2-1\right)^2 \coth ^{-1}(\xi_0)+5 \xi_0\right)}$ & $ K_{lo}=\frac{24}{5 \left(9 \xi_0^5 \csc ^{-1}(\xi_0)-3 \xi_0 \sqrt{\xi_0^2-1} \left(3 \xi_0^2+2\right)\right)}$\\ 
\hline
\end{tabular}
\caption{The aspect-ratio-dependent stresslet coefficients that appear in the far-field stresslet disturbance (see equation (4.1) in the text).}
\label{tab:coeff}
\end{table}
\begin{figure}
  \centering
\begin{subfigure}[b]{0.49\textwidth}
\centering
  \includegraphics[scale=0.16]{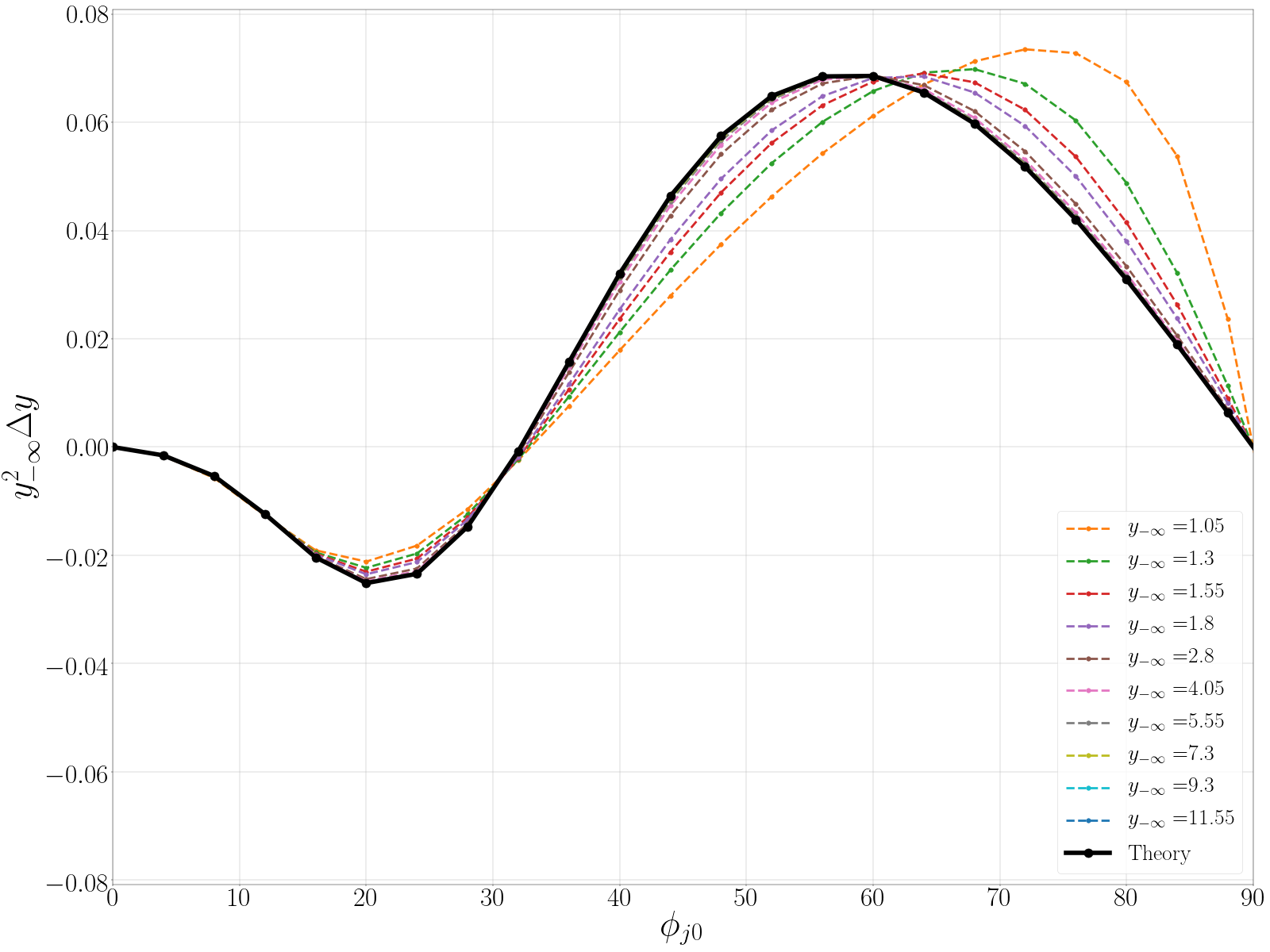}
\caption{}
\end{subfigure}
\hfill
\begin{subfigure}[b]{0.49\textwidth}
\centering
  \includegraphics[scale=0.16]{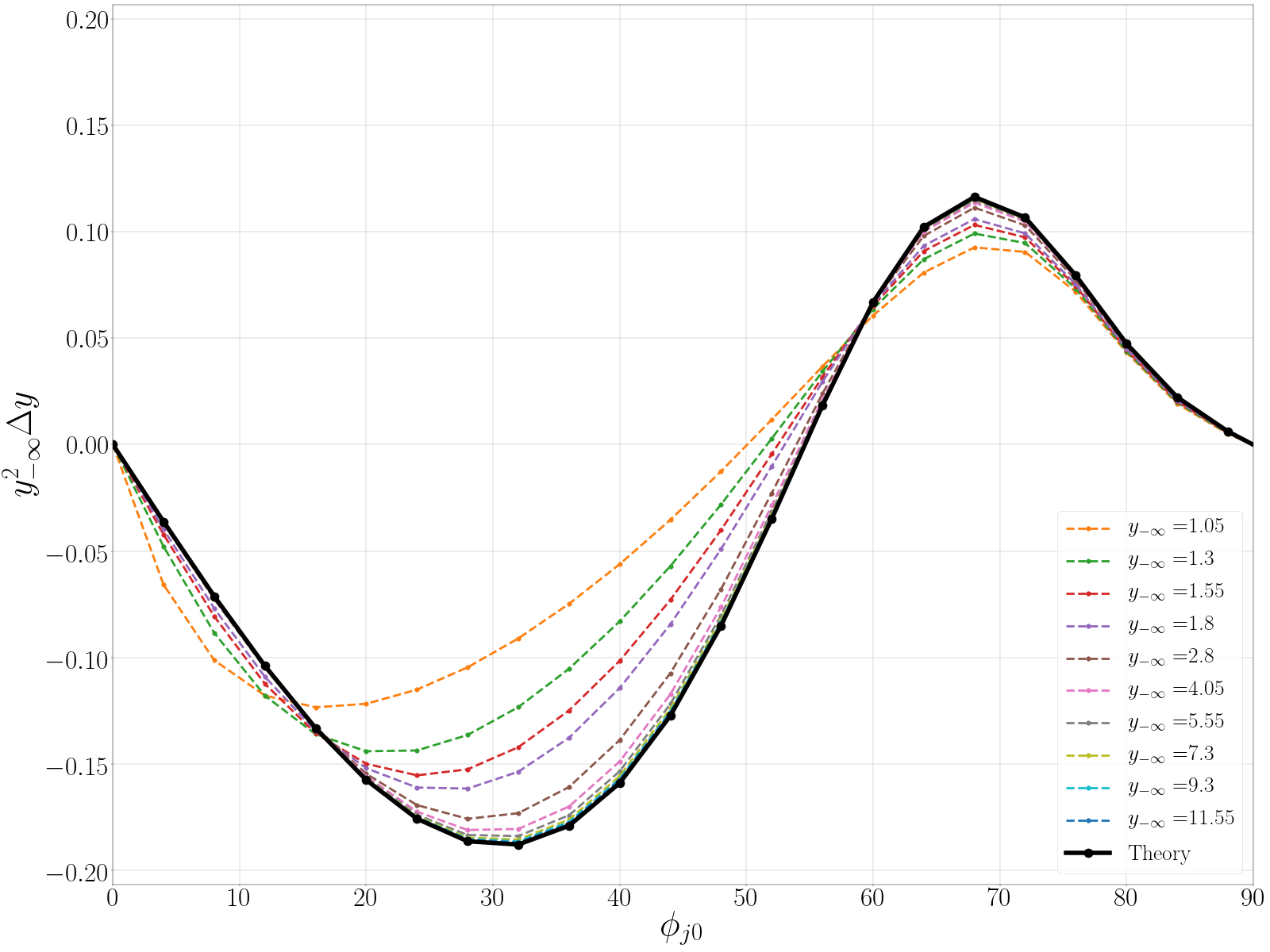}
\caption{}
\end{subfigure}
  \caption{Lateral displacements in the gradient direction for fluid elements in the flow-gradient plane, as a function of the initial orientation($\phi_{j0}$). (a) A prolate spheroid ($\xi_0=1.01(\kappa=7.12)$), and (b) an oblate spheroid ($\xi_0=1.01(\kappa=0.14)$); both spheroids are in the tumbling orbit. The lateral displacement for $\phi_{j0} \in (\pi/2,\pi)$ may be obtained as $\Delta y(\phi_{j0}) = -\Delta y(\pi-\phi_{j0})$.}
\label{fig:latdisp_tumb_plot}
\end{figure}
\begin{figure}
  \centering
\begin{subfigure}[b]{0.49\textwidth}
\centering
  \includegraphics[scale=0.16]{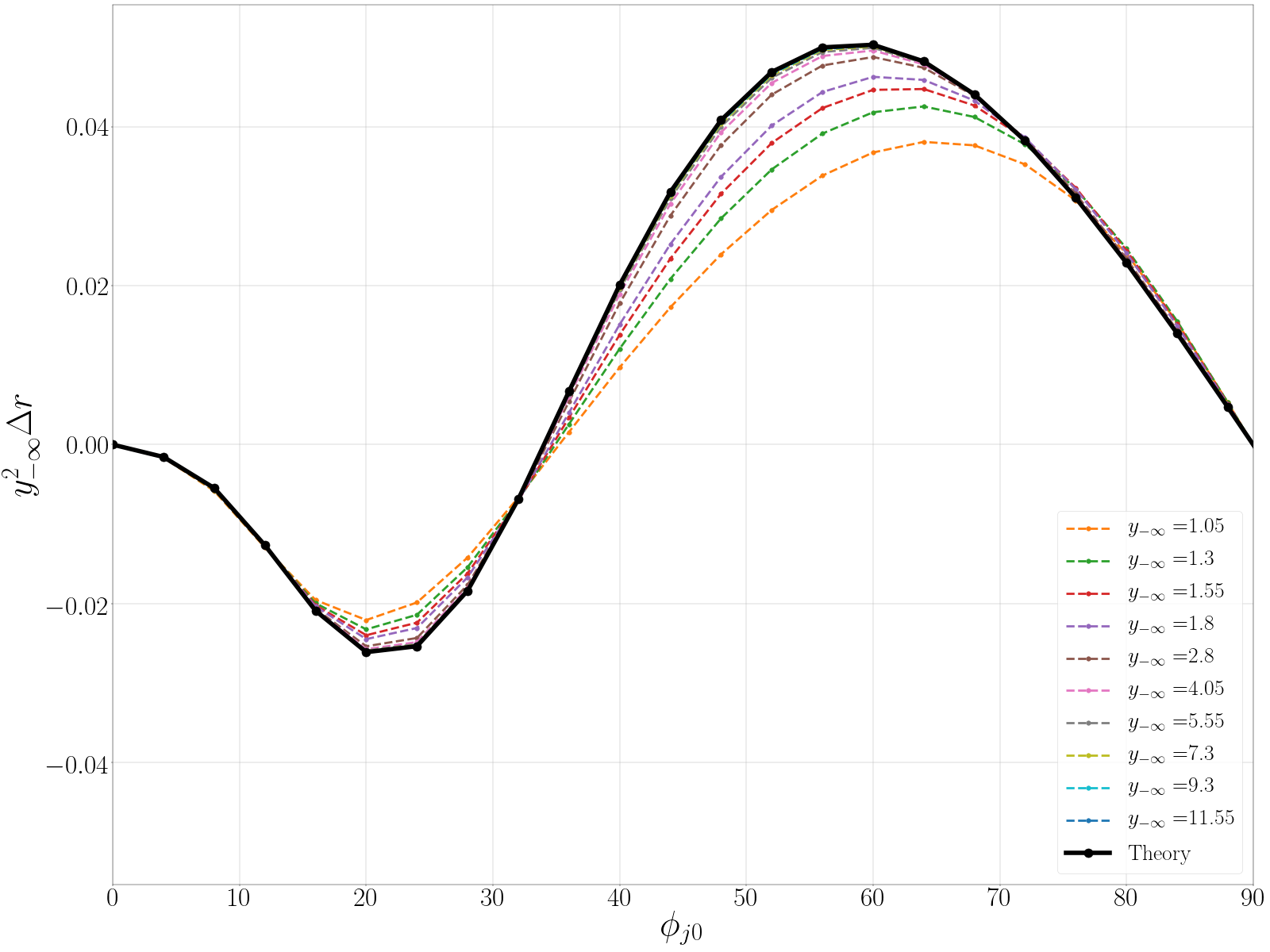}
\caption{}
\end{subfigure}
\hfill
\begin{subfigure}[b]{0.49\textwidth}
\centering
  \includegraphics[scale=0.16]{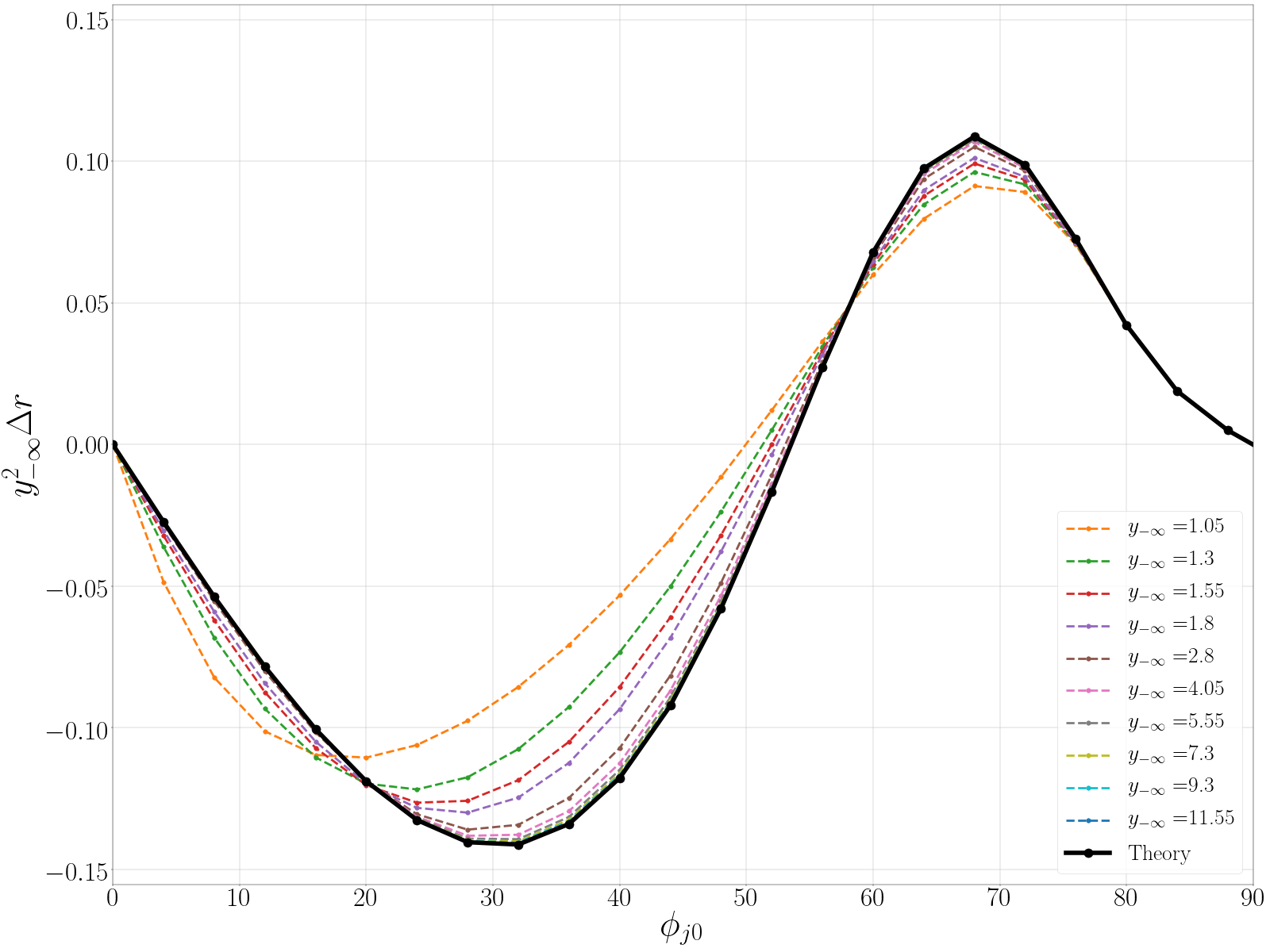}
\caption{}
\end{subfigure}
  \caption{The total lateral displacements for off-plane fluid elements($\Theta=\pi/9$) as a function of the initial orientation($\phi_{j0}$). (a) A prolate spheroid ($\xi_0=1.01(\kappa=7.12)$), and (b) an oblate spheroid ($\xi_0=1.01(\kappa=0.14)$); both spheroids are in the tumbling orbit. The lateral displacement for $\phi_{j0} \in (\pi/2,\pi)$ may be obtained as $\Delta y(\phi_{j0}) = -\Delta y(\pi-\phi_{j0})$.}
\label{fig:latdisp_tumb_z_plot}
\end{figure}
Figure \ref{fig:latdisp3D_plot} shows the re-scaled total lateral displacements for off-plane fluid elements ($\Theta =\pi/9$), and for prolate and oblate spheroids in a non-tumbling orbit ($C = 10$). Note that for this generic case, it is necessary to consider $\phi_{j0}$ in the larger interval ($0,\pi$), with the displacements in the interval ($\pi,2\pi$) again obtained via a reflection transformation. In all cases (figures \ref{fig:latdisp_tumb_plot}, \ref{fig:latdisp_tumb_z_plot} and \ref{fig:latdisp3D_plot}), the numerical displacement curves approach the analytical far-field ones in the limit $(y_{-\infty}^2 + z_{-\infty}^2)^{1/2} \rightarrow \infty$.
\begin{figure}
  \centering
\begin{subfigure}[b]{0.49\textwidth}
\centering
  \includegraphics[scale=0.16]{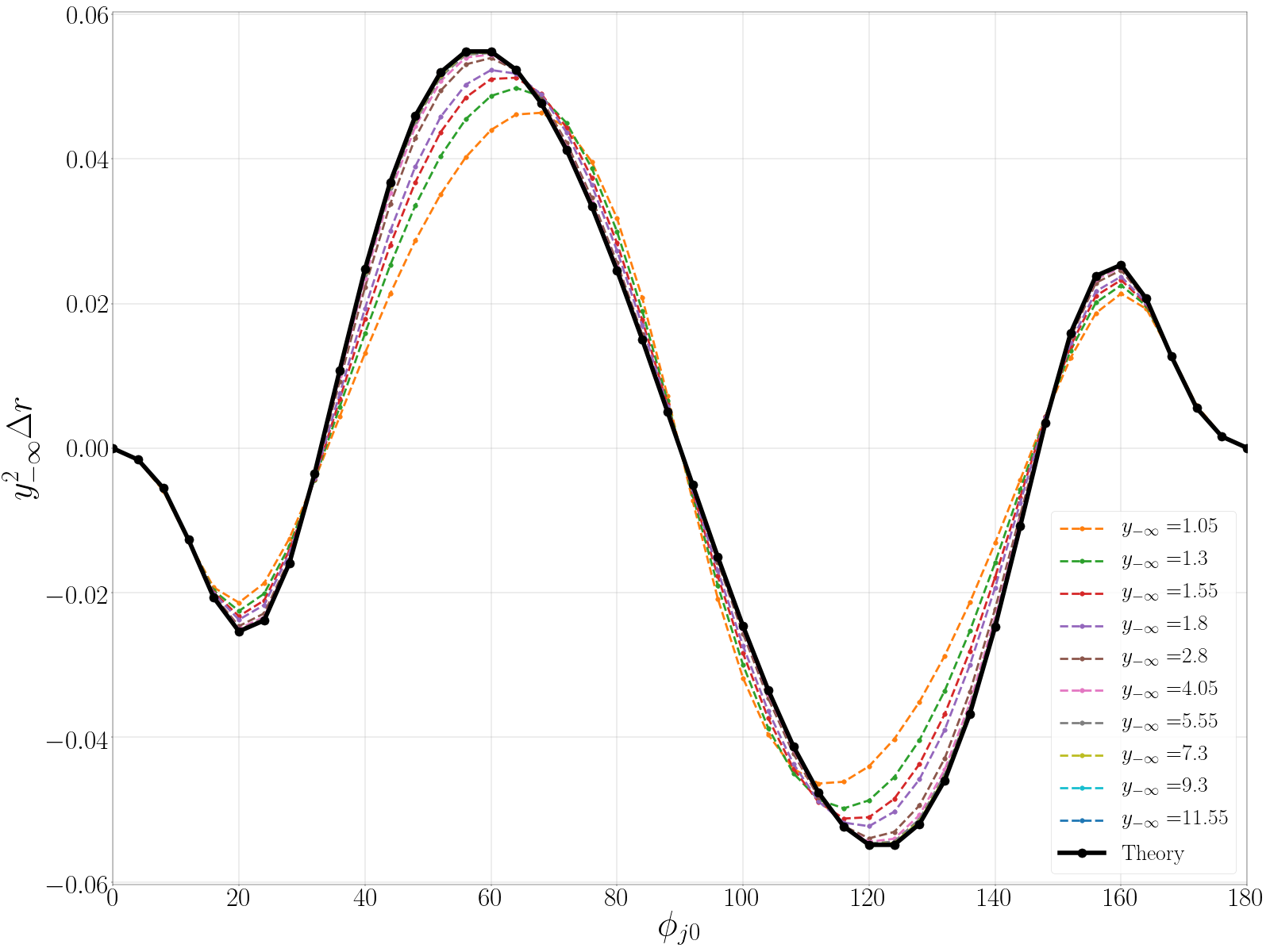}
\caption{}
\end{subfigure}
\hfill
\begin{subfigure}[b]{0.49\textwidth}
\centering
  \includegraphics[scale=0.16]{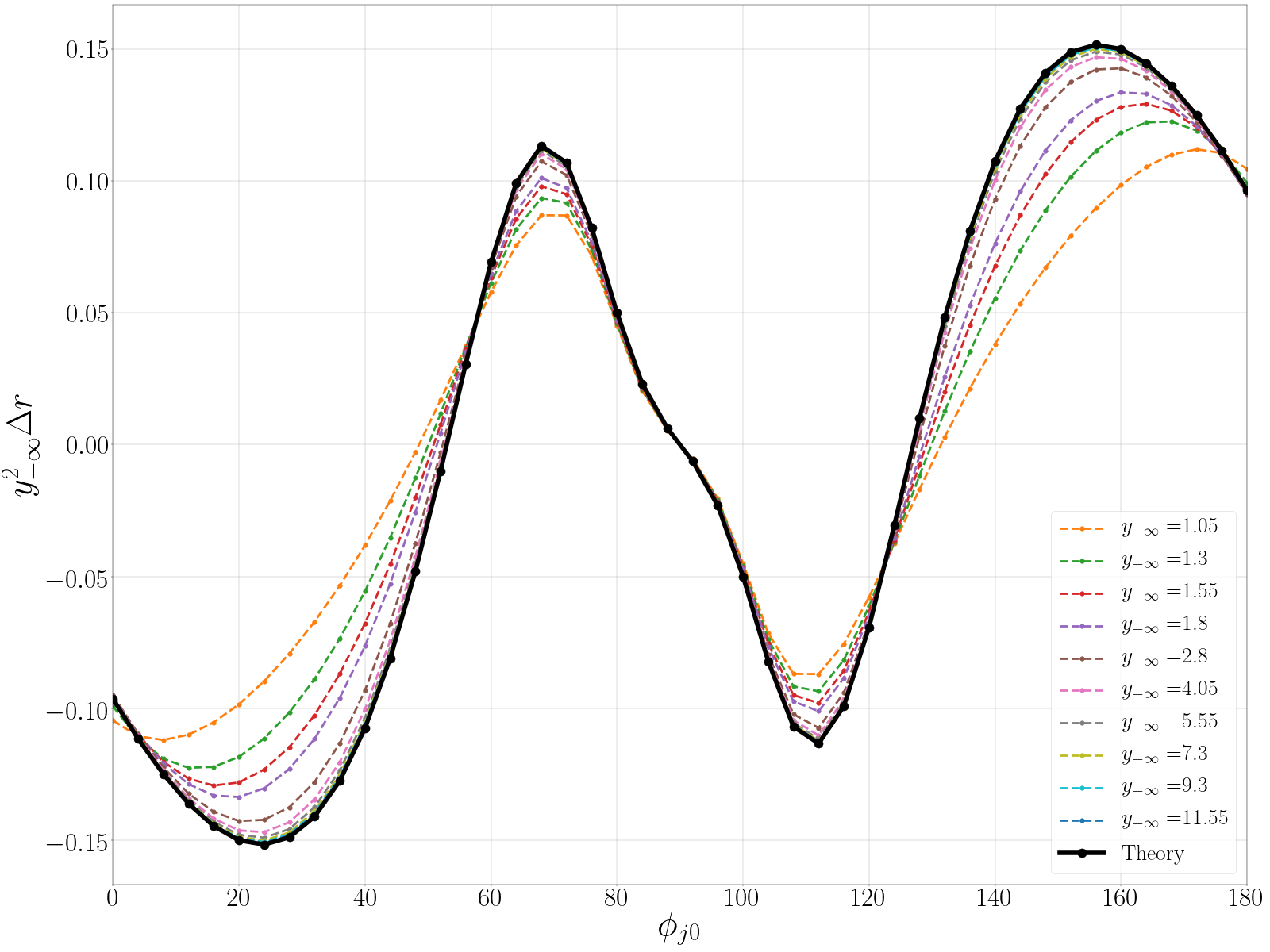}
\caption{}
\end{subfigure}
  \caption{The total lateral displacements for off-plane fluid elements($\Theta=\pi/9$) as a function of the initial orientation($\phi_{j0}$). (a) A prolate spheroid ($\xi_0=1.01(\kappa=7.12)$), and (b) an oblate spheroid ($\xi_0=1.01(\kappa=0.14)$). Both spheroids are in a Jeffery orbit with $C=10$. The lateral displacement for $\phi_{j0} \in (\pi,2\pi)$ may be obtained as $\Delta y(\phi_{j0}) = -\Delta y(2\pi-\phi_{j0})$.}
\label{fig:latdisp3D_plot}
\end{figure}
%

\subsubsection{The separatrix of a tumbling spheroid}\label{sec:separatrix}
We end this subsection on the regular pathlines by analyzing the approach of the separatrices, marked in red in figure 7, towards the flow axis - the approach occurs in both the upstream and downstream directions for $\phi_{j0} = 0,\pi/2$, and in either of these two directions for other $\phi_{j0}$'s. For the spinning spheroid, as seen earlier((\ref{eq:spin_sepsys_1}) and (\ref{eq:spin_sepsys_2})), the rate of approach of the far-field separatrix\,($y \propto x^{-3/2}$) remains identical to that of a sphere, with only a pre-factor that is a function of aspect ratio. The analysis for the analogously defined separatrices in figure 7 is more involved owing to the spheroid rotation. In the far-field ($r \gg 1$), one may, however, exploit a separation of time scales, with the fast scale corresponding to the Jeffery period  (of $O(\dot{\gamma}^{-1})$ for $\kappa \sim O(1)$) and the slow time scale characterizing the rate of change of position of the fluid element. Note that, unlike the regular pathlines for large $y_{-\infty}$ analyzed above where this latter time scale remains $O(\dot{\gamma}^{-1})$, for the separatrices, the approach towards the flow axis implies that $y$ is asymptotically small and the time scale of convection by the ambient shear is therefore correspondingly large. Note also that the magnitude of this slow time scale, being a function of $y$, is not known apriori. Keeping this in mind, we write the governing equations as:
\begin{eqnarray}
 && \frac{d\boldsymbol{x}}{dT}=\boldsymbol{u}(\boldsymbol{x},\boldsymbol{p}(\tau))\quad \label{eq:vel}\\
 && \frac{d\boldsymbol{p}}{d\tau}=\boldsymbol{\omega}\cdot\boldsymbol{p}+\frac{\kappa^2-1}{\kappa^2+1}\left(\mathsfbi{E}.\boldsymbol{p}-\boldsymbol{p}\left(\mathsfbi{E}\colon\boldsymbol{pp}\right)\right). \label{eq:ori}
\end{eqnarray}
where we have (formally) identified $\tau$ and $ T$ as the fast and slow time variables, respectively. The identification implies that $\boldsymbol{p}$ only evolves on the fast time scale, while $\boldsymbol{x}$ is a function of both $\tau$ and $T$. One may expand $\boldsymbol{x}$ in the usual manner $\boldsymbol{x}(T,\tau)=\boldsymbol{X}(T,\tau)+\tilde{\boldsymbol{x}}(\boldsymbol{X},T,\tau)$  with the fast variable satisfying $1/T_j \int_0^{T_j}\tilde{\boldsymbol{x}}d\tau=0$, with $T$ fixed, and $T_j$ being the Jeffery period. The slow and fast contributions of $\boldsymbol{x}$ may further be expanded in the following manner: $\boldsymbol{X}(T,\tau)=\boldsymbol{X}_0(T) + \boldsymbol{X}_1(T,\tau)+\cdots$ and $\tilde{\boldsymbol{x}}= \boldsymbol{\tilde{x}}_0+\boldsymbol{\tilde{x}}_1+\cdots $, respectively, where we have anticipated the leading order position to only evolve on the slow time scale. Since there isn't a readily available estimate of the slow time scale apriori, we don't introduce $\varepsilon$ above,  but nevertheless assume the higher order terms in the above expansions to be asymptotically smaller; thus, $\tilde{\boldsymbol{x}} \ll \boldsymbol{X}, \boldsymbol{X}_1 \ll \boldsymbol{X}_0$, and so on. We restrict ourselves to calculating the leading order slow contribution $\boldsymbol{X}_0(T)$ below (although, note that $\tilde{x}_0$ is responsible for the small-scale wiggles on the separatrices seen in figure 8 and figure 13.)

The far-field form may now be obtained by recalling the approximations used in the context of the spinning spheroid in section \ref{sec:farfield}. Thus, in the far-field at leading order, we only retain the ambient flow in the $x$-component, and the octupolar contribution in the $y$-component. One may now write the scalar components of the two-time-scale equations as:
\begin{eqnarray}
\frac{dX_0}{dT}+\frac{\partial\tilde{x}_0}{\partial \tau}&=&Y_0+\tilde{y}_0\\\label{eq:full_dynamics_x}
\frac{dY_0}{dT}+\frac{\partial\tilde{y}_0}{\partial \tau}&=&-\frac{15}{4}d^5\xi_0^3L^2_c\kappa^2(\kappa^2-1)]\left[\left(K_{ap/o}-2K_{tp/o}+K_{lp/o}\right)\kappa^2 \right. \\ \nonumber
&&\left. + 2\left(K_{tp/o}-K_{ap/o}\right)\cot^2\left(\frac{\kappa\tau}{\kappa^2+1}\right)\right]\left[\kappa^2+\cot^2\left(\frac{\kappa\tau}{\kappa^2+1}\right)\right]^{-3} X_0^{-4}\label{eq:full_dynamics_y}
\end{eqnarray}
where $L_c=\bar{\xi_0}(\xi_0)$ for a prolate(an oblate) spheroid. Averaging (\ref{eq:full_dynamics_x}) and (\ref{eq:full_dynamics_y}) over $\tau$, with $T$ fixed, and using the periodicity of the fast contributions, one obtains the equations governing $X_0$ and $Y_0$ as:
\begin{eqnarray}
\frac{dX_0}{dT} &=& Y_0 \label{eq:slowsep_x} \\
\frac{dY_0}{dT} &=& -\frac{d^5\xi_0^3L_c^2}{32X_0^4(1+\kappa)^2} \kappa\left[K_{ap/o} (-3 - 30\kappa + 9 \kappa^2) \right. \nonumber \\
\label{eq:slowsep_y}
&&\left. + K_{lp/o} (7 - 5 \kappa(2 + \kappa)) - 4 K_{tp/o} (5 + \kappa (-2 + 5 \kappa))\right],
\end{eqnarray}
The analytical far-field form may now be obtained by taking the ratio of (\ref{eq:slowsep_x}) and (\ref{eq:slowsep_y}). Reverting to the original notation, one obtains the far-field separatrices (without the superimposed small-scale wiggles) in the following form:
\begin{eqnarray}\label{eq:tumb_sep1}
y_p&=&4\sqrt{\frac{\left(K_{ap} \left(-9 \kappa^2+30 \kappa +3\right)+4 K_{tp} (\kappa  (5 \kappa -2)+5)+K_{lp} (5 \kappa  (\kappa +2)-7)\right)}{3\kappa(\kappa +1)^2\,x^3}},\\
\label{eq:tumb_sep2}
y_o&=&4\sqrt{\frac{\kappa  \left(K_{ao} \left(-9 \kappa^2+30 \kappa +3\right)+4 K_{to} (\kappa  (5 \kappa -2)+5)+K_{lo} (5 \kappa  (\kappa +2)-7)\right)}{3(\kappa +1)^2\,x^3}}.
\end{eqnarray}
\begin{figure}
\centering
\begin{subfigure}[b]{0.48\textwidth}
\centering
\includegraphics[scale=0.17]{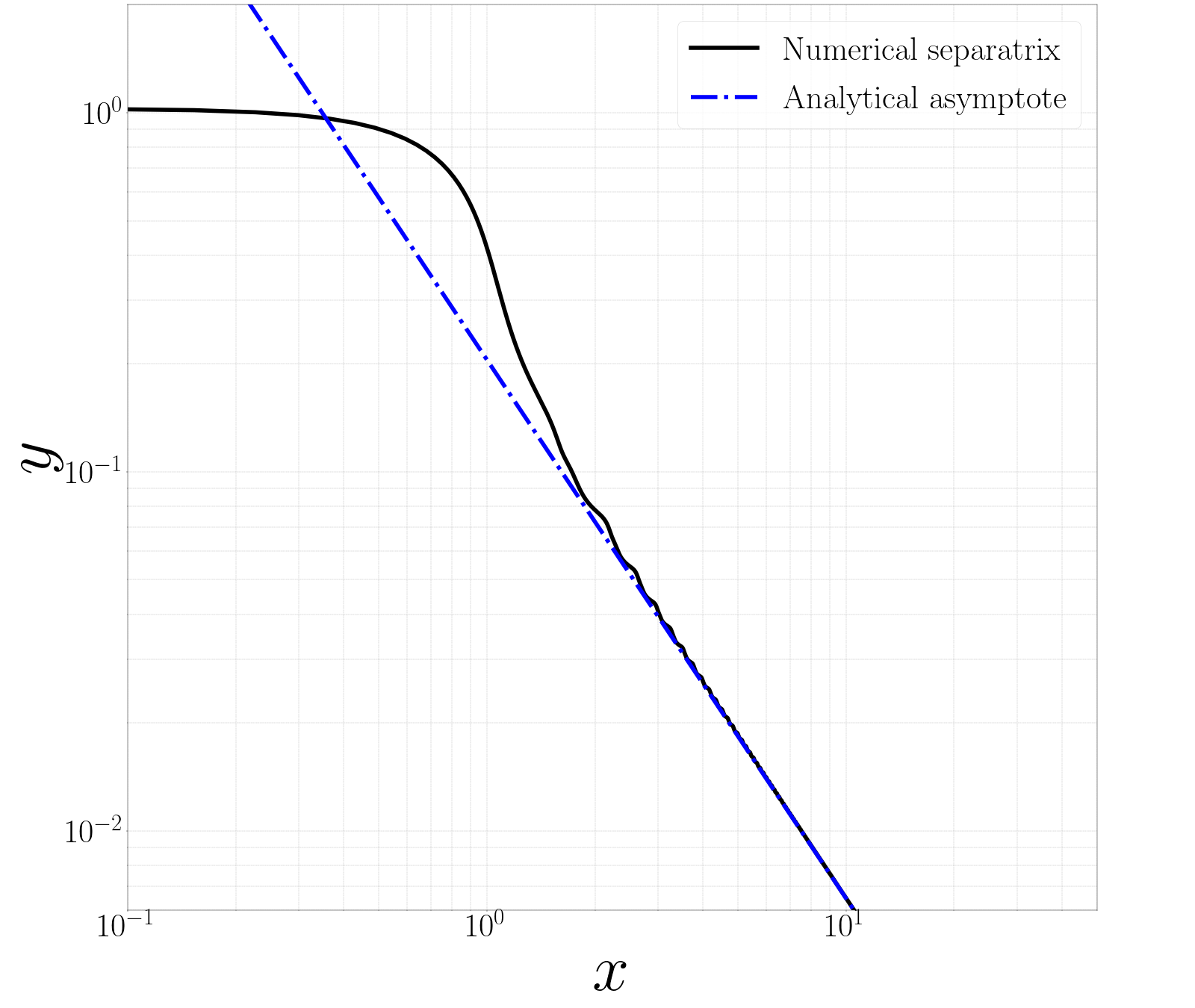}
\caption{Prolate-$\phi_{j0}=\pi/2$}
\end{subfigure}
\begin{subfigure}[b]{0.48\textwidth}
\centering
\includegraphics[scale=0.18]{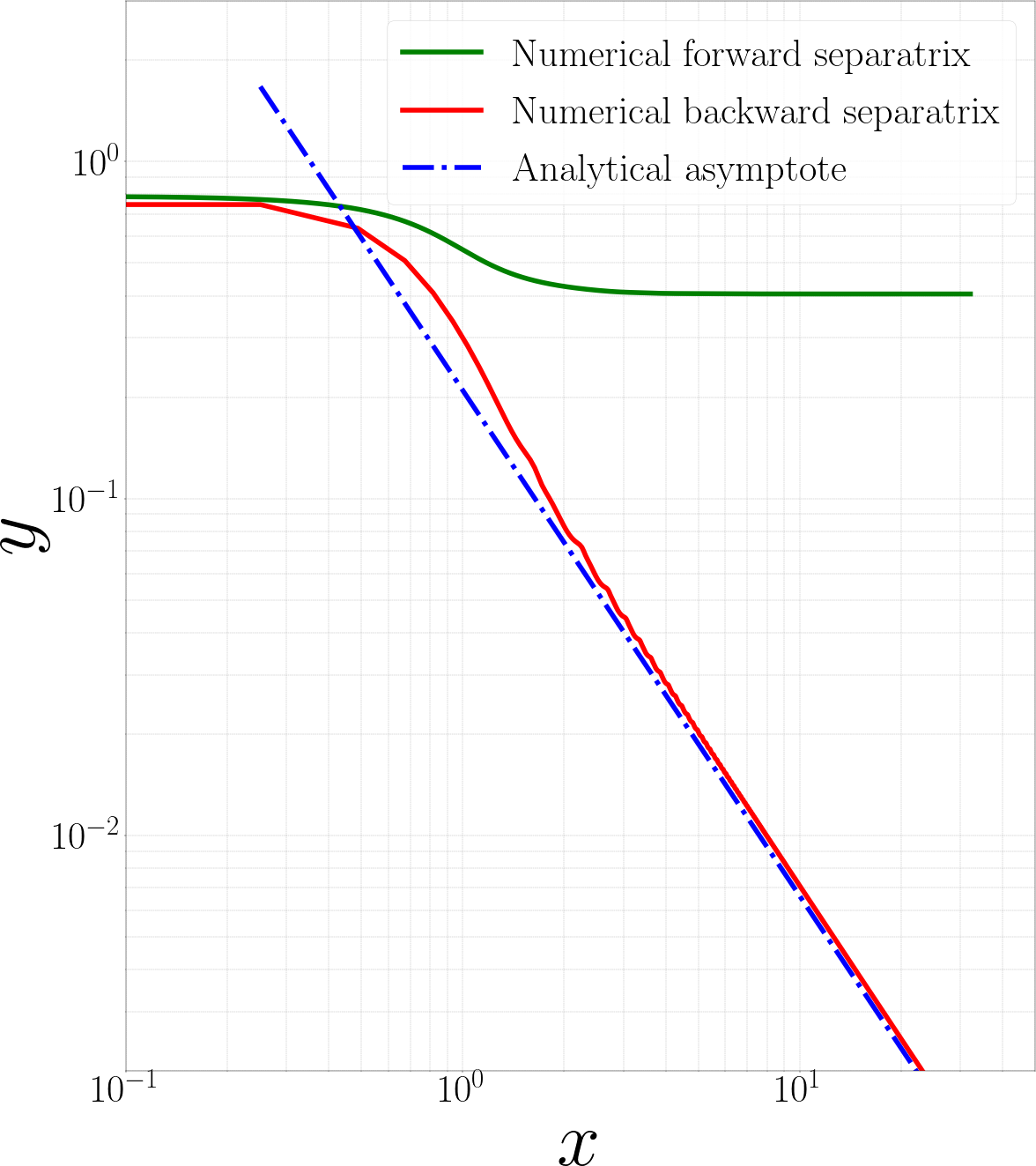}
\caption{Prolate-$\phi_{j0}=\pi/3$}
\end{subfigure}
\begin{subfigure}[b]{0.48\textwidth}
\centering
\includegraphics[scale=0.17]{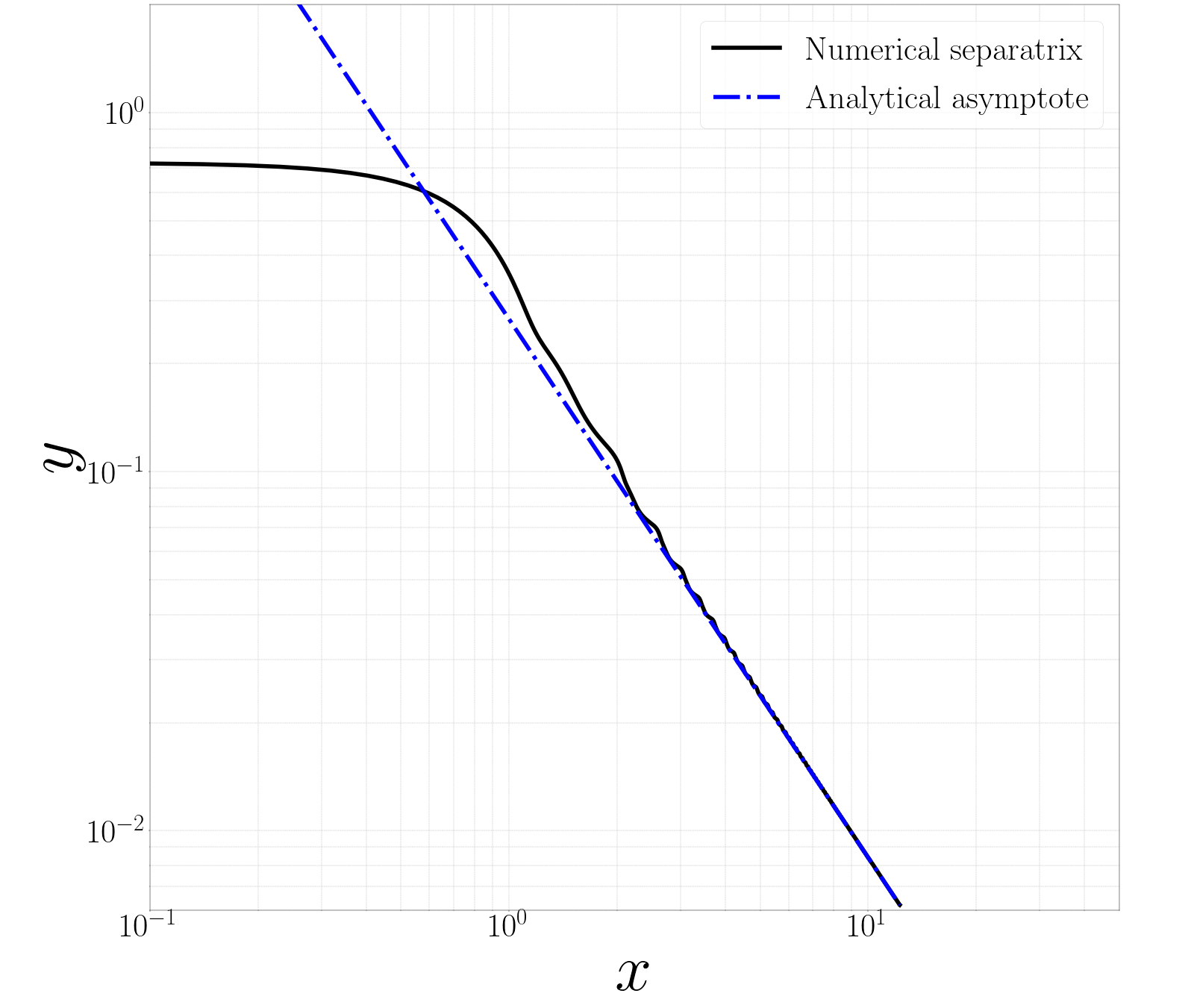}
\caption{Oblate-$\phi_{j0}=\pi/2$}
\end{subfigure}
\begin{subfigure}[b]{0.48\textwidth}
\centering
\includegraphics[scale=0.18]{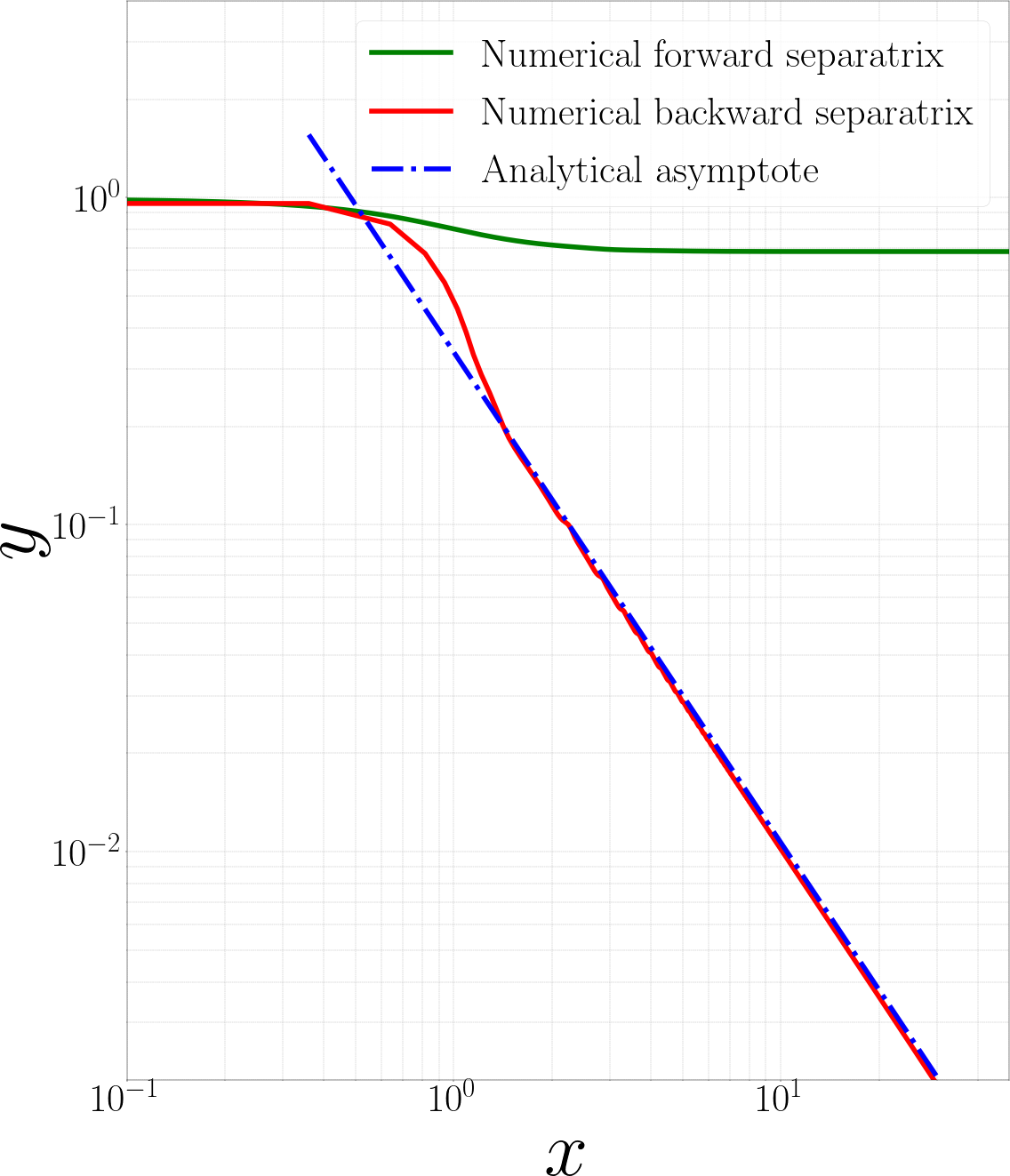}
\caption{Oblate-$\phi_{j0}=\pi/6$}
\end{subfigure}
\caption{Comparison between the numerical and analytical separatrix branches for a tumbling prolate spheroid with $\xi_0=1.15 (\kappa =2.025)$: (a) $\phi_{j0} = \pi/2$ (fore-aft symmetric) (b) $\phi_{j0} =\pi/3$(fore-aft asymmetric) and for an oblate spheroid with $\xi=1.15(\kappa=0.49)$ :(c) $\phi_{j0}=\pi/2$(fore-aft symmetric) (d) $\phi_{j0}=\pi/6$(fore-aft asymmetric).}
\label{fig:separatrixt}
\end{figure}
again exhibiting a far-field decay of $x^{-3/2}$. Figure \ref{fig:separatrixt} shows the comparison between both fore-aft symmetric (for $\phi_{j0} = \pi/2$) and asymmetric (for $\phi_{j0} = \pi/3(\pi/6$ for oblate)) separatrices obtained numerically, and the analytical predictions above. The numerical separatrices are obtained by choosing an initial location ($y_0$) on the gradient axis, and progressively refining this based on the final outcome of the integrated trajectory (escape to $x \rightarrow \pm \infty$, or crossing of the flow axis at a finite $x$). Notwithstanding the small-scale wiggles that decrease in amplitude with increasing $x$, a good comparison is obtained with  (\ref{eq:tumb_sep1}) and (\ref{eq:tumb_sep2}). 

\subsection{The singular pathlines}\label{sec:closed}
Thus far, we have dealt with the regular open pathlines, which are (except for $\phi_{j0} = 0,\pi/2$) fore-aft asymmetric generalizations of the open streamlines around a sphere. That these do not constitute the entire set of pathlines may be seen from the fact that all of the in-plane separatrices in figure 7 intersect the gradient axis at an ordinate value (say, $y^{sep}_0$) greater than $y_{min}$, the value corresponding to the intersection of the spheroid surface with $x = 0$; the $y_{min}$ value depends on $\phi_{j0}$, and ranges between $\bar{\xi}_0/\xi_0$ and unity, these limiting values corresponding to flow and gradient-aligned prolate spheroids, respectively. In-plane pathlines spanning the interval ($y_{min},y^{sep}_0$) have therefore not been accounted for; a similar argument may be made for the off-plane pathlines. The pathlines in these intervals are the analogs of closed streamlines for the sphere case, and the naive expectation is perhaps that the strictly periodic closed streamlines for a sphere would transform to aperiodic but bounded pathlines for a non-spinning spheroid. Interestingly, as will be seen below, this is not the case.

To examine pathlines in the interval ($y_{min}, y^{sep}_0$), we consider the case $\phi_{j0} = \pi/2 (y_{min} = 1)$. The regular (in-plane) pathline configuration analogous to that shown in figure \ref{fig:latdisp_phi}, but for a prolate spheroid with $\xi_0=2(\kappa = 1.15)$, is shown in figure \ref{fig:sing0}. The regular pathlines are bounded below by a fore-aft symmetric separatrix that asymptotes to the flow axis at upstream and downstream infinity, and intersects the gradient axis at $y_0^{sep} \approx 1.11621$.  As seen in figures \ref{fig:closed_sep}b-e, the pathlines `below' the separatrix that intersect the gradient axis in the interval ($1,y_0^{sep}$) are again open, in that they eventually asymptote to finite $y$ values in the upstream and downstream directions, but loop around the spheroid a certain number of times before doing so. Importantly, there appears to be no pattern to the number of loops; this number varies in a seemingly random (but sensitive) fashion as $y_0$ is decreased below $y^{sep}_0$ even by a very small amount. For instance, the number of loops for the pathlines in figure \ref{fig:sing1} is 1 and in \ref{fig:sing2} it is 2, but that in figure \ref{fig:sing3} is 23, while that in figure \ref{fig:sing4} is again 2; the dramatic increase in the number of loops in figure \ref{fig:sing3} occurs despite a change in $y_0$ that is $O(10^{-5})$! The fact that the pathlines crossing the gradient axis below the separatrix nevertheless asymptote to finite gradient offsets at upstream and downstream infinity obviously implies a crossing of the separatrix at some point. This crossing does not violate uniqueness since the spheroid orientation, at the time of crossing, is different for the separatrix and any of these looped trajectories (which we term the singular open pathlines); an argument along the same lines also shows that such looped trajectories, asymptoting to different upstream and downstream gradient offsets, is not inconsistent with the reversibility of the Stokes equations either.

\subsubsection{The residence time distributions}
To characterize this seemingly random dependence of the number of loops of the pathlines above, we define an appropriate residence time ($\mathcal{D}t$), taken to be the total time that a fluid element spends in a certain neighbourhood of the spheroid. Unless otherwise specified, this  neighbourhood  is taken to be the $x$-interval ($-20,20$) in the figures to follow; the qualitative nature of the findings reported below is independent of the particular choice of neighbourhood, provided the $x$-value chosen is large enough compared to the spheroid dimensions. Figures \ref{fig:res_tumb} and \ref{fig:res_spin} depict the variation in the residence time defined above. The residence times are plotted as a function of $y_0$, over a range that brackets $y_0^{sep}$, for both a tumbling prolate spheroid (with $\phi_{j0} = \pi/2$) and a spinning one. In the former case, we see a discontinuous change in the nature of the residence time curve as one moves across the separatrix. $\mathcal{D}t$ exhibits an initial smooth increase for the regular pathlines corresponding to $y_0 > y_0^{sep}$, but there is an abrupt transition to a seemingly random dependence for the singular pathlines with $y_0 < y_0^{sep}$. Figure \ref{fig:res_spin} shows the analogous residence time plot for the spinning spheroid, in which case the residence time varies in a smooth manner with $y_0$, attaining a maximum at the closed streamline(corresponding to $y'_0$ in the inset) that exactly spans the $x$-interval under consideration. The subsequent decrease for $y_0 < y_0'$ corresponds to choosing $\mathcal{D}t$ as half the period of the closed streamlines that no longer span the chosen $x$-interval, but instead intersect the flow axis at a pair of points, within the interval (-20,20), and symmetrically placed about the origin.
\begin{figure}
  \begin{center}
  \begin{subfigure}[b]{0.98\textwidth}
  \centering
  \includegraphics[scale=0.35]{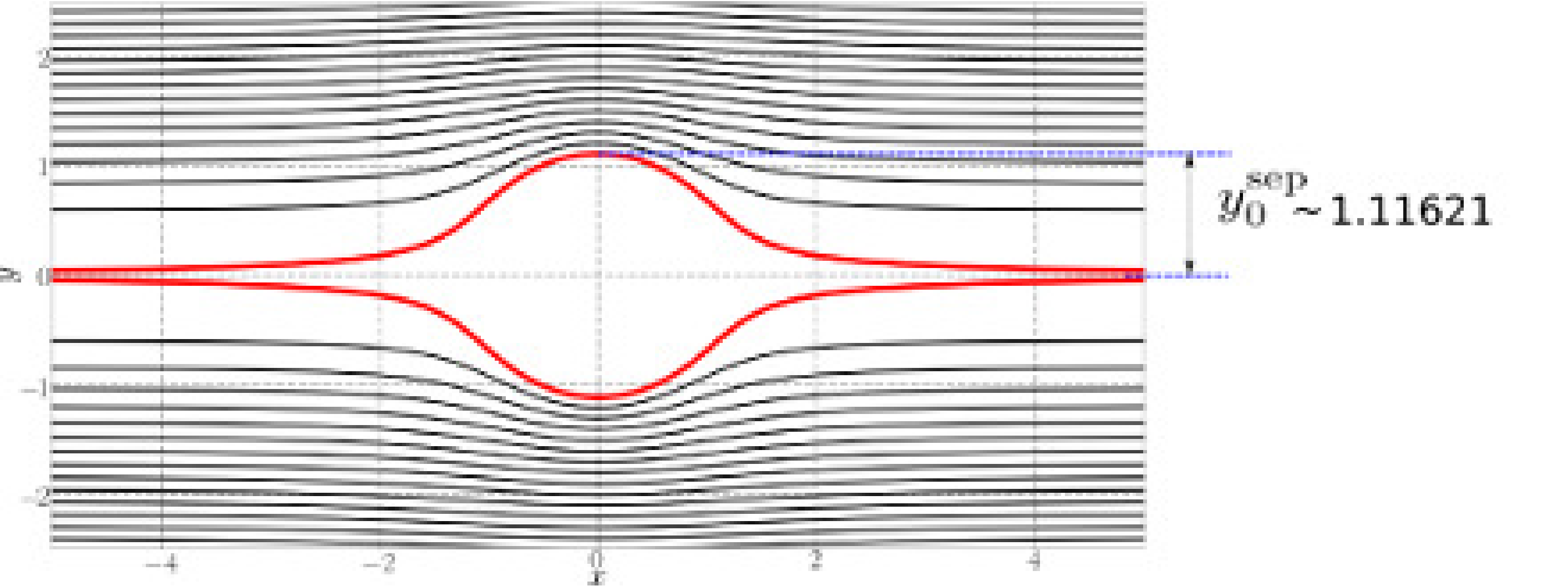}
  \caption{}
   \label{fig:sing0}
  \end{subfigure}
  \begin{subfigure}[b]{.48\textwidth}
  \centering
  \includegraphics[scale=0.22]{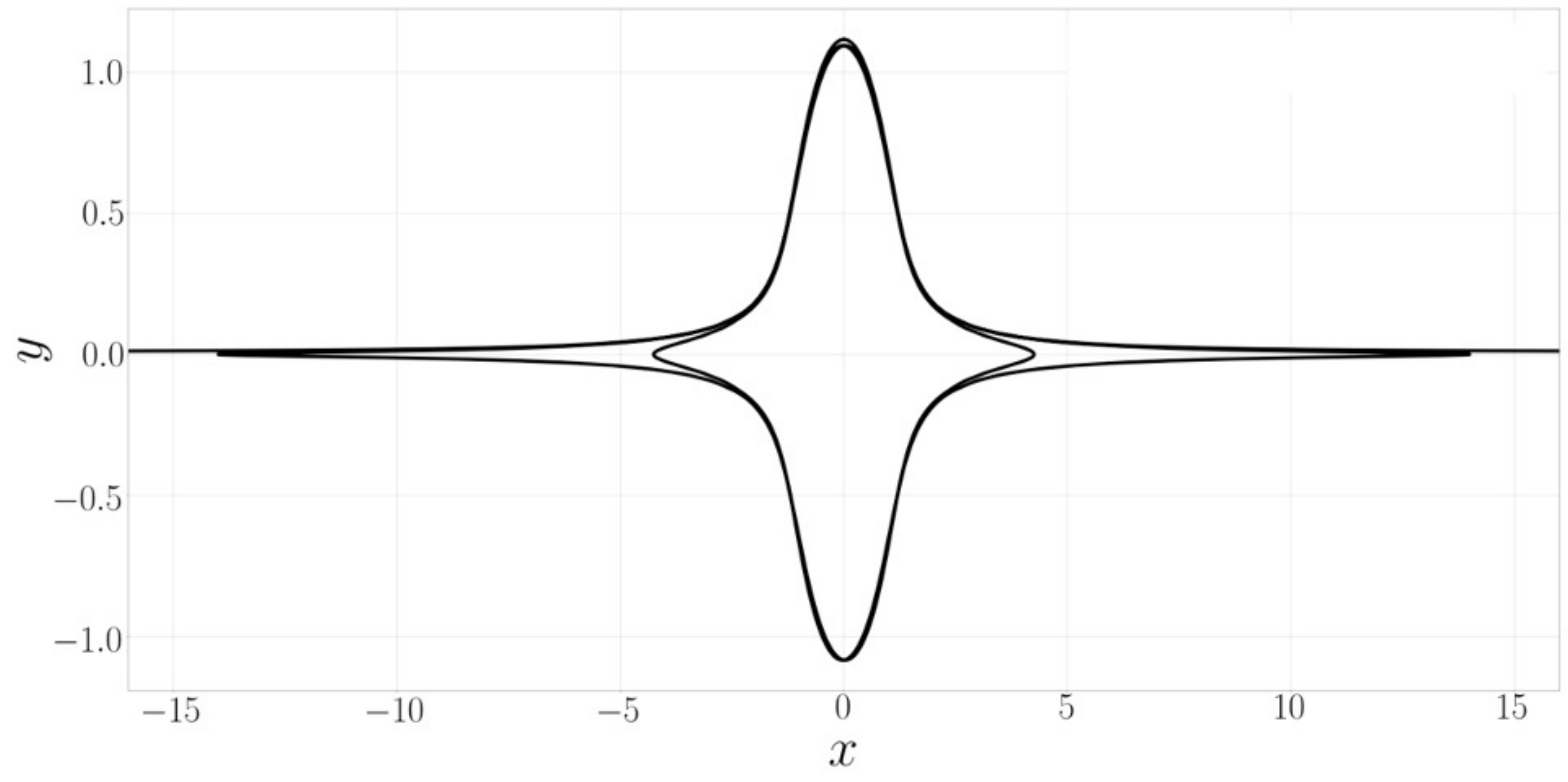}
  \caption{$y_0 = y_0^{\rm sep}-10^{-5}$}
  \label{fig:sing1}
  \end{subfigure}
  \begin{subfigure}[b]{.48\textwidth}
  \centering
  \includegraphics[scale=0.22]{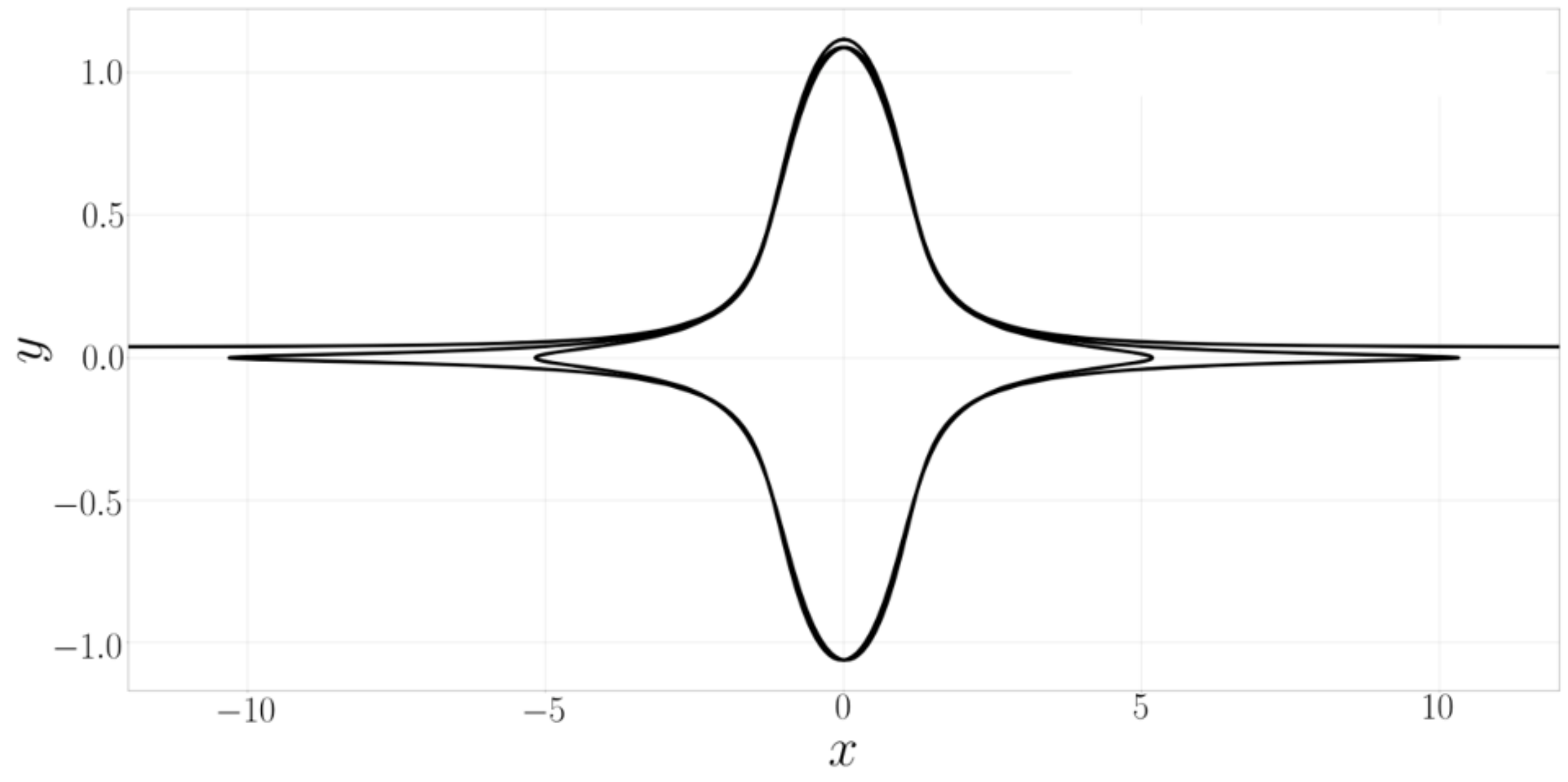}\\
  \caption{$y_0 = y_0^{\rm sep}-4\times 10^{-5}$}
  \label{fig:sing2}
  \end{subfigure}
  \begin{subfigure}[b]{.48\textwidth}
  \centering
  \includegraphics[scale=0.22]{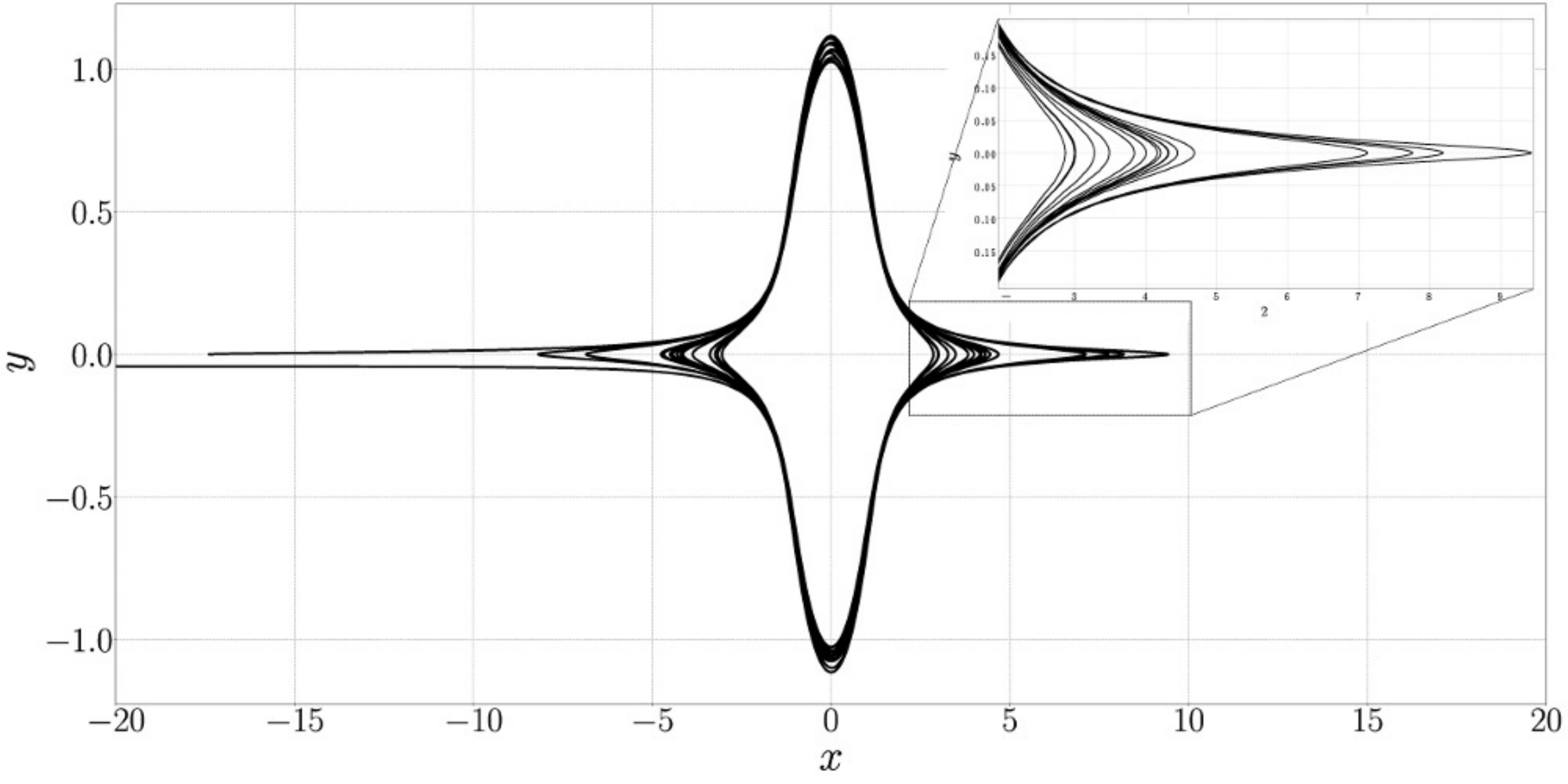}
  \caption{$y_0 = y_0^{\rm sep}-10\times 10^{-5}$}
  \label{fig:sing3}
  \end{subfigure}
  \begin{subfigure}[b]{.48\textwidth}
  \centering
  \includegraphics[scale=0.22]{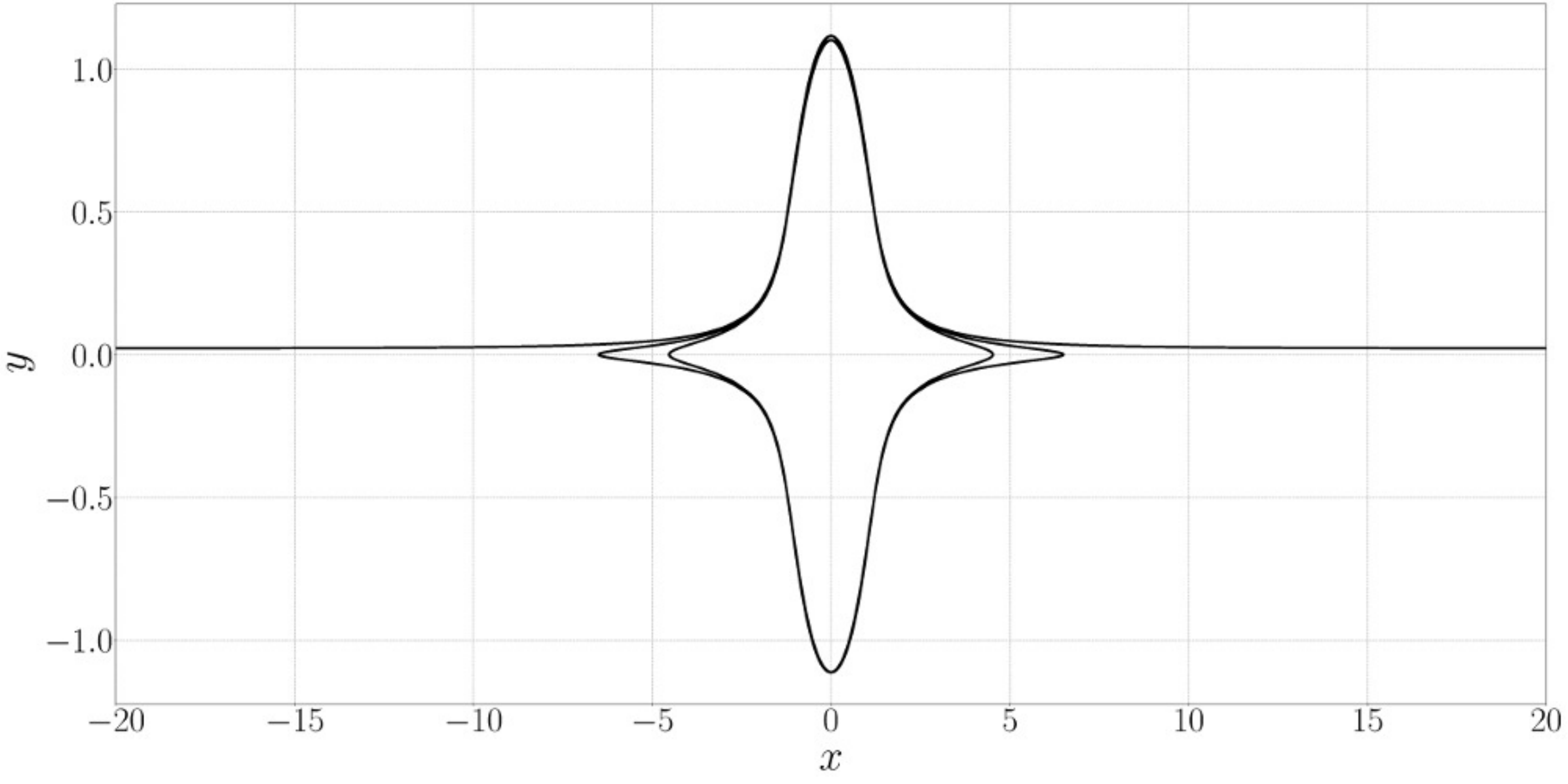}\\
  \caption{$y_0 = y_0^{\rm sep}-20\times 10^{-5}$}
  \label{fig:sing4}
  \end{subfigure}
  \end{center}
  \caption{The open pathline configuration for a tumbling prolate spheroid ($\xi0 =2.0(\kappa=1.15)$) with $\phi_{j0} = \pi/2$. (b)-(e) correspond to singular open pathlines that cross the gradient axis just below the separatrix(shown in red, with $y_0=y_0^{\rm sep}$) at four different $y_0$'s.}
\label{fig:closed_sep}
\end{figure}
\begin{figure}
\begin{center}
\begin{subfigure}[b]{0.45\textwidth}
\centering
  \includegraphics[scale=0.25]{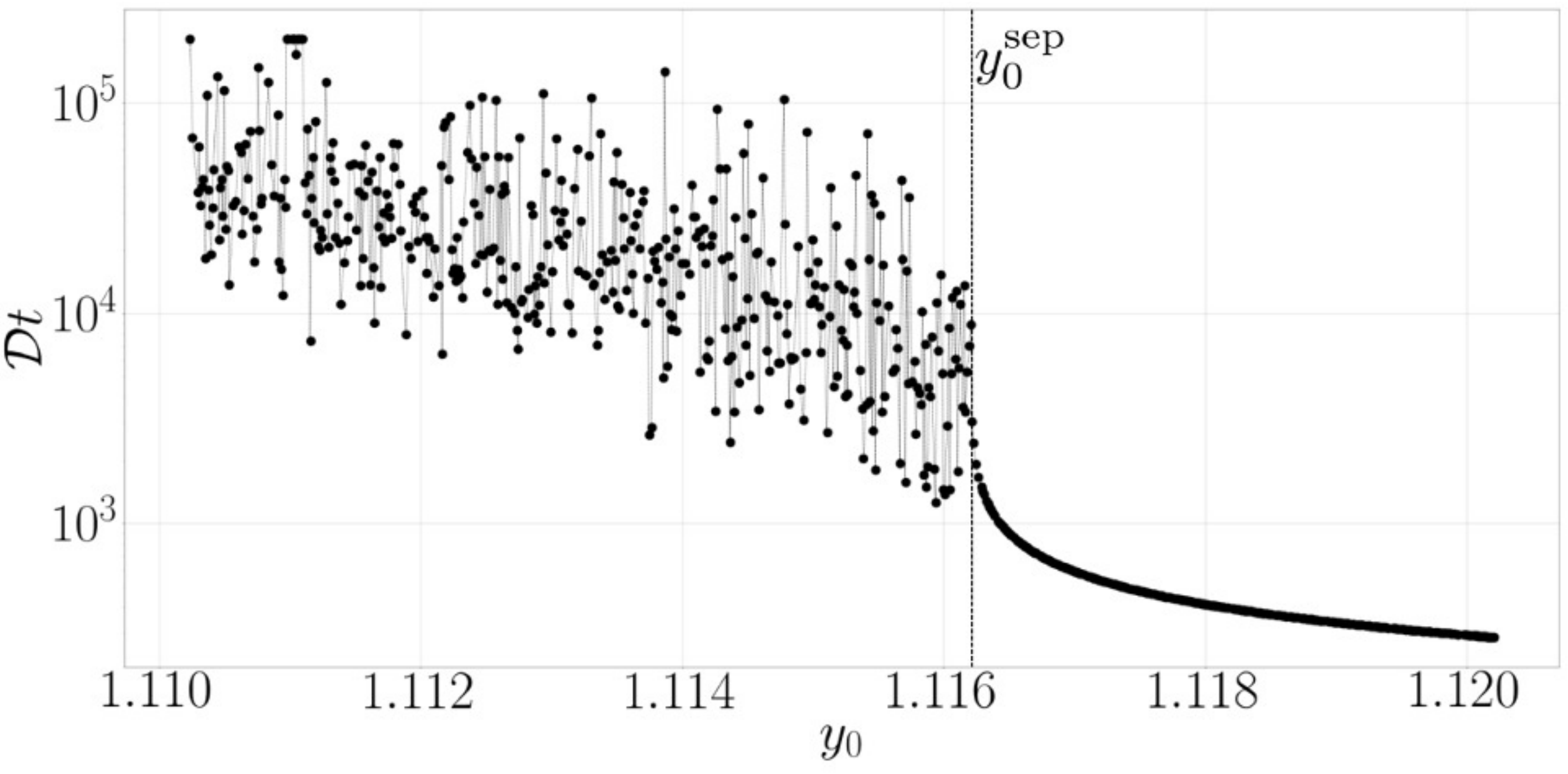}
\caption{Tumbling}
\label{fig:res_tumb}
\end{subfigure}
\hfill
\begin{subfigure}[b]{0.45\textwidth}
\centering
  \includegraphics[scale=0.25]{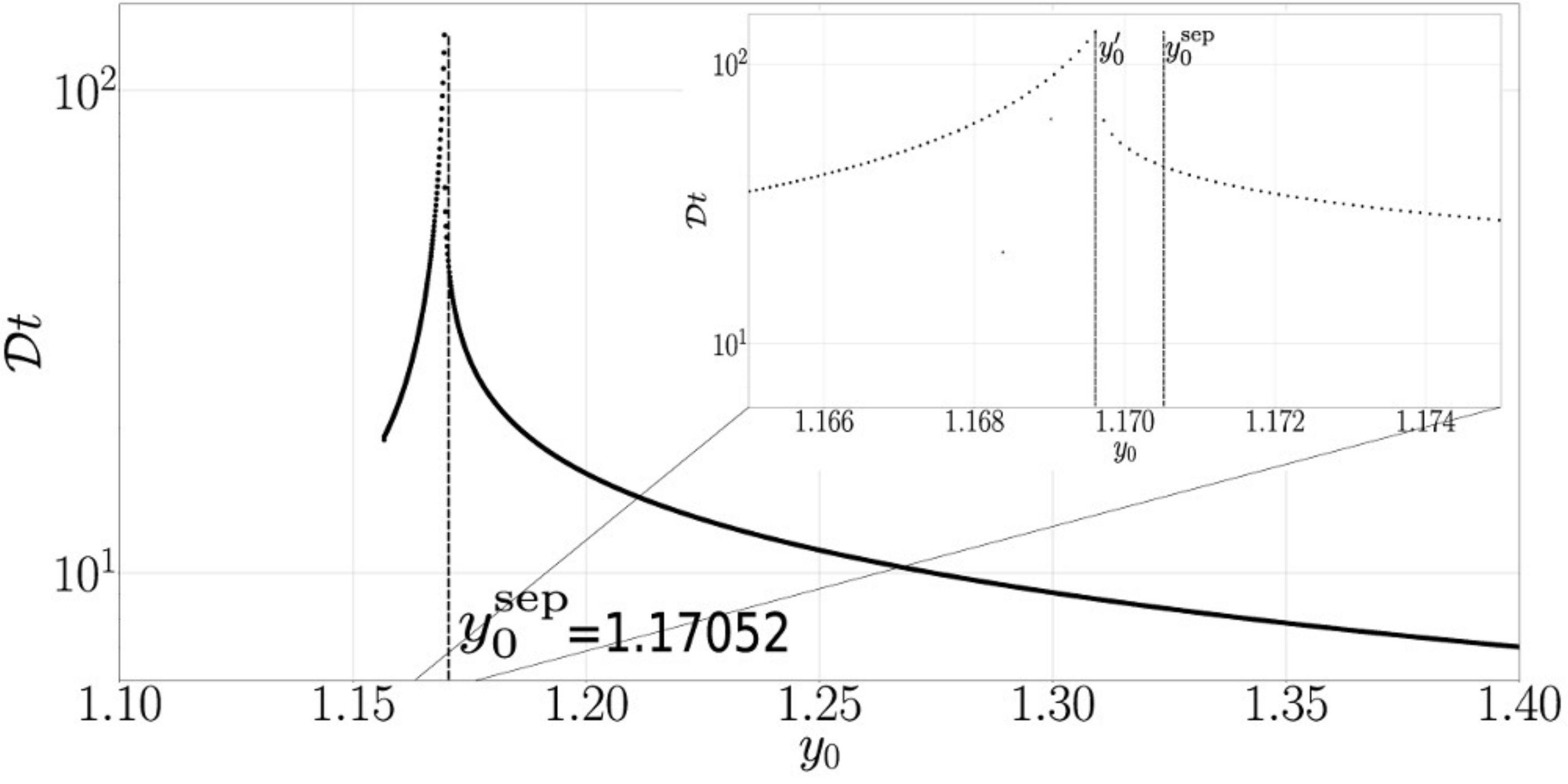}
\caption{Spinning}
\label{fig:res_spin}
\end{subfigure}
  \end{center}
  \caption{Variation of the residence time of a fluid element for a prolate spheroid($\xi_0=2.0(\kappa=1.15)$): (a) in the tumbling orbit with $\phi_{j0} = \pi/2$, (b) in the spinning mode; the ordinate of the separatrix $y_0^{sep}$ is shown by the dashed line; $y_0'$ corresponds to the closed orbit spanning the $x$-interval ($-20,20$).}
\label{fig:turns_timeperiod}
\end{figure}

\begin{figure}
  \begin{center}
  \begin{subfigure}[b]{.98\textwidth}
  \centering
  \includegraphics[scale=0.35]{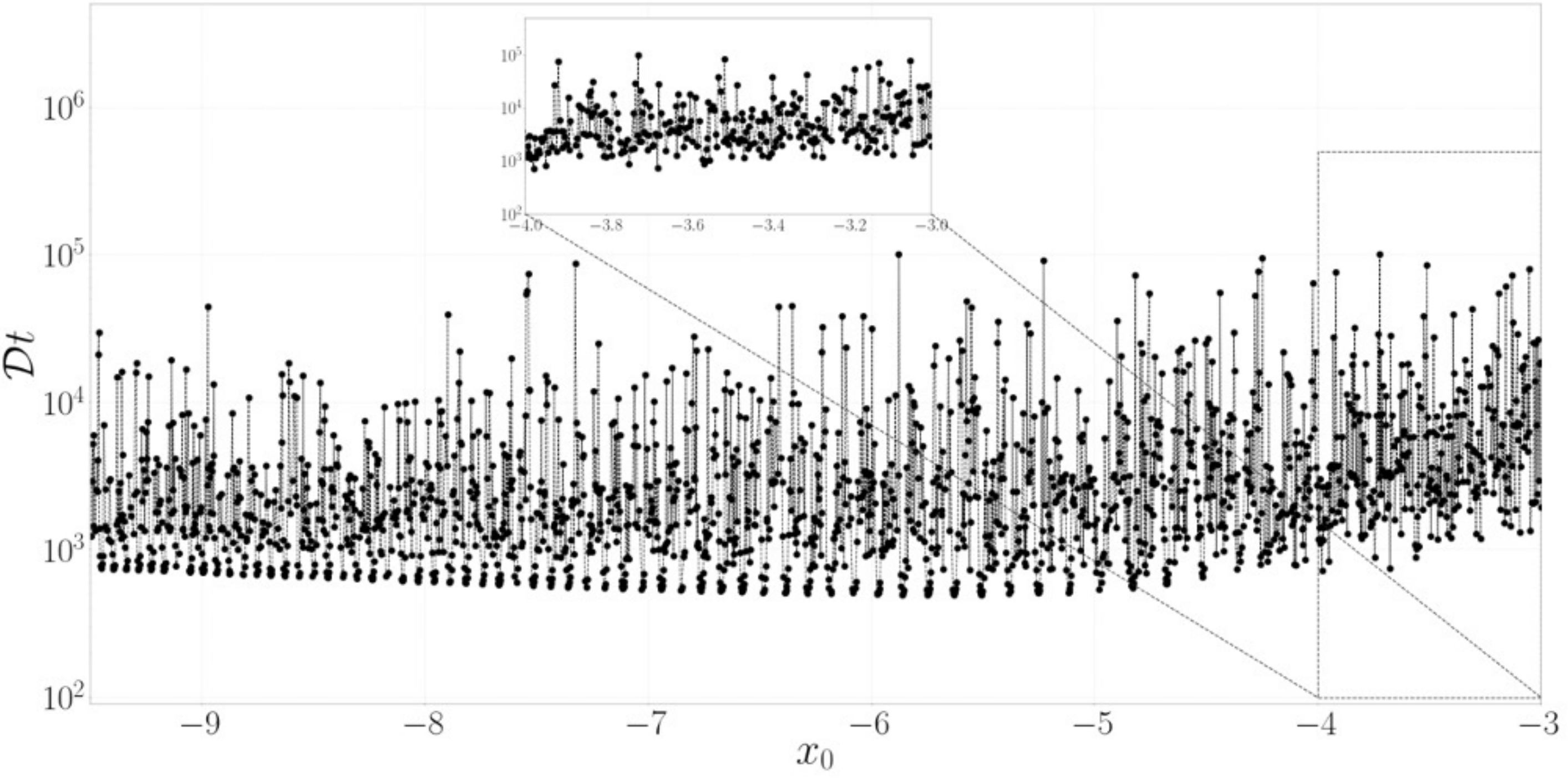}
  \caption{}
  \label{fig:closed_time}
  \end{subfigure}
  \begin{subfigure}[b]{.48\textwidth}
  \centering
  \includegraphics[scale=0.22]{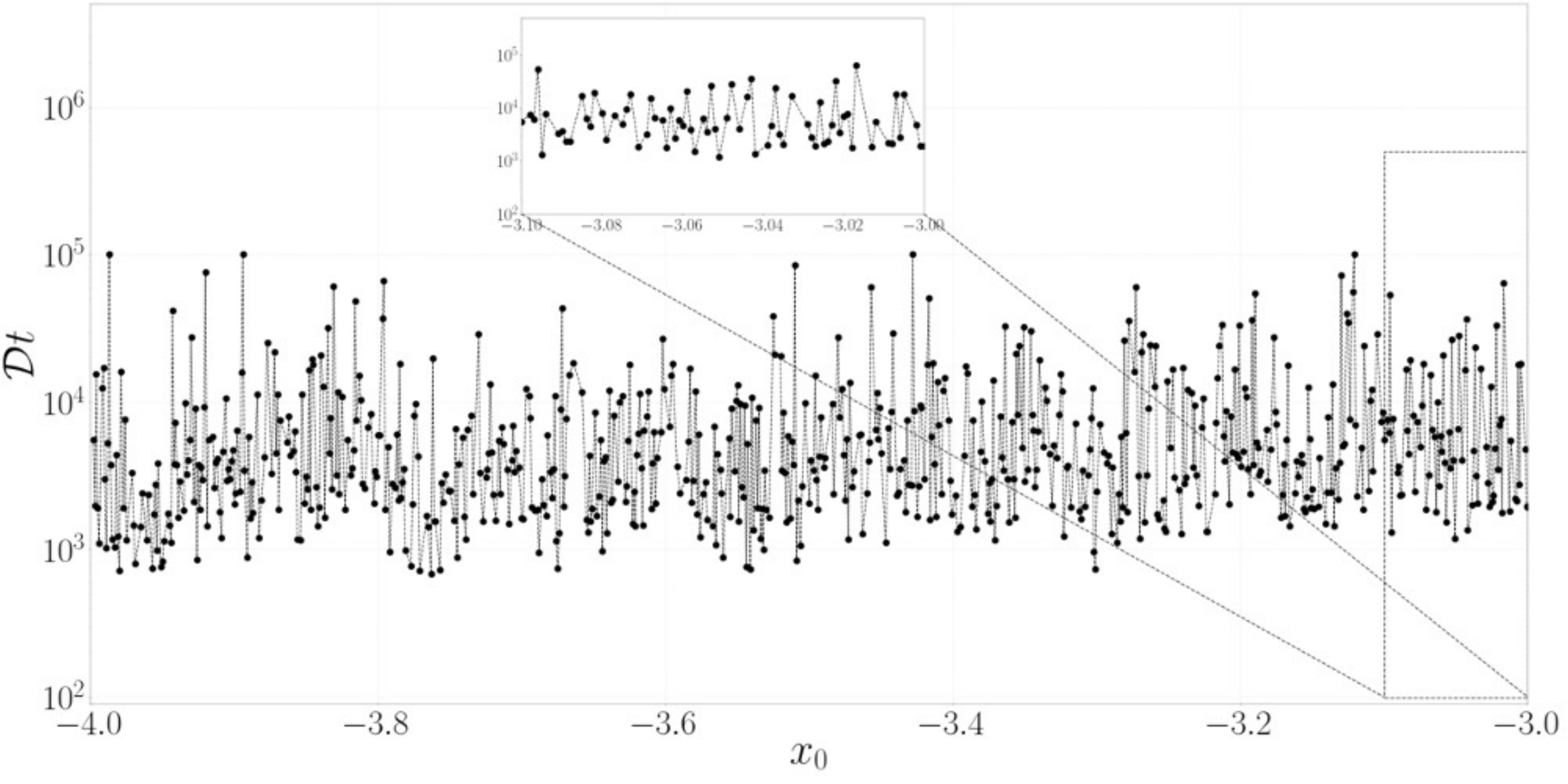}
  \caption{}
  \end{subfigure}
  \begin{subfigure}[b]{.48\textwidth}
  \centering
  \includegraphics[scale=0.22]{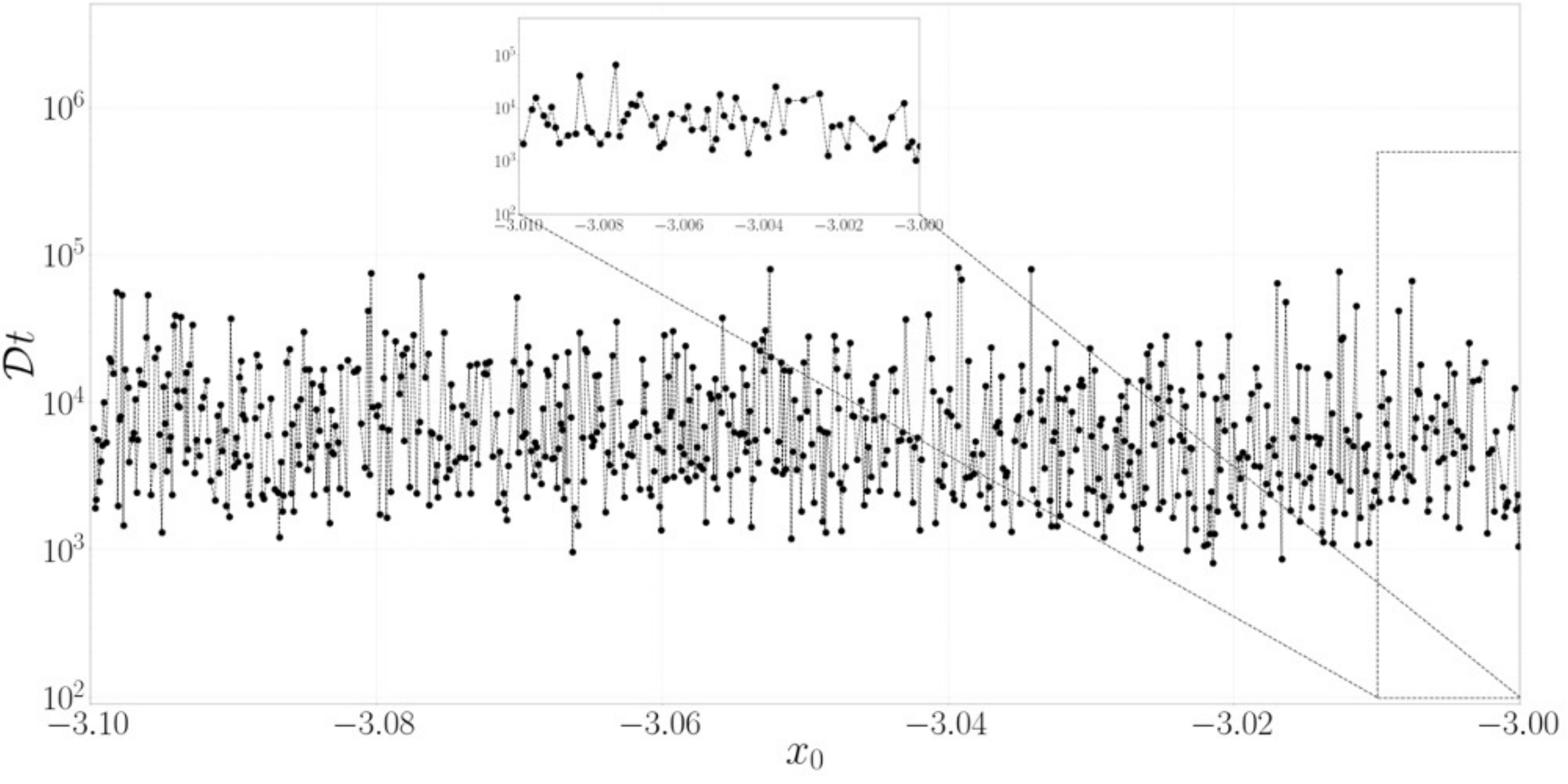}
  \caption{}
  \end{subfigure}
  \begin{subfigure}[b]{.48\textwidth}
  \centering
  \includegraphics[scale=0.22]{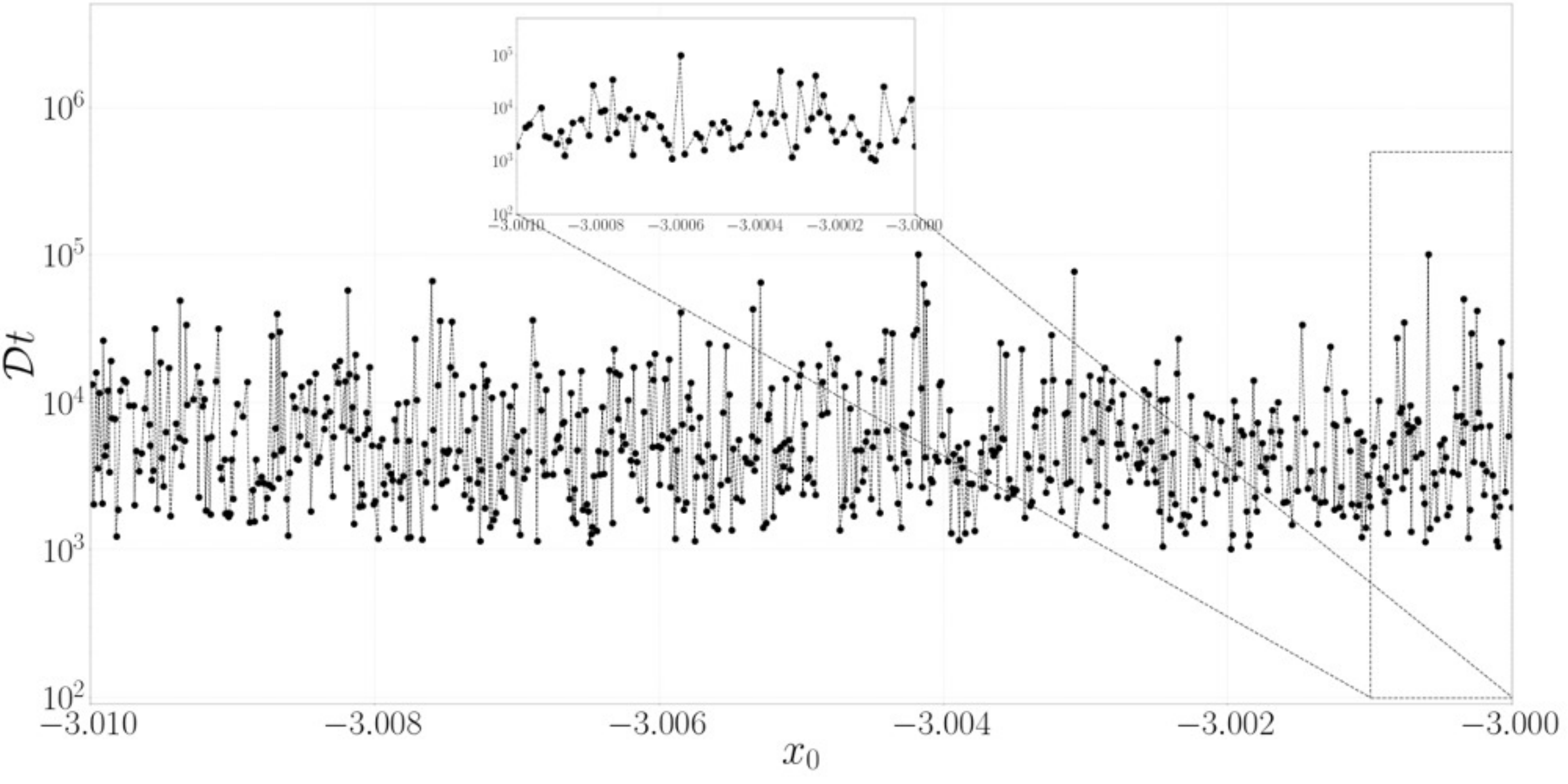}
  \caption{}
  \end{subfigure}
  \begin{subfigure}[b]{.48\textwidth}
  \centering
  \includegraphics[scale=0.22]{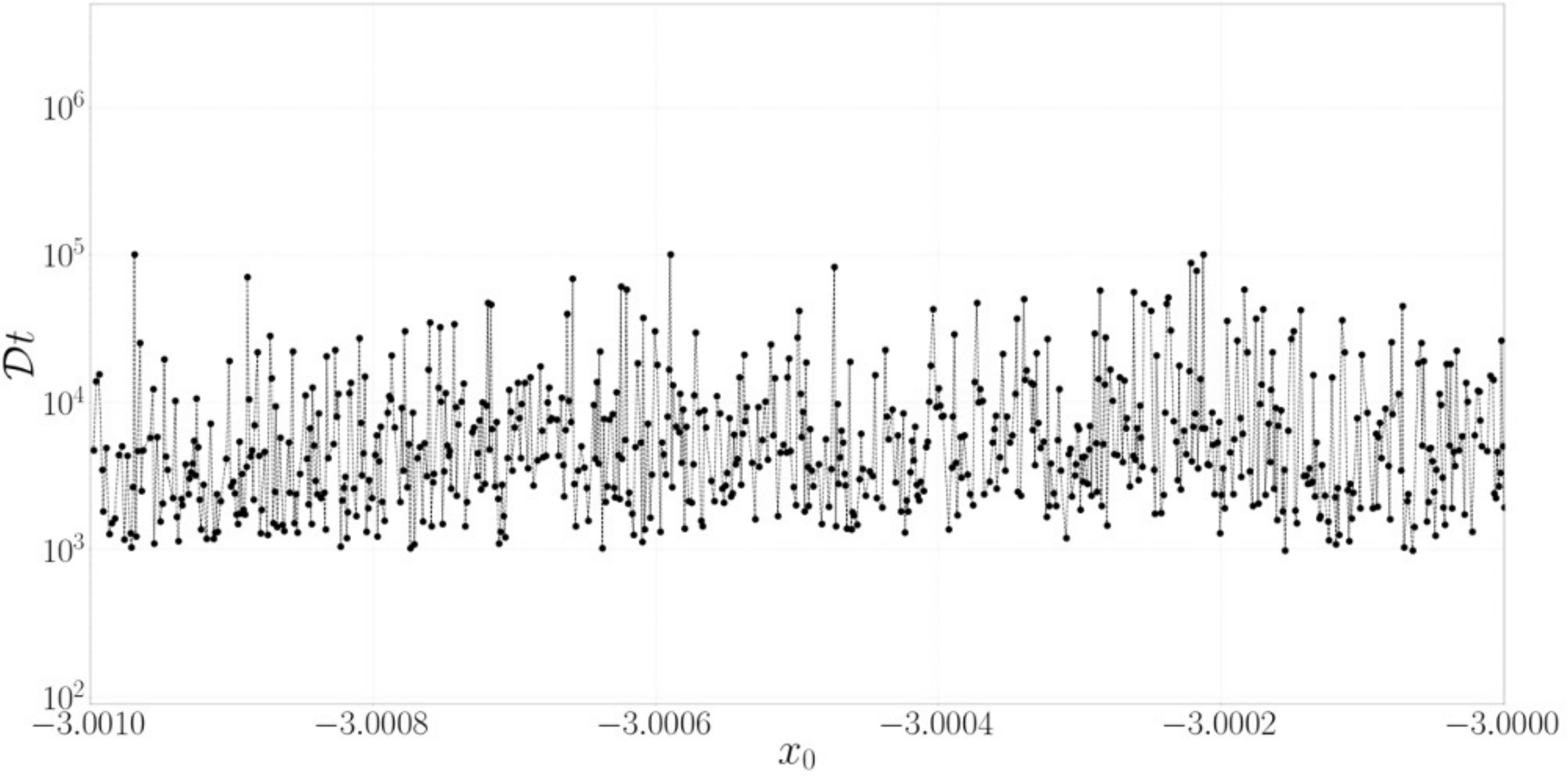}
  \caption{}
  \end{subfigure}
  \end{center}
  \caption{The sensitive dependence of the residence time of a fluid element, for a tumbling prolate spheroid ($\xi_0=2.0(\kappa=1.13)$) with $\phi_{j0}=0$, plotted as a function of the `initial condition' (defined as a point on the negative $x$-axis). Figure (a) depicts the dependence in the entire interval (-9.5,-3), while figures (b)-(e) show the dependence of the residence time  in the intervals (-4,-3), (-3.1,3), (-3.01,3) and (-3.001,3), respectively.}
\label{fig:closed_time_res}
\end{figure}

\begin{figure}
  \begin{center}
  \begin{subfigure}[b]{.48\textwidth}
  \centering
  \includegraphics[scale=0.3]{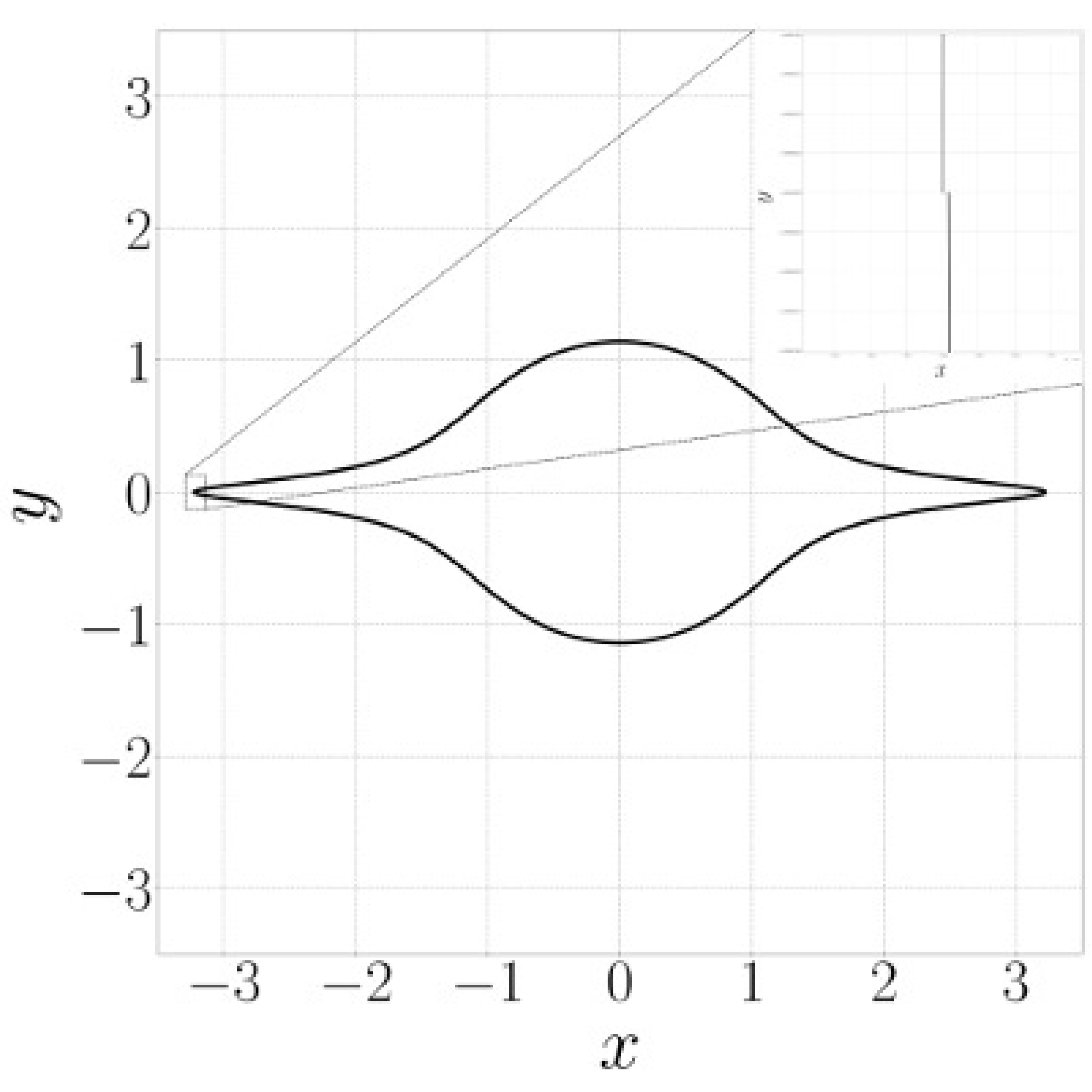}
  \caption{$x_0=-3.2214088$}
  \end{subfigure}
  \begin{subfigure}[b]{.48\textwidth}
  \centering
  \includegraphics[scale=0.3]{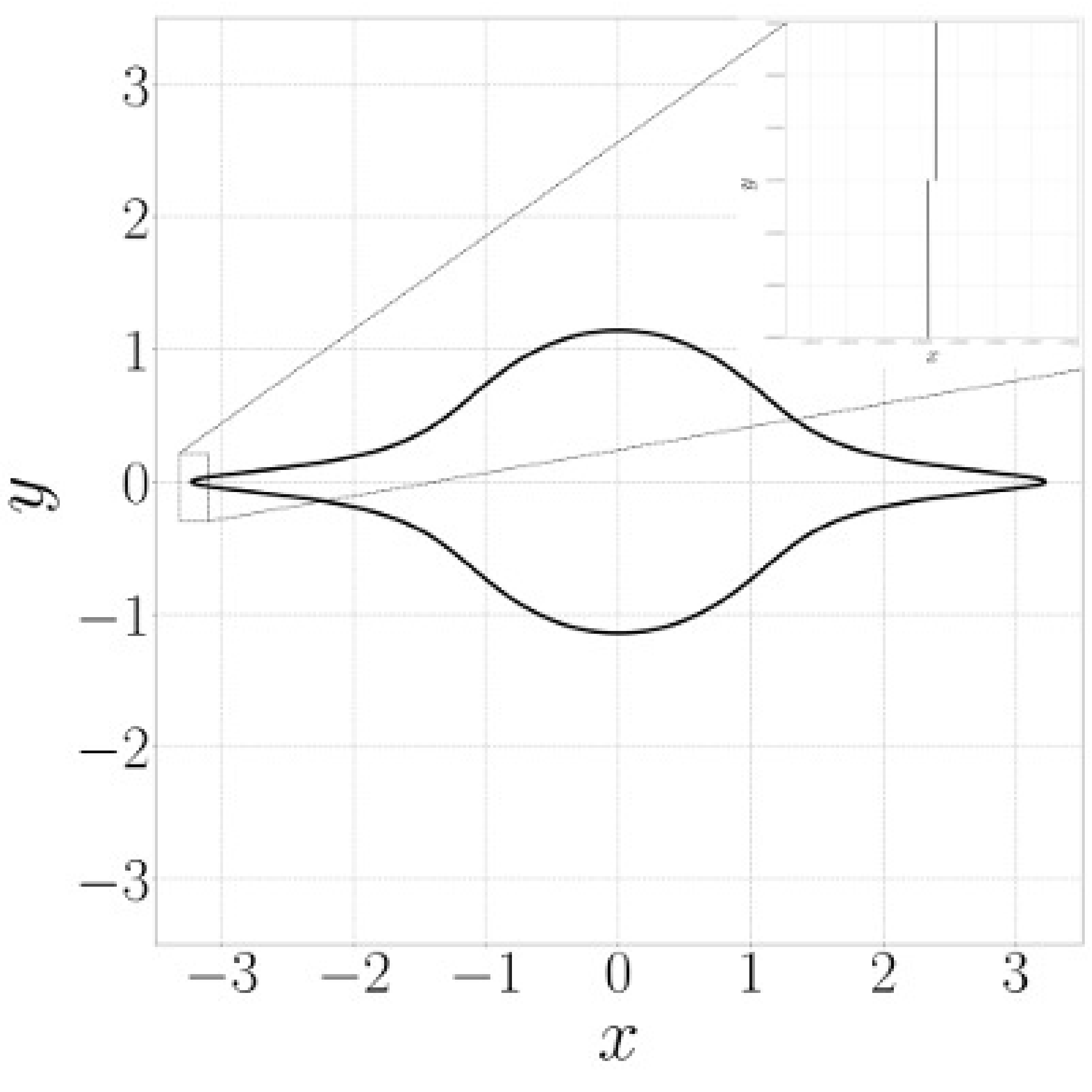}
  \caption{$x_0=-3.2214089$}
  \end{subfigure}
  \begin{subfigure}[b]{.48\textwidth} 
  \end{subfigure}
  \end{center}
  \caption{Existence of periodic pathlines for the tumbling prolate spheroid($\xi_0=4.0(\kappa=1.03)$- a single-loop periodic orbit with an approximate period of  $40 T_j$: (a) final point displaced to the right of the initial point $(-3.2214088,0)$ and (b) final point displaced to the left of the initial point $(-3.2214089,0)$.}
\label{fig:periodic_KAM}
\end{figure}

We now examine the irregular variation of the residence time encountered in figure \ref{fig:res_tumb} in greater detail. For this purpose, we consider the same spheroid as in Figure \ref{fig:closed_sep}, again in its tumbling orbit ($\theta_j = \pi/2$). The consideration of such a spheroid with an aspect ratio near unity ($\xi_0$=2; $\kappa = 1.15$) serves the additional purpose of illustrating the singular nature of the spherical particle limit from the perspective of the streamline/pathline topology. Figure \ref{fig:closed_time} shows the variation of the residence time of the singular open pathlines which pass through the negative $x$-axis within the interval $x_0 \in (-9.5,-3)$. The residence time is obtained in the manner defined above, via forward and backward time integrations starting from a specific initial point in the aforementioned interval, and until the pathline leaves the interval (-20,20). Similar to figure \ref{fig:res_tumb}, the extremely irregular dependence of the residence time on $x_0$ is readily apparent (note that each of the pathlines intersects the $x$-axis multiple times, and in principle, it is possible that a pair of chosen initial conditions end up being part of the same singular pathline, although this should, in principle, occur with an infinitesimal probability; we have verified that this is not the case by monitoring the upstream and downstream coordinates, and ensuring that these are different for each of the initial points chosen). The plots that follow in figures \ref{fig:closed_time_res}b-e show the variation in the residence time at progressively finer resolutions by focussing on increasingly small subsets of the $x_0$-interval considered above. These figures correspond to 1000 initial conditions in the $x_0$-intervals (-4,-3), (-3.1,-3), (-3.01,3) and (-3.001,3), respectively. Thus, in going from figure \ref{fig:closed_time_res}a to e, the length scale of the interval under consideration has decreased by four orders of magnitude, but the irregular dependence of the residence time clearly persists down to the smallest scales. This fractal dependence of the residence time measure is a signature of chaotic scattering, and is suggestive of the existence of a chaotic saddle in the region around the spheroid\citep{bleher-prl,bleher, aref-pomphrey-letter,aref-pomphrey,aref-stremler,faisst}.
The chaotic saddle above includes an infinite number of unstable periodic orbits (and bounded aperiodic ones), whose stable and unstable manifolds intersect to form a Cantor-like set. The existence of single-looped periodic orbits may, for instance, be inferred by considering two neighbouring pathlines, integrated over the approximate duration of a single loop, as shown in figures \ref{fig:periodic_KAM}a and b. The pair of initial points in the figures correspond to $x_0=-3.2217088$ and $x_0=-3.2214089$ on the negative $x$-axis, and one observes the subsequent intersections with the (negative) $x$-axis to occur opposite sides of the initial point in the two cases, implying that the pair of pathlines shown bracket a single-loop closed orbit between them. This argument may be extended in an obvious manner to show the existence of closed pathlines with multiple loops.

\begin{figure}
  \begin{center}
  \begin{subfigure}[b]{.48\textwidth}
  \centering
  \includegraphics[scale=0.22]{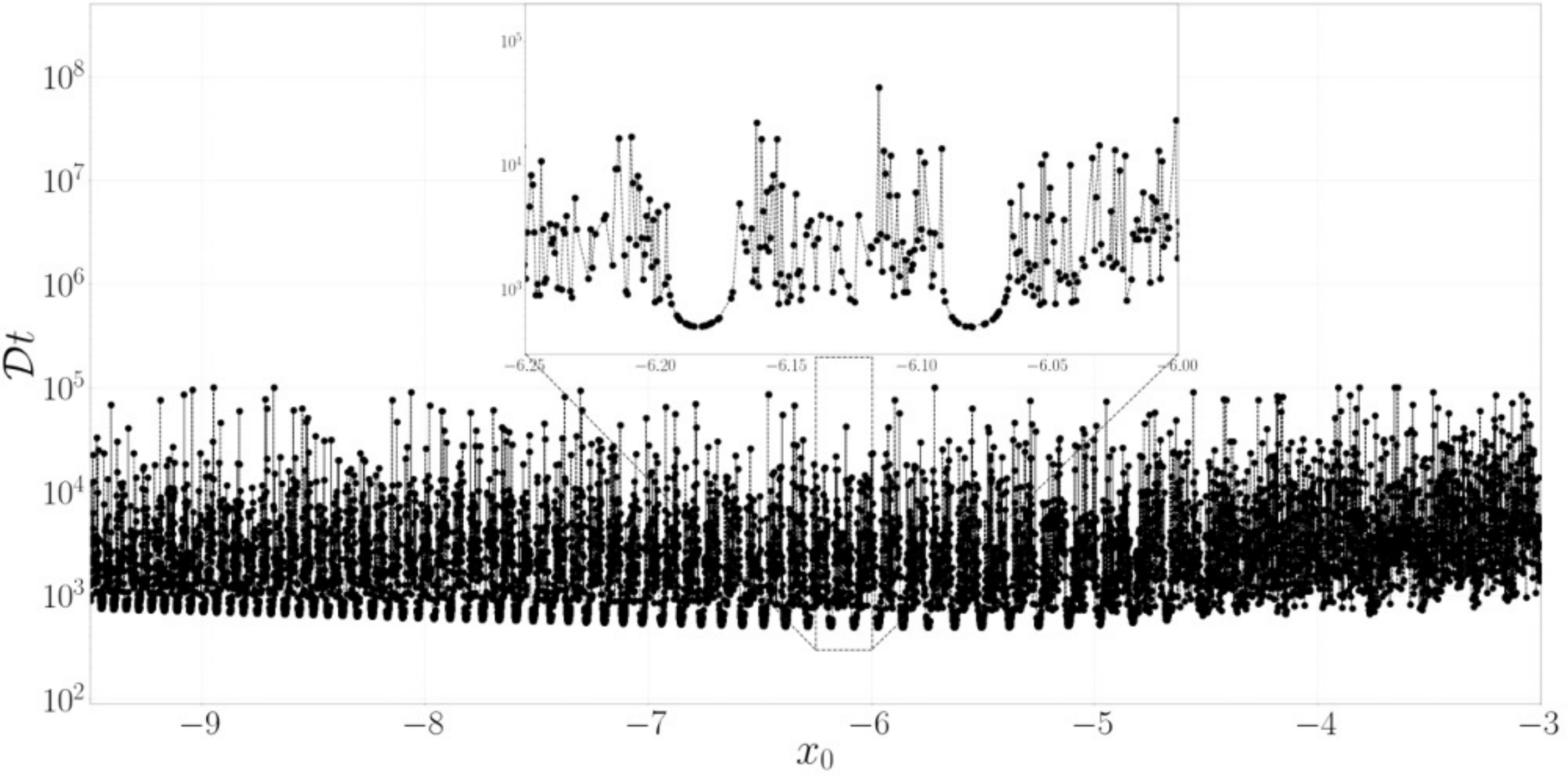}
  \caption{$C=\infty$}
  \end{subfigure}
  \begin{subfigure}[b]{.48\textwidth}
  \centering
  \includegraphics[scale=0.22]{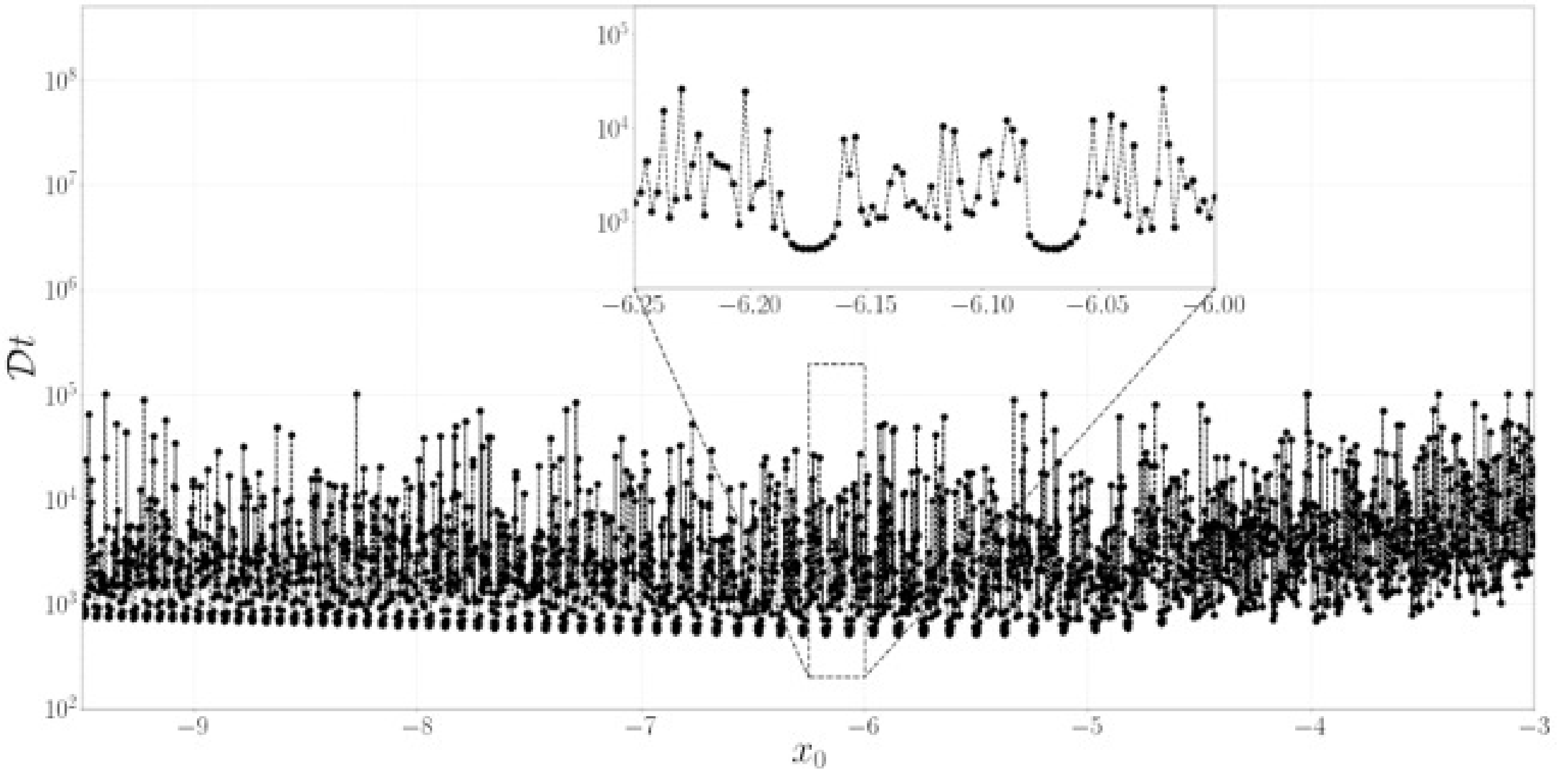}
  \caption{$C=5.0$}
  \end{subfigure}
  \begin{subfigure}[b]{.48\textwidth}
  \centering
  \includegraphics[scale=0.22]{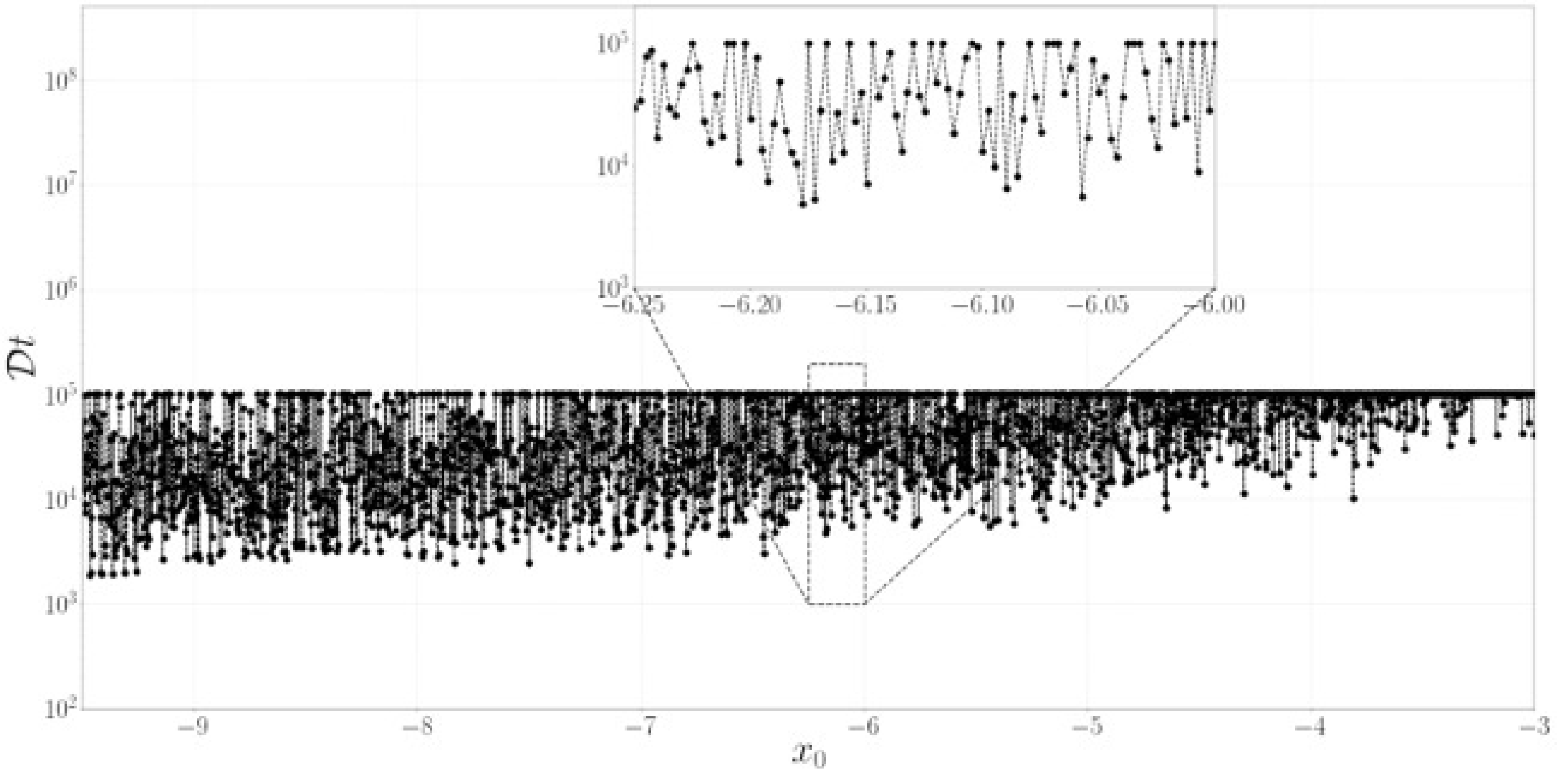}
  \caption{$C=0.5$}
  \end{subfigure}
  \begin{subfigure}[b]{.48\textwidth}
  \centering
  \includegraphics[scale=0.22]{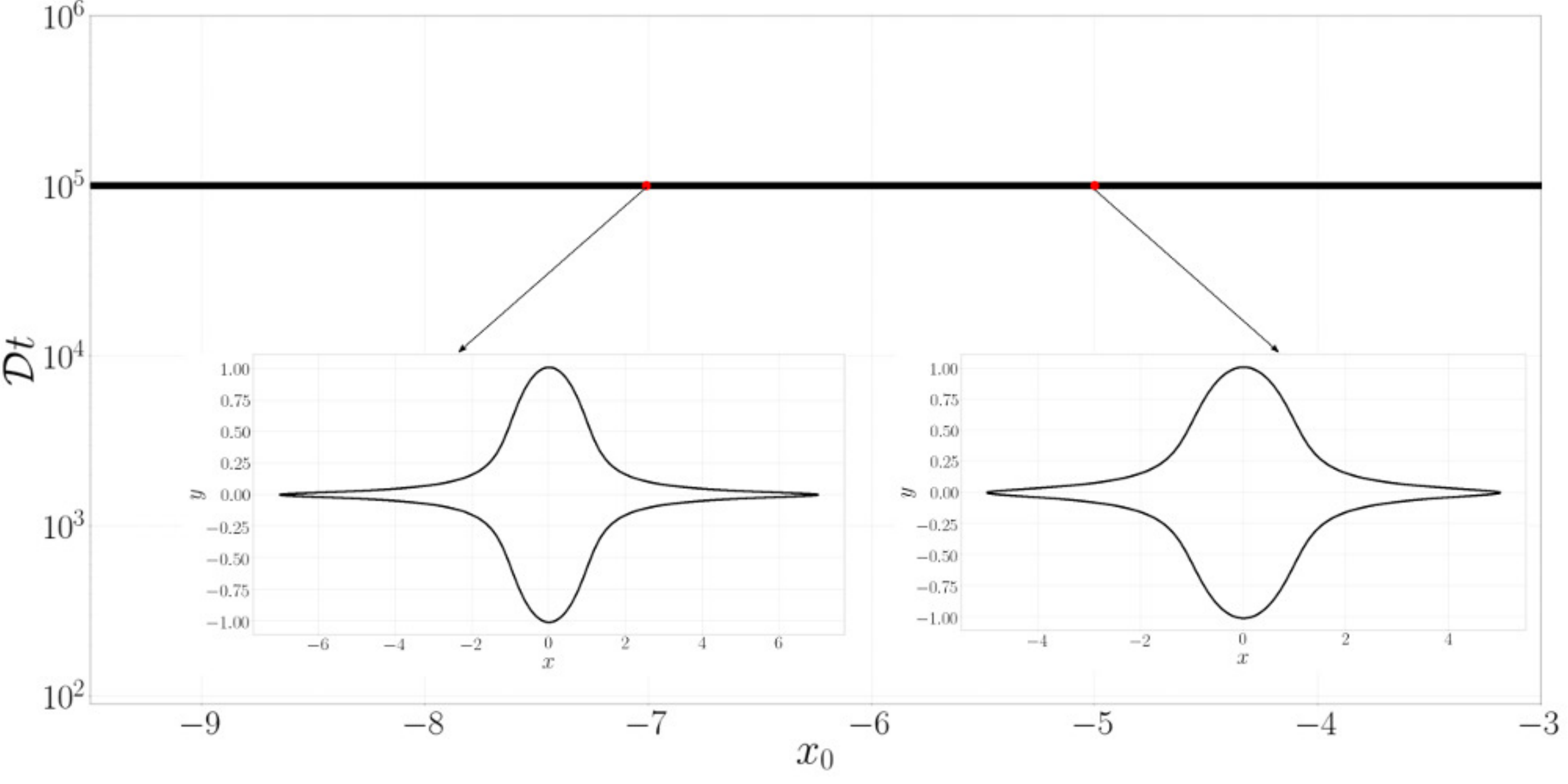}
  \caption{$C=0.05$}
  \end{subfigure}
  \begin{subfigure}[b]{.96\textwidth}
  \centering
  \includegraphics[scale=0.35]{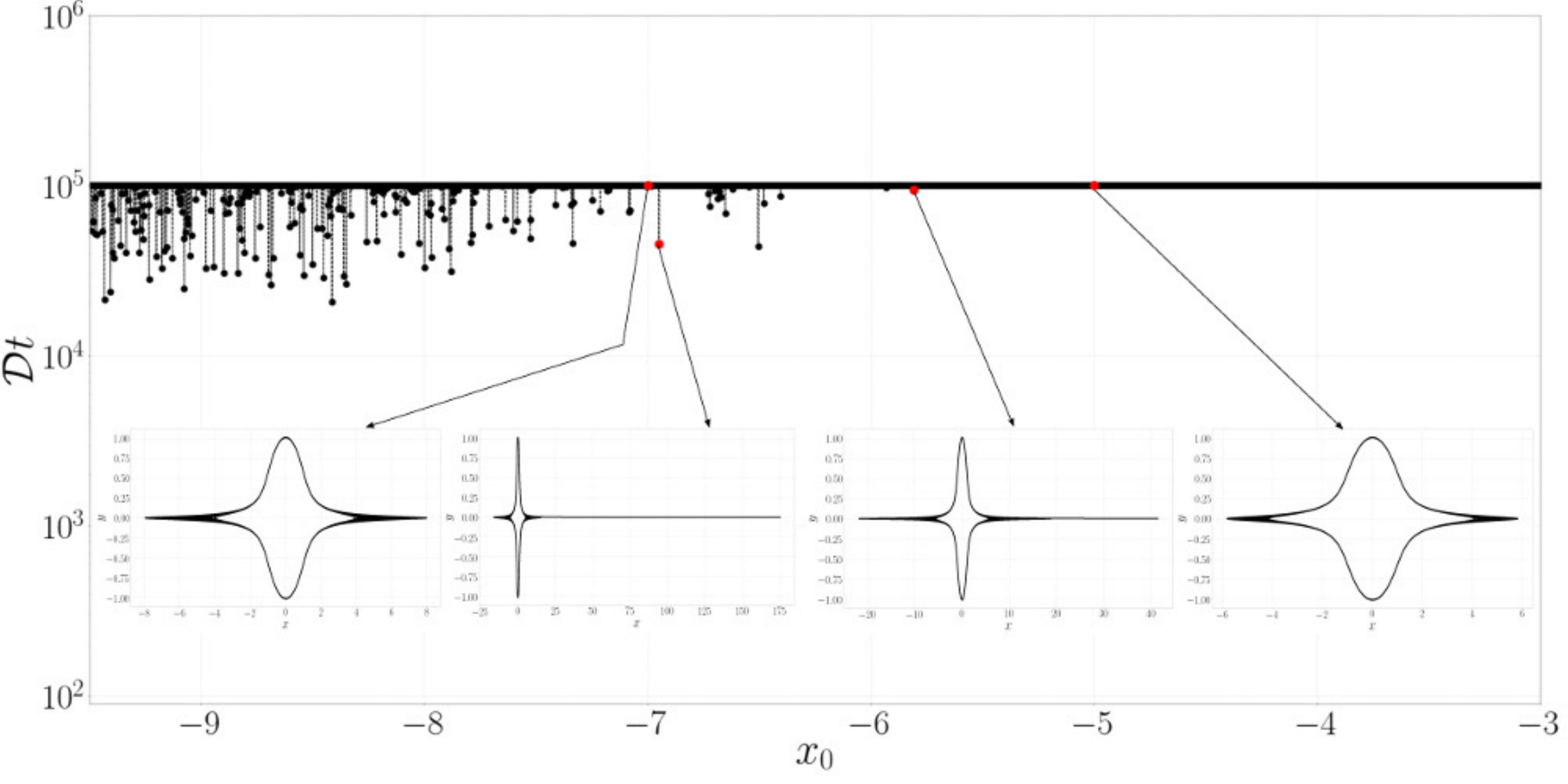}
  \caption{$C=0.2$}
  \end{subfigure}
  \end{center}
  \caption{Sensitive dependence of the residence time of fluid pathlines, on initial conditions, for a prolate spheroid($\xi_0=2.0(\kappa=1.13)$) in finite-$C$ precessional orbits (a) $C=\infty$, (b) $C=5.0$, (c) $C=0.5$, (d) $C=0.05$ and (e) $C=0.2$. The insets in figure 18e shows that the highest residence times (plateau points) correspond to bounded pathlines, while those off the plateau lead to pathlines that open up over times shorter than the integration time($8000 T_j$); the insets in 18d confirm that all of the pathlines for this case are bounded. }
\label{fig:closed3D_time}
\end{figure}

\begin{figure}
  \begin{subfigure}[b]{.48\textwidth}
  \centering
  \includegraphics[scale=0.22]{./figures_lowres/prolate3D_xi0_2_C_0_2_x0_-3_-9_5.pdf}
  \caption{$t_{max}=8000T_j$}
  \end{subfigure}
  \begin{subfigure}[b]{.48\textwidth}
  \centering
  \includegraphics[scale=0.22]{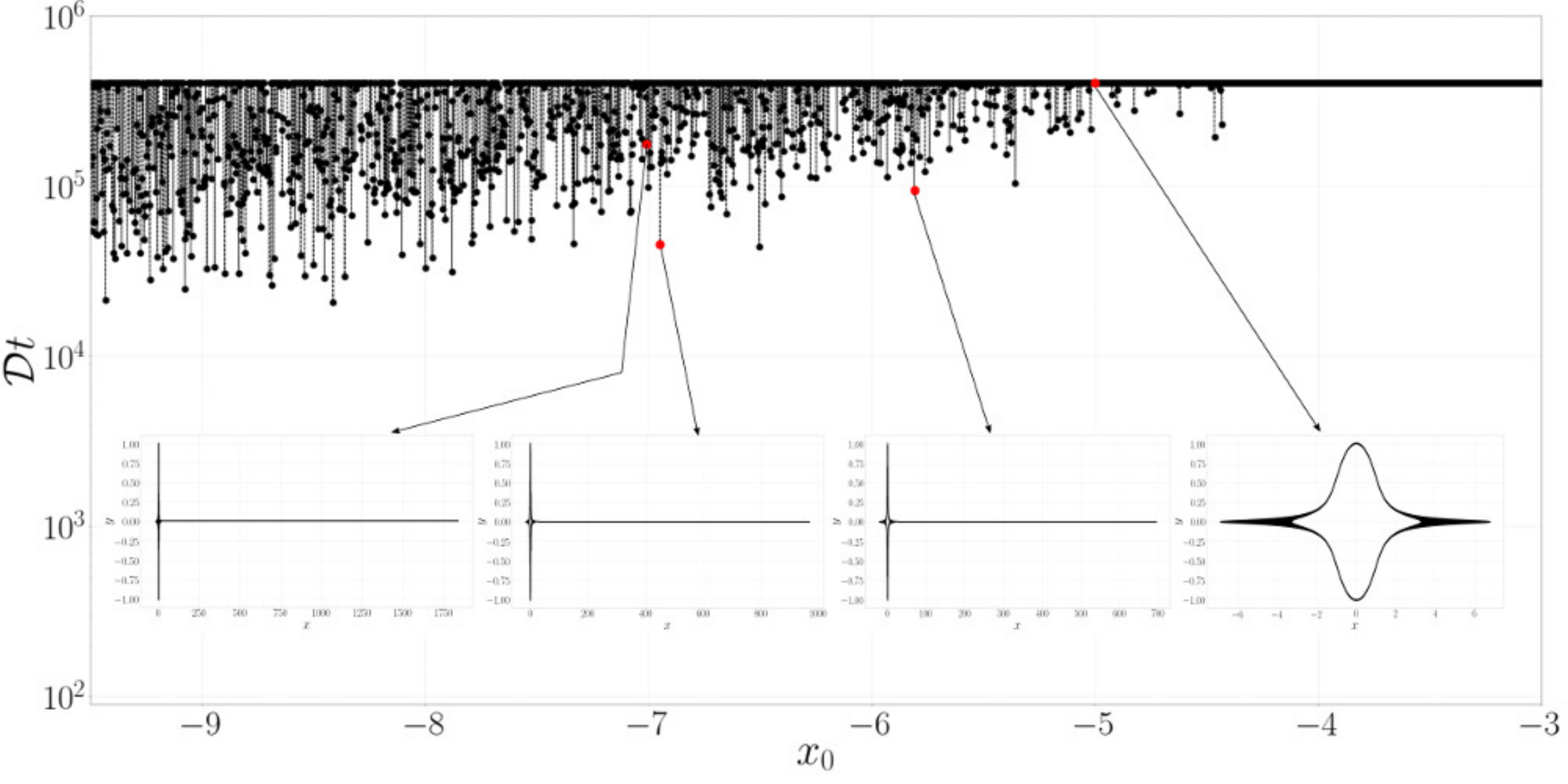}
  \caption{$t_{max}=32000T_j$}
  \end{subfigure}
  \caption{The residence time of fluid pathlinse for a prolate spheroid($\xi_0=2(\kappa=1.13)$), in Jeffery orbit corresponding to $C=0.2$, for two different integration times: (a) $t_{max}=8000T_j$, (b) $t_{max}=32000T_j$. The fraction of singular open pathlines increases significantly for $t_{max} = 32000 T_j$; sample pathlines are shown for points on the plateau as well as for ones off it.}
\label{fig:closed3D_long}
\end{figure}

Having established the signatures of chaotic scattering in the residence time plot for a tumbling spheroid, in figures \ref{fig:closed3D_time}a-d, we consider analogous residence time plots for the spheroid in other finite-$C$ precessional orbits. The spheroid is the same as that in figure \ref{fig:closed_time_res}, and the orbits chosen correspond to $C = \infty$(already shown), 5, 0.5, 0.1 and 0.05, respectively, with the initial point again restricted to the interval (-9.5,-3) on the negative $x$-axis. The plots appear to show a progressive decrease in the extent of irregularity with decreasing $C$. One reason for this decrease is that, for lower $C$'s, there appear regular intervals between those that correspond to chaotic scattering, and this interlacing behavior will be seen in more detail in the next section; note that this interlacing tendency is, in fact, already present for $C = \infty$, although not evident owing to the much smaller scale (see the inset of figure \ref{fig:closed3D_time}a which presents a magnified view). The decrease in irregularity is also because, for the smaller $C$'s, an increasing fraction of pathlines do not open for the duration of the numerical integration($t_{max} = 8000 T_j$); examples of such bounded pathlines are shown in figures \ref{fig:closed3D_time}d and e. Such bounded pathlines, arising as an artifact of the finite integration duration, begin to populate an upper horizontal plateau with residence times equal to the maximum integration time; in fact, figure 18(d) for $C$ = 0.05 consists entirely of pathlines that do not open over the duration of the integration. However, as shown in figure \ref{fig:closed3D_long},  the fraction of initial conditions leading to bounded pathlines for the case $C$ = 0.2 decreases perceptibly for an integration duration ($t_{max} = 32000 T_j$) four times the one chosen for figure \ref{fig:closed3D_time}. Thus, the fraction of bounded pathlines appears to depend on the duration of the numerical integration. Although impossible to ascertain by numerical means, the limiting integrable case of a spinning spheroid\,($C$ = 0) is a singular one, with a spheroid rotating in an orbit with $C$ small but finite nevertheless leading to chaotic scattering; the associated large residence times are, however, inaccessible to numerics.

\subsubsection{Local graphs of the invariant manifolds and the chaotic saddle}
\begin{figure}
  \begin{subfigure}[b]{.48\textwidth}
  \centering
  \includegraphics[scale=0.2]{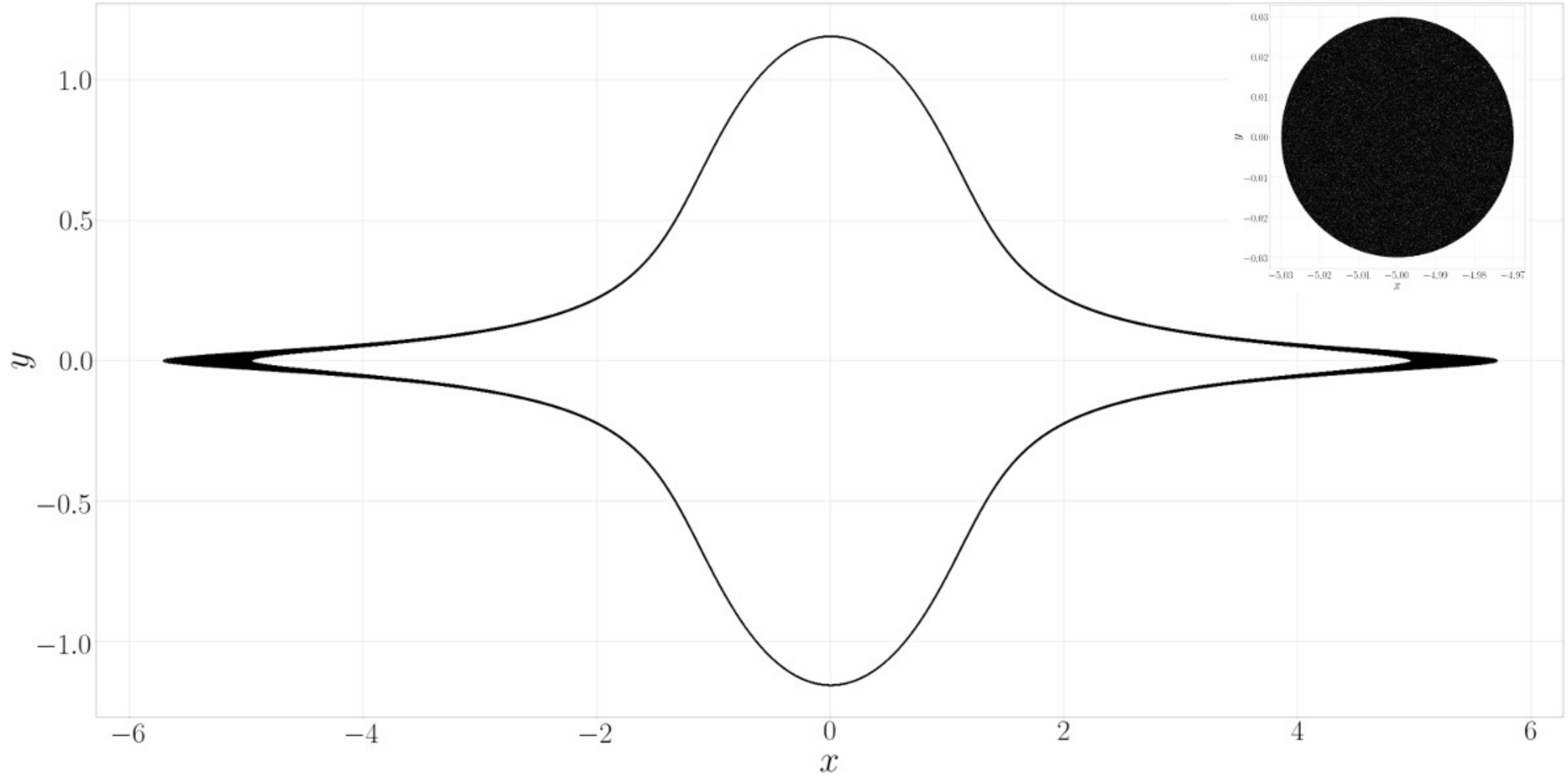}
  \caption{$t=-2000 T_j$}
  \end{subfigure}
  \begin{subfigure}[b]{.48\textwidth}
  \centering
  \includegraphics[scale=0.038]{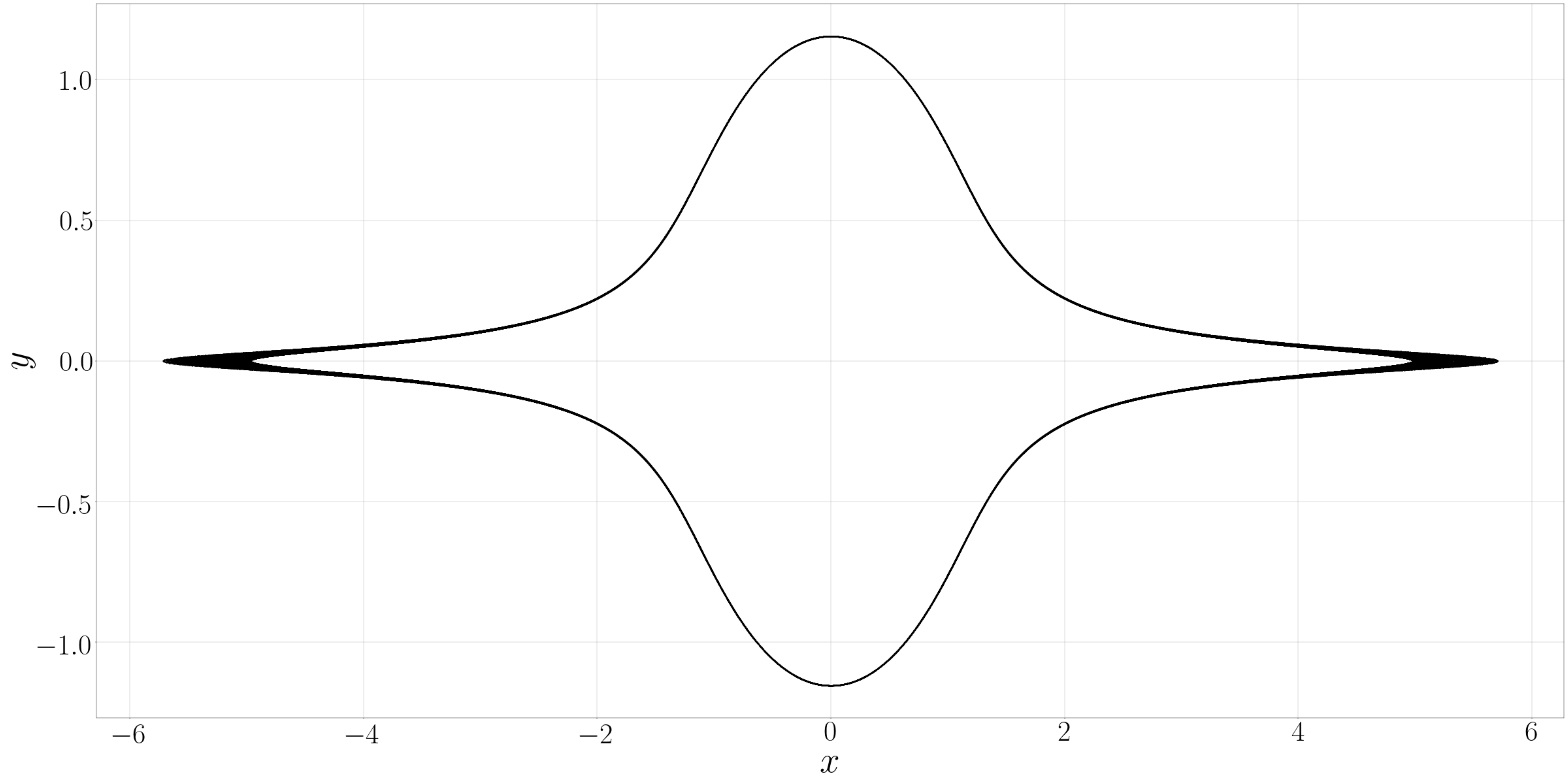}
  \caption{$t=2000 T_j$}
  \end{subfigure}
  \begin{subfigure}[b]{.48\textwidth}
  \centering
  \includegraphics[scale=0.2]{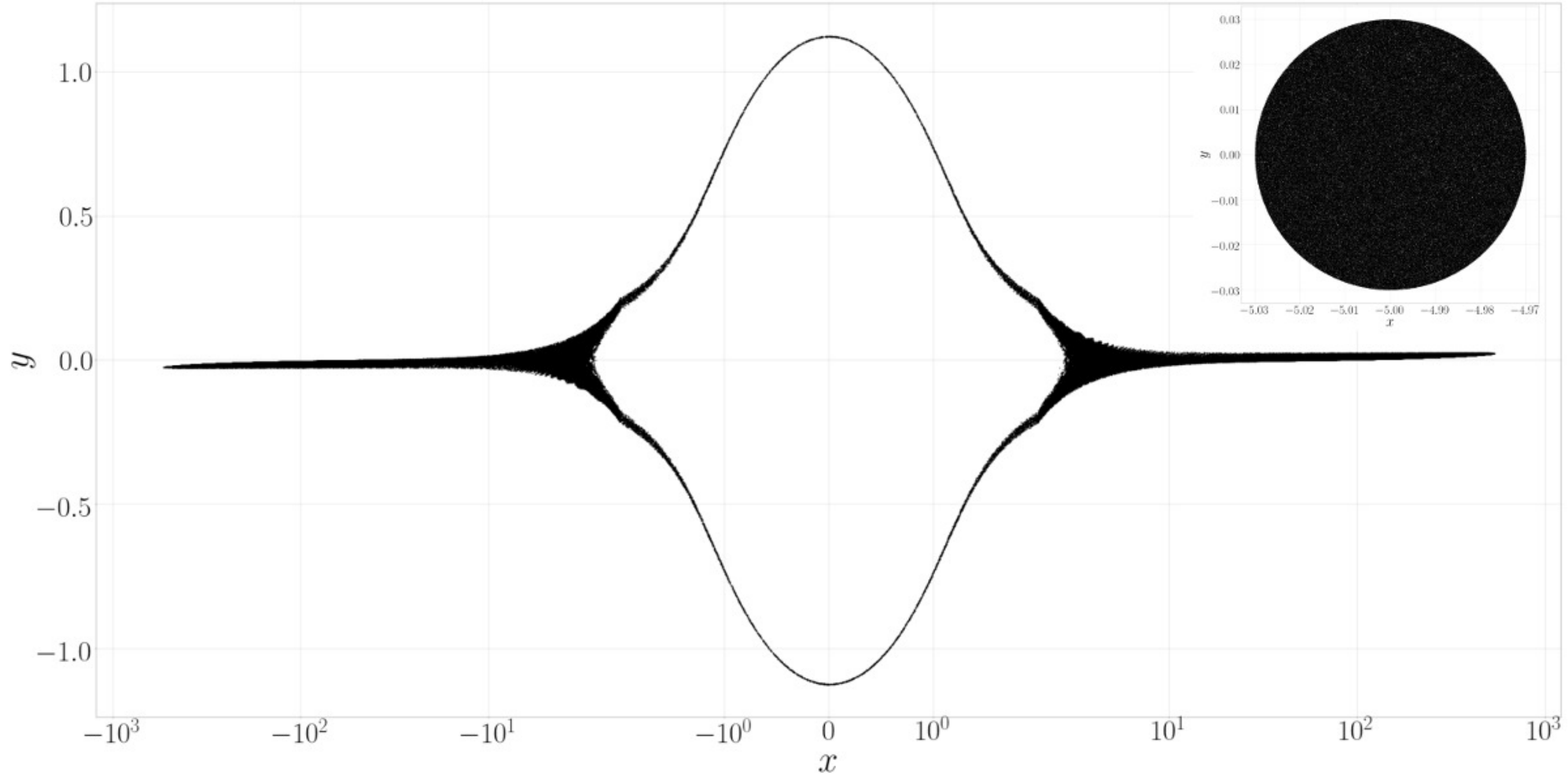}
  \caption{$t=-2000 T_j$}
  \end{subfigure}
  \hfill
  \begin{subfigure}[b]{.48\textwidth}
  \centering
  \includegraphics[scale=0.2]{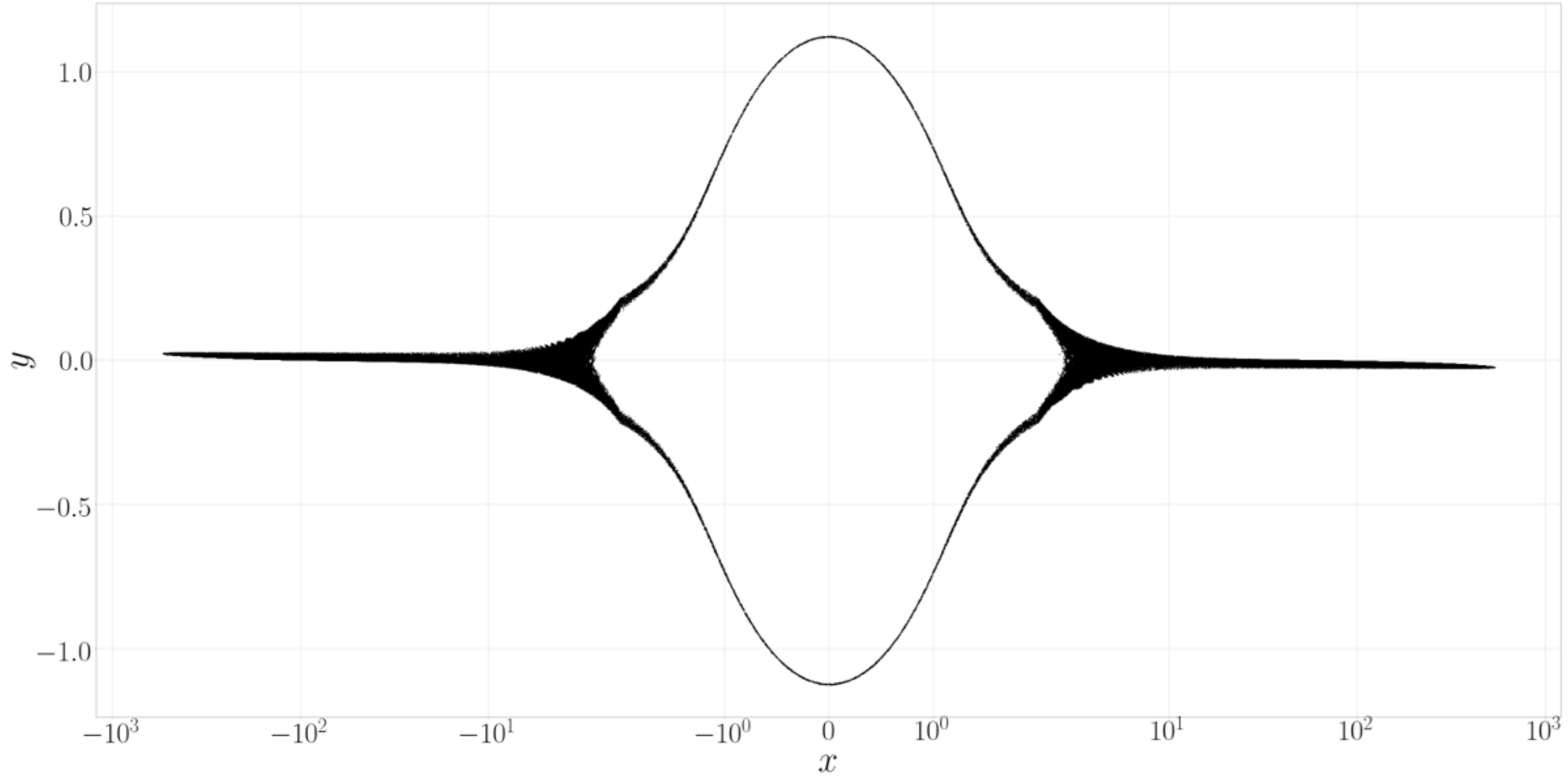}
  \caption{$t=2000 T_j$}
  \end{subfigure}
  \caption{The evolution of an initially circular blob of radius = 0.03, located at (-5,0,0)(shown in the inset) for a sphere and a prolate spheroid with $\kappa = 1.03(\xi_0=4.0)$ rotating in a Jeffery orbit with $C =$ 20: (a,c)forward integration (b,d)backward integration.}
\label{fig:blob0}
\end{figure}

\begin{figure}
  \centering
  \begin{subfigure}[b]{.32\textwidth}
  \centering
  \includegraphics[scale=0.15]{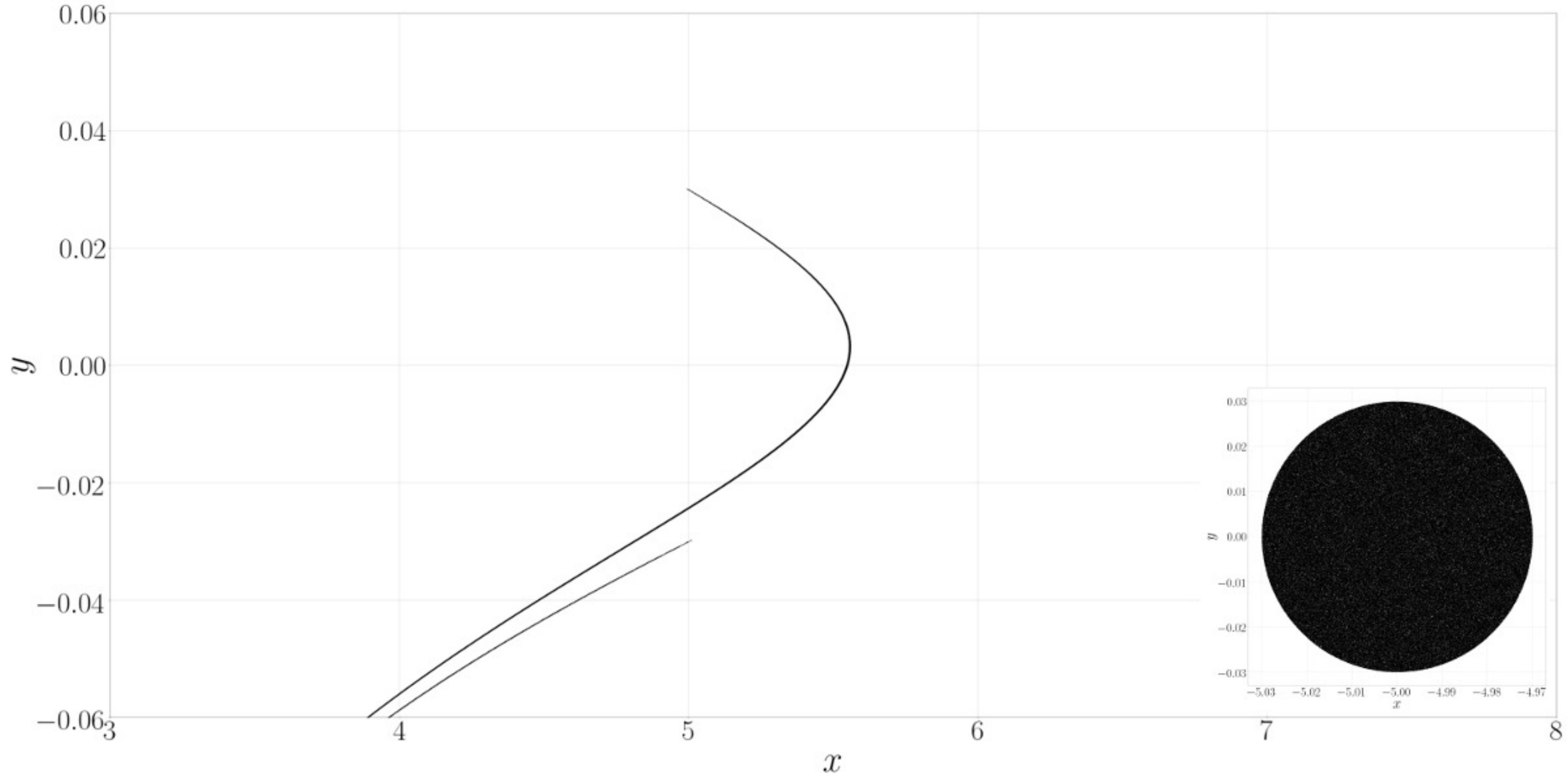}
  \caption{$t=50 T_j$}
  \end{subfigure}
  \begin{subfigure}[b]{.32\textwidth}
  \centering
  \includegraphics[scale=0.15]{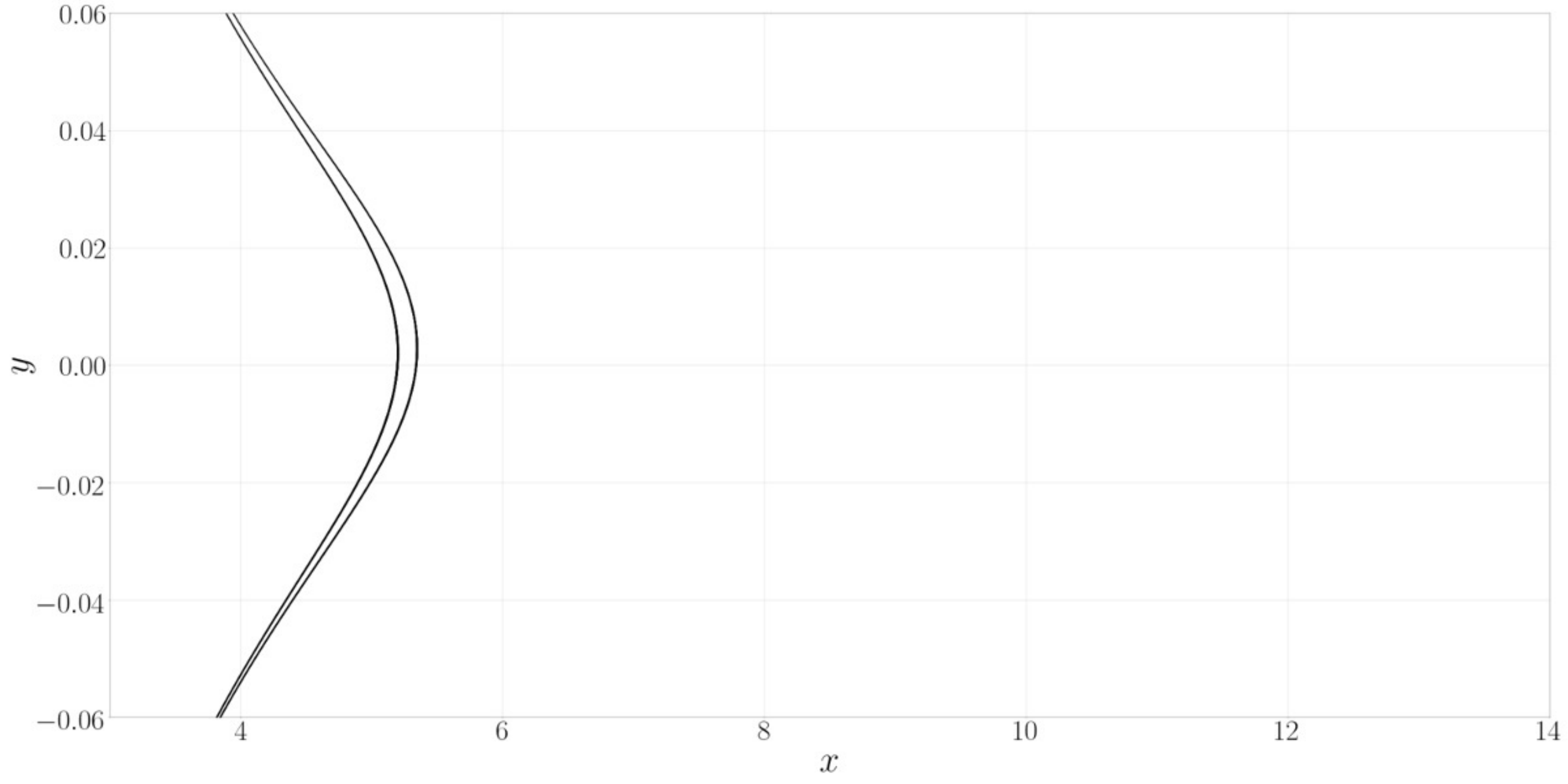}
  \caption{$t=68 T_j$}
  \end{subfigure}
  \begin{subfigure}[b]{.32\textwidth}
  \centering
  \includegraphics[scale=0.15]{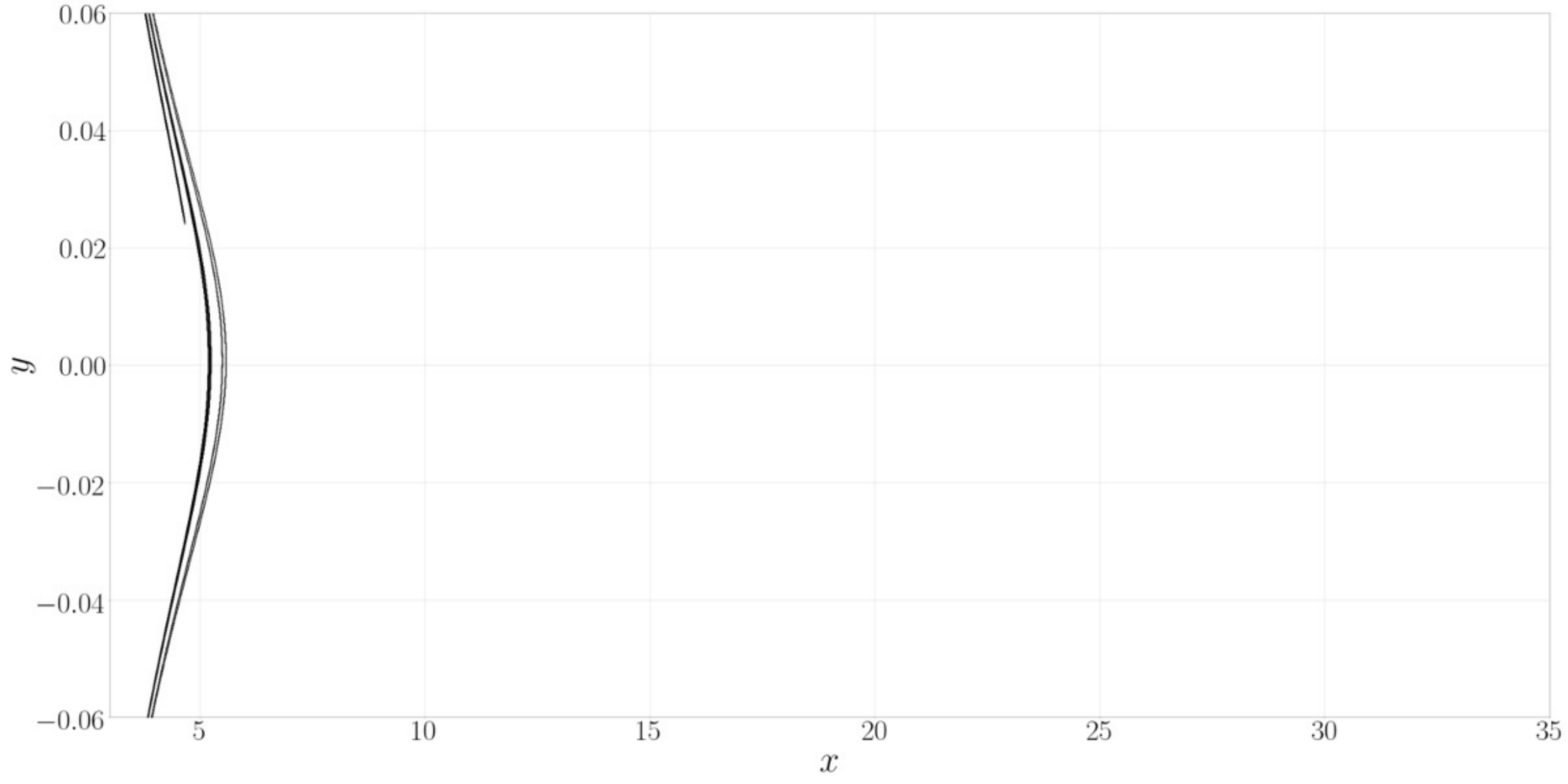}
  \caption{$t=200 T_j$}
  \end{subfigure}
  \begin{subfigure}[b]{.32\textwidth}
  \centering
  \includegraphics[scale=0.15]{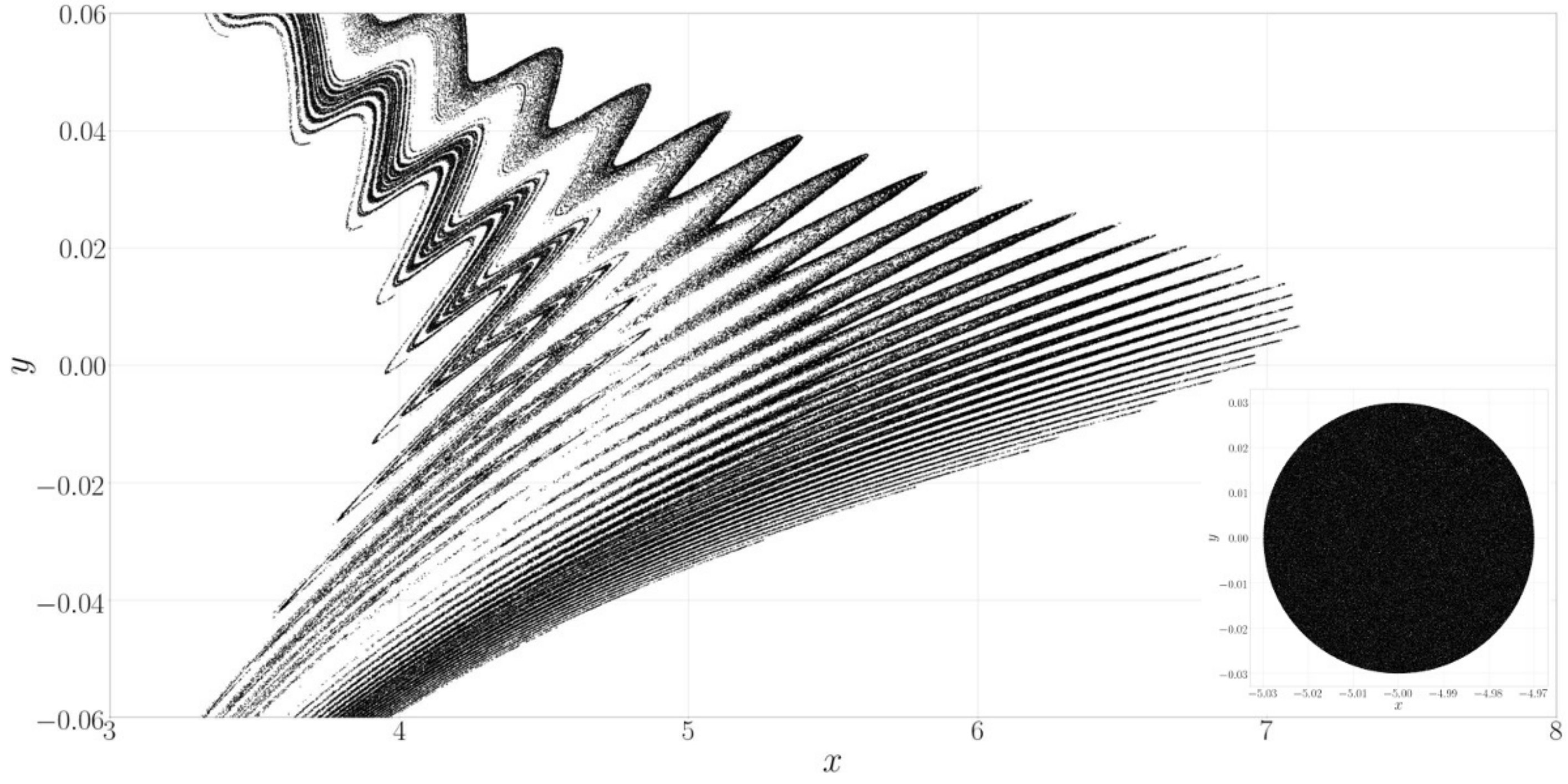}
  \caption{$t=50 T_j$}
  \end{subfigure}
  \begin{subfigure}[b]{.32\textwidth}
  \centering
  \includegraphics[scale=0.15]{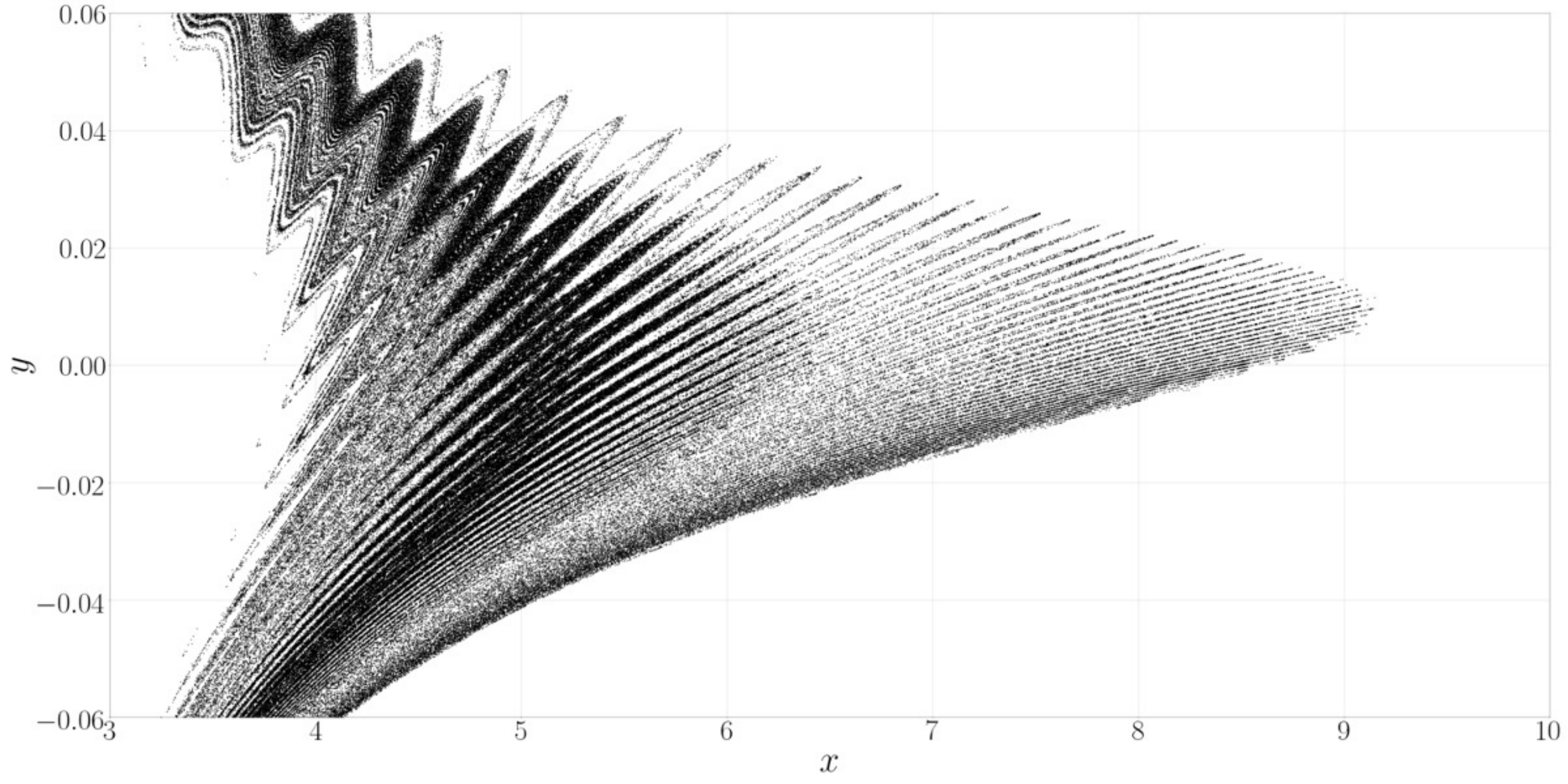}
  \caption{$t=68 T_j$}
  \end{subfigure}
  \begin{subfigure}[b]{.32\textwidth}
  \centering
  \includegraphics[scale=0.15]{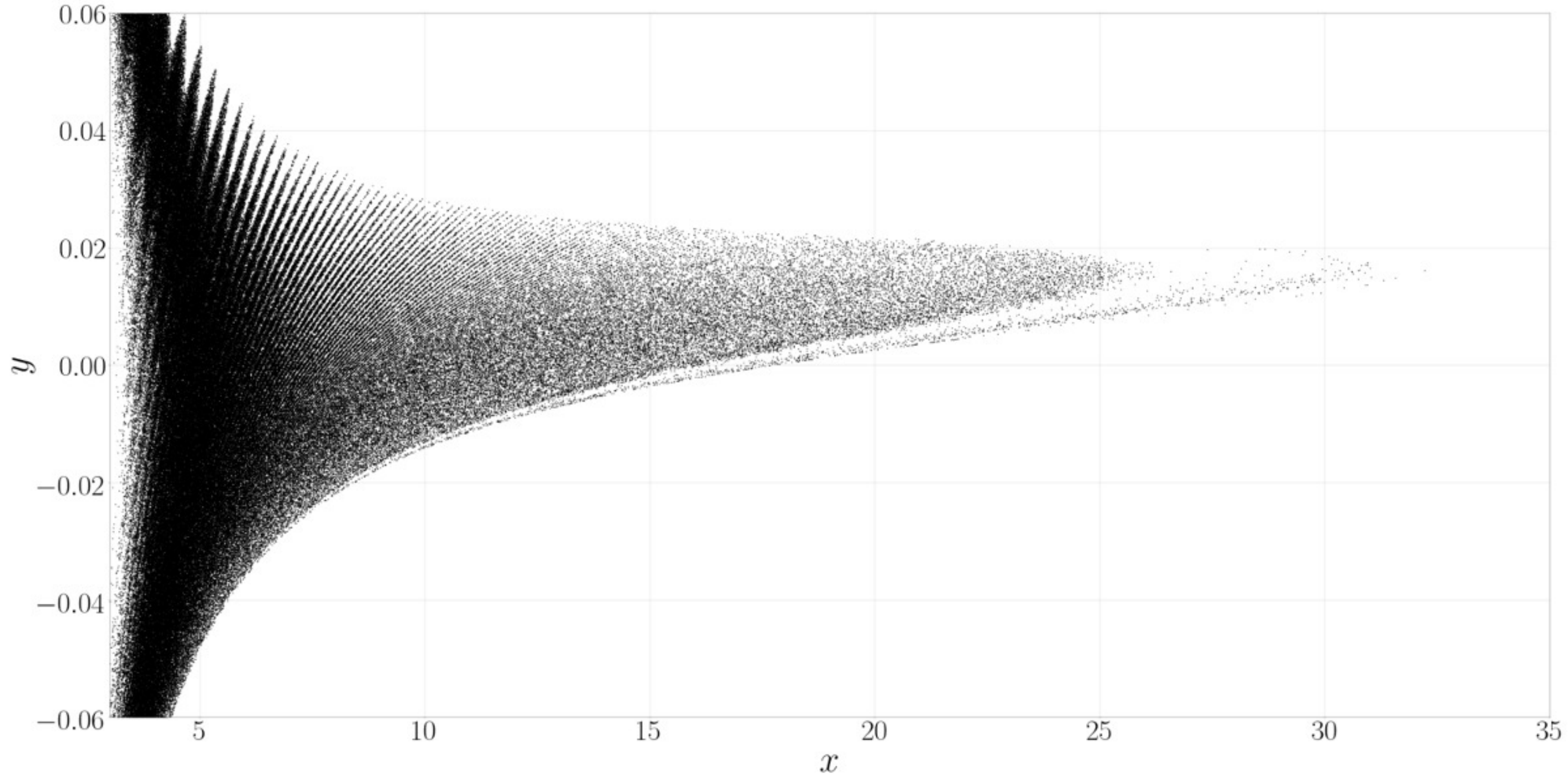}
  \caption{$t=200 T_j$}
  \end{subfigure}
  \caption{A comparison of the evolution of an initially circular blob for a sphere and a prolate spheroid ($\xi_0=4(\kappa=1.03)$) in a precessional orbit with $C$ = 20. The evolving blob, for the spheroid, traces out the unstable manifold, of the underlying chaotic saddle, for large times.}
\label{fig:blob}
\end{figure}

Graphical representations of the unstable and stable manifolds of the chaotic saddle are another qualitative signature of chaos. Here, we numerically evolve an initially circular blob of `dye' \citep{thiffeault} located along the negative $x$-axis on account of being advected by the fluid motion induced by the rotating spheroid. A forward integration in time must trace out the unstable manifold of the chaotic saddle for sufficiently long times, while a backward-in-time integration of the equations yields the stable manifold. Figure \ref{fig:blob0} compares the evolution of the aforementioned circular blob (made up of $10^6$ randomly distributed initial points) for a sphere and for a (nearly spherical) prolate spheroid with $\xi_0 = 4 (\kappa = 1.03)$; the spheroid is chosen to rotate in a Jeffery orbit with $C = 20(\phi_{j0} = 0)$. Figure \ref{fig:blob0}a and b show that the blob for the case of a sphere, although significantly distorted for large times owing to differential convection, has, nevertheless a finite spatial extent - the distorted blob lies within the interval (-6,6) for all time. This is because the initial blob has been chosen to lie entirely within the region of closed streamlines, and is therefore always bounded by the largest closed streamline that passes through one of the initial points on its periphery. In contrast, figures \ref{fig:blob0}c and d show that the blob, for a spheroid, is sheared out to arbitrarily large distances in the flow direction as time goes to plus and minus infinity on account of the singular open pathlines discussed above (note the differing horizontal scales in figures \ref{fig:blob0}a and b, and in figures \ref{fig:blob0}c and d). While figure \ref{fig:blob0} highlights the singular role of non-sphericity with regard to the long-time temporal evolution of a blob, which in turn has implications for the rate of mass/heat transport from the spheroid(see conclusions section), it is figure \ref{fig:blob} that highlights the striking differences between the sphere and spheroid cases even for finite times. The magnified views in figure \ref{fig:blob} clearly show the fine-scaled features(wiggles) that emerge in the evolving blob, only for the spheroid, on account of the convoluted nature of the underlying unstable manifold of the chaotic saddle. In fact, in figure 21f, the initial distribution of points has been sheared out to an extent that the small-scale wiggles associated with the underlying chaotic saddle are no longer well resolved.

\begin{figure}
  \centering
  \begin{subfigure}[b]{.32\textwidth}
  \centering
  \includegraphics[scale=0.15]{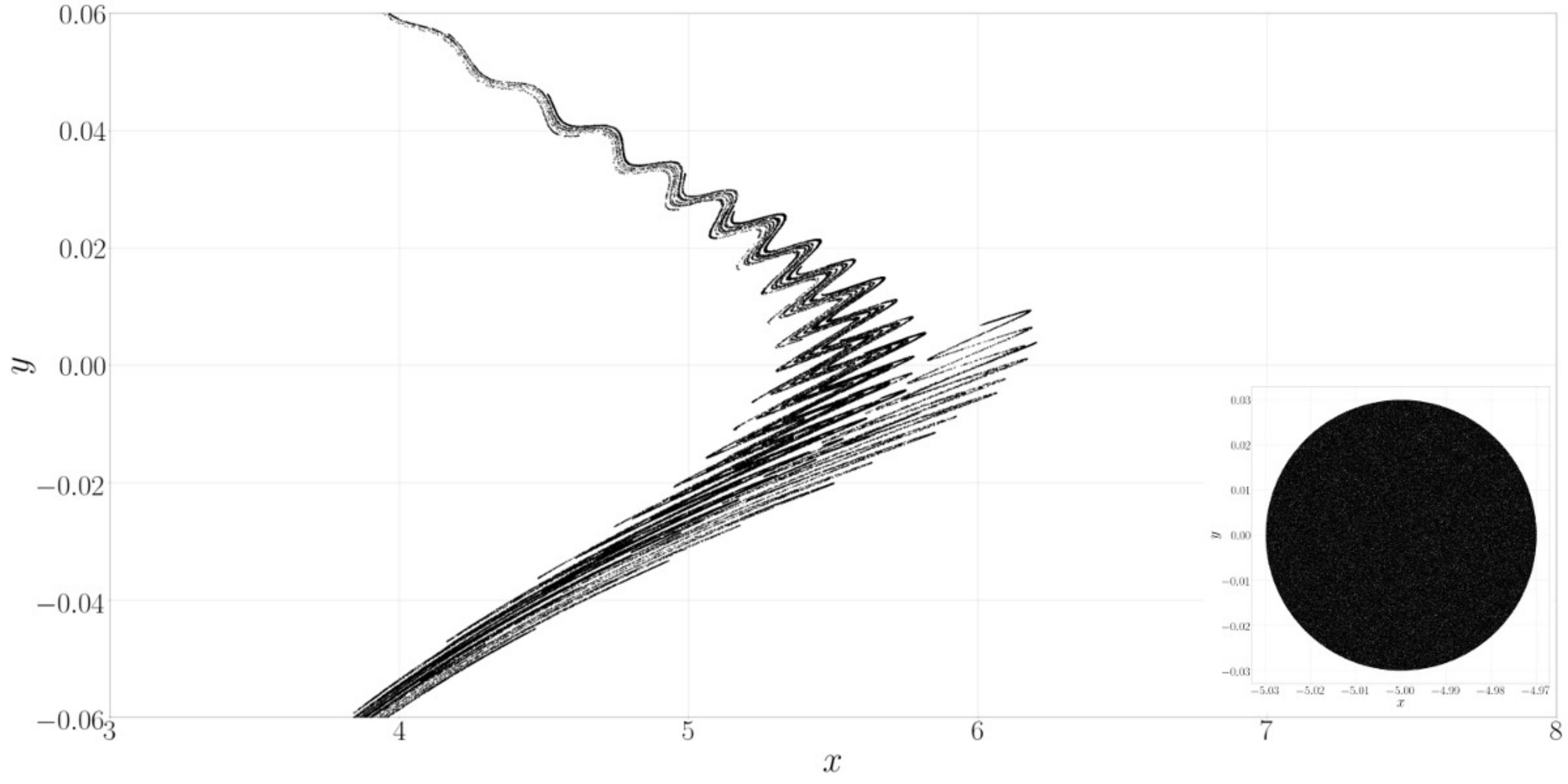}
  \caption{$t=50 T_j$}
  \end{subfigure}
  \begin{subfigure}[b]{.32\textwidth}
  \centering
  \includegraphics[scale=0.15]{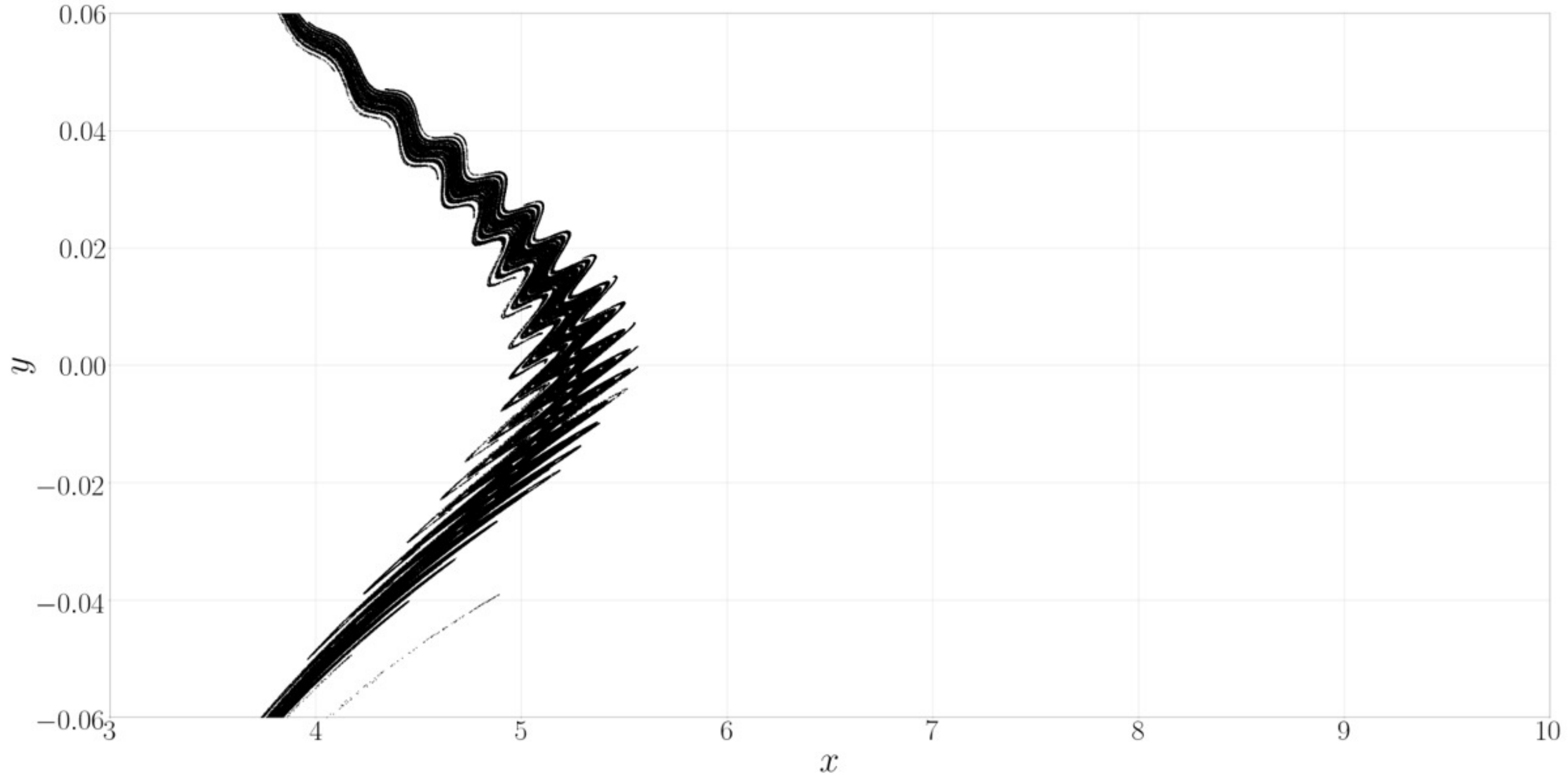}
  \caption{$t=68 T_j$}
  \end{subfigure}
  \begin{subfigure}[b]{.32\textwidth}
  \centering
  \includegraphics[scale=0.15]{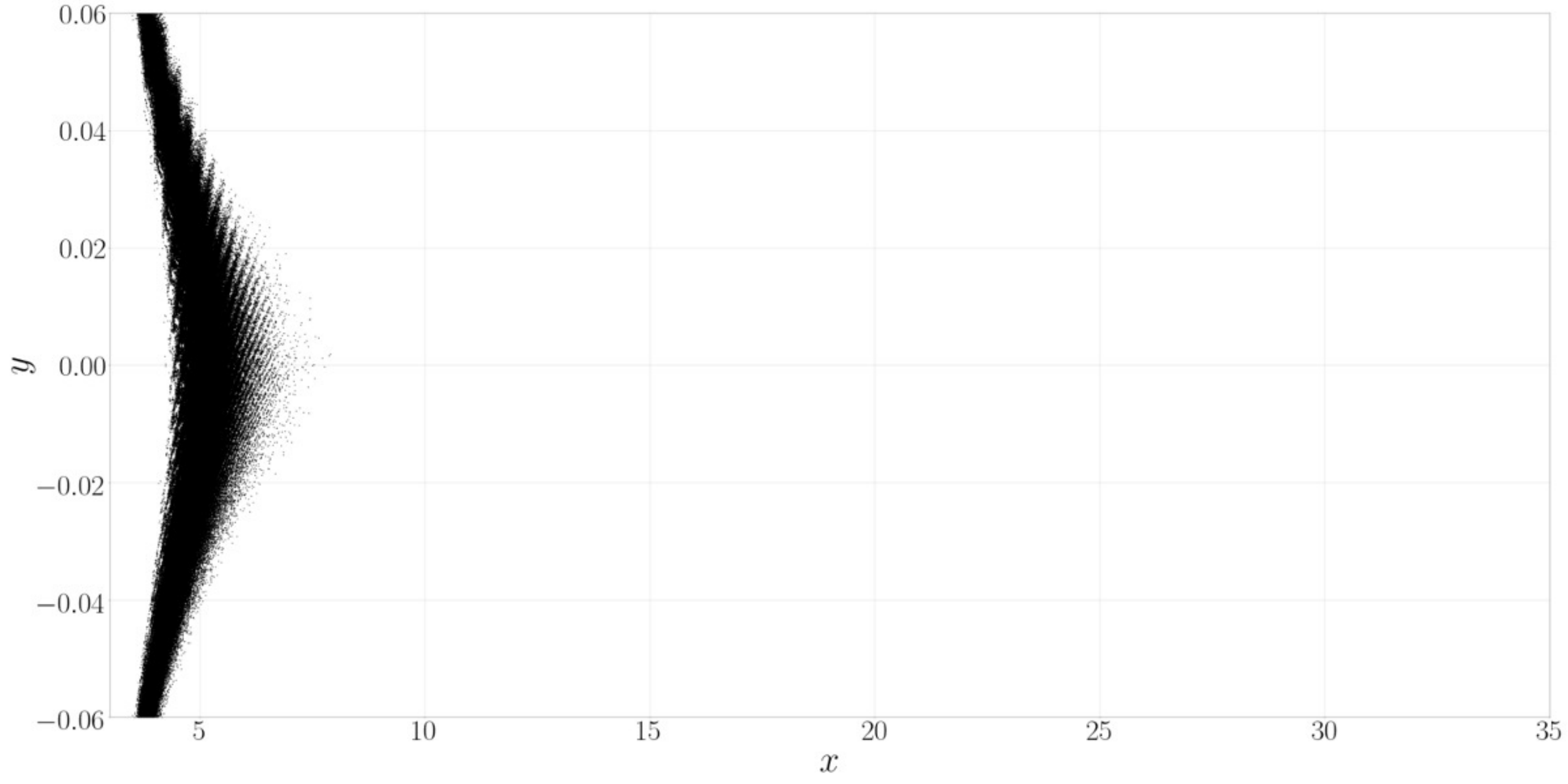}
  \caption{$t=200 T_j$}
  \end{subfigure}
  \begin{subfigure}[b]{.32\textwidth}
  \centering
  \includegraphics[scale=0.15]{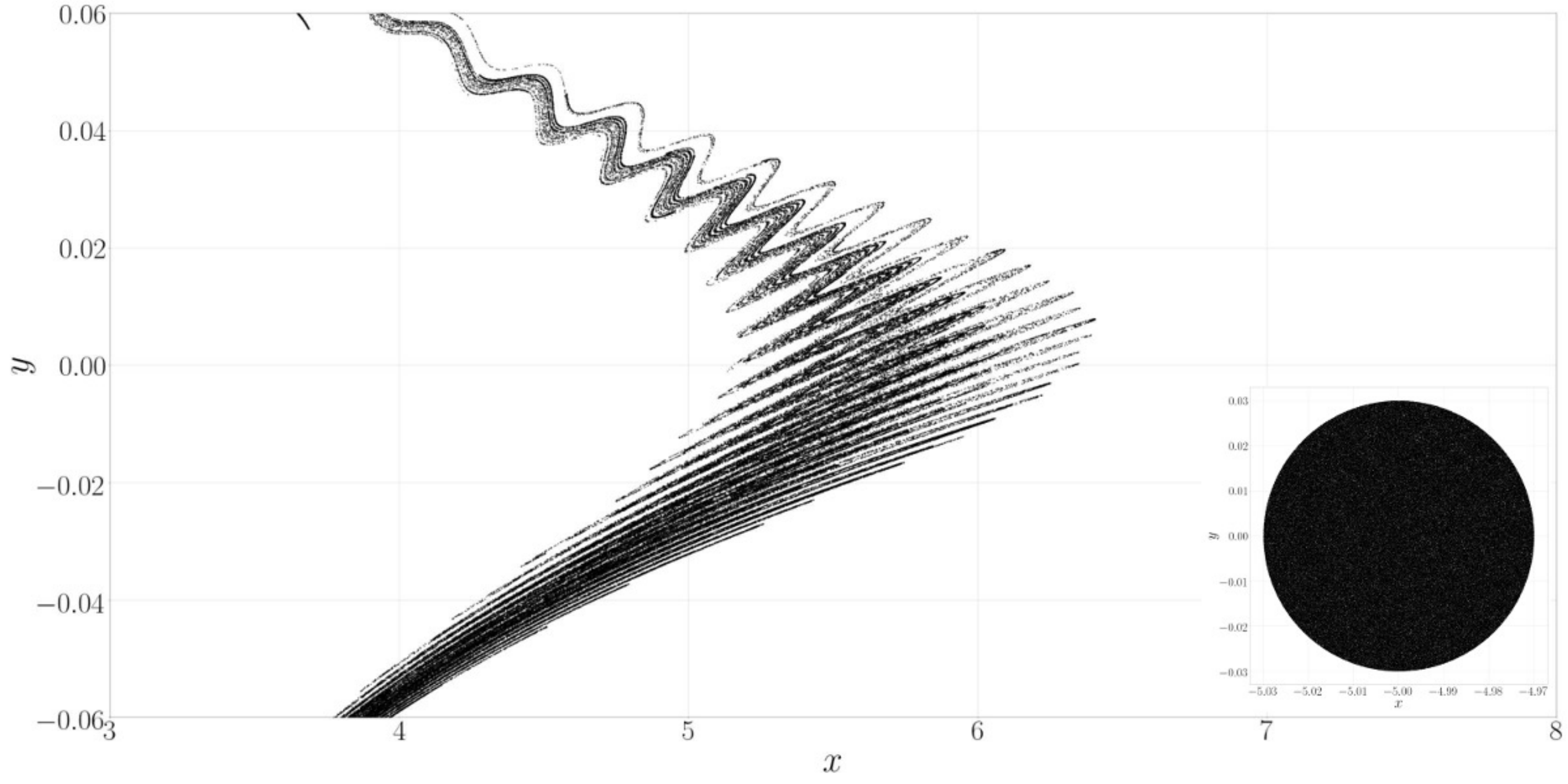}
  \caption{$t=50 T_j$}
  \end{subfigure}
  \begin{subfigure}[b]{.32\textwidth}
  \centering
  \includegraphics[scale=0.15]{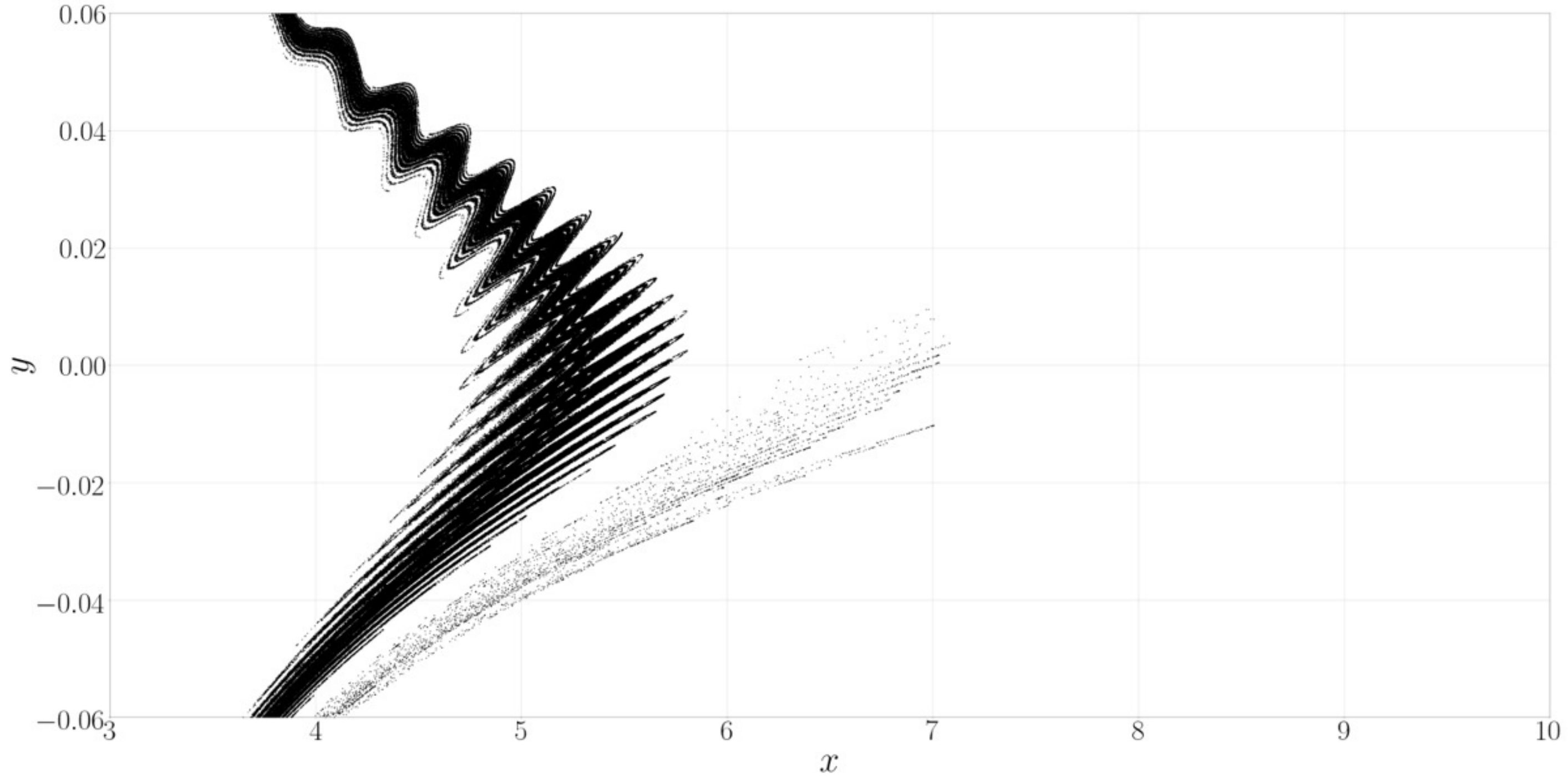}
  \caption{$t=68 T_j$}
  \end{subfigure}
  \begin{subfigure}[b]{.32\textwidth}
  \centering
  \includegraphics[scale=0.15]{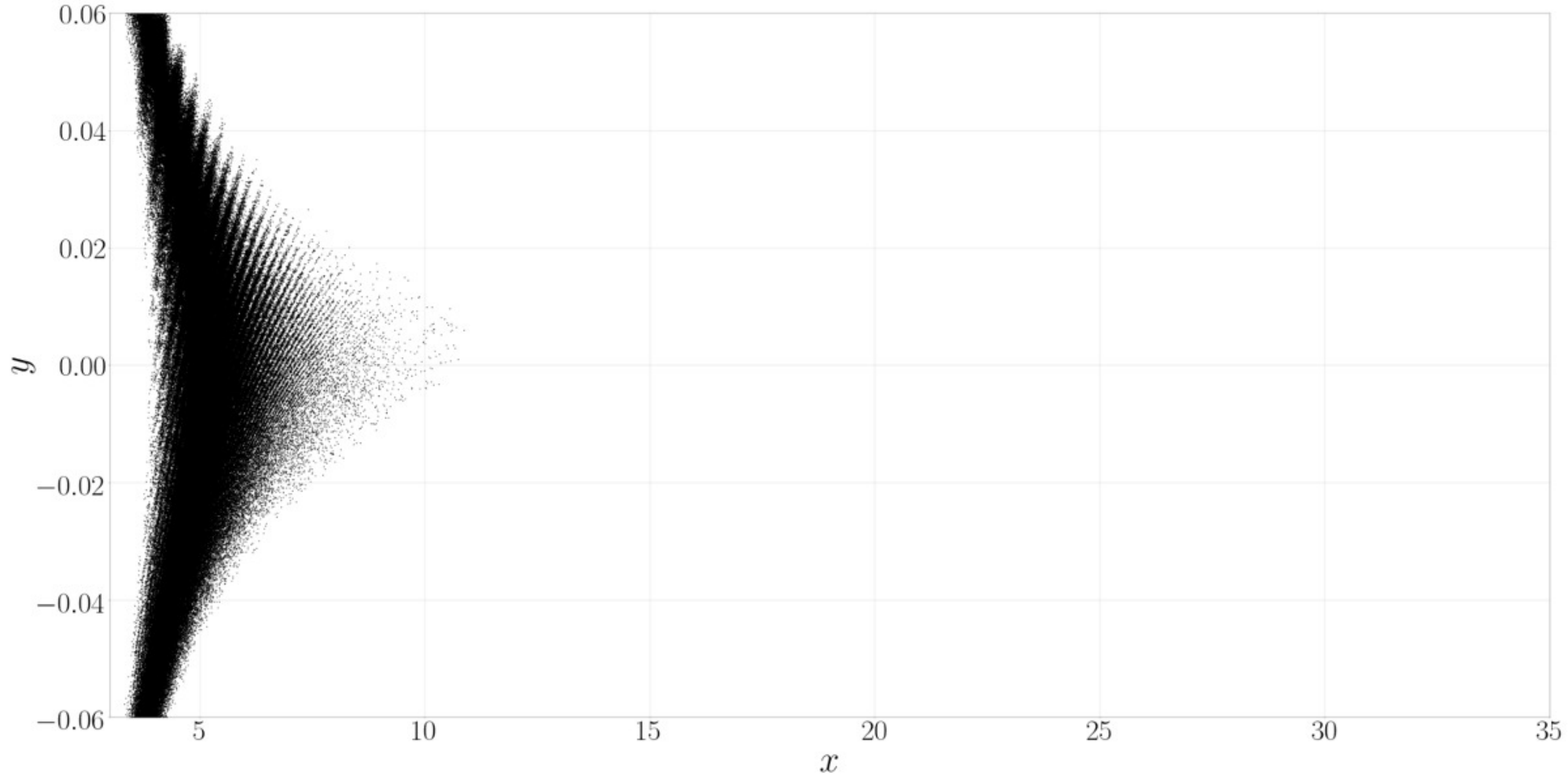}
  \caption{$t=200 T_j$}
  \end{subfigure}
  \begin{subfigure}[b]{.32\textwidth}
  \centering
  \includegraphics[scale=0.15]{./figures_lowres/prolate3DC20_xi0_4_50Tj_umanifod_mag.pdf}
  \caption{$t=50 T_j$}
  \end{subfigure}
  \begin{subfigure}[b]{.32\textwidth}
  \centering
  \includegraphics[scale=0.15]{./figures_lowres/prolate3DC20_xi0_4_68Tj_umanifod_mag.pdf}
  \caption{$t=68 T_j$}
  \end{subfigure}
  \begin{subfigure}[b]{.32\textwidth}
  \centering
  \includegraphics[scale=0.15]{./figures_lowres/prolate3DC20_xi0_4_200Tj_umanifod_mag.pdf}
  \caption{$t=200 T_j$}
  \end{subfigure}
  \caption{A comparison of evolving blobs for prolate spheroids in a precessional orbit with $C$ = 20:(a,b,c)$\xi_0=8.0(\kappa=1.008)$, (d,e,f)$\xi_0=6.0(\kappa=1.01)$ and (g,h,i)$\xi_0=4.0(\kappa=1.03)$. The scale of the wiggles, and the horizontal spread of the distorted blob, increase with increasing aspect ratio.}
\label{fig:blob_compare}
\end{figure}

\begin{figure}
\centering
  \begin{subfigure}[b]{.96\textwidth}
  \centering
  \includegraphics[scale=0.35]{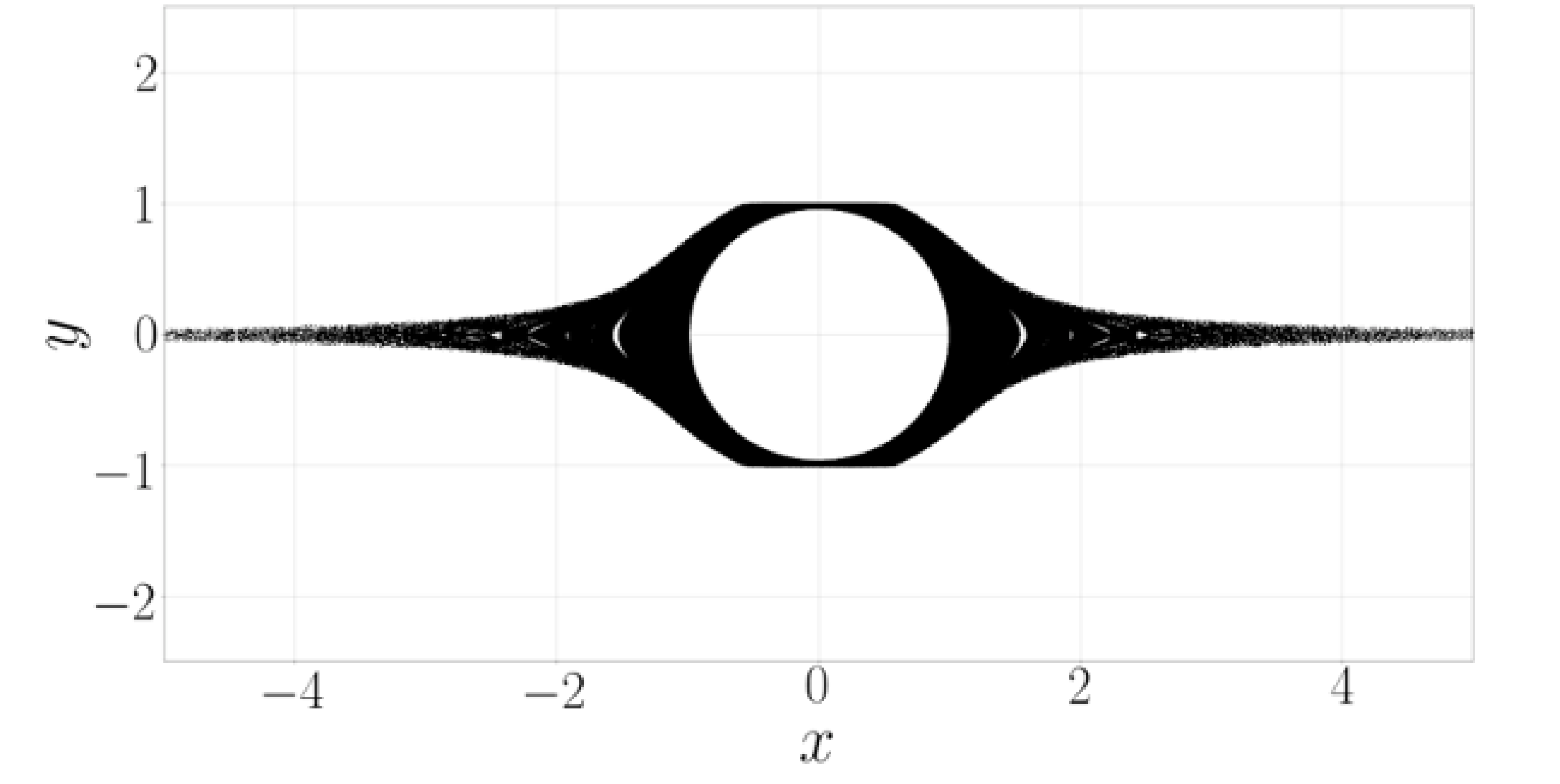}
  \caption{}
  \end{subfigure}
  \begin{subfigure}[b]{.96\textwidth}
  \centering
  \includegraphics[scale=0.35]{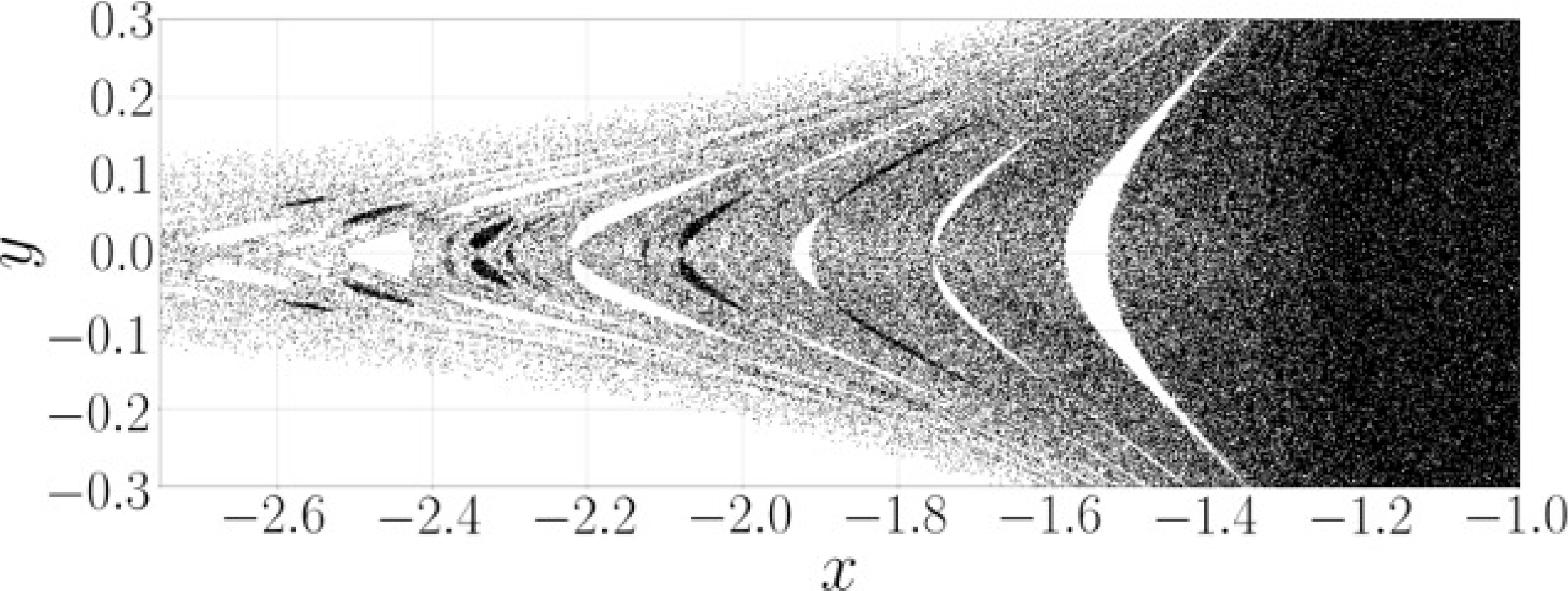}
  \caption{}
  \end{subfigure}
  \begin{subfigure}[b]{.96\textwidth}
  \centering
  \includegraphics[scale=0.35]{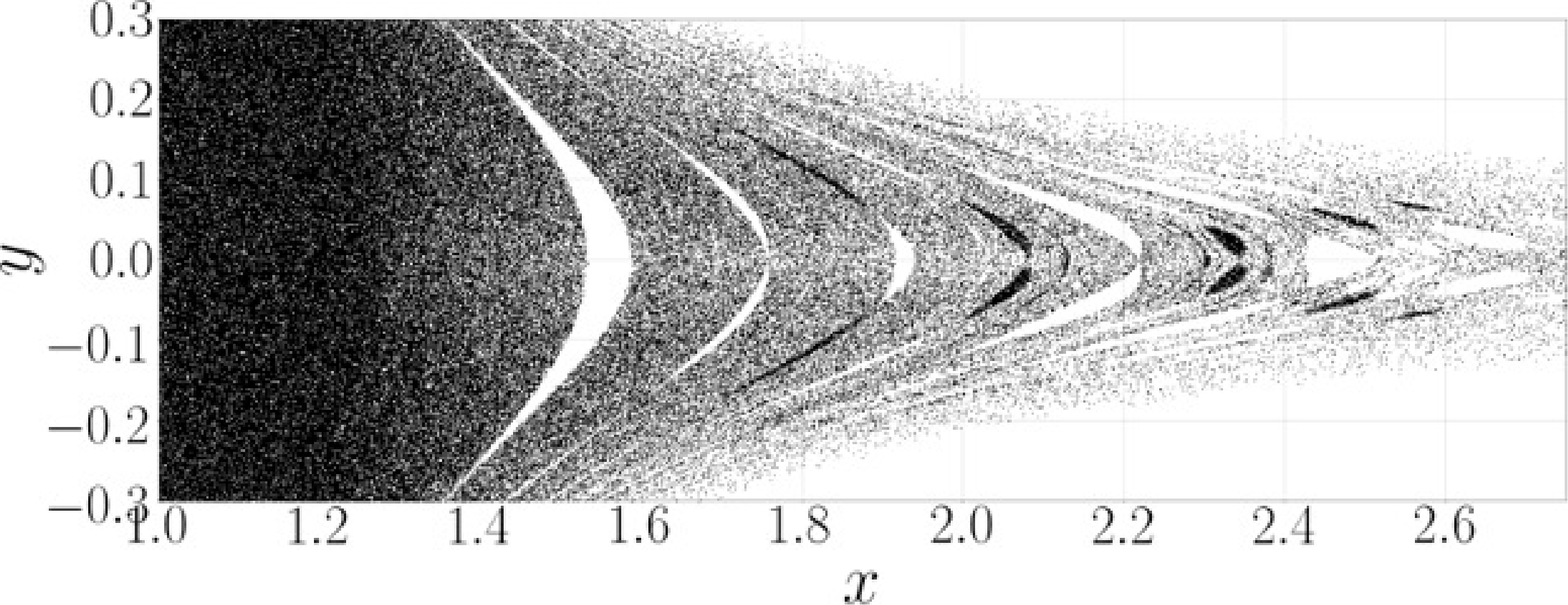}
  \caption{}
  \end{subfigure}
\caption{Local graph of the initial configuration of the chaotic saddle, plotted for a prolate spheroid with $\xi_0 = 4 (\kappa=1.03)$ in $C=20$: (a)Full view, (b)Left magnified view, (c)Right magnified view.}
\label{fig:blob_saddle}
\end{figure}

Figure \ref{fig:blob_compare} shows evolving blobs for spheroids of different aspect ratios, and here, the scale of the aforementioned wiggles is seen to becomes larger with increasing aspect ratio. Further, the horizontal extent of the sheared blob, for large times, also increases substantially for high aspect ratio spheroids, implying a decrease in the `loopiness' of the singular open pathlines. In the next section we will see a similar behavior emerge from a different perspective, wherein the dependence of the residence time (of fluid pathlines) is analyzed from the point of view of an `impact parameter', defined as the gradient and/or vorticity coordinates of a fluid element far upstream.
Finally, in Figure \ref{fig:blob_saddle}, we give a representation of the invariant chaotic saddle, its projection onto the flow-gradient plane and in the neighborhood of the rotating spheroid, obtained in the following manner. The representation corresponds to a near-sphere, with $\xi_0 = 4 (\kappa = 1.03)$, and it therefore makes sense to relate this representation to the corresponding invariant set for a sphere, the latter being the region of closed streamlines contained within the separatrices above and below the flow axis. With this in mind, we initialize 18 million fluid elements uniformly distributed in a rectangular domain spanning the intervals (-5,5) and (-1,1) in the $x$ and $y$-directions, and integrate these over a duration of $1000 T_j$ both forward and backward in time. Those initial points that remain within a composite region, consisting of an ellipse for $|x| < 1.5$ with semi-minor axis equal to 1.143 (the y-coordinate of the sphere-separatrix at x = 0) and semi-major axis equal to 1.5, and the far-field form of the sphere-separatrix $(y = \pm \sqrt{1/3}x^{-3/2})$ for $1.5 < x < 5$, are plotted in the said figure; the composite region above is regarded as an approximation of the actual invariant region between the sphere-separatrices. The fine-scaled structure evident in Figure 23 is a representation of the invariant set responsible for the chaotic scattering signatures detailed above. Further, the emergent non-trivial foliated structure is also indicative of the singular nature of the spherical limit. For a sphere, the spatial extent of the initial distribution would have remained unchanged, since fluid elements within this region remain so regardless of the duration of integration.

\subsubsection{The fractal dimension of the chaotic saddle}
\begin{figure}
  \begin{center}
  \begin{subfigure}[b]{.48\textwidth}
  \centering
  \includegraphics[scale=0.23]{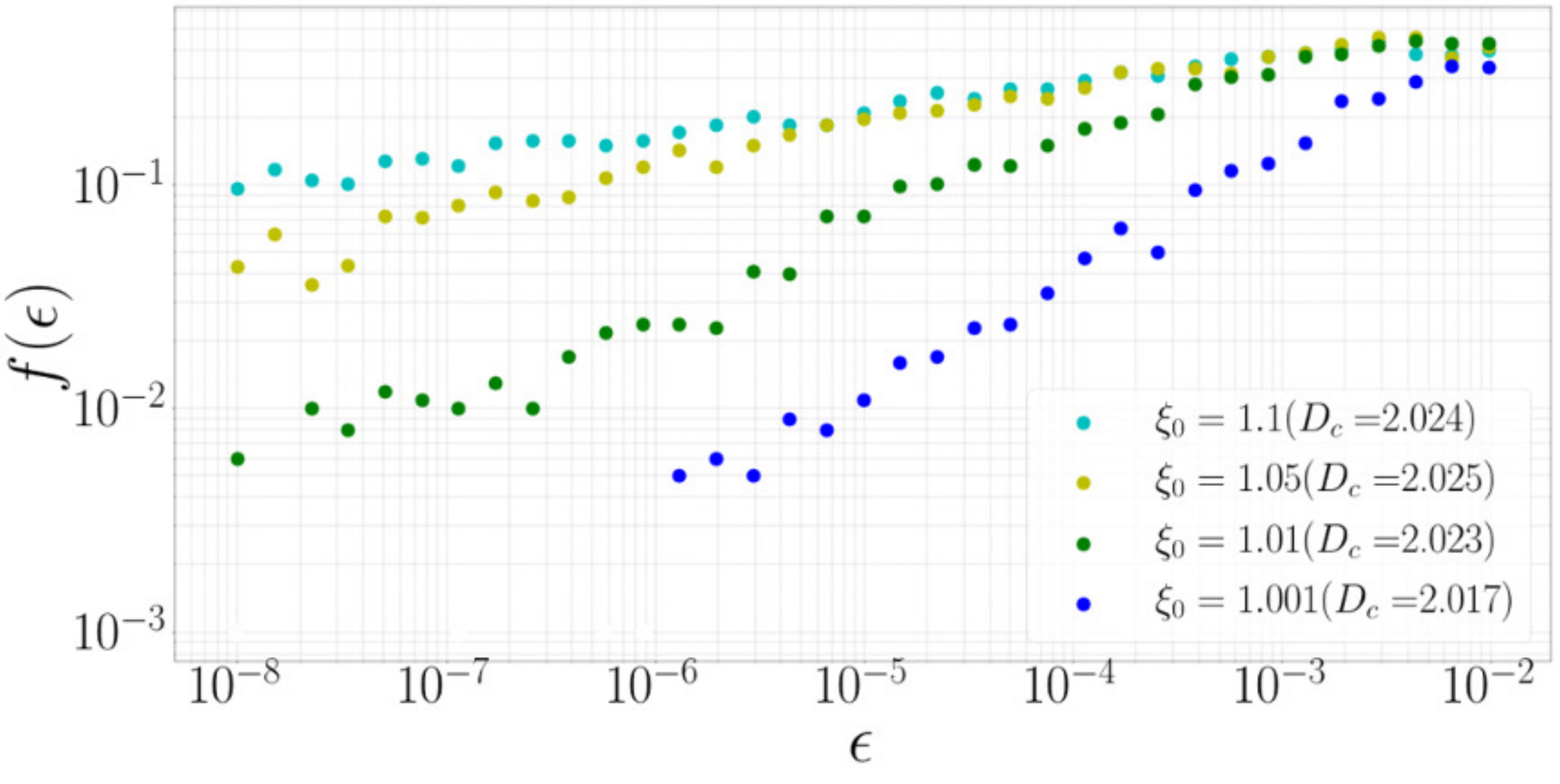}
  \caption{}
  \end{subfigure}
  \begin{subfigure}[b]{.48\textwidth}
  \centering
  \includegraphics[scale=0.23]{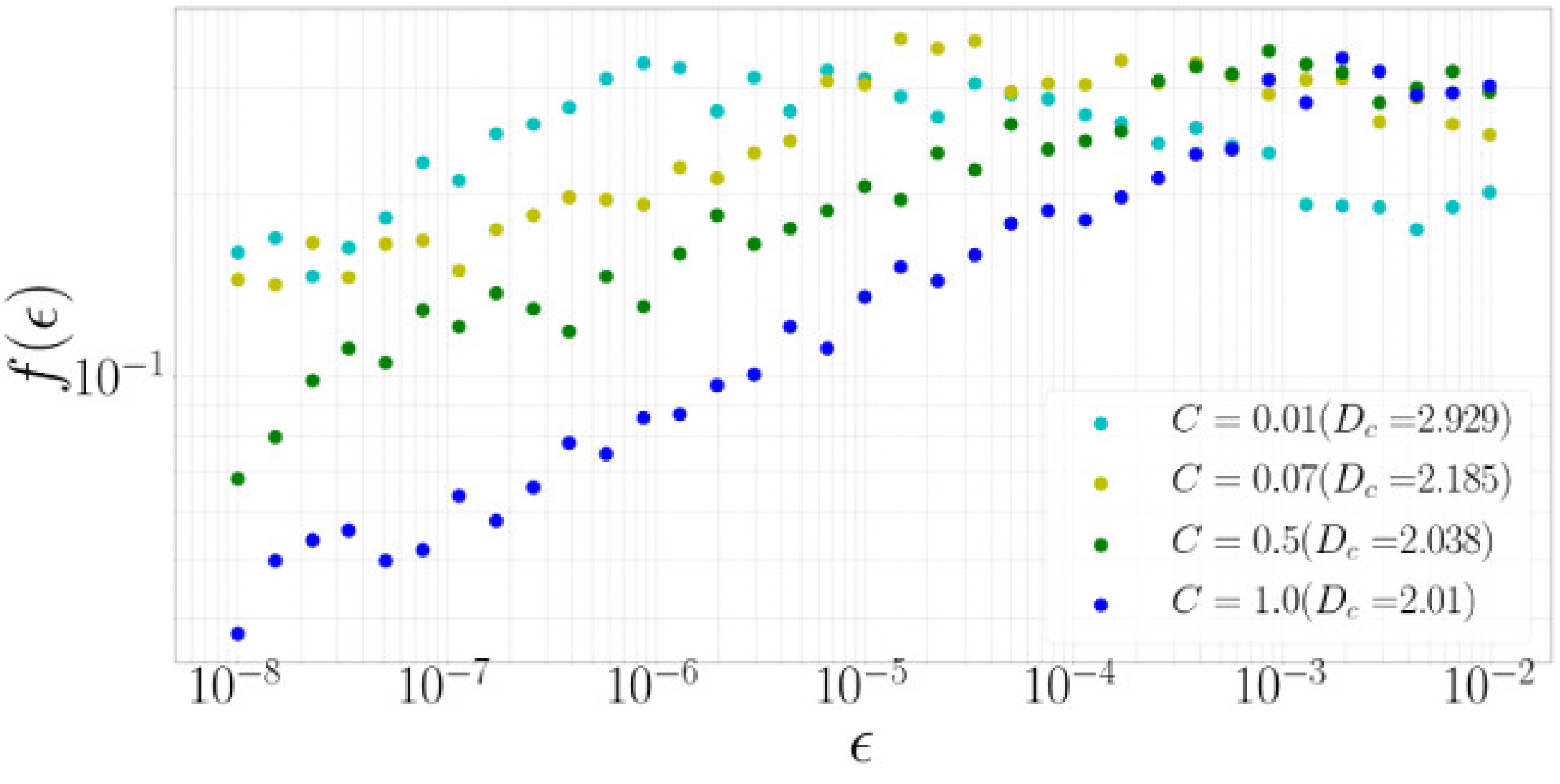}
  \caption{}
  \end{subfigure}
  \end{center}
  \caption{Fractal dimension of the chaotic saddle for (a) prolate spheroids of different aspect ratios in the precessional orbit of $C$=20 and (b) prolate spherois of $\xi_0=1.05$ in different Jeffery orbits.}
\label{fig:fractal}
\end{figure}

The chaotic saddle may be characterized quantitatively in terms of its fractal dimension. While there are quite a few ways by which one may obtain the fractal dimension, herein we adopt the methodology of \citep{bleher} by determining the uncertainty dimension. In this method a large number($N_0$) of initial conditions are chosen randomly, and each of these conditions is perturbed by a small amount ($\epsilon$), and is marked uncertain if the corresponding output states are markedly different. In the present case, one chooses several pairs of original and perturbed upstream gradient offsets in the interval $0 \le y_{-\infty} < y^{sep}_{-\infty}$ with $x=-20,z=0,\phi_{j,-\infty}=0$ and see whether the integrated final pairs of states also have similarly signed $y$ component. Those pairs\,($N$) that don't are deemed uncertain, and one then calculates the uncertainty probability as $f(\epsilon) = N/N0$. For small $\epsilon$, this uncertainty probability scales as $f(\epsilon)\sim\epsilon^{\alpha}$, and the uncertainty dimension can be defined as $D_s=D-\alpha$ with $D$ being the dimension of the dynamical system. Figures \ref{fig:fractal}a and b  show the variation of the uncertainty dimension with respect to tumbling prolate spheroids of different aspect ratios and prolate spherois of $\xi_0=1.05$ in different Jeffery orbits, respectively, where it is observed that as aspect ratio tends towards unity or the Jeffery constant($C$) drifts towards zero(spinning orbit) the fractal dimension shifts towards an integer value.

\section{Transition from regular to chaotic pathlines: Boundary of the chaotic saddle}\label{sec:transition}
Having established the signatures of chaotic scattering for the fluid pathlines around any non-spinning spheroid, we now proceed to analyze the pathlines from the perspective used in earlier efforts(for instance, see \cite{bleher-prl,bleher}). This is done by plotting the output variable, which is still the residence time, as a function of an appropriate input one, the focus being on the singular regions corresponding to apparent divergences of the residence time. Such regions would appear scale-dependent in a numerical investigation, on account of the true divergences being localized on a Cantor set of vanishing measure \citep{jung-particle,ziemniak,skufca,eckhardt-pipe-annual}. The  input variable must correspond to the analog of the impact parameter used in the chaotic scattering literature\citep{bleher}. In our case, this is the upstream gradient offset($y_{-\infty}$) of a fluid element in the plane of shear; or, both the gradient and vorticity offsets ($y_{-\infty}$ and $z_{-\infty}$) of an element off the flow-gradient plane. In addition one, of course, needs to specify the particular Jeffery orbit, and the phase of rotation(within a given orbit) at the initial instant; the latter is again done via the azimuthal angle which is now denoted as $\phi_{j-\infty}$. Note that these upstream offsets have already been used to analyze the regular open pathlines above. The difference here is that the offsets are used as a common basis to analyze both the regular and singular pathlines; the latter group of pathlines was examined earlier, only as a function of their points of intersection with the flow axis, in order to establish the signatures associated with chaotic scattering. One may now analyze the transition from the regular to the singular open pathlines, as a function of the aforementioned upstream offsets, and thereby, analyze the boundary of the chaotic saddle (what has been termed `the edge of chaos' in the literature; see \cite{skufca}). The latter has been the subject of earlier investigations in the context of two-dimensional Hamiltonian problems\citep{jung}.

\begin{figure}
  \begin{subfigure}[b]{.48\textwidth}
  \centering
  \includegraphics[scale=0.25]{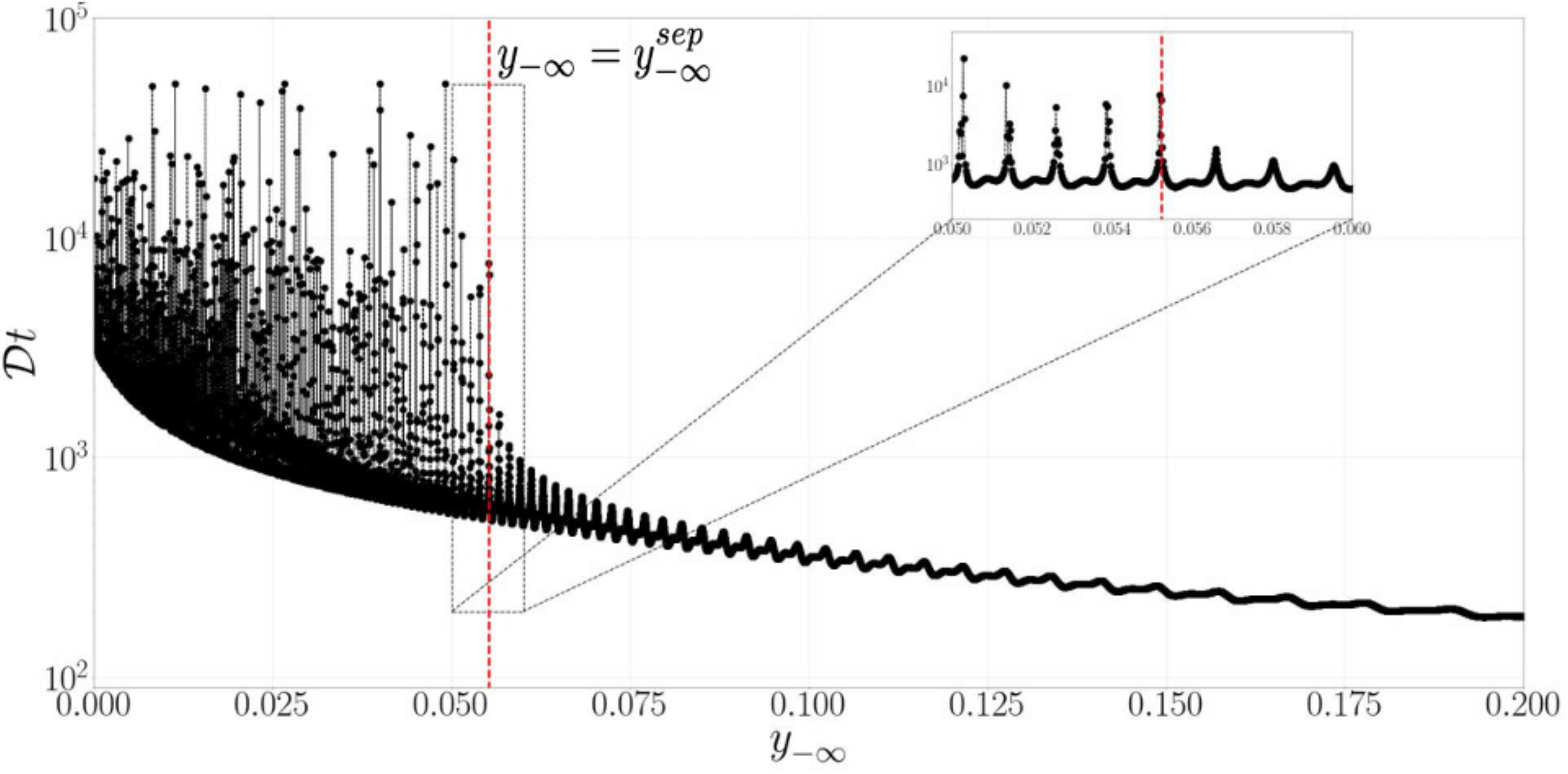}
  \caption{}
  \end{subfigure}
  \begin{subfigure}[b]{.48\textwidth}
  \centering
  \includegraphics[scale=0.25]{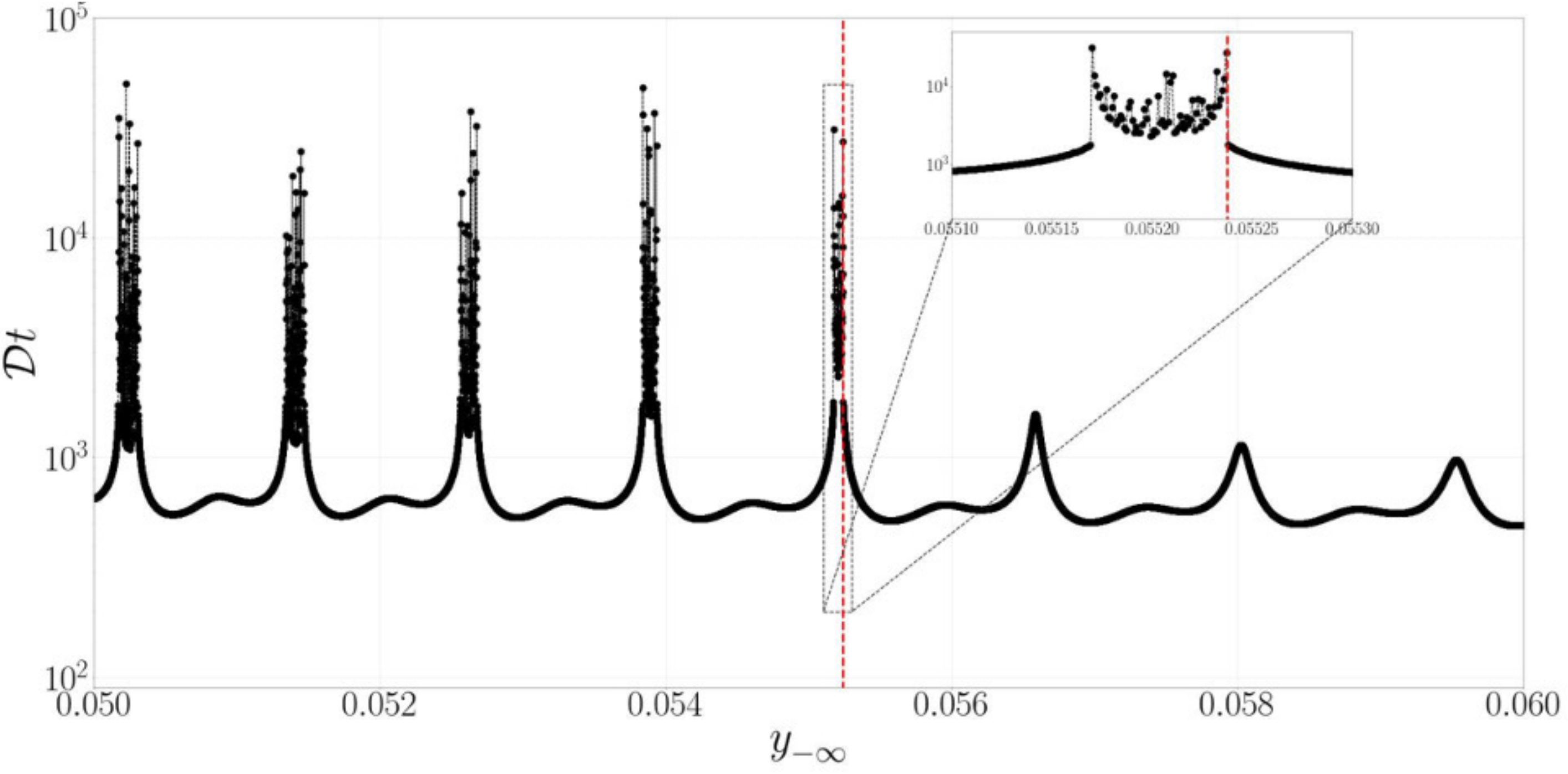}
  \caption{}
  \end{subfigure}
  \begin{subfigure}[b]{.96\textwidth}
  \centering
  \includegraphics[scale=0.25]{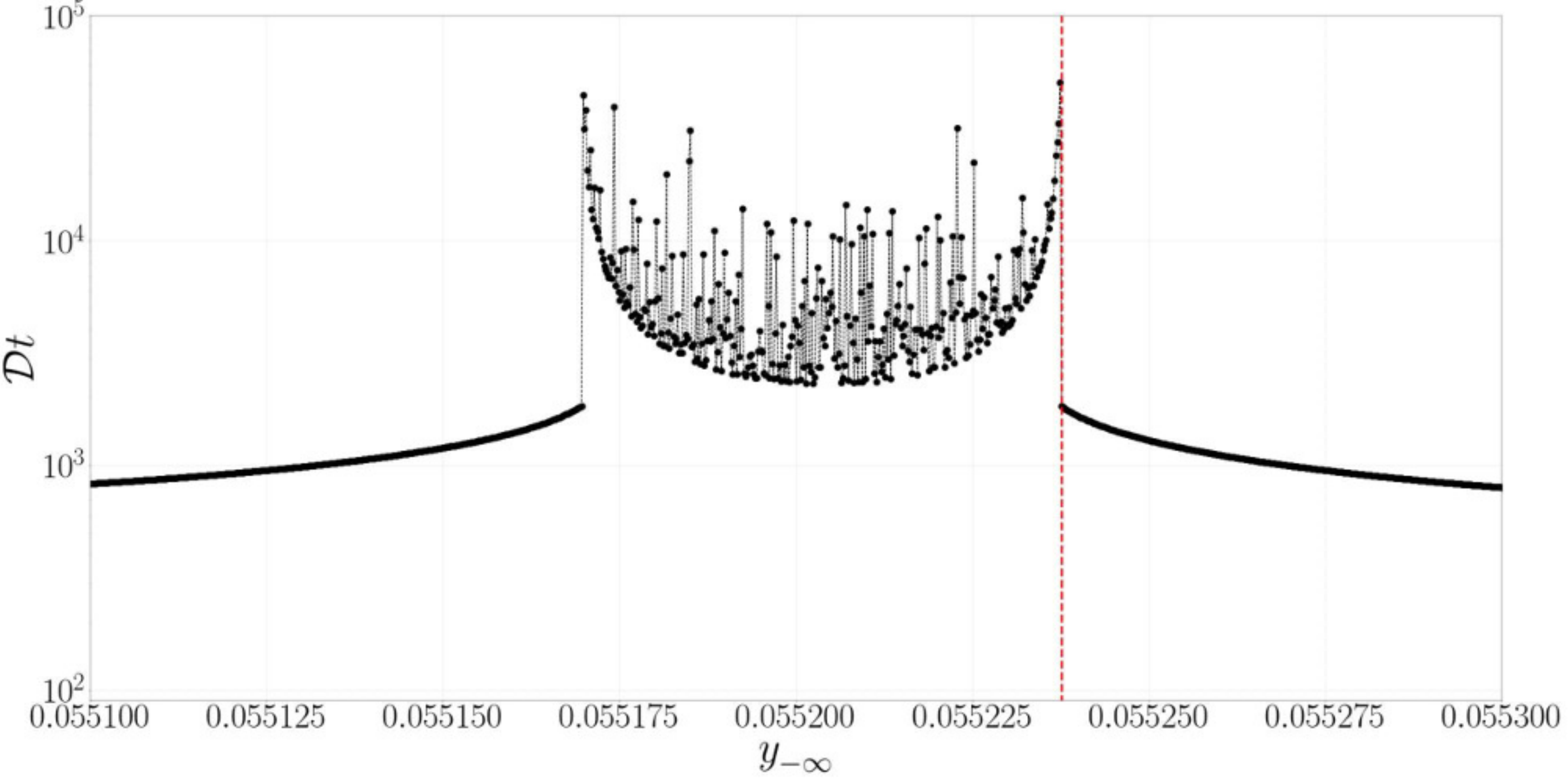}
  \caption{}
  \end{subfigure}
  \caption{Variation of the residence time for a tumbling prolate spheroid($\xi_0=2.0(\kappa=1.13)$), with $\phi_{j,-\infty}=0$, as a function of the upstream gradient offset of the fluid pathline (`upstream' here corresponds to $x = -20$). The red vertical line in all figures denotes the separatrix($ y^{sep}_{-\infty}$) that separates the regular($y_{-\infty} > y^{sep}_{-\infty}$) region from the one that includes intervals of chaotic scattering($y_{-\infty} < y^{sep}_{-\infty}$).}
\label{fig:chaos_boundary0}
\end{figure}

\begin{figure}
  \begin{subfigure}[b]{.48\textwidth}
  \centering
  \includegraphics[scale=0.25]{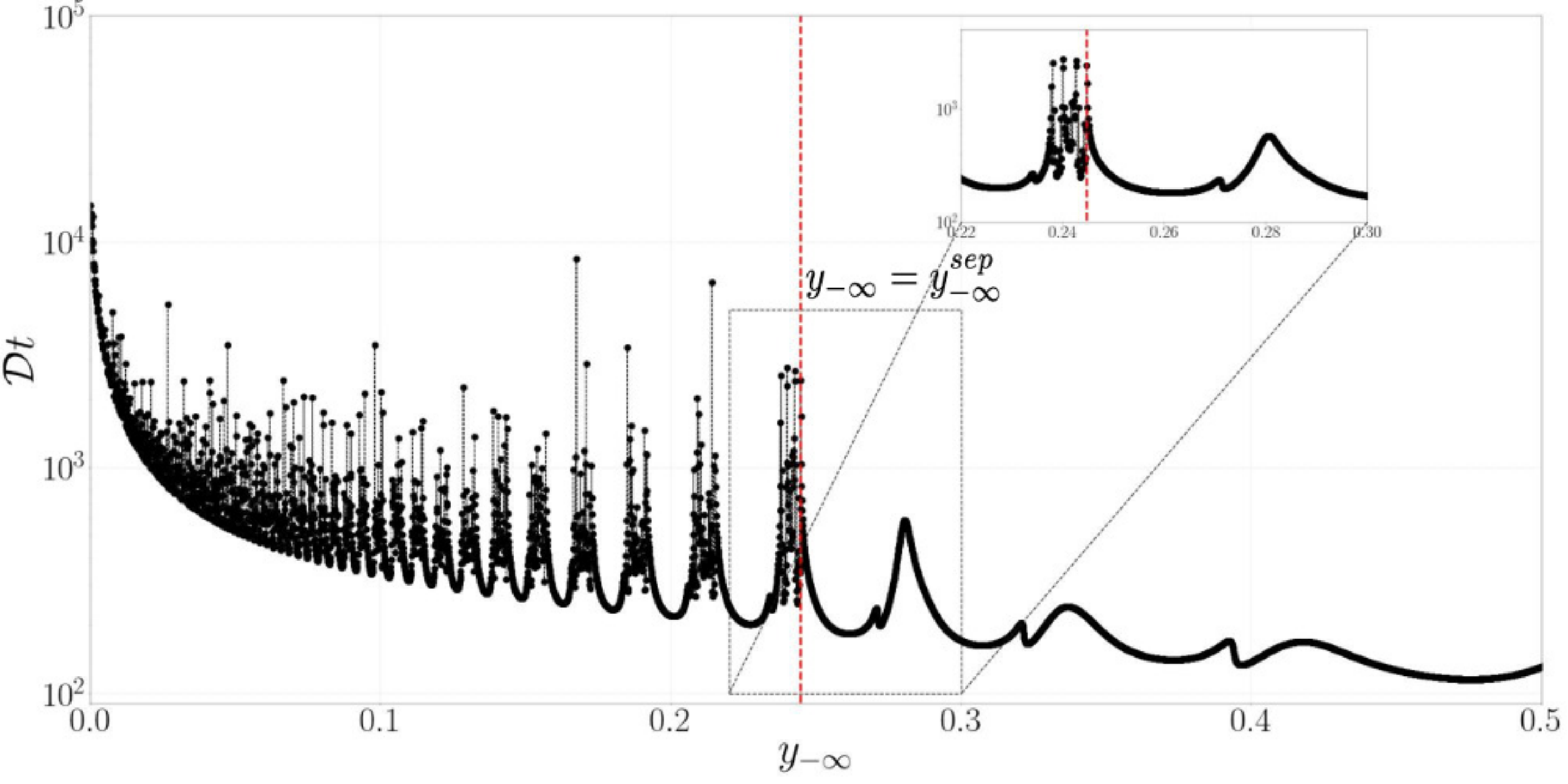}
  \caption{}
  \end{subfigure}
  \begin{subfigure}[b]{.48\textwidth}
  \centering
  \includegraphics[scale=0.25]{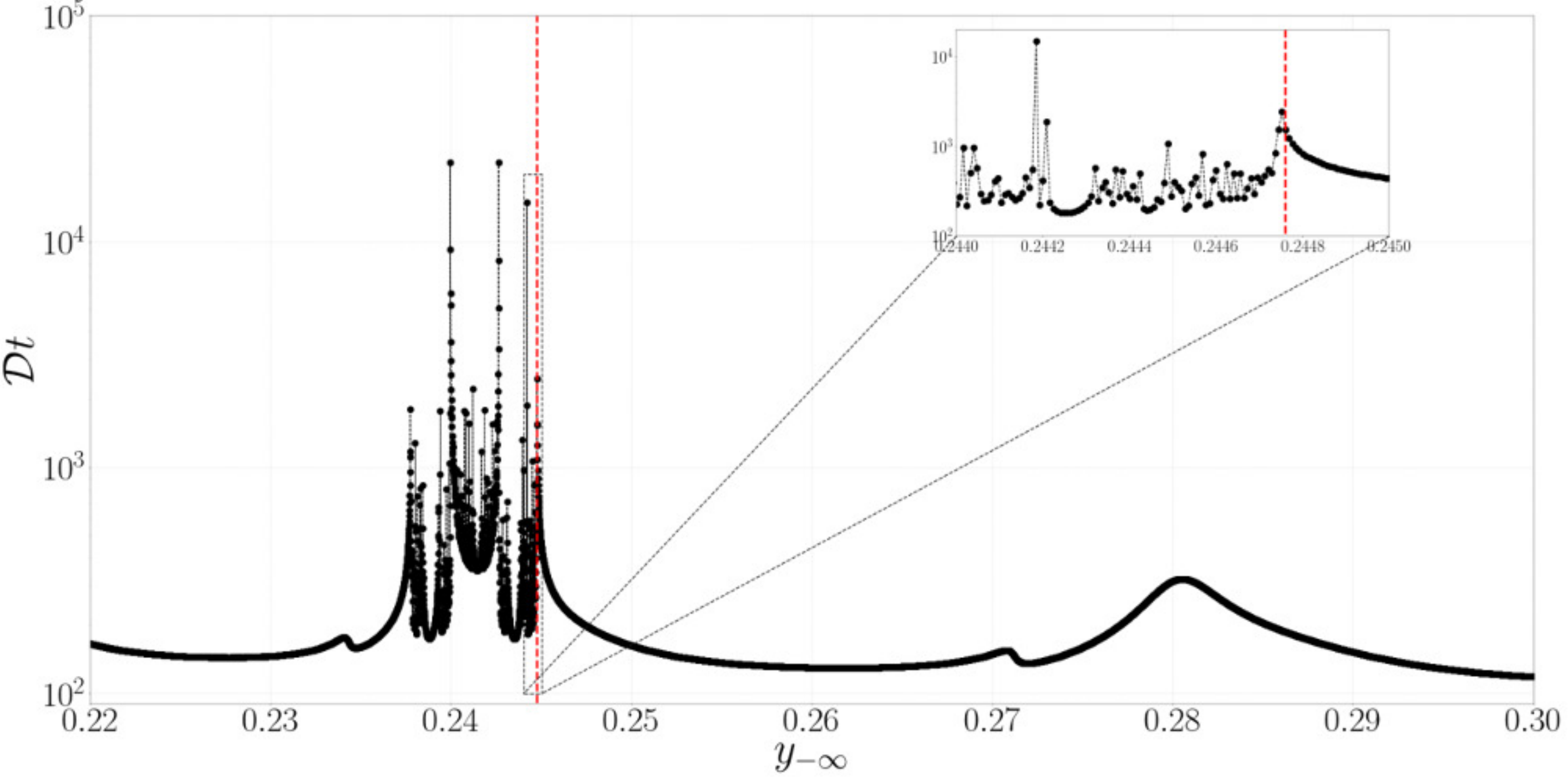}
  \caption{}
  \end{subfigure}
  \begin{subfigure}[b]{.96\textwidth}
  \centering
  \includegraphics[scale=0.25]{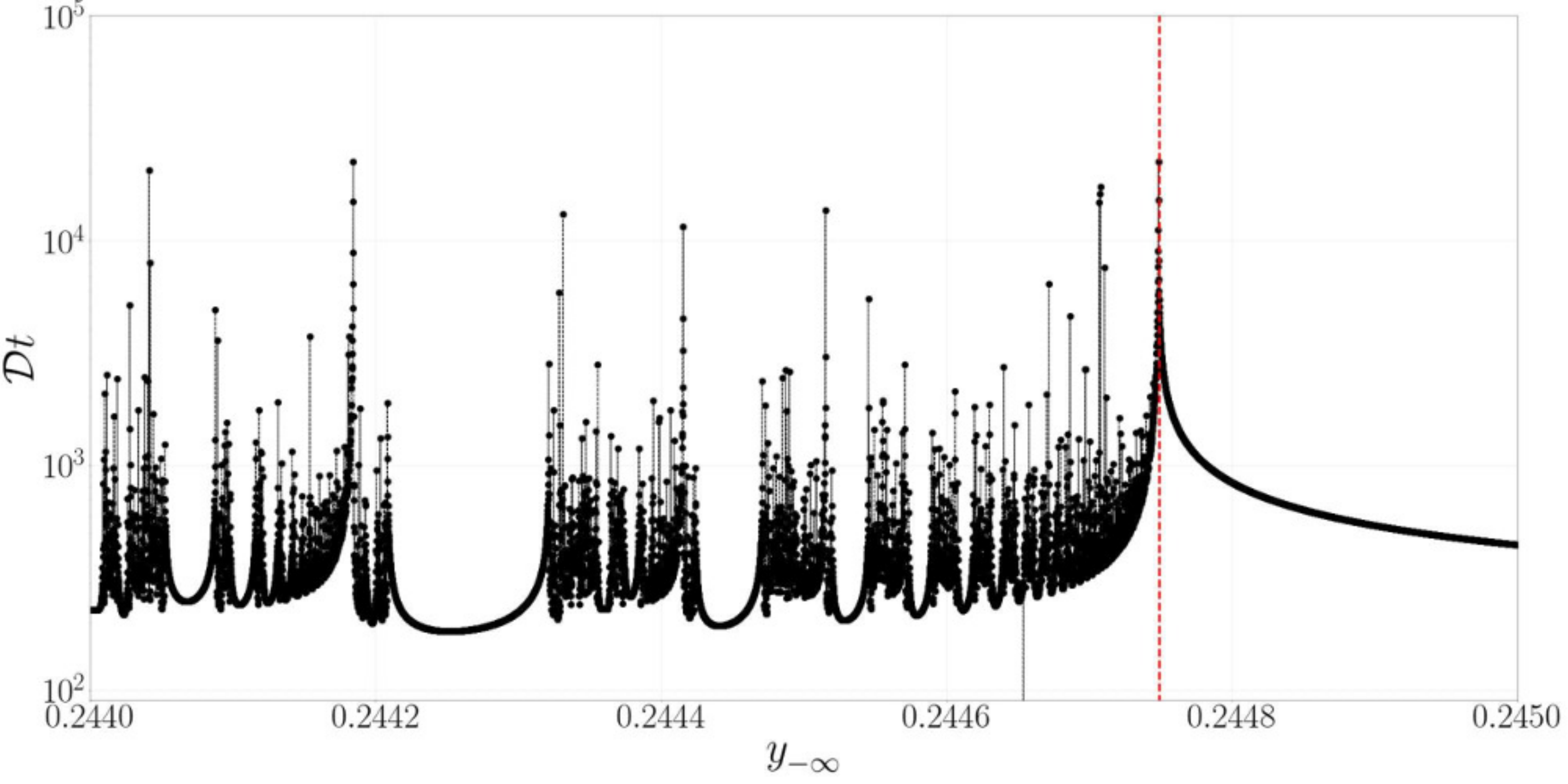}
  \caption{}
  \end{subfigure}
  \caption{Variation of the residence time for a tumbling prolate spheroid($\xi_0=1.05(\kappa=3.28)$), with $\phi_{j,-\infty}=0$, as a function of the upstream gradient offset of the fluid pathline (`upstream' here corresponds to $x = -20$). The red vertical line in all figures denotes separatrix($ y^{sep}_{-\infty}$) that separates the regular($y_{-\infty} > y^{sep}_{-\infty}$) region from the one that includes intervals of chaotic scattering($y_{-\infty} < y^{sep}_{-\infty}$).}
\label{fig:chaos_boundary1}
\end{figure}

In figures \ref{fig:chaos_boundary0} and \ref{fig:chaos_boundary1} we plot the residence time as a function of $y_{-\infty} (z_{-\infty} = 0)$ for tumbling prolate spheroids with $\xi_0=2.0(\kappa=1.13)$ and $\xi_0=1.05(\kappa=3.28$), respectively; $\phi_{j,-\infty} = 0$ for both cases. The residence time curves exhibit a small-scale modulation superposed on an underlying monotonic increase with decreasing $y_{-\infty}$. The reason for the latter increase is obvious. The modulation leading to a non-monotonic dependence on $y_{-\infty}$ arises owing to the spheroid orientation being a function of time. For $y_{-\infty}$'s not too different, a fluid element crossing the gradient-vorticity plane, at the instant that the spheroid is close to a vertical orientation, will do so at a higher $y_0$, leading to a reduced residence time. With decreasing $y_{-\infty}$, the residence time, for a given change in $y_{-\infty}$,  changes by a greater amount relative to the fixed Jeffery period. This leads to the wavelength of the modulation in Figures \ref{fig:chaos_boundary0} and \ref{fig:chaos_boundary1}(and the ones thereafter) decreasing with $y_{-\infty}$. Note that the absolute wavelength scales in proportion to the Jeffery period, and is therefore greater for the larger aspect ratio spheroid ($\xi_0$ = 1.05). The red vertical line in each plot denotes the ordinate of the separatrix ($y^{sep}_{-\infty}$) that demarcates the regular regions from those that exhibit intervals of chaotic scattering; this demarcation is evident from the magnified views in figures \ref{fig:chaos_boundary0}b and \ref{fig:chaos_boundary1}b, which clearly contrast the smooth peak(s) for $y_{-\infty} > y^{sep}_{-\infty}$ with the irregular dependence that ensues for $y < y^{sep}_{-\infty}$. Note that it is $\phi_{j,-\infty}$ that is now fixed at 0, and the value of $\phi_{j0}$ emerges during the course of the integration; one finds $\phi_{j0} = 111.8^o$ and  $168.6^o$ for the separatices corresponding to $\xi_0=2.0$ and $\xi_0=1.05$, respectively. Thus, the separatrices marked by the red vertical lines in figures \ref{fig:chaos_boundary0} and \ref{fig:chaos_boundary1} correspond to fore-aft asymmetric separatrices with finite upstream offsets - keeping in mind the pathline configurations in Figure 7, these separatrices correspond to $\phi_{j0} \in (\pi/2,\pi)$, and are obtained from those in Figure 7 by reflecting about the gradient axis. Similar to Figure 15a, the dependence for $y_{-\infty}$'s less than $y^{sep}_{-\infty}$ isn't entirely irregular. Instead, there are intervals of regular dependence separated by chaotic `bursts', with the relative sizes of the regular and chaotic intervals being clearly sensitive to the aspect ratio.  Figures \ref{fig:chaos_boundary0}b,c and \ref{fig:chaos_boundary1} b,c correspond to magnified views, with each view corresponding to $10^4$ initial conditions ($y_{-\infty}$'s), and highlight the interlacing of regular and chaotic intervals down to the smallest (numerically) resolved scales; in other words, what appears as an interval of chaotic dependence at a given coarse resolution contains smaller intervals of regular intervals within, and this appears to continue ad infinitum. Figures \ref{fig:chaos_boundary0}c and \ref{fig:chaos_boundary1}c suggest qualitatively different transitions from regular to chaotic scattering for the two aspect ratios. The transition for the spheroid with a near-unity aspect ratio($\xi_0=2$) in Figure \ref{fig:chaos_boundary0}c appears to be discontinuous, with $\mathcal{D}t$ approaching a finite limit for $y_{-\infty} \rightarrow y^{sep+}_{-\infty}$, but appearing to diverge for  $y_{-\infty} \rightarrow y^{sep-}_{-\infty}$;  in contrast, the transition for the larger aspect ratio spheroid appears to involve a divergence of the residence time regardless of the direction of approach towards the separatrix offset. Despite this difference in appearance, one expects $\lim_{y_{-\infty} \rightarrow y^{sep+}_{-\infty}}\mathcal{D}t$ to be finite, since this corresponds to the finite residence times of the fore-aft asymmetric separatrices mentioned above in the interval (-20,20).

\begin{figure}
  \begin{subfigure}[b]{.48\textwidth}
  \centering
  \includegraphics[scale=0.25]{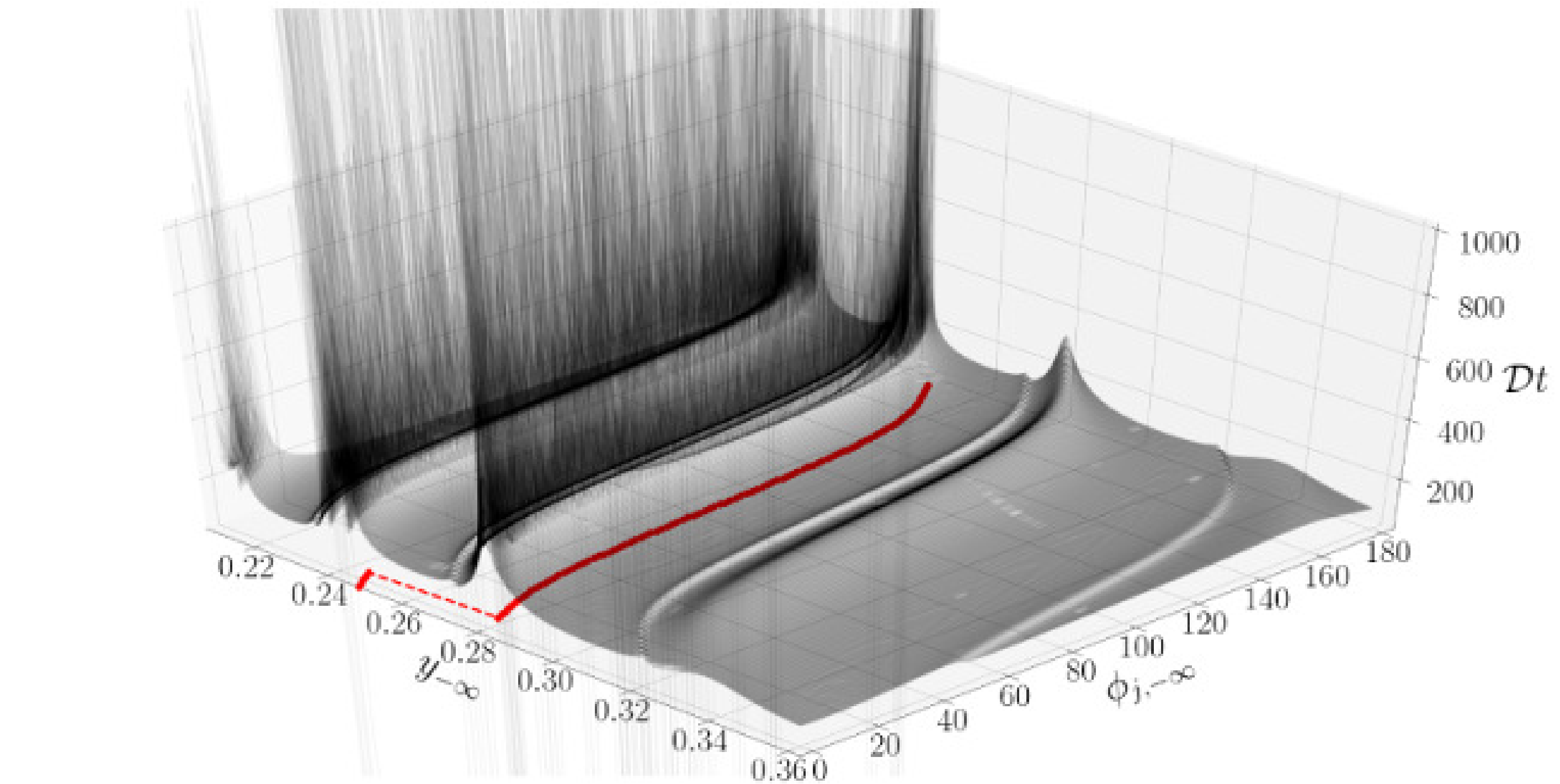}
  \caption{}
  \end{subfigure}
  \begin{subfigure}[b]{.48\textwidth}
  \centering
  \includegraphics[scale=0.25]{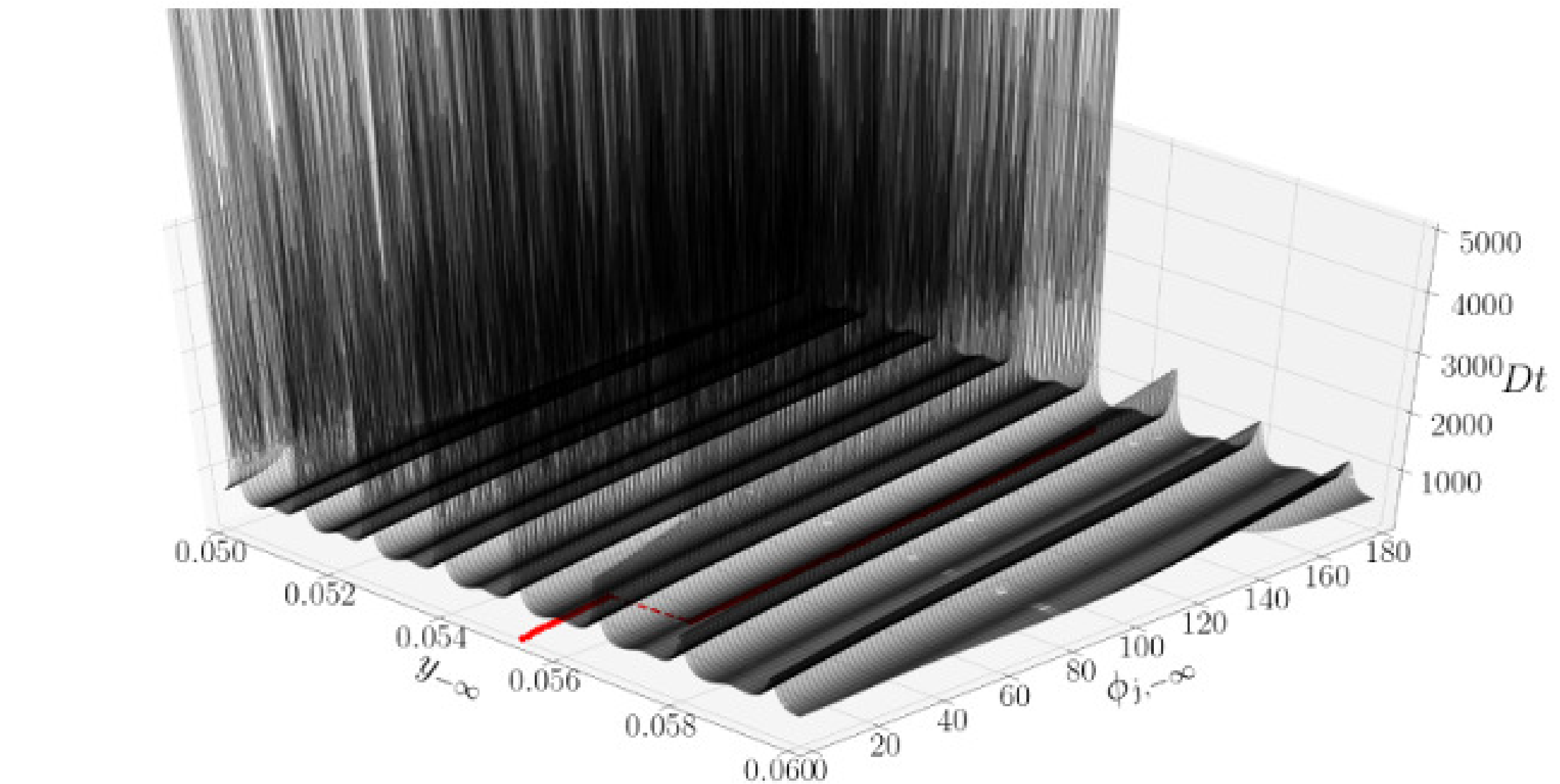}
  \caption{}
  \end{subfigure}
  \caption{The residence time surface plotted on the $y_{-\infty}-\phi_{j,-\infty}$ plane, for  tumbling spheroids with(a) $\xi_0=1.05(\kappa=3.28)$) and (b) $\xi_0=2(\kappa=1.15)$), as a function of the upstream gradient offset of the fluid pathline (`upstream' here corresponds to $x = -20$) and $\phi_{j,-\infty}$. The red separatrix curves mark the transition from regular to chaotic scattering in both cases.}
\label{fig:res_surf}
\end{figure}

The separatrix offsets in Figures \ref{fig:chaos_boundary0} and \ref{fig:chaos_boundary1} arise for $\phi_{j,-\infty} = 0$. A different offset will result for another choice of $\phi_{j,-\infty}$. To access all possible separatrices, one needs to choose $\phi_{j,-\infty}$'s spanning the interval ($0,\pi$) for a given large negative $x$. The maximum among all such separatrix offsets would then determine the onset of chaotic scattering in the flow-gradient plane. The restriction of a large negative $x$ is necessary so that the aforementioned maximum separatrix offset is independent of $x$; although, this maximum will occur at an $x$-dependent $\phi_{j,-\infty}$. Thus, in other to characterize the onset of chaotic scattering in the flow-gradient plane, one ought to have a plot, as depicted in Figure \ref{fig:res_surf}, of the residence-time surface as a function of both $y_{-\infty}$ and $\phi_{j,-\infty}$. 
The onset of chaotic scattering is demarcated by a critical curve $y^{sep}_{-\infty}(\phi_{j,-\infty})$ in this plane. The residence-time surface is smooth on one side of this separatrix curve, but one expects a jagged irregular distribution of peaks on the other side; the latter irregular dependence has indeed been observed in the context of the laminar-turbulent transition\citep{eckhardt-pipe-annual}. Note that, provided $x$ is sufficiently large, the set of $y^{sep}_{-\infty}$ values is invariant to a change in $x$. Instead, and as mentioned above, each $y^{sep}_{-\infty}$ now corresponds to a different $\phi_{j,-\infty}$, the relation between the two $\phi_{j,-\infty}$'s being determined by the angle through which the spheroid rotates as the fluid element translates from the first to the second value of $x$\,(this translation is almost entirely due to the ambient simple shear). This `juggling' of $y^{sep}_{-\infty}$'s will, of course, change the separatrix curve in the $y_{-\infty},\phi_{j,-\infty}$ plane. In both figures \ref{fig:res_surf}a and b, the separatrix curve has a step discontinuity; as will be seen below, this jump arises due to the abrupt appearance of a chaotic burst on a previously regular peak.

\begin{figure}
  \begin{subfigure}[b]{.48\textwidth}
  \centering
  \includegraphics[scale=0.24]{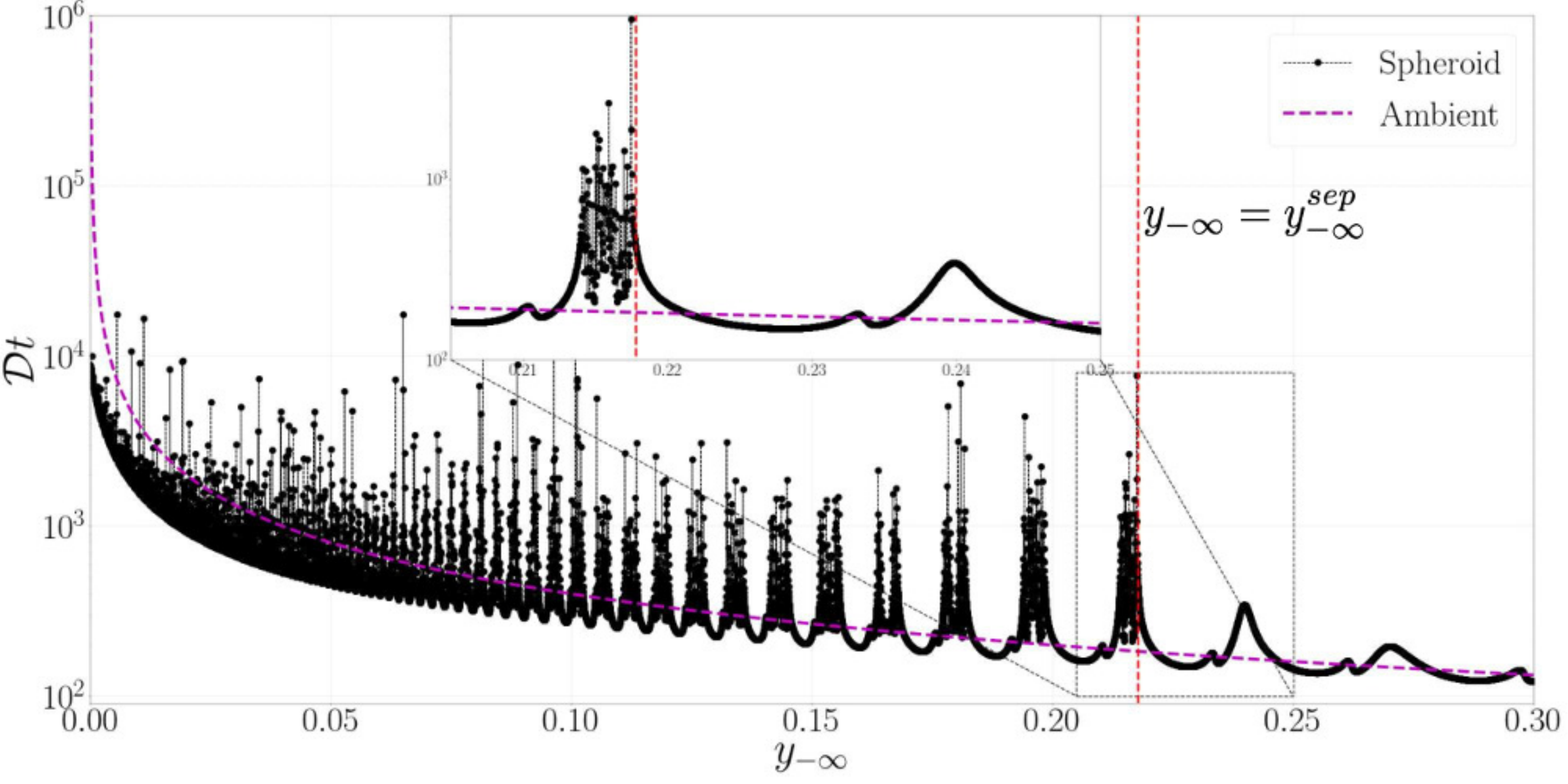}
  \caption{$\xi_0=1.1$}
  \end{subfigure}
  \begin{subfigure}[b]{.48\textwidth}
  \centering
  \includegraphics[scale=0.24]{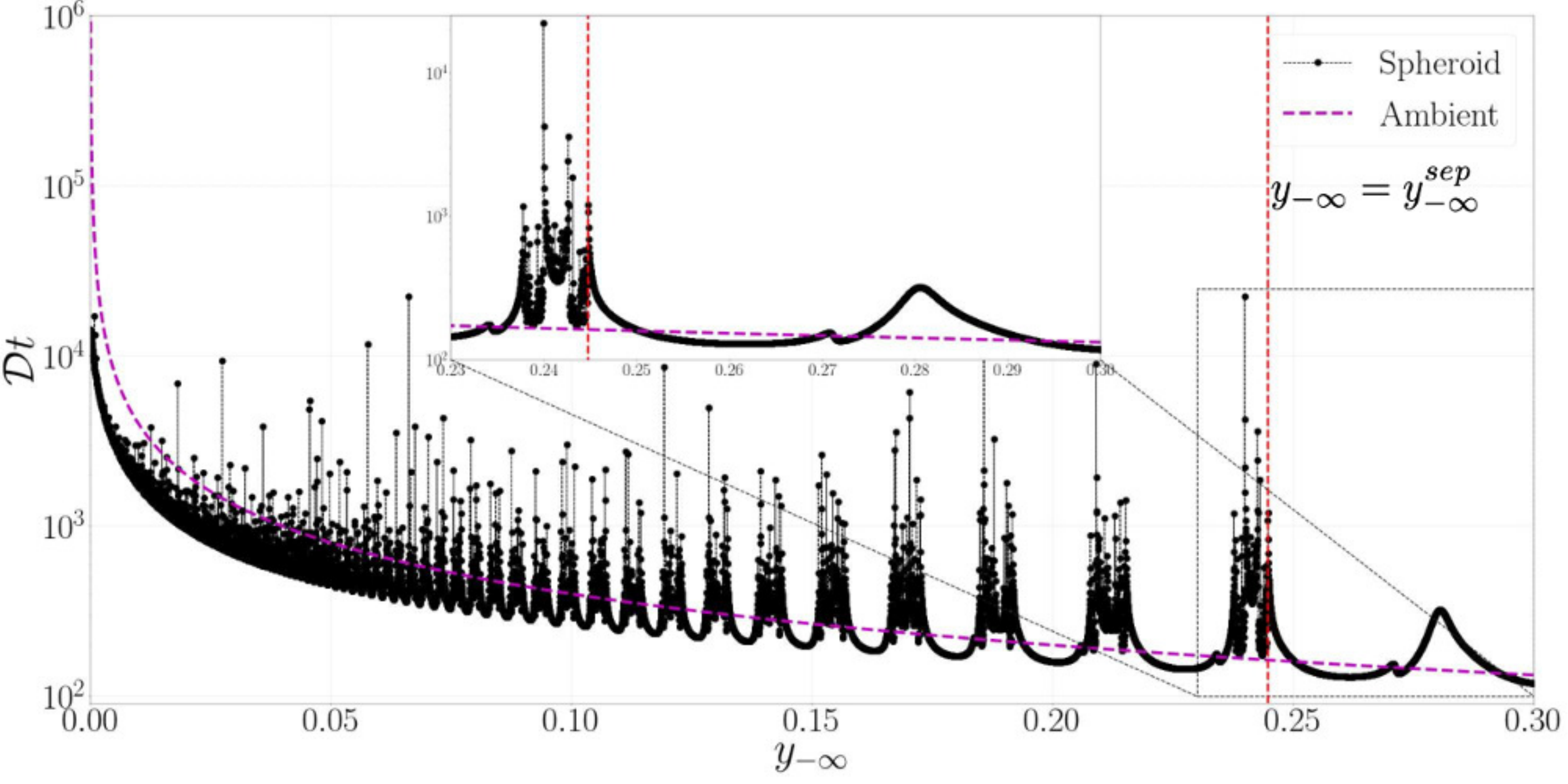}
  \caption{$\xi_0=1.05$}
  \end{subfigure}
  \begin{subfigure}[b]{.48\textwidth}
  \centering
  \includegraphics[scale=0.24]{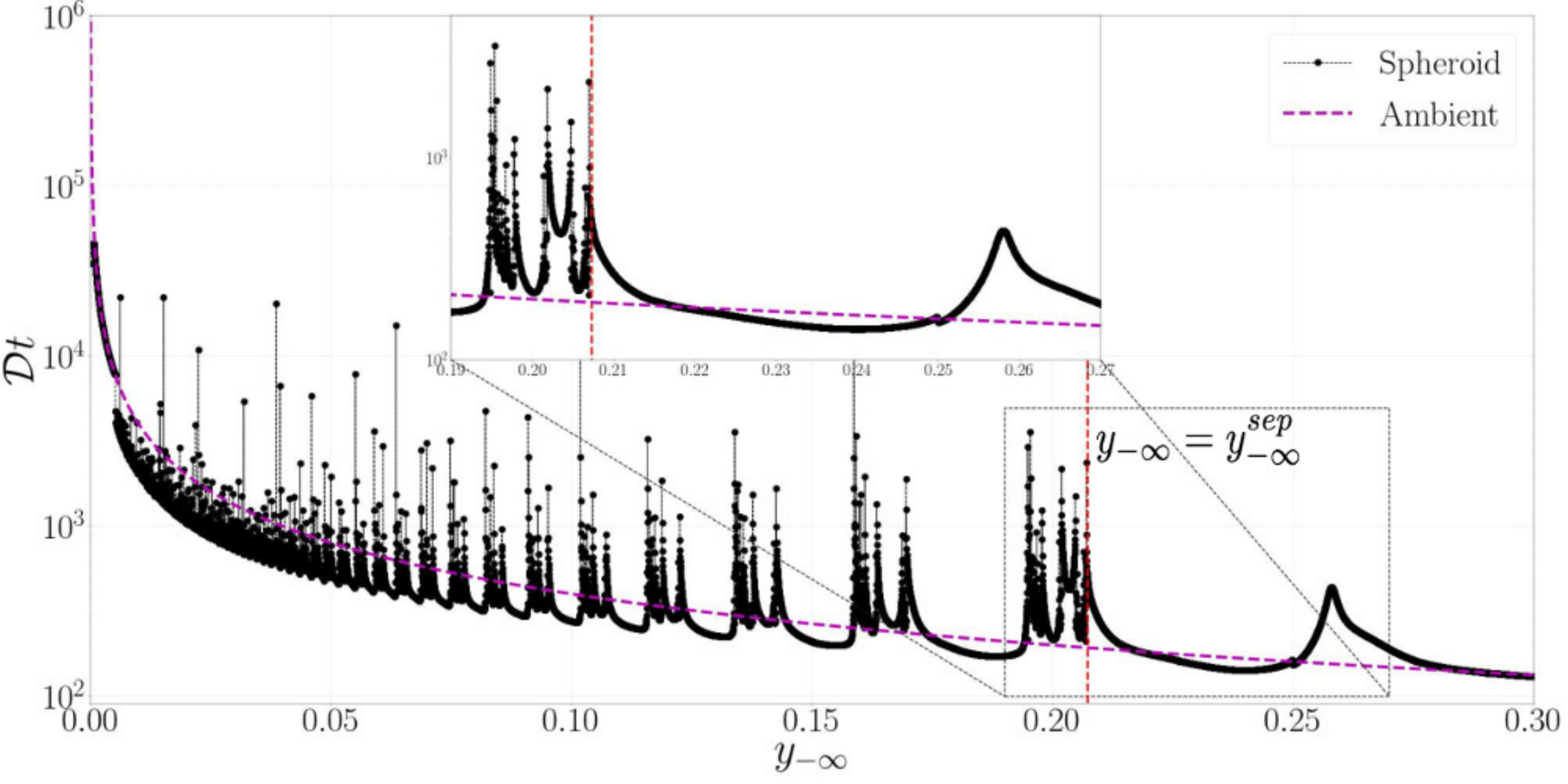}
  \caption{$\xi_0=1.01$}
  \end{subfigure}
  \begin{subfigure}[b]{.48\textwidth}
  \centering
  \includegraphics[scale=0.24]{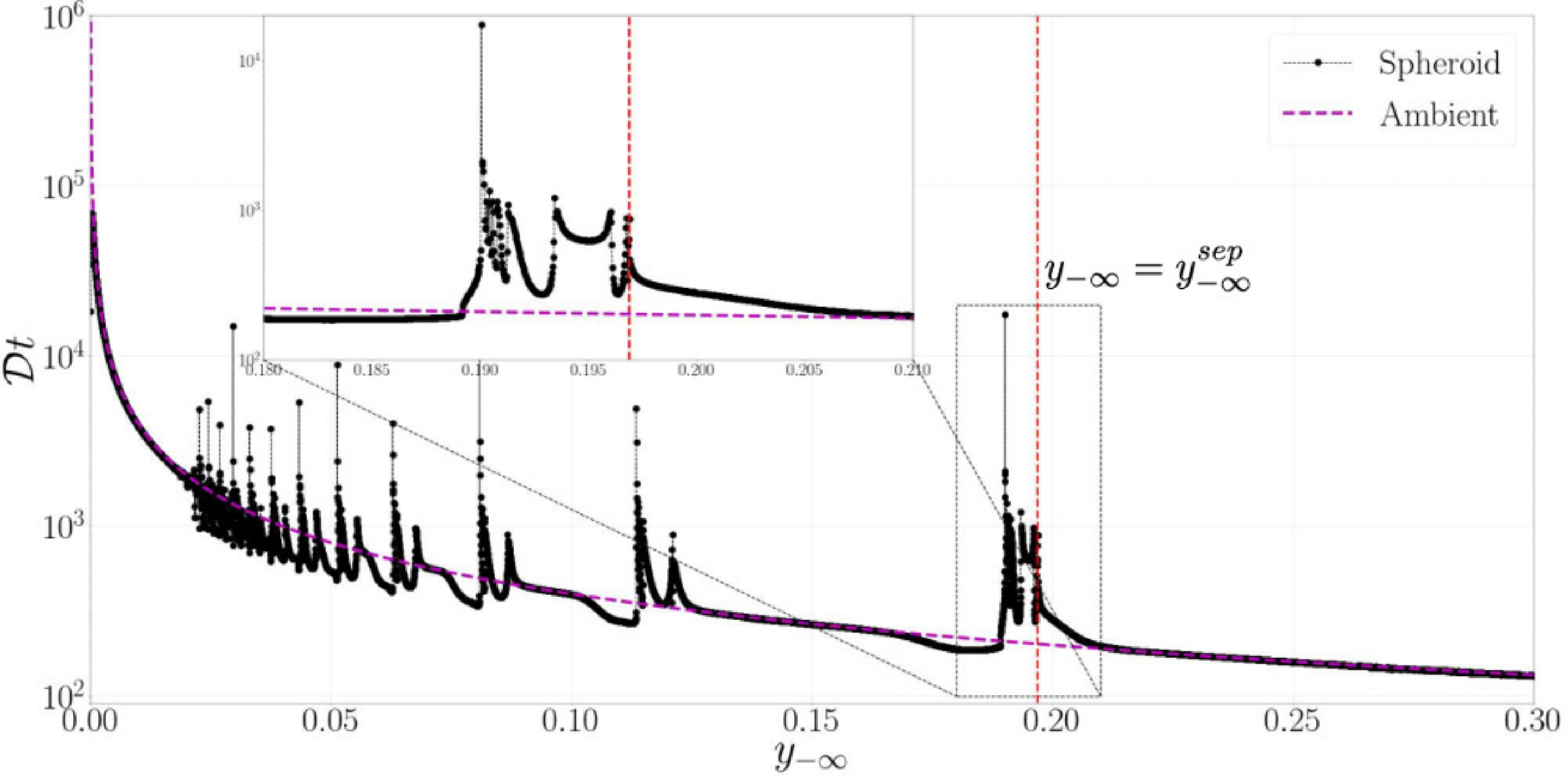}
  \caption{$\xi_0=1.001$}
  \end{subfigure}
  \begin{subfigure}[b]{.96\textwidth}
  \centering
  \includegraphics[scale=0.45]{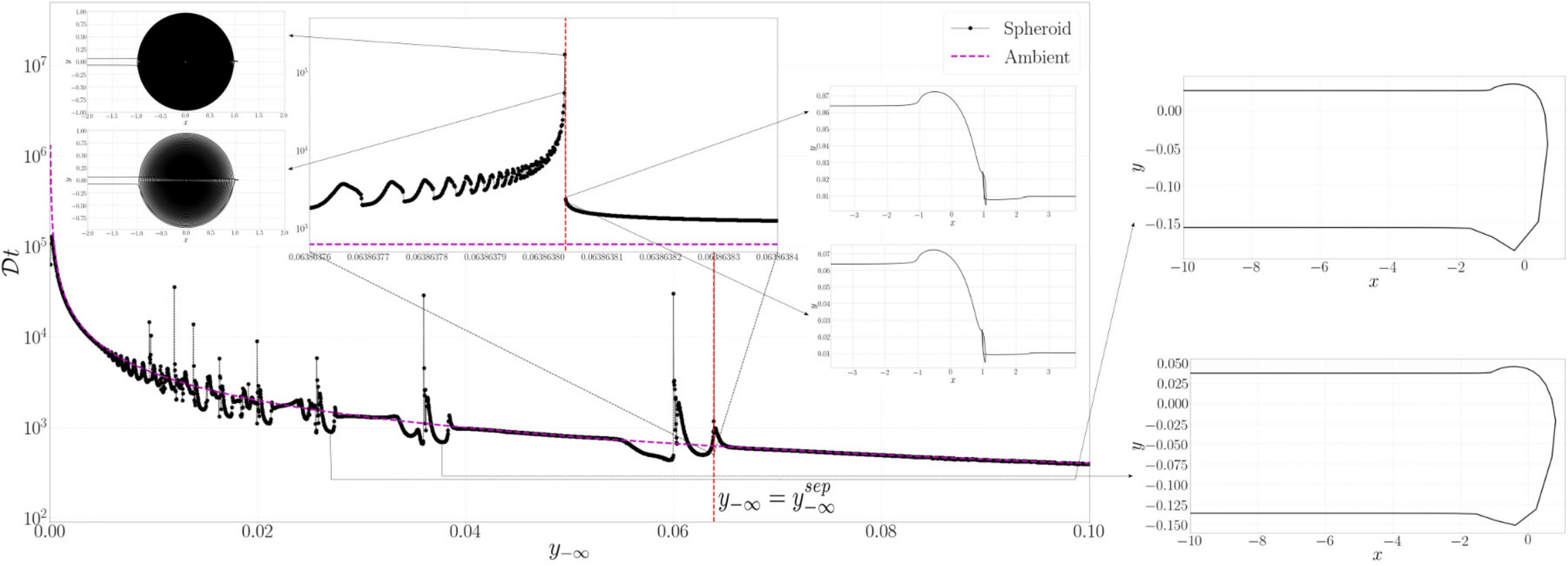}
  \caption{$\xi_0=1.0001$}
  \end{subfigure}
  \caption{Variation of the residence time for tumbling prolate spheroids with $\xi_0=1.1,1.05,1.01,1.001,1.0001(\kappa=2.4,3.28,7.12,22.68,70.7)$) with $\phi_{j,-\infty}=0$. The red vertical line separates the regular and chaotic regions, the magenta curve denotes the residence time estimate based on the ambient simple shear flow. The insets in Figure \ref{fig:chaos_boundary2}e show sample pathlines for upstream offsets less and greater than that of the separatrix ($y^{sep}_{-\infty}$); those below the separatrix, and within the chaotic burst intervals, loop around the spheroid a large number of times (the two insets on the left), while the other pathlines are open (the insets on the right).}
\label{fig:chaos_boundary2}
\end{figure}

Having characterized the residence time surface, we return to the specific case of $\phi_{j,-\infty} = 0$, and now compare the manner in which the signatures of chaotic scattering disappear as one approaches the limit of an integrable system (note that the term `integrable' is used here in a loose sense, to denote the regular nature of the trajectories, and the resulting smooth dependence on initial conditions; the problem of a spheroid in a simple shear flow is evidently non-Hamiltonian). The two limiting integrable cases are that of a spinning spheroid of any aspect ratio, the streamline topology for which has already discussed in detail in section \ref{sec:spinning}, and an infinitely slender prolate spheroid ($\kappa \rightarrow \infty$ or $\xi_0 \rightarrow 1$) in any of the finite-$C$ Jeffery orbits. We examine the latter limit first. The integrability in this case arises because viscous slender body theory\citep{batchelor_sbt} shows that the time dependent disturbance velocity field associated with a slender fiber (responsible for the non-integrability of this system), in a region around it of order its length, scales as the inverse of the logarithm of its (large) aspect ratio. Thus, at leading logarithmic order, a fluid element is merely convected by the ambient simple shear flow. Such an approximation has been earlier used to analyze the effect of polymeric stresses on fiber motion in simple shear flow; in particular, to analyze the viscoelasticity-induced drift across (meridional) Jeffery orbits for large Deborah numbers\citep{harlen_koch}. In this limit, the polymeric stress at a point is approximated based on the integrated effect of velocity gradients experienced by a fluid element (containing polymer molecules) along an ambient flow streamline that convects the element upto the point of interest. Figures \ref{fig:chaos_boundary2}a-e show a comparison of the residence time distributions for tumbling prolate spheroids of increasing aspect ratio, approaching the slender fiber limit; $\kappa$ ranges from 2.4 ($\xi_0$ = 1.1) in figure \ref{fig:chaos_boundary2}a to 70.7 ($\xi_0$ = 1.0001) in figure \ref{fig:chaos_boundary2}e. Note that the length scale characterizing the undulations in the residence time curves, induced by the spheroid orientation dynamics, increases with increasing $\kappa$. As mentioned earlier, this is due to an increase in the Jeffery period. Each of the residence time curves is overlaid on a smooth curve that plots the residence time estimated based on convection by the ambient shear alone and that is therefore proportional to $1/y^{-\infty}$. For the smaller aspect ratios, notwithstanding the chaotic bursts, the ambient-flow-based estimate, ends up overestimating the residence time. This is because accounting for the disturbance velocity field leads to the fluid element being advected to larger $y$ in the vicinity of the rotating spheroid, in turn leading to a faster (local) convection by the ambient flow. For the largest aspect ratios (figures \ref{fig:chaos_boundary2}d and e), the ambient-flow-based estimate is quite accurate, validating to some extent a slender-body-theory-based analysis that neglects of the disturbance velocity field as far as the fluid pathlines is concerned. From figures \ref{fig:chaos_boundary2}d and e, it is also evident that the separatrix offset, $y^{sep}_{-\infty}$, that marks the onset of the chaotic bursts, eventually decreases with increasing aspect ratio, again emphasizing the dominance of the ambient flow, and thence, the approach towards an integrable limit $(\kappa = \infty)$. This approach is, however, a singular one, with the intermittent chaotic bursts corresponding to large departures of the residence time from the $O(1/y^{-\infty})$ ambient-flow-based estimate even for the largest $\kappa$ shown. As shown in figure \ref{fig:chaos_boundary2}e, these departures correspond to a `trapping' of the fluid pathline for a long time in the vicinity of the rotating spheroid. Such trapped pathlines execute a large number of loops around the spheroid; in contrast, the pathlines above the separatix, and those in the regular intervals for $y_{-\infty} < y^{sep}_{-\infty}$, are open. Interestingly, for the largest $\kappa$, the dips of the residence time curve below the ambient-flow-based estimate correspond to regular reversing pathlines that do not loop around the spheroid (see inset in Figure \ref{fig:chaos_boundary2}e). Finally, note that the signatures of chaotic scattering are also not readily apparent for the largest $\kappa$. Thus, although we have marked a separatrix offset in this case, it is possible that the irregular dependence of the residence time, in a manner resembling the smaller $\kappa$'s, now occurs on scales smaller than those resolved.

\begin{figure}
\begin{subfigure}[b]{.96\textwidth}
\centering
\includegraphics[scale=0.35]{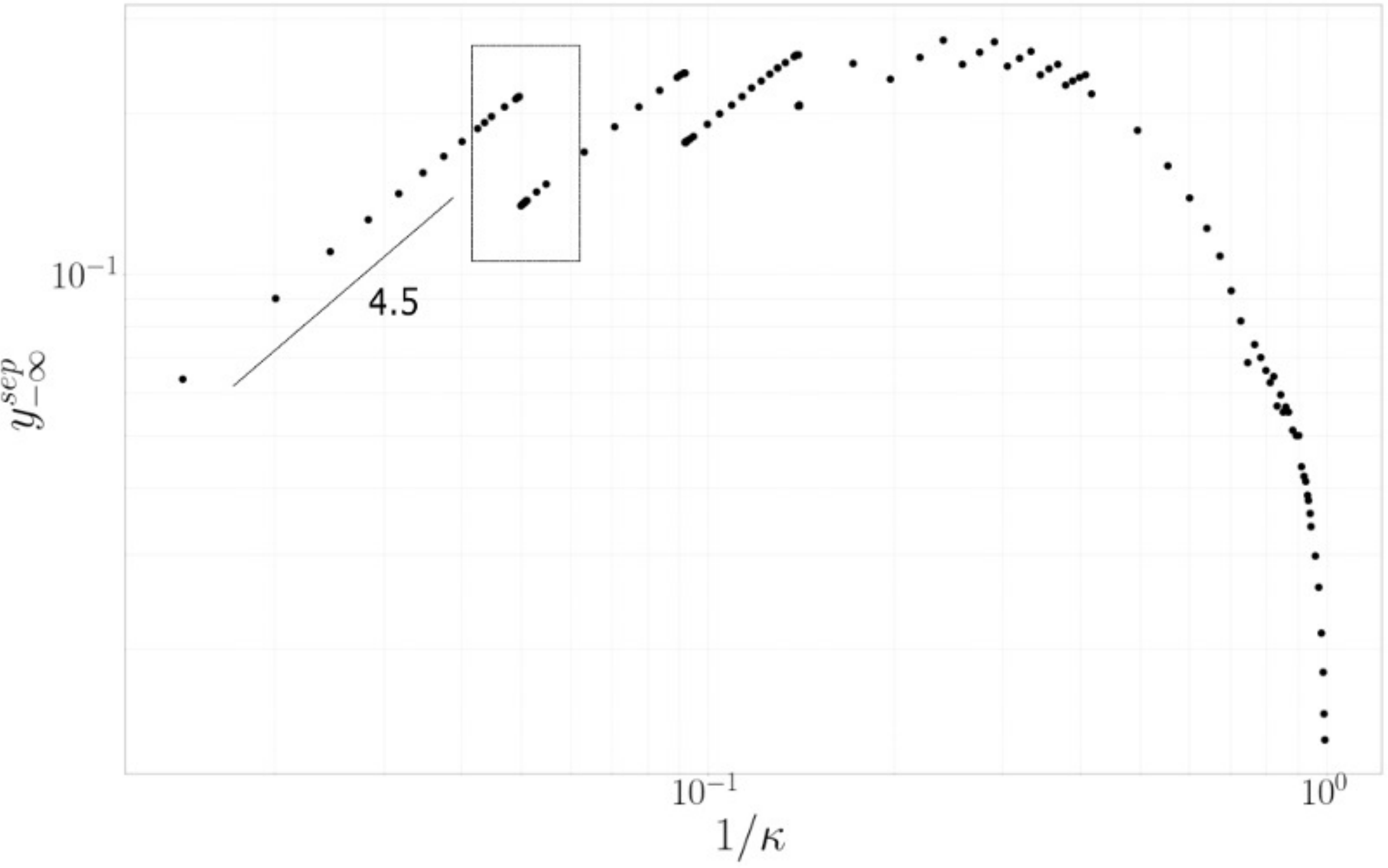}
\caption{}
\label{fig:chaos_boundary3a}
\end{subfigure}
\begin{subfigure}[b]{.96\textwidth}
\centering
\includegraphics[scale=0.4]{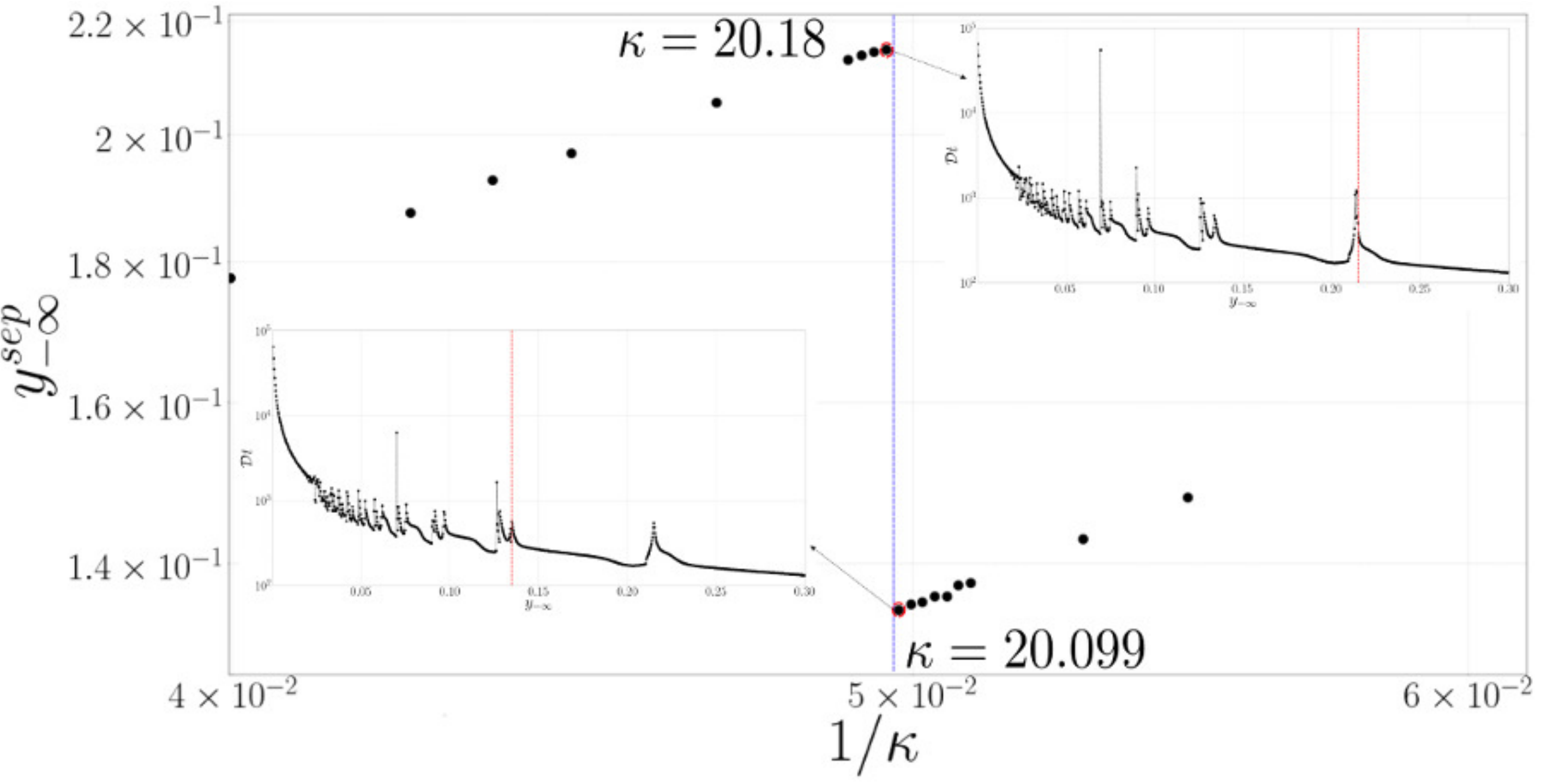}
\caption{}
\label{fig:chaos_boundary3b}
\end{subfigure}
\caption{(a) The upstream gradient offset of the separatrix, $y^{sep}_{-\infty}$, for tumbling spheroids, as a function of the spheroid aspect ratio, (b) Residence time plots for the (circled) aspect ratios $\kappa=$20.18 and 20.099, in the close vicinity and on either side of a $y^{sep}_{-\infty}$ jump.}
\label{fig:chaos_boundary3}
\end{figure}

Based on the residence time distributions shown in figure \ref{fig:chaos_boundary2}, and those for other aspect ratios, figure \ref{fig:chaos_boundary3}a plots the upstream (gradient) offset of the separatrix ($y^{sep}_{-\infty}$), marking the onset of chaotic scattering, as a function of the spheroid aspect ratio. Although the separatrix offset appears to eventually decrease monotonically for the largest aspect ratios with $y^{sep}_{-\infty} \sim \kappa^{-4.5}$, for $\kappa \rightarrow \infty$, the variation for moderately large aspect ratios is non-monotonic on account of a series of seemingly discontinuous jumps in $y^{sep}_{-\infty}$ at certain aspect ratios (similar to the jumps seen earlier in figures \ref{fig:res_surf}a and b). To examine this further, we plot in figure \ref{fig:chaos_boundary3}b the residence time distributions for a pair of aspect ratios on either side of a particular $y^{sep}_{-\infty}$ jump. It is seen from the distributions that the jump arises from the sudden appearance of a chaotic burst in a previously smooth peak (located just above the separatrix) with increasing aspect ratio. Note that, even for a slender fiber, there would still be fluid elements that approach the fiber to within a distance of order its diameter, and which would then suffer a strong interaction (since the disturbance field isn't small in this region). But, the fraction of fluid elements that undergo such a strong interaction must approach zero as $\kappa^{-1}$, owing to the smallness of the `collisional' cross section which scales as the product of the fiber length and diameter. Thus, the upstream gradient offset of fluid elements that undergo a strong interaction is expected to scale as $\kappa^{-1}$. While $y^{sep}_{-\infty}$ in figure \ref{fig:chaos_boundary3}a asymptotes to zero more rapidly, as $\kappa^{-4.5}$, the numerical values of the offsets remain larger than the spheroid diameter for the largest aspect ratios examined. We have verified that the separatrix offsets for oblate spheroids exhibit a similar series of jumps for $\kappa$ decreasing to zero\,(not shown). Unlike the prolate case, the oblate spheroid disturbance field remains finite in the limit $\kappa \rightarrow 0$, and one might therefore expect the separatrix offset for a flat disk to also remain finite. Over the range of aspect ratios examined, however, we find separatrix offsets for oblate and prolate spheroids to remain comparable in magnitude.


\begin{figure}
  \begin{subfigure}[b]{.48\textwidth}
  \centering
  \includegraphics[scale=0.24]{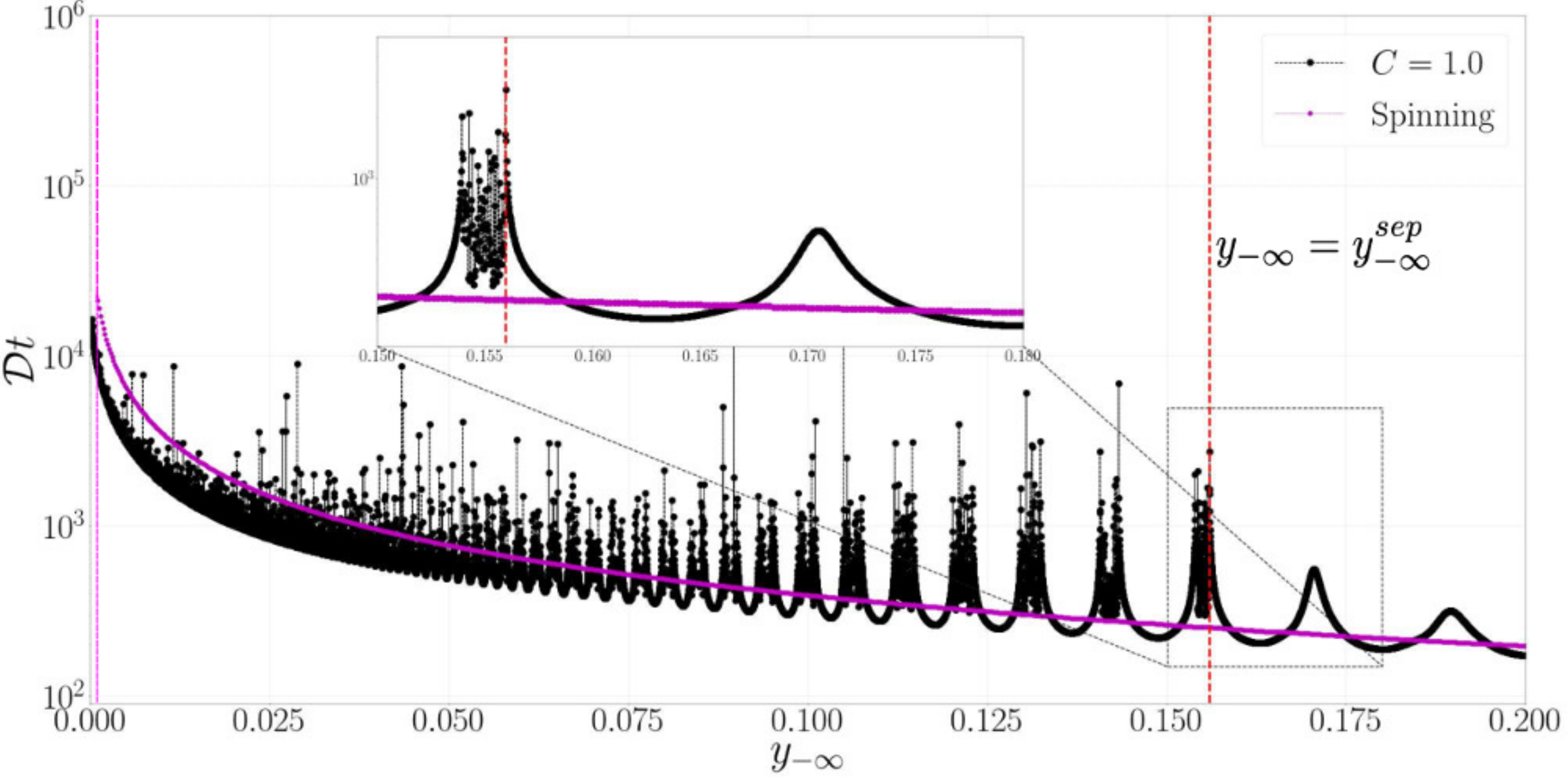}
  \caption{$C=1, t=4000T_j$}
  \end{subfigure}
  \begin{subfigure}[b]{.48\textwidth}
  \centering
  \includegraphics[scale=0.24]{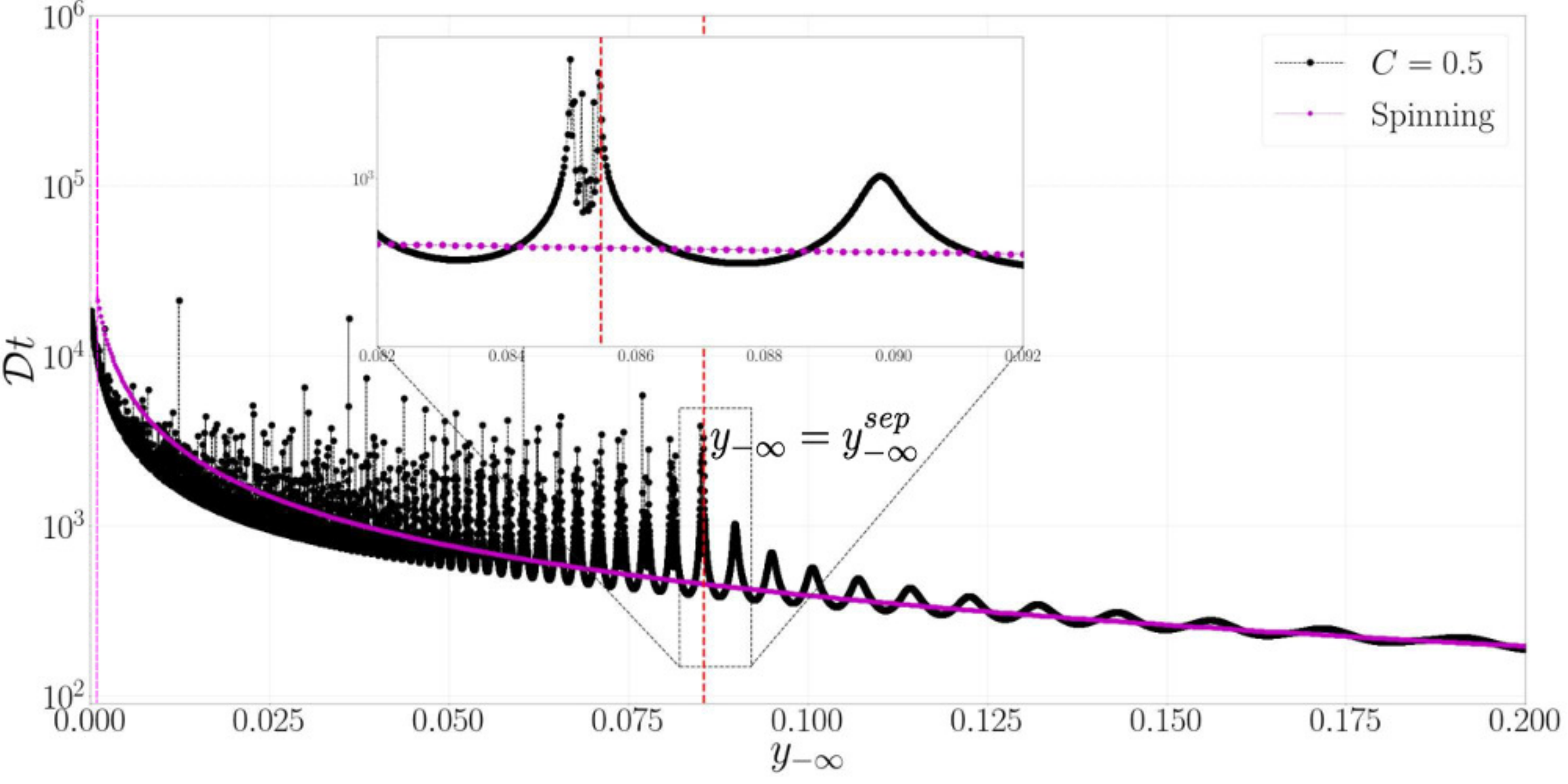}
  \caption{$C=0.5, t=4000T_j$}
  \end{subfigure}
  \begin{subfigure}[b]{.48\textwidth}
  \centering
  \includegraphics[scale=0.24]{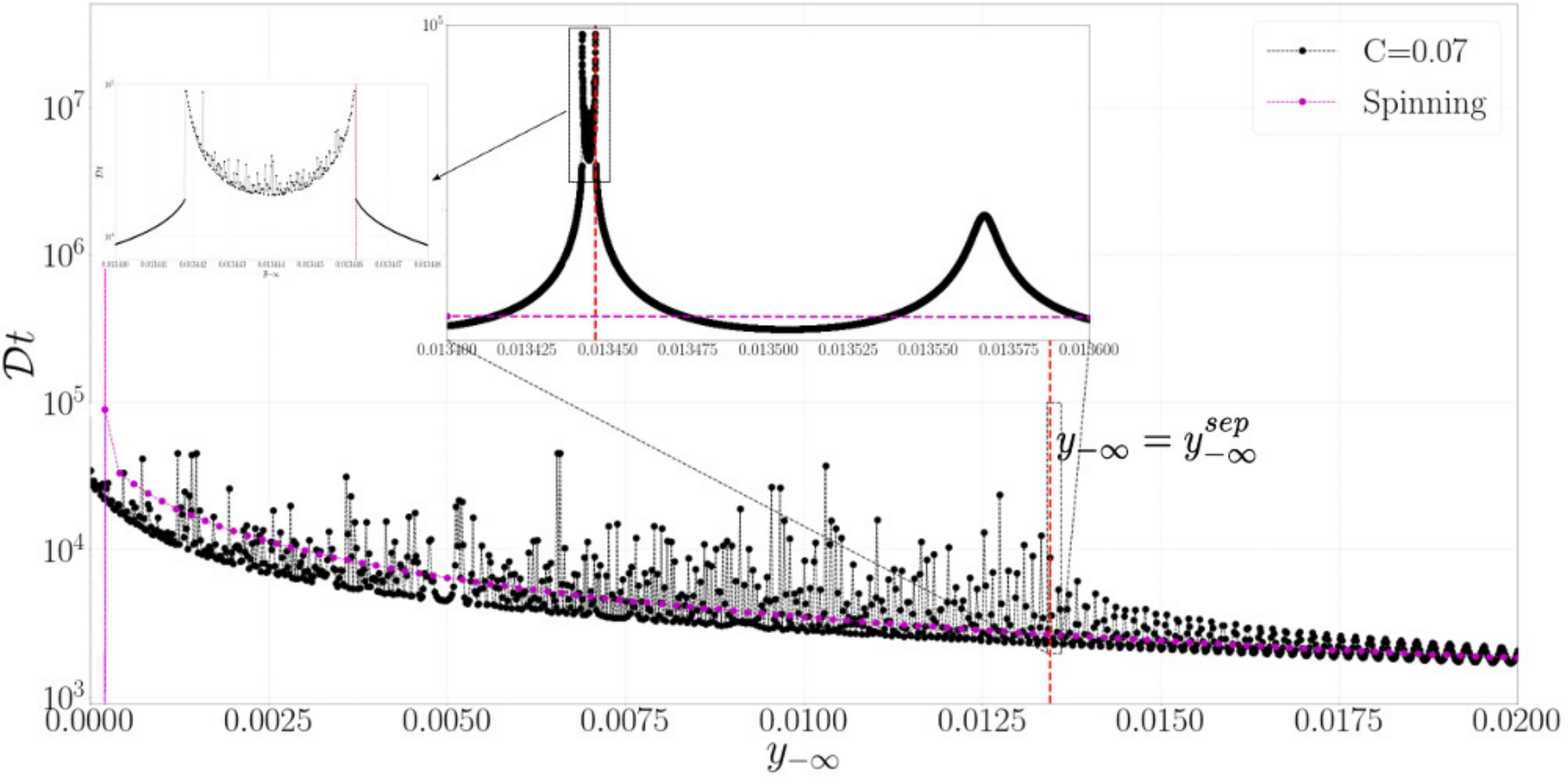}
  \caption{$C=0.07, t=10000T_j$}
  \end{subfigure}
  \begin{subfigure}[b]{.48\textwidth}
  \centering
  \includegraphics[scale=0.24]{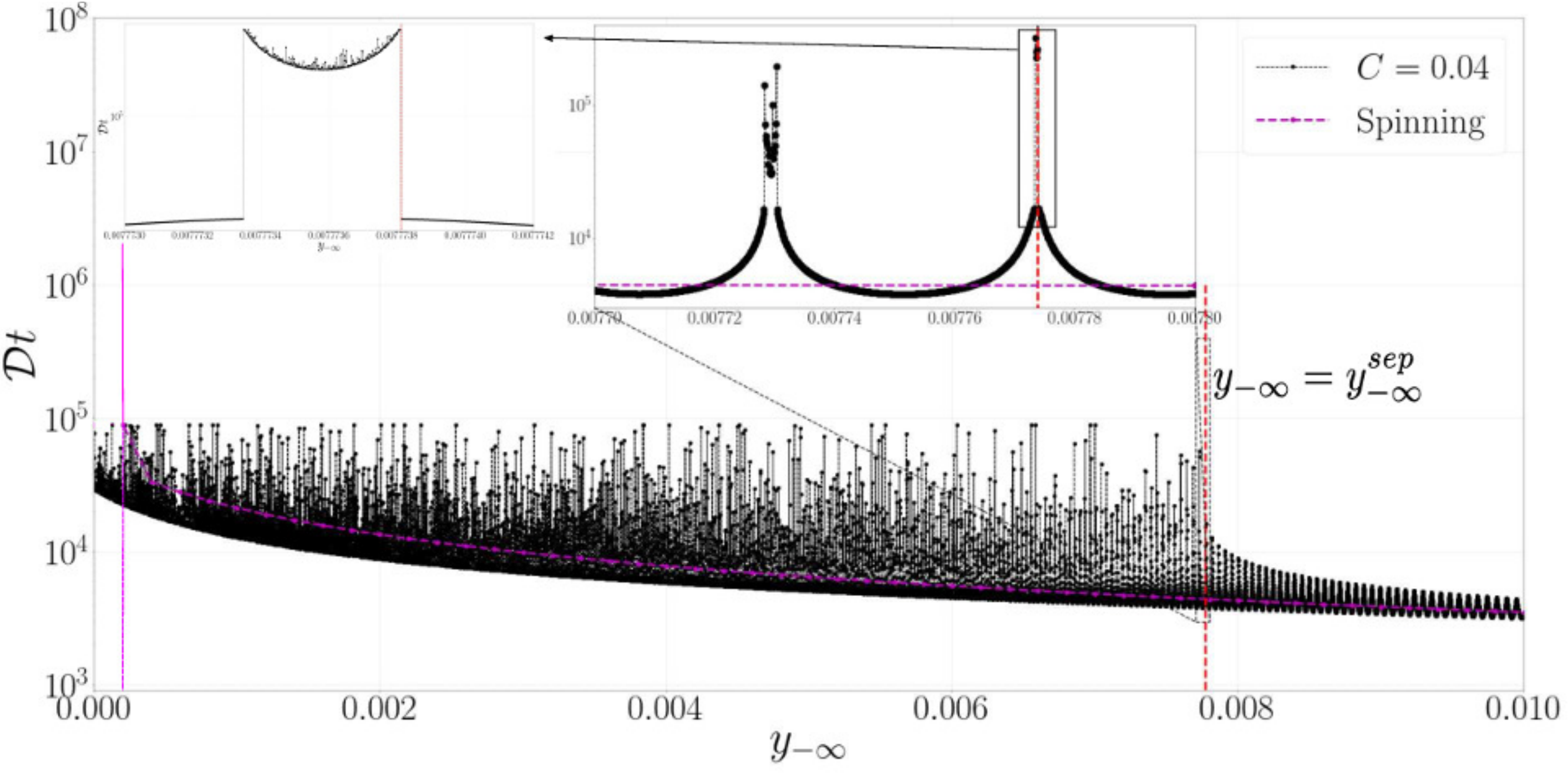}
  \caption{$C=0.04, t=10000T_j$}
  \end{subfigure}
  \hfill
  \begin{subfigure}[b]{.96\textwidth}
  \centering
  \includegraphics[scale=0.45]{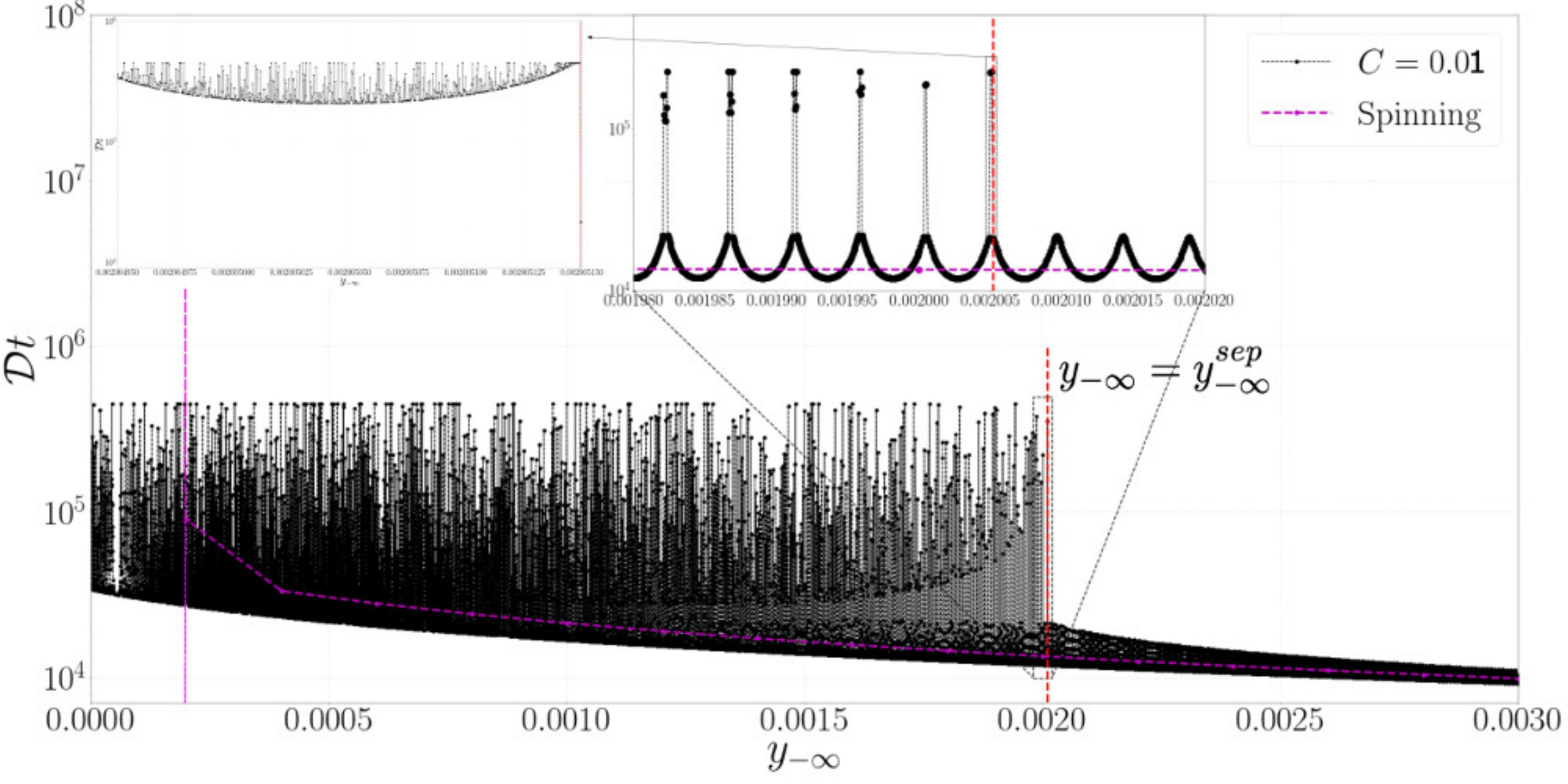}
  \caption{$C=0.01, t=20000T_j$}
  \end{subfigure}
  \caption{Variation of the residence time, for prolate spheroid with $\xi_0=1.05(\kappa=3.28)$ in Jeffrey orbits $C=1.0,0.5,0.07,0.04,0.01$ with $\phi_{j,-\infty}=0$. Red vertical line in each figure separates the regular and chaotic regions, the magenta curve denotes the residence time for the spinning($C=0$) spheroid, with the magenta vertical line denoting the separatrix for the spinning orbit.}
\label{fig:chaos_boundary4}
\end{figure}

\begin{figure}
\centering
\includegraphics[scale=0.3]{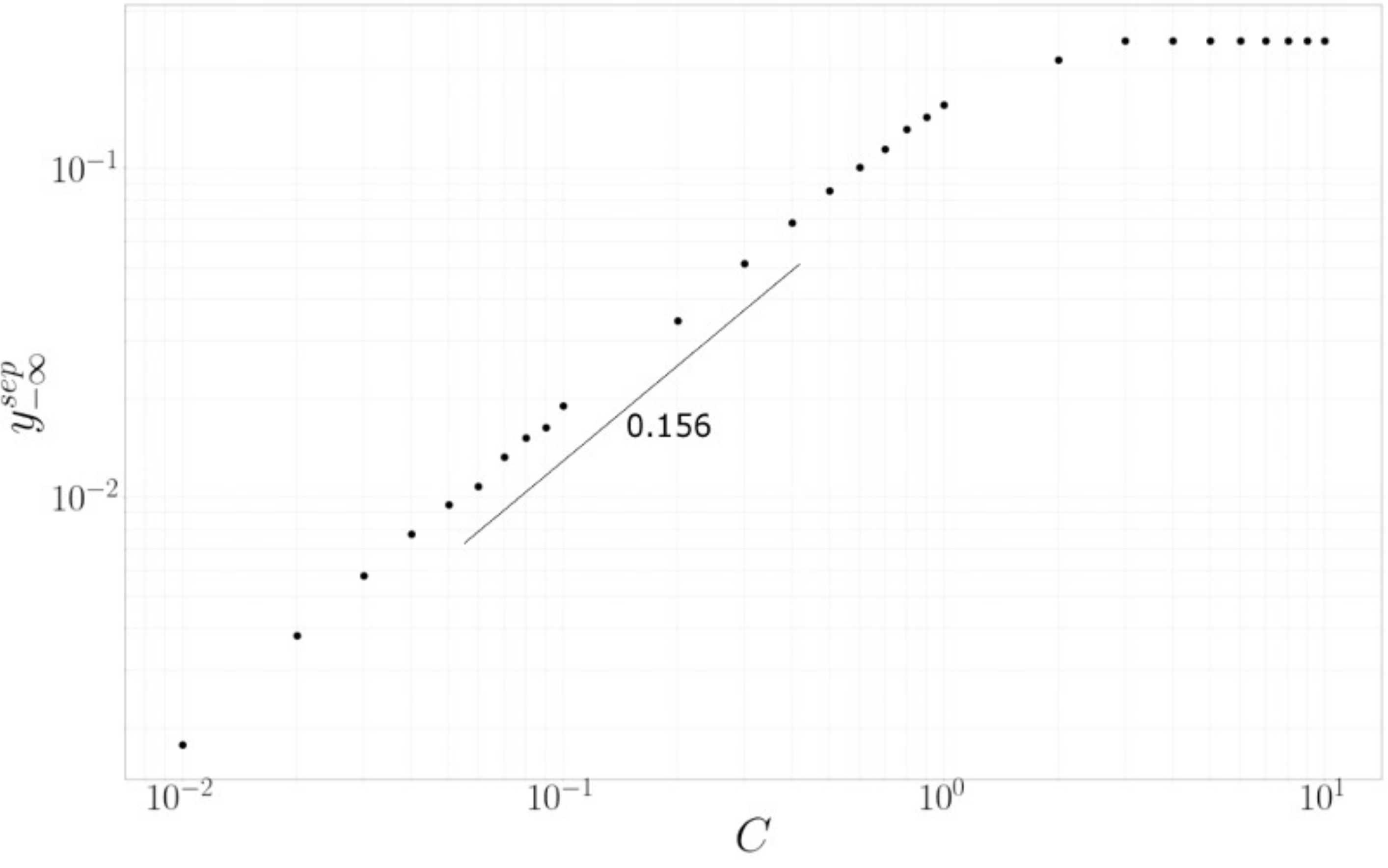}
\caption{The gradient offset of the separatrix, $y^{sep}_{-\infty}$, plotted as a function of the Jeffery orbit constant $C$, for a prolate spheroid with $\xi_0 = 1.05 (\kappa=3.28)$.}
\label{fig:chaos_boundary5}
\end{figure}

Next, in figure \ref{fig:chaos_boundary4}, we present a comparison of the residence time distributions as a function of the orbit constant $C$, for a prolate spheroid of $\xi=1.05(\kappa=3.28)$. The sequence of orbit constants chosen is $C$=1, 0.5,  0.1, 0.04, 0.01, and thus approaches the other integrable limit (a spinning spheroid: $C = 0$) mentioned above. Similar to figure \ref{fig:chaos_boundary2}, and for purposes of comparison, we plot the residence time distribution corresponding to a spinning spheriod, of the same aspect ratio, in each of the sub-figures. For a spinning spheroid, and for the finite $x$-interval (-20,20) under consideration, the residence time remains finite regardless of $y_{-\infty}$. Since closed streamlines correspond to an infinite residence time, the spinning spheroid curve shown terminates at the separatrix($y^{sep(spin)}_{-\infty}$); this is unlike figure15b where it was continued to the closed orbits within, by taking the residence time to be equal to the orbital half-periods. Although for the aspect ratio examined in Figure \ref{fig:chaos_boundary4}, there remains a large difference between the offsets of the actual separatrix ($y^{sep}_{-\infty}$) and the spinning-spheroid separatrix ($y^{sep(spin)}_{-\infty}$) down to the smallest $C$'s, the location of the separatrix for the smallest $C$'s nevertheless begins to correlate with a rather abrupt increase in the residence times, in a manner resembling the spinning-spheroid distribution. Unlike the slender fiber limit examined above, figure \ref{fig:chaos_boundary5} shows that the separatrix offset starts from an order unity value (specific to $\kappa = 1.05$) for large $C$, and decreases monotonically for $C$ going to zero, implying a similar decrease of the size of the chaotic region as one approaches the spinning-spheroid limit.

\begin{figure}
  \begin{center}
  \begin{subfigure}[b]{.96\textwidth}
  \centering
  \includegraphics[scale=0.45]{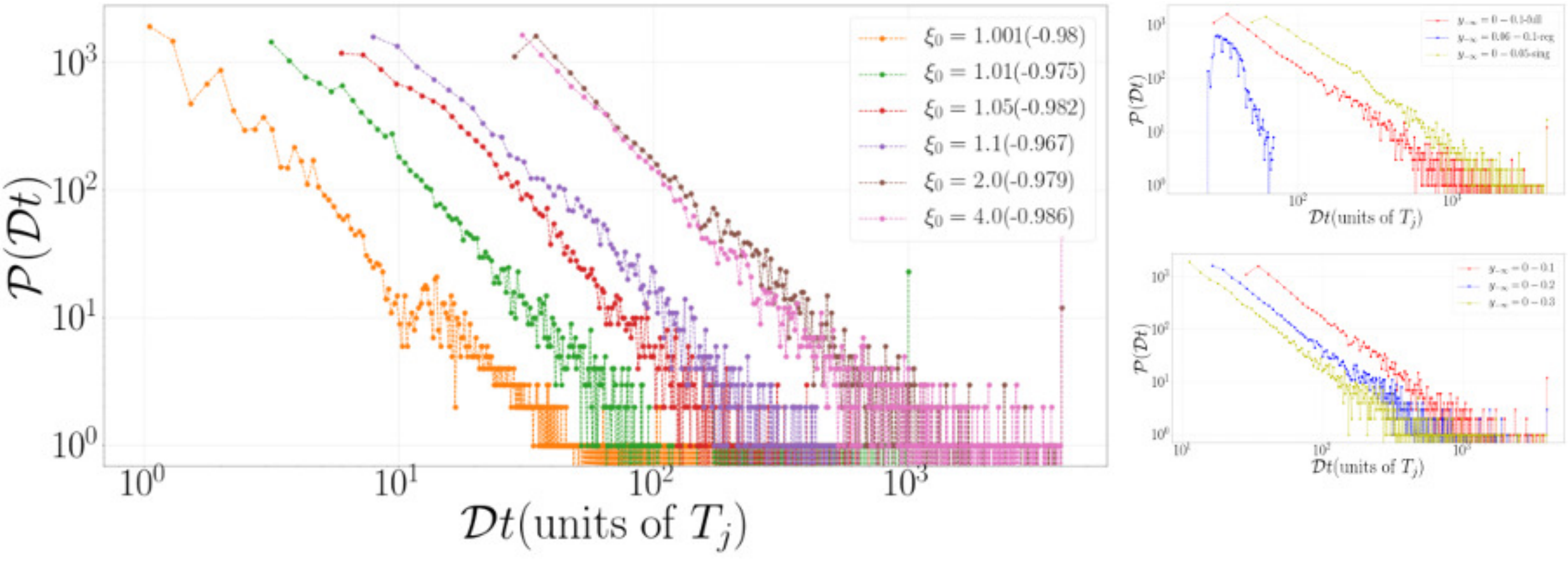}
  \end{subfigure}
  \end{center}
  \caption{The main figure shows the residence time probability densities for tumbling prolate spheroids of different aspect ratios; the slopes of the exponential tails appear within brackets. The top figure on the right shows the total probability density for $\xi_0 = 2$, and the component densities corresponding to regular\,($y^{sep}_{-\infty},y_{-\infty}$) and singular ($0,y^{sep}_{-\infty}$) pathlines. The bottom figure on the right demonstrates the insensitivity of the tail to the interval of upstream offsets considered.}
\label{fig:pr}
\end{figure}

Figure \ref{fig:pr} shows the residence time probability densities for tumbling prolate spheroids of decreasing aspect ratios: $\kappa(\xi_0)$=22.38(1.001), 7.12(1.01), 3.28(1.05), 2.4(1.1), 1.15(2) and 1.03(4). The exponential tails, whose slopes are a measure of the average residence time along the singular pathlines in the vicinity of the spheroid, are readily evident. The top figure on the right shows three probabililty densities corresponding to regular pathlines in ($y^{sep}_{-\infty},y_{-\infty}$), singular pathlines in ($0,y^{sep}_{-\infty}$), and all of the pathlines in ($0,y_{-\infty}$). The tail, corresponding to asymptotically long residence times, is seen to be entirely controlled by the singular pathlines. The lower figure on the right shows the invariance of the exponential tail to the interval of upstream offsets considered; clearly, an increase in $y_{-\infty}$ only amounts to considering a greater number of regular pathlines, with shorter residence times, leaving the tails unaffected.

\section{Conclusion}

In this study, we have examined the topology of the fluid pathlines induced by freely rotating neutrally buoyant spheroids, both prolate and oblate, and of an arbitrary aspect ratio, in simple shear flow. The fluid pathline configuration is analyzed as a function of the particular spheroid orbit, as defined by the orbit constant $C$. In the spinning or log-rolling ($C$ = 0) orientations, one has a steady scenario, and the fluid pathlines are the same as streamlines. Further, the topology of the streamline configuration is identical to that already known for the limiting cases of a sphere or a cylinder. There are two distinct groups of streamlines, open and closed, and these are separated by a surface of limiting streamlines, termed separatrices, that `closes at infinity'. For spinning prolate spheroids, various measures relating to the streamline configuration, including the size of the closed streamline region, the orbital periods of the individual closed streamlines, are bounded between those for a sphere and a cylinder (see figure 5).

On the other hand, when the spheroid is in any of the precessional orbits, including the tumbling one($0<C\leq\infty$), almost all pathlines are open, in that they come from and eventually go to infinity. Nevertheless, the open pathlines may be divided into two groups. The first group are the regular open pathlines which come from and go to infinity without looping around the precessing spheroid. The residence time for these pathlines, in a certain neighborhood of the spheroid, is a smooth function of their initial coordinates defined as their offsets, far upstream, in the gradient and vorticity directions. One may therefore regard these regular pathlines as trivial generalizations of the open pathlines around a sphere or a cylinder. In general, these pathlines are fore-aft asymmetric (with respect to the gradient vorticity plane), and may therefore be characterized via the net displacements in the gradient or vorticity directions (see figure \ref{fig:latdisp_phi}). For the most slender spheroids, there is the appearance of regular reversing pathlines that are absent for a torque-free sphere or a cylinder.

\begin{figure}
  \begin{center}
  \includegraphics[scale=0.35]{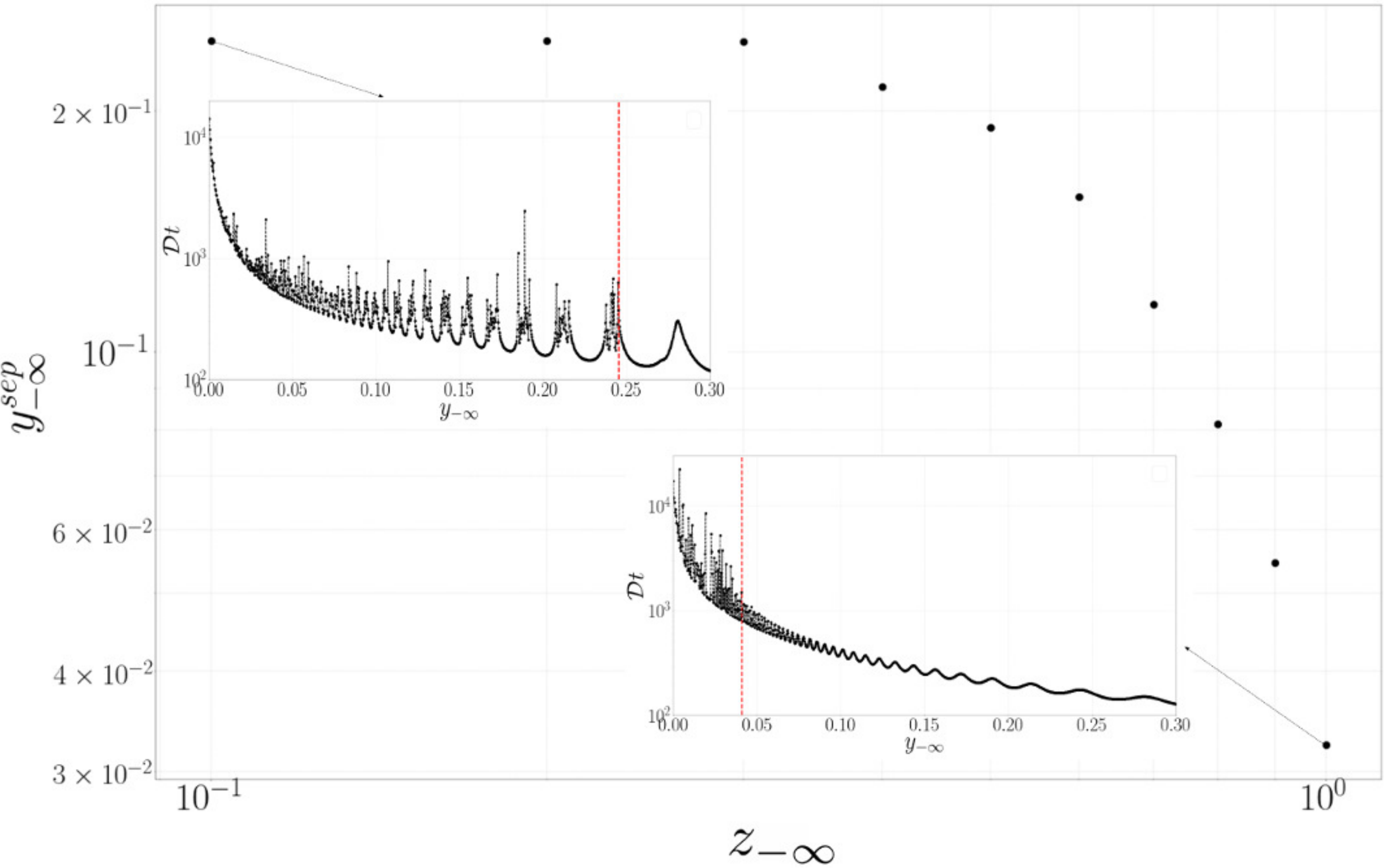}
  \end{center}
  \caption{The gradient offset of the separatrix, $y^{sep}_{-\infty}$, plotted as a function of the vorticity offset $z_{-\infty}$, for a prolate spheroid with $\xi_0 = 1.05 (\kappa=3.28)$, with the sample residence time distributions for $z_{-\infty}=$ 0.1 and 1.0}
\label{fig:zinf}
\end{figure}

The most important results of this manuscript concern the second group of open pathlines, the singular ones, that loop around the precessing spheroid on their way from upstream to downstream infinity. These differ in a profound manner from the closed pathlines around a freely rotating cylinder or a sphere. It is observed that the number of loops, or alternatively, the residence time of these pathlines in a certain neighborhood of the spheroid, has an extremely sensitive, seemingly random, dependence on the initial offset, with this sensitivity persisting until the smallest (numerically) resolved scales. Such a fractal dependence on initial coordinates is a defining characteristic of chaotic scattering as investigated earlier in other scenarios that include the laminar-to-turbulent transition in plane Couette flow\citep{skufca}, pipe flow\citep{eckhardt-pipe-annual,faisst,schneider}, dynamics of point vortices\citep{aref-pomphrey-letter,aref,eckhardt-vortex}, to name a few. The infinitely sensitive variation of the residence time found here implies the existence of a chaotic saddle, and thence, of a set of singularities on a Cantor-like set in the vicinity of the rotating spheroid; for the particular case of simple shear flow examined here (see final paragraph below), this singular set might extend to infinity along the flow axis. The singular set refers to the set of bounded pathlines with infinite residence times, and includes both an infinite number of periodic orbits, and an uncountably infinite number of aperiodic ones. It is, of course, numerically impossible to exactly locate any of these pathlines, but their existence may be inferred in a manner similar to that in Figure 17. We have investigated the nature of the local graph of the unstable manifold by numerically mimicking the experiment of \cite{thiffeault}, whereby the fractal structure is readily evident(figure \ref{fig:blob_compare}); a partial representation of the invariant chaotic saddle, for a near-sphere, is shown in figure \ref{fig:blob_saddle}. We have also investigated in detail the  transition from regular to chaotic scattering as a function of the underlying physical parameters that include the upstream gradient offset, the orbit constant and the spheroid aspect ratio. This included looking at the nature of approach to the known integrable limits - that of an arbitrary aspect ratio spheroid in the spinning orbit (figure \ref{fig:chaos_boundary5}), and that of an infinitely slender prolate spheroid in any orbit (figure \ref{fig:chaos_boundary3}). An aspect not investigated in detail is the variation in the spatial extent of the chaotic saddle as one moves away from the flow-gradient plane, that is, as a function of the upstream vorticity offset of the pathlines. Figure \ref{fig:zinf} presents the results of a limited investigation along these lines, and shows that (1) the scenario, for a fixed non-zero $z_{-\infty}$, remains similar to the inplane case, and (2) that the separatrix offset $y^{sep}_{-\infty}$ decreases with increasing $z_{-\infty}$.

One needs to emphasize the distinction between Hamiltonian and non-Hamiltonian settings here. In the former case, one can relate the onset of chaos to the onset of bound orbit sequences that correspond to infinite yet non-recurring sequences of reflection from the Hamiltonian peaks (as indicated by \cite{eckhardt-aref2,eckhardt-vortex}). It is also not difficult to identify a critical energy level, based on the governing Hamiltonian, that signals the onset of the aforementioned bound sequences. There doesn't seem to exist any such analogy in our case, and thus, no obvious way to predict the offset of the separatrices that mark the onset of chaotic scattering. The latter is true for the other non-Hamiltonian problems too (the laminar-turbulent transition, for example).

As indicated in the Introduction, the findings of the present investigation might have  profound implications for transport problems in disperse multiphase systems. In what follows, we briefly discuss three instances, in roughly increasing order of complexity. The first is the transport of heat or mass from particles in shearing flows. As originally shown by Acrivos \citep{acrivos,poe_acrivos}, the rate of scalar transport from a freely rotating sphere in a planar linear flow has an exceptional character; planar linear flows form a one-parameter subset of the general family of linear flows (see discussion in the final paragraph below, for the values that this parameter takes), with the limiting values of the parameter corresponding to planar extension and solid-body rotation. The dimensionless rate of transport, as characterized by the Nusselt number (Nu), does not increase indefinitely with the Peclet number (\emph{Pe}). Such an increase would occur if, at large \emph{Pe} corresponding to the convectively dominant limit, the transport were to occur via a thin boundary layer, as is the case for a translating sphere; see \cite{acrivos_goddard}. Instead, \emph{Nu} saturates at an order unity value dependent on flow-type, the saturation arising due to the transport eventually being limited by diffusion across closed streamlines that surround the sphere in all of the planar linear flows (except planar extension). The singular alteration of the trajectory topology owing to a deviation of the particle from sphericity, found here, must manifest as qualitatively different \emph{Nu} v/s \emph{Pe }curves for a sphere and spheroid in a given planar linear flow. In contrast to a sphere, for large enough \emph{Pe}, one expects the transport from a freely rotating spheroid to be dominated by the set of singular open pathlines which convect heat or mass away in an efficient manner. This should avoid the diffusion limitation for a sphere, leading to a continued growth of \emph{Nu} with \emph{Pe}, with the nature of this growth dependent on the residence time distributions of fluid elements obtained here (for instance, those shown in figure \ref{fig:chaos_boundary0}); any scaling exponent characterizing the large-\emph{Pe }asymptotics of \emph{Nu}, would be a sensitive function of aspect ratio via the residence time distributions. For spheroids with near-unity aspect ratios (the weakly non-integrable cases), in particular, one expects \emph{Nu} to exhibit an intermediate plateau on account of diffusion limitation, similar to a sphere. For sufficiently large \emph{Pe}, however, \emph{Nu} will increase in a manner determined by the fraction of fluid pathlines with residence times shorter than a characteristic diffusion time through the (hypothetical) envelope of closed streamlines. Recent efforts \citep{sub_koch_2005,subramanian_koch_2006,deepak_2018_2} have examined the singular effect of weak inertia-induced convection on the originally diffusion-limited behavior of \emph{Nu} in the Stokesian limit. The above discussion suggests that departure from sphericity  again have a singular effect, manifesting as a convective enhancement for large \emph{Pe}. The convective flow in the inertial case may be derived using a straightforward perturbation expansion in the particle Reynolds number, but that for the non-spherical case is non-trivial, and even for a near-sphere would require the notion of lobe dynamics. The latter has been used earlier to describe the efficient (non-diffusive) exchange of material between the irrotational volume entrained by a vortex ring, or a pair of counter-rotating point vortices, and the ambient\citep{rom-kedar}.

A second problem concerns the rheology of dilute viscoelastic suspensions of anisotropic (axisymmetric) particles which sensitively depends on the orientation dynamics of these particles in an imposed shear. The degenerate nature of the orientation distribution, arising from the existence of closed (Jeffery) orbits in the Stokesian limit \citep{leal, dabade1, marath2018} is well known, and implies that the orientation distribution, in a viscoelastic suspending medium, crucially depends on a viscoelasticity-induced drift across orbits. This drift is governed by the polymeric stress field. For small Deborah numbers (\emph{De}), the polymeric stress may be determined using an ordered fluid expansion, and is therefore only a function of the local velocity gradient associated with the Stokesian velocity field \citep{madival}. But, for large \emph{De}, the relaxation times are long compared to the characteristic residence time of a fluid element in the vicinity of the rotating particle, and one expects the polymeric stresses to therefore arise from the integrated effects of the velocity gradients seen by a polymer molecule as it is convected along a fluid pathline. For small polymer concentrations, these pathlines will have a near-Newtonian (Stokesian) character, and the non-trivial pathline topology identified here implies that the polymeric stress field, and thence the particle drift, might be crucially influenced by the stresses that develop along the singular open pathlines. In fact, on account of the non-integrable nature of the pathline topology, one expects the spatial variation of polymeric stresses to have a singular character; that is to say, the polymeric stresses at points located arbitrarily close to each other might differ by a finite amount, on account of an analogous difference in the residence times of fluid pathines arriving at these points. This in turn points to a probabilistic (rather than deterministic) formulation of the particle-polymer-molecule interaction where, rather than determining the polymer stress field at each point in the domain, one attempts to instead determine the probability distribution of polymeric stresses that develops in the neighborhood of a single point (within the domain influenced by the chaotic saddle). For large aspect ratios, the above problem admits a well known simplification, on account of the disturbance velocity field being logarithmically small (as discussed in section \ref{sec:transition}). For finite \emph{De}, this simplification allows one to approximate the polymeric stresses by integrating a functional of the Stokesian velocity gradients along an ambient streamline instead of the actual fluid pathline. This simplification has been exploited in earlier calculations (for instance, see \cite{harlen_koch}). While such a simplification is certainly valid for steady velocity fields, as for instance, the flow around a sedimenting spheroid with an unchanging orientation in the Stokesian limit (see section 5 in \cite{dabade1}), the scenario in an ambient shear flow might be more complicated even in the slender fiber limit. As shown in figure \ref{fig:chaos_boundary2}, the approach to this limit has a singular character - while the residence time, based on the convection by the ambient shear alone, does give an accurate estimate for the regular intervals, for sufficiently large aspect ratios, the occasional chaotic bursts nevertheless lead to very pronounced departures of the residence time from the ambient-shear-based envelope. It would be of interest to examine quantitatively the consequences of such departures for the polymeric stresses that develop along the heavily looped pathlines.

The third problem which, in fact, motivated the examination of the fluid pathline topology in the first place concerns pair-spheroid interactions. As mentioned in the introduction, one expects the fluid pathline topology around a single particle to have some bearing on the nature of pair-particle interactions. This is certainly true for a spherical particle, where the streamline and pair-particle-pathline topologies are identical. Thus, the findings here imply that pair-spheroid trajectories are likely to have a similar singular character. As indicated in \cite{marath_dwivedi}, one would ideally want to characterize pair-spheroid interactions via a phase-averaged scattering kernel that would relate the pre- and post-interaction orbit constants. The findings here would, however, imply that even an infinitesimal change in the input orbit constant would likely lead to an order unity change in the post-interaction orbit constant. In other words, the scattering kernel would be singular (in fact, such singular kernels in the context of reacting molecules constituted the very first pieces of evidences of chaotic scattering, see \cite{gottdiener,noid}). Therefore, similar to the polymer-spheroid problem above, instead of the exact pair-interaction kernel, one would again want to describe even individual pair-interactions in a probabilistic framework.

Although a canonical flow-type from the fluid mechanical viewpoint, owing to its significance to rheology, for instance, the choice of simple shear flow does appear inconvenient from the dynamical systems perspective. As is well known (for instance, see \citep{marath_linear,subramanian_2006POF}), simple shear flow may be regarded as a member of the aforementioned one-parameter family of planar linear flows. The parameter here is $\alpha$ (say), with $(1+\alpha)/(1-\alpha)$ being proportional to the ratio of extension to vorticity. Simple shear, corresponding to $\alpha = 0$, is then the threshold flow, with equal magnitudes of extension and vorticity (leading to straight streamlines), separating the hyperbolic planar linear flows with open streamlines ($\alpha > 0$), from the elliptic ones with closed streamlines ($\alpha < 0$). If one now considers a spinning spheroid in a hyperbolic linear flow, from the results of section \ref{sec:spinning}, one expects, similar to a sphere, a trajectory topology where separatrices demarcate the closed from the open streamlines; for instance, see figure 34a above(figure 6 in \cite{subramanian_2006POF} which sketches the trajectory topology for a sphere). Importantly, for a hyperbolic linear flow, the closed streamline region, projected onto the flow-gradient plane, has a finite spatial extent for any spheroid aspect ratio, and the aforementioned separatrices are now heteroclinic trajectories connecting connecting a pair of saddle points(see figure 34a). If the analogy with a sphere is exact, then the closed streamline envelope has a finite volume, and the saddle points will  correspond to the points of intersection, of a fixed-point curve, with the flow-gradient plane. All of these fixed points have a saddle-like character, with the curve reducing to a fixed-point circle for the case of a sphere. The existence of such heteroclinic connections highlights the structurally unstable trajectory topology for a spinning spheroid. One may now interpret the deviation of the spheroid, from the spinning orbit to a precessional one corresponding to a small but finite $C$, to constitute a perturbation that renders the system non-integrable. Importantly, one expects the non-integrability of the perturbed system to manifest as transverse intersections of the stable and unstable manifolds associated with the two saddle points(see figure 34b), and the existence of such intersections can be rigorously demonstrated via the construction of a Melnikov function \citep{wiggins}, as has been done earlier in the fluid mechanical context \citep{kawakami,angilella}. 
\begin{figure}
  \begin{center}
  \includegraphics[scale=0.5]{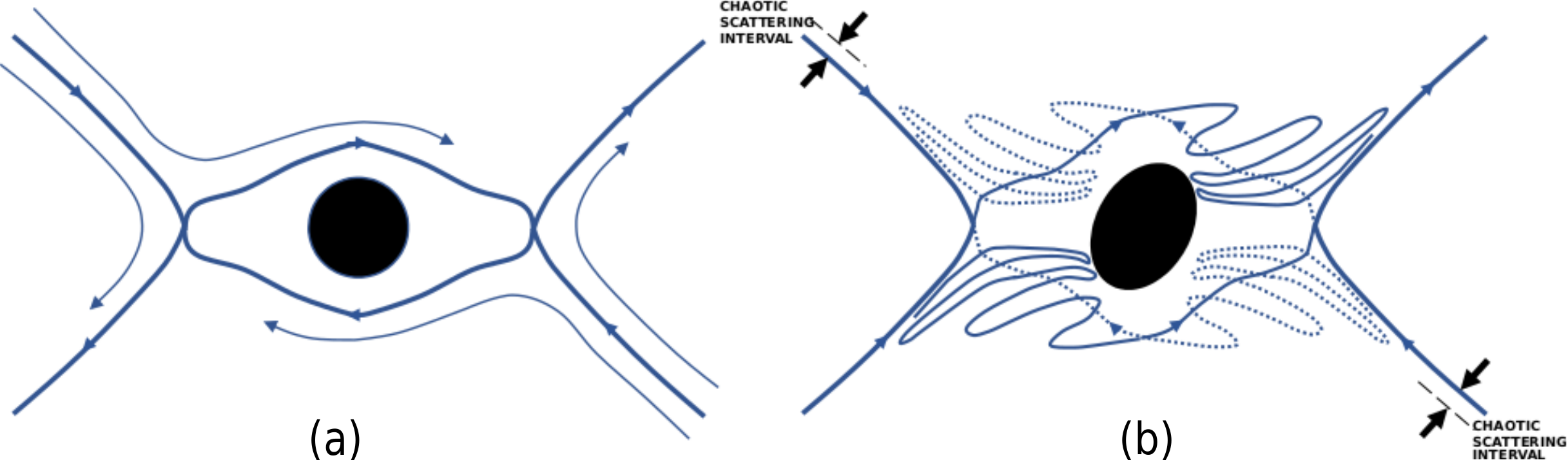}
  \end{center}
  \caption{Smooth and chaotic separatrix in the hyperbolic flow over a sphere(left) and a spheroid(right)}
\label{fig:hyper}
\end{figure}
As sketched in Figure \ref{fig:hyper}, the implication of these intersections is that intervals of chaotic scattering must replace the original separatrix for the spheroid rotating in any precessional orbit. The trajectory topology in Figure \ref{fig:hyper}, although less symmetric, nevertheless resembles the pathline topology analyzed by \cite{rom-kedar}, wherein the non-integrability arose from the time dependence imposed by an oscillatory planar extension acting on a counter-rotating vortex pair. Although related, the emphasis in this work was not on the chaotic scattering aspects, but instead on the exchange of fluid across the (infinitely) convoluted intersecting manifolds. Figure \ref{fig:hyper} suggests that, from a dynamical systems viewpoint, the non-integrability for the simple shear flow examined here is perhaps best interpreted as the singular limiting case, for $\alpha \rightarrow 0^+$, of a hyperbolic linear flow, when the saddle points recede to infinity. The nature of chaotic scattering in hyperbolic linear flows, and its relation to the chaotic scattering in the simple shear flow configuration examined here, will be taken up in a future communication.

\bibliographystyle{jfm}
\bibliography{spheroid_manuscript}

\end{document}